\documentclass[a4paper,fleqn,usenatbib]{mnras}

\usepackage{newtxtext,newtxmath}

\usepackage[T1]{fontenc}
\usepackage{ae,aecompl}

\usepackage{graphicx}	
\usepackage{amsmath}	
\usepackage{amssymb}	

\title[The TIMER Project]{Time Inference with MUSE in Extragalactic Rings (TIMER): Properties of the Survey and High-Level Data Products}

\author[D. A. Gadotti \& The TIMER Team]{
Dimitri A. Gadotti,$^{1,2}$\thanks{E-mail: dgadotti@eso.org}
Patricia S\'anchez-Bl\'azquez,$^{3}$
Jes\'us Falc\'on-Barroso,$^{4,5}$
\newauthor
Bernd Husemann,$^{6}$
Marja K. Seidel,$^{7}$
Isabel P\'erez,$^{8,9}$
Adriana de Lorenzo-C\'aceres,$^{4,5}$
\newauthor
Inma Martinez-Valpuesta,$^{4,5}$
Francesca Fragkoudi,$^{10,11}$
Gigi Leung,$^{6}$
Glenn van de Ven,$^{1,6}$
\newauthor
Ryan Leaman,$^{6}$
Paula Coelho,$^{12}$
Marie Martig,$^{6,13}$
Taehyun Kim,$^{14,15}$
\newauthor
Justus Neumann,$^{2,16}$
Miguel Querejeta$^{1,17}$
\\
$^{1}$European Southern Observatory, Karl-Schwarzschild-Str. 2, D-85748 Garching bei M\"unchen, Germany\\
$^{2}$European Southern Observatory, Casilla 19001, Santiago 19, Chile\\
$^{3}$Departamento de F\'isica Te\'orica, Universidad Aut\'onoma de Madrid, E-28049 Cantoblanco, Spain\\
$^{4}$Instituto de Astrof\'isica de Canarias, 38205 La Laguna, Tenerife, Spain\\
$^{5}$Departamento de Astrof\'isica, Universidad de La Laguna, 38206 La Laguna, Tenerife, Spain\\
$^{6}$Max-Planck-Institut f\"ur Astronomie, K\"onigstuhl 17, D-69117 Heidelberg, Germany\\
$^{7}$Carnegie Observatories, 813 Santa Barbara St., CA 91101 Pasadena, USA\\
$^{8}$Departamento de F\'isica Te\'orica y del Cosmos, Universidad de Granada, Facultad de Ciencias (Edificio Mecenas), E-18071 Granada, Spain\\
$^{9}$Instituto Universitario Carlos I de F\'isica Te\'orica y Computacional, Universidad de Granada, E-18071 Granada, Spain\\
$^{10}$GEPI, Observatoire de Paris, PSL Research University, CNRS, Place Jules Janssen, 92195 Meudon, France\\
$^{11}$Max-Planck-Institut f\"ur Astrophysik, Karl-Schwarzschild-Str. 1, D-85748 Garching bei M\"unchen, Germany\\
$^{12}$Instituto de Astronomia, Geof\'isica e Ci\^encias Atmosf\'ericas, Universidade de S\~ao Paulo, Rua do Mat\~ao, 1226, 05508-090 S\~ao Paulo-SP, Brazil\\
$^{13}$Astrophysics Research Institute, Liverpool John Moores University, IC2 Liverpool Science Park, 146 Brownlow Hill, L3 5RF Liverpool, UK\\
$^{14}$Korea Astronomy and Space Science Institute,  34055 Daejeon, Korea\\
$^{15}$Astronomy Program, Department of Physics and Astronomy, Seoul National University, 08826 Seoul, Korea\\
$^{16}$Leibniz-Institut f\"ur Astrophysik Potsdam, An der Sternwarte 16, D-14480 Potsdam, Germany\\
$^{17}$Observatorio Astron\'omico Nacional (IGN), Alfonso XII 3, E-28014 Madrid, Spain
}

\pubyear{2018}

\begin{document}
\label{firstpage}
\pagerange{\pageref{firstpage}--\pageref{lastpage}}
\maketitle

\begin{abstract}
The {\bf T}ime {\bf I}nference with {\bf M}USE in {\bf E}xtragalactic {\bf R}ings (TIMER) project is a survey with the VLT-MUSE integral-field spectrograph of 24 nearby barred galaxies with prominent central structures ({\it e.g.}, nuclear rings or inner discs). The main goals of the project are: {\bf (i)} estimating the cosmic epoch when discs of galaxies settle, leading to the formation of bars; {\bf (ii)} testing the hypothesis whereby discs in more massive galaxies are assembled first; and {\bf (iii)} characterising the history of external gas accretion in disc galaxies. We present details on the sample selection, observations, data reduction, and derivation of high-level data products, including stellar kinematics, ages and metallicities. We also derive star formation histories and physical properties and kinematics of ionised gas. We illustrate how this dataset can be used for a plethora of scientific applications, {\it e.g.}, stellar feedback, outflows, nuclear and primary bars, stellar migration and chemical enrichment, and the gaseous and stellar dynamics of nuclear spiral arms, barlenses, box/peanuts and bulges. Amongst our first results -- based on a few selected galaxies --, we show that the dynamics of nuclear rings and inner discs is consistent with the picture in which they are formed by bars, that the central few hundred parsecs in massive disc galaxies tend to show a pronounced peak in stellar metallicity, and that nuclear rings can efficiently prevent star formation in this region. Finally, we present evidence that star-bursting nuclear rings can be fed with low-metallicity gas from low-mass companions.
\end{abstract}


\section{Introduction and Project Goals}

In the nearby Universe, discs of massive spiral and lenticular galaxies show orderly dynamics, dominated by differential rotation and a relatively smooth rotation curve. To some extent, this results from the fact that discs are sufficiently massive today and thus their self-gravity becomes a dominant factor on their dynamics. However, this has not always been the case. At earlier cosmic epochs, at redshifts $z\gtrsim1-2$, discs were characterised by turbulent dynamics and a clumpy, irregular structure \citep[see, {\it e.g.},][]{ForGenLeh06,GenTacEis06,GenBurBou08,LawSteErb07,ElmBouElm08,LawSteErb09}. When and how did this transition happen? Or in other words, when do galaxy discs settle dynamically and how? The main goal of the TIMER project is to answer these questions. If we are to understand galaxy formation and evolution we have to look for those answers. In addition, in the same framework of the downsizing picture of galaxy formation and evolution \citep[see, {\it e.g.},][]{CowSonHu96,ThoMarSch10}, \citet{SheMelElm12} suggest that the most massive discs will settle and form bars first. The TIMER project will be able to test this hypothesis.

One possible path for answering these questions is to study the dynamics of large samples of disc galaxies at a number of redshift bins and assess directly how their dynamics evolve. A number of studies have gone down this path, with the help of powerful instruments such as KMOS and SINFONI at the {\it Very Large Telescope} (VLT), which provide 2D kinematics measurements on rest-frame optical wavelengths for galaxies at $z\sim1-2$ \citep[{\it e.g.},][]{ShaGenFor08,ForGenBou09,EpiTasAmr12,WisForWuy15}. Those studies make use of nebular emission lines -- typically H$\alpha$ from warm, ionised gas clouds -- since obtaining direct measurements of stellar kinematics from absorption lines is still very challenging in this redshift regime. They generally conclude that a large fraction of the massive, star-forming galaxies at those redshifts are rotationally supported. \citet{WisForWuy15} found that 93 per cent of galaxies at $z\sim1$ and 74 per cent at $z\sim2$ are rotationally supported, with a continuous velocity gradient and $v_{\mathrm{rot}}/\sigma_0 > 1$, where $v_{\mathrm{rot}}$ is the rotational velocity corrected for inclination, and $\sigma_0$ is the velocity dispersion measured at outer regions. Using stricter criteria they found the fraction of disc galaxies to be 70 per cent at $z\sim1$ and 47 per cent at $z\sim2$. They also found that resolved, non-rotating galaxies are found primarily at low stellar masses, consistent with downsizing. Nevertheless, in contrast to local discs, these discs show high velocity dispersion and are thus turbulent and thick \citep[see also, {\it e.g.},][for similar results from observations of the cold molecular gas]{TacNerGen13}. We note that \citet{RodHamFlo17} found that only a third of the $z\sim1$ galaxies are isolated, virialised discs, pushing to lower redshifts the cosmic epoch of disc settling. They have used the same data as \citet{WisForWuy15} but employed different criteria in their analysis, which highlights the enduring difficulties confronted in studies of distant galaxies.

Another possible path that can shed light on this issue is to look for the ``archeological'' evidence for the dynamical settling of discs in nearby barred galaxies. About 2/3 of disc galaxies in the local universe are barred \cite[see, {\it e.g.},][]{EskFroPog00,MenSheSch07,MasNicHoy11,ButSheAth15}. Theoretical studies, both analytical and numerical, suggest that bars can only form after the dynamical settling of the disc or at least part of it. In addition, the formation of the bar will often happen quickly, a few 10$^8$\,yr after the disc settles, if the disc is unstable to the formation of the bar. Furthermore -- similarly quickly after the formation of the bar -- the non-axisymmetric potential introduced by the bar produces tangential forces across the disc, which cause the cold gas in the interstellar medium (ISM) to shock and lose angular momentum and funnel down to the central region along the leading edges of the bar, traced by dust lanes. This is seen in both theoretical and observational work \citep[see, {\it e.g.},][and references therein]{ComGer85,Ath92b,SelWil93,PinStoTeu95,SakOkuIsh99,RauSal00,RegTeu03,RegTeu04,Ath05b,SheVogReg05,Woz07,Kna07,Gad09a,KimSeoSto12,ColDebErw14,EmsRenBou15,SorBinMag15,FraAthBos16,QueMeiSch16,PerMarRui17}. When the gas transitions from the region where x1 orbits dominate (i.e, orbits parallel to the bar) to the more central region where x2 orbits dominate (perpendicular to the bar), due to the gas collisional nature, stars in (near) circular orbits are formed, which often end up in a nuclear ring \citep[see, {\it e.g.},][see also  \citealt{ButCom96} and \citealt{MaoBarHo01} and the pioneering works of \citealt{SerPas65,SerPas67}]{vanEms98,FalBacBur06,SarFalDav06,KunEmsBac10,PelKutvan12}. Further, not only nuclear rings, but other central stellar structures, such as inner discs (also known as pseudo-bulges), are also thought to form from gas brought to the centre by the bar. Inner discs may also -- just as the major galaxy disc -- host nuclear bars and nuclear spiral arms.

Therefore, the oldest stars in such bar-built structures hold fossil evidence of the time the bar first brought gas to the centre: the ages of those oldest stars tell us directly how long ago was the first bar-driven gas accretion event. This in turn gives us an estimate of the time of both bar formation (the age of the bar) and disc settling. This is the path chosen with TIMER.

In many galaxies, the stellar population in the ring, or in other bar-built structures, is superposed on an underlying stellar component (e.g., the disc) that may have a different star formation history. Since our goal is to measure the oldest stars in the bar-built structures, we are faced with the challenge of separating the stellar population of those structures from the underlying stellar component. This is not an easy undertaking and the team is currently working on different approaches. One possibility is to use spectral decomposition techniques \citep[e.g.,][]{CocFabSag18} to explore the fact that the kinematical properties of the bar-built and underlying structures may be noticeably different. In fact, as we will show below, bar-built structures rotate faster than the underlying disc at the same radius. Another possibility is to interpolate the stellar properties of the underlying component just outside the bar-built structures to derive their properties at the position of the bar-built structures. These developments will be presented elsewhere.

In \citet{GadSeiSan15} we offered a proof of concept with data obtained during the Science Verification campaign of the Multi-Unit Spectroscopic Explorer (MUSE), the latest integral field spectrograph on the VLT. We observed the central region of NGC\,4371, a lenticular galaxy in the core of the Virgo cluster with a stellar mass of $M_\star\approx10^{10.8}\,{\rm M}_\odot$ \citep{GalTreMar10}. We derived the stellar age distributions and star formation histories of the nuclear ring and other stellar structures in NGC\,4371 and found that the oldest stars in the nuclear ring have ages of about 10\,Gyr. In fact, the whole central region of the galaxy is vastly dominated by very old stars, which avoids the issue discussed above of how to separate the stellar population of the ring from the underlying stellar population. This means that the bar was formed about 10\,Gyr ago, with the disc settling shortly before. This is consistent with the large stellar mass of this galaxy and the expectation from the downsizing picture that the most massive discs settle first. Our structural analysis indicates that the mass of the disc in NGC\,4371 is $\approx3\times10^{10}\,{\rm M}_\odot$. Interestingly, this suggests that the turbulent, thick discs observed at $z\sim2$ are able to form bars \citep[see also][for an observational study on barred galaxies up to $z\sim2$]{SimMelLin14}. In addition, we found no star formation in the field covered by MUSE, indicating that environmental processes in the cluster have quenched the galaxy.

TIMER is the full-fledged form of our Science Verification project, encompassing 24 galaxies in total with a range of physical properties. Apart from its main scientific goals, as described above, the TIMER project has great legacy value. As we will make clear below, our data has high signal-to-noise ratio (SNR), and because of the superb capabilities of MUSE (such as its fine spatial sampling and broad wavelength coverage), the amount of physical information and detail in the TIMER data cubes is unprecedented for its sample. This allows a number of further studies, {\it e.g.}, on the properties of nuclear and primary bars (stellar population content, kinematics and dynamics, and star formation) and the ISM (metal content, shocks, kinematics and dynamics, and ionisation states). In addition, while the oldest stars in bar-built nuclear structures tell us about the first bar-driven gas accretion event onto the nuclear region, the complete star formation histories of those structures can reveal other gas accretion events as more recent peaks in star formation activity. This information is fundamental to understand the observed properties of the central regions of galaxies and their evolution. Furthermore, some of our data cubes reveal in exquisite detail outflows driven either by Active Galactic Nuclei (AGN) or stellar feedback.

This paper is organised as follows. In the next section we describe the selection of the TIMER sample, while in Sect. \ref{sec:obs} we explain in detail our observations and data reduction. One of the main goals of this paper is to describe, illustrate and discuss the derivation of our high-level data products to a level of detail such that our forthcoming studies do not necessarily have to go through it in detail. This is done in Sects. \ref{sec:kin}, \ref{sec:stelpop} and \ref{sec:emli}. Section \ref{sec:kin} corresponds to products related to the stellar kinematics, while Sect. \ref{sec:stelpop} corresponds to the stellar population content, namely, ages, chemical content and star formation histories. In Sect. \ref{sec:emli} we are concerned with the fluxes and kinematics from emission lines. A brief illustration of the potential of the TIMER dataset for scientific exploration is given in Sect. \ref{sec:sci}, where we analyse the data presented in the previous sections and discuss our first results. Finally, we conclude in Sect. \ref{sec:conc}, presenting an outlook on the short- and long-term exploration of the dataset, and an outline of our plans for future papers. Throughout TIMER -- unless otherwise noted -- we use a Hubble constant of ${\rm H_0}=67.8\,\rm{km}\,\rm{s}^{-1}\,\rm{Mpc}^{-1}$ and $\Omega_{\rm m}=0.308$ in a universe with flat topology \citep{AdeAghArn15}.

\section{Sample Selection}
\label{sec:sample}

\begin{figure*}
\begin{center}
	\includegraphics[trim=1cm 9.5cm 2.9cm 9.5cm, clip=true, width=0.5\columnwidth]{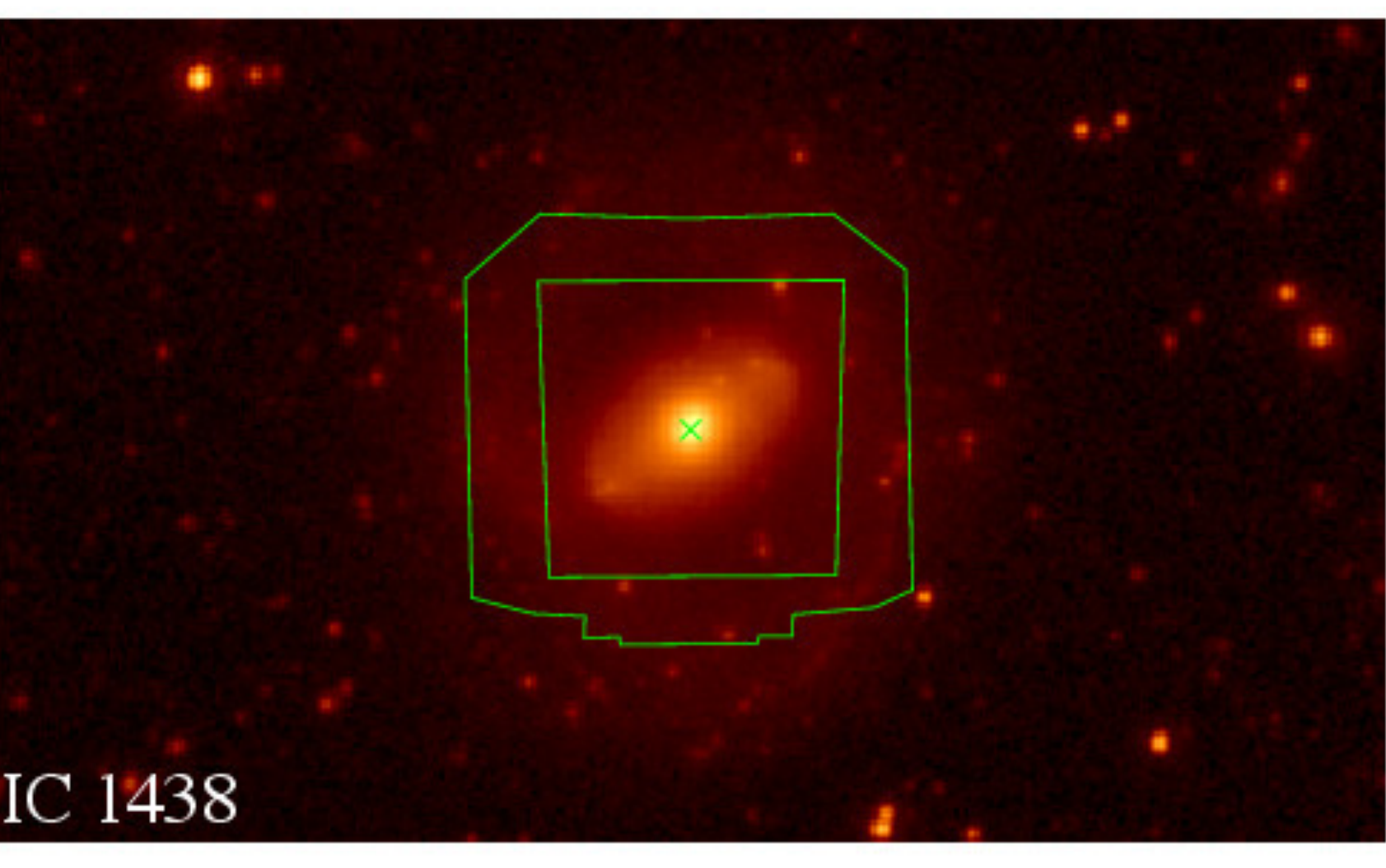}
	\includegraphics[trim=1cm 9.5cm 2.9cm 9.5cm, clip=true, width=0.5\columnwidth]{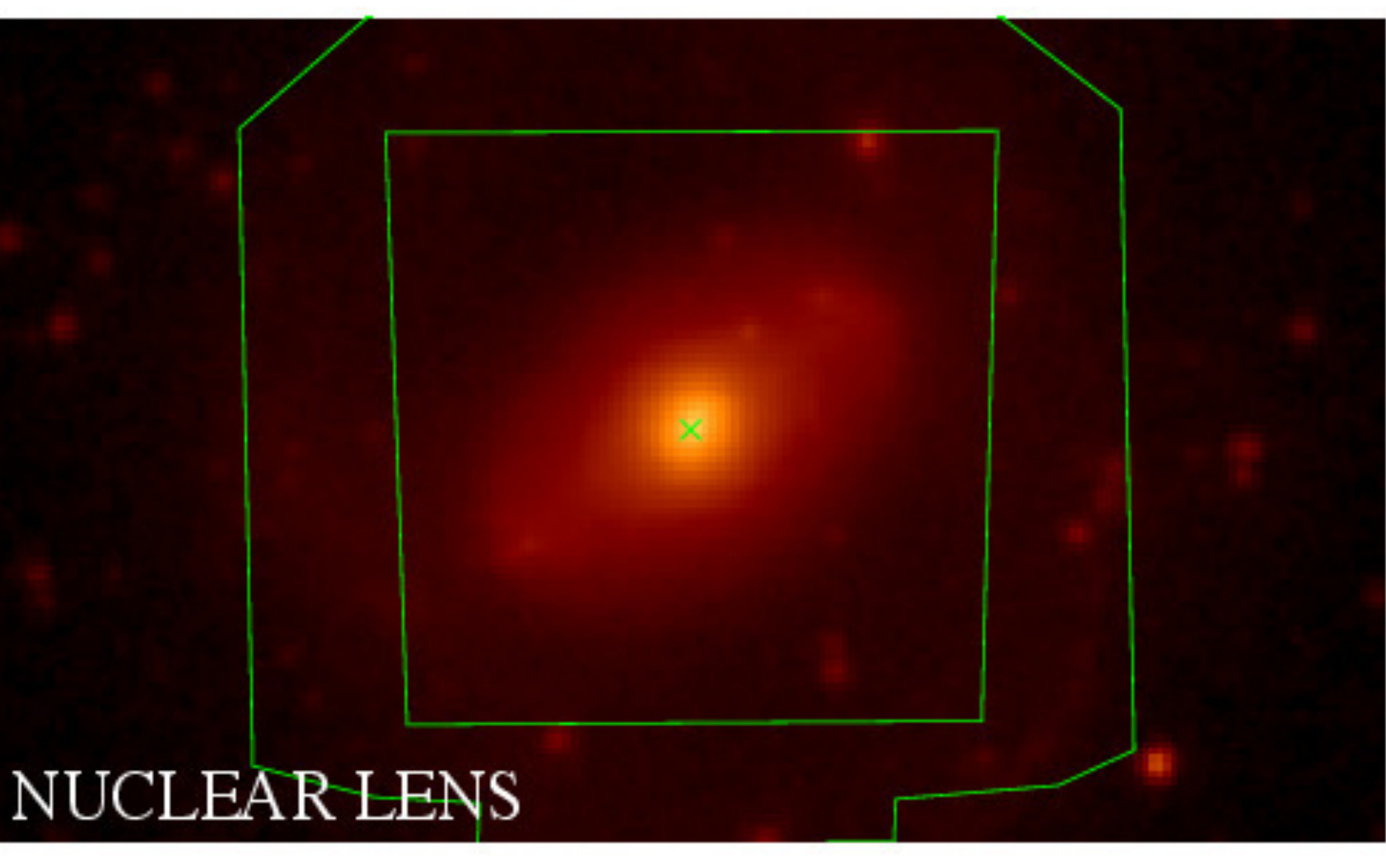}
	\includegraphics[trim=1cm 9.5cm 2.9cm 9.5cm, clip=true, width=0.5\columnwidth]{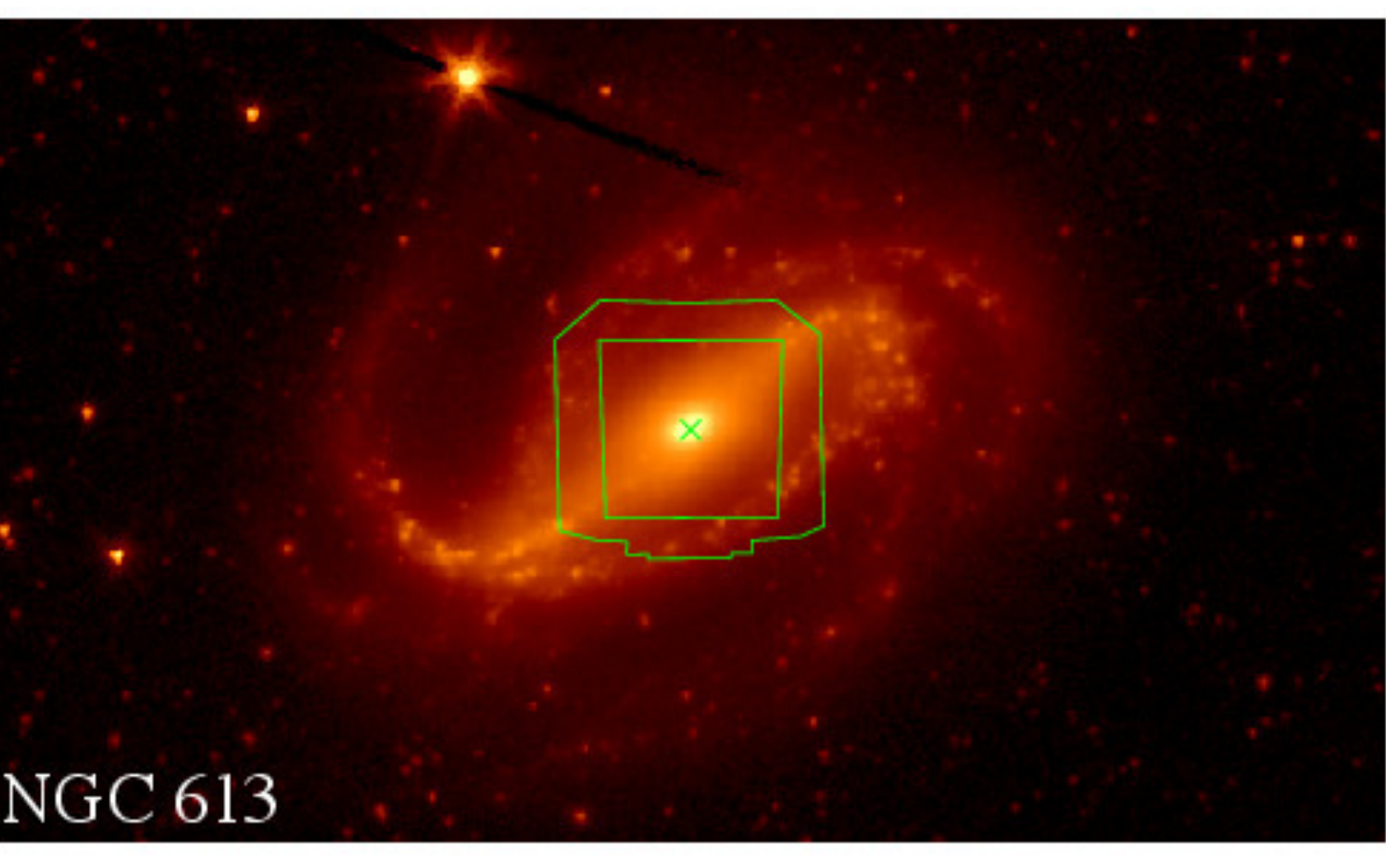}
	\includegraphics[trim=1cm 9.5cm 2.9cm 9.5cm, clip=true, width=0.5\columnwidth]{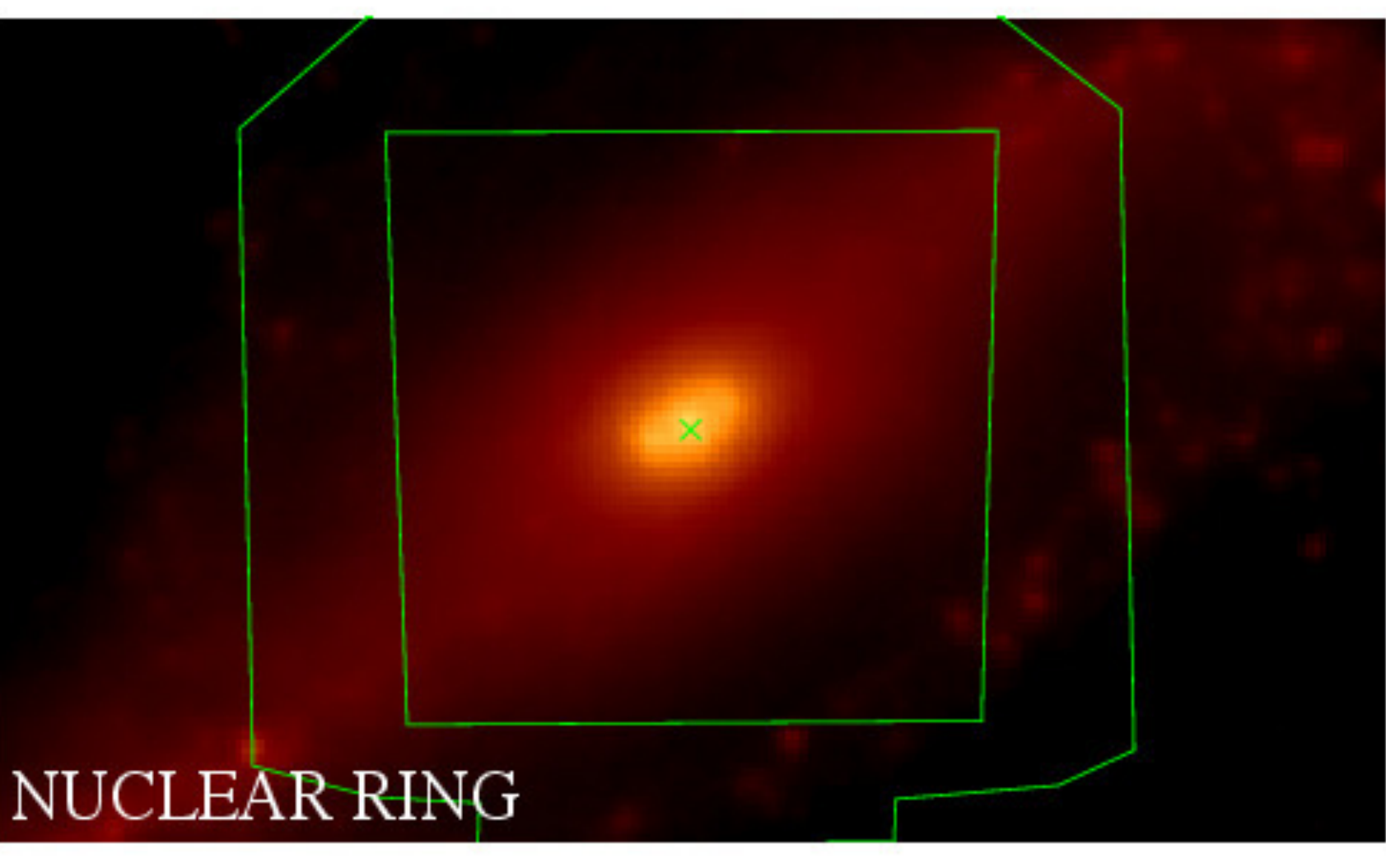}\\
	\includegraphics[trim=1cm 9.5cm 2.9cm 9.5cm, clip=true, width=0.5\columnwidth]{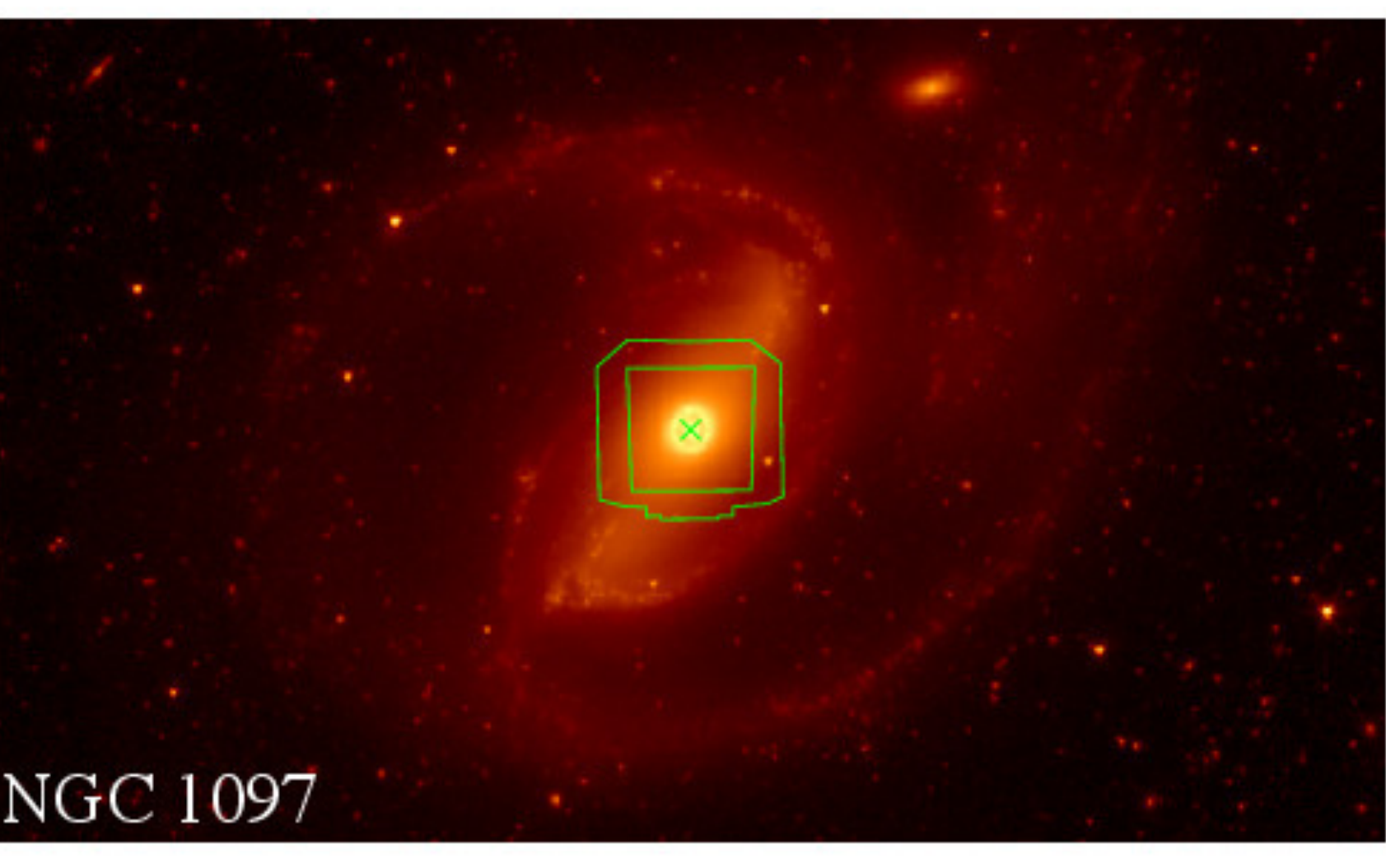}
	\includegraphics[trim=1cm 9.5cm 2.9cm 9.5cm, clip=true, width=0.5\columnwidth]{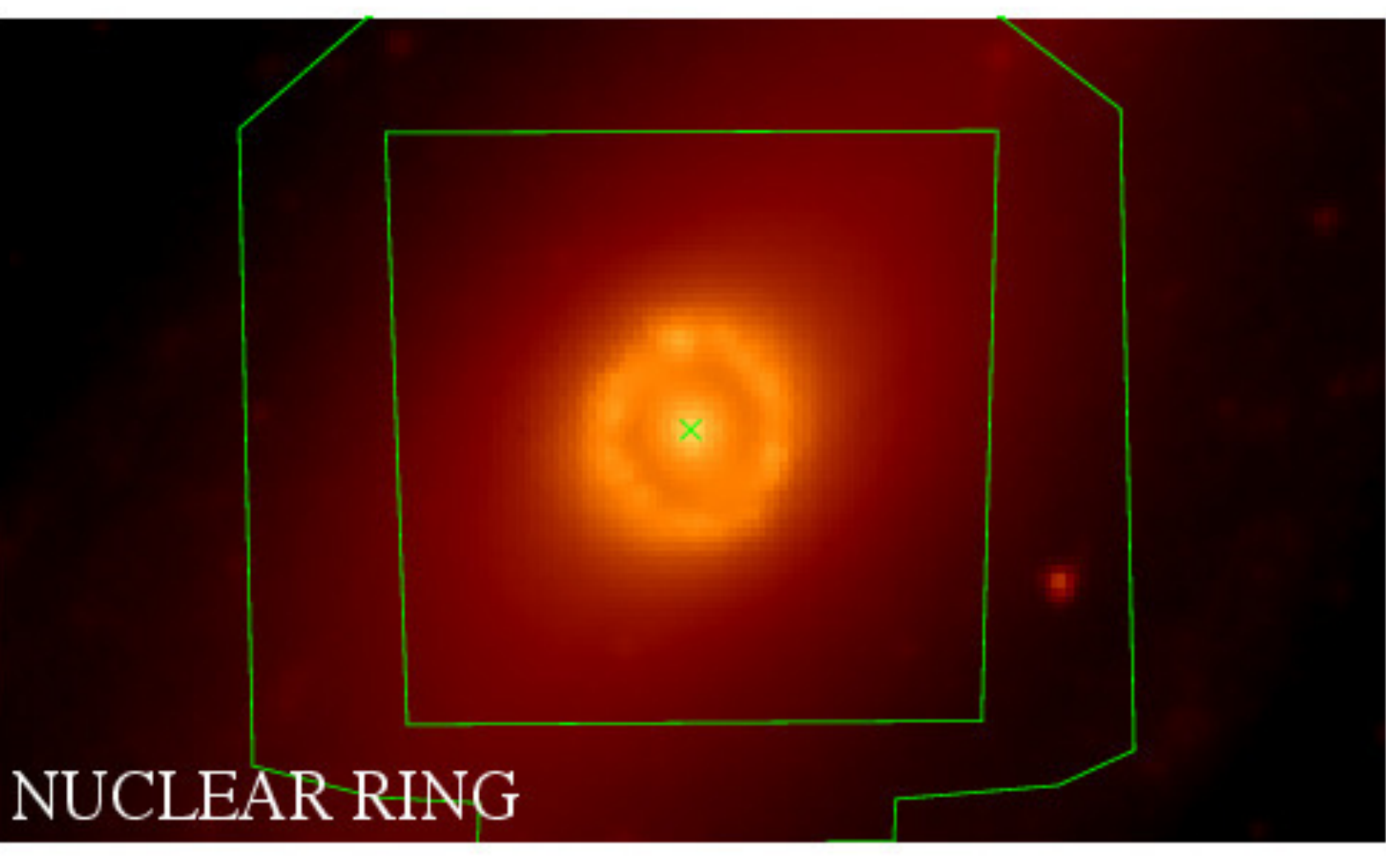}
	\includegraphics[trim=1cm 9.5cm 2.9cm 9.5cm, clip=true, width=0.5\columnwidth]{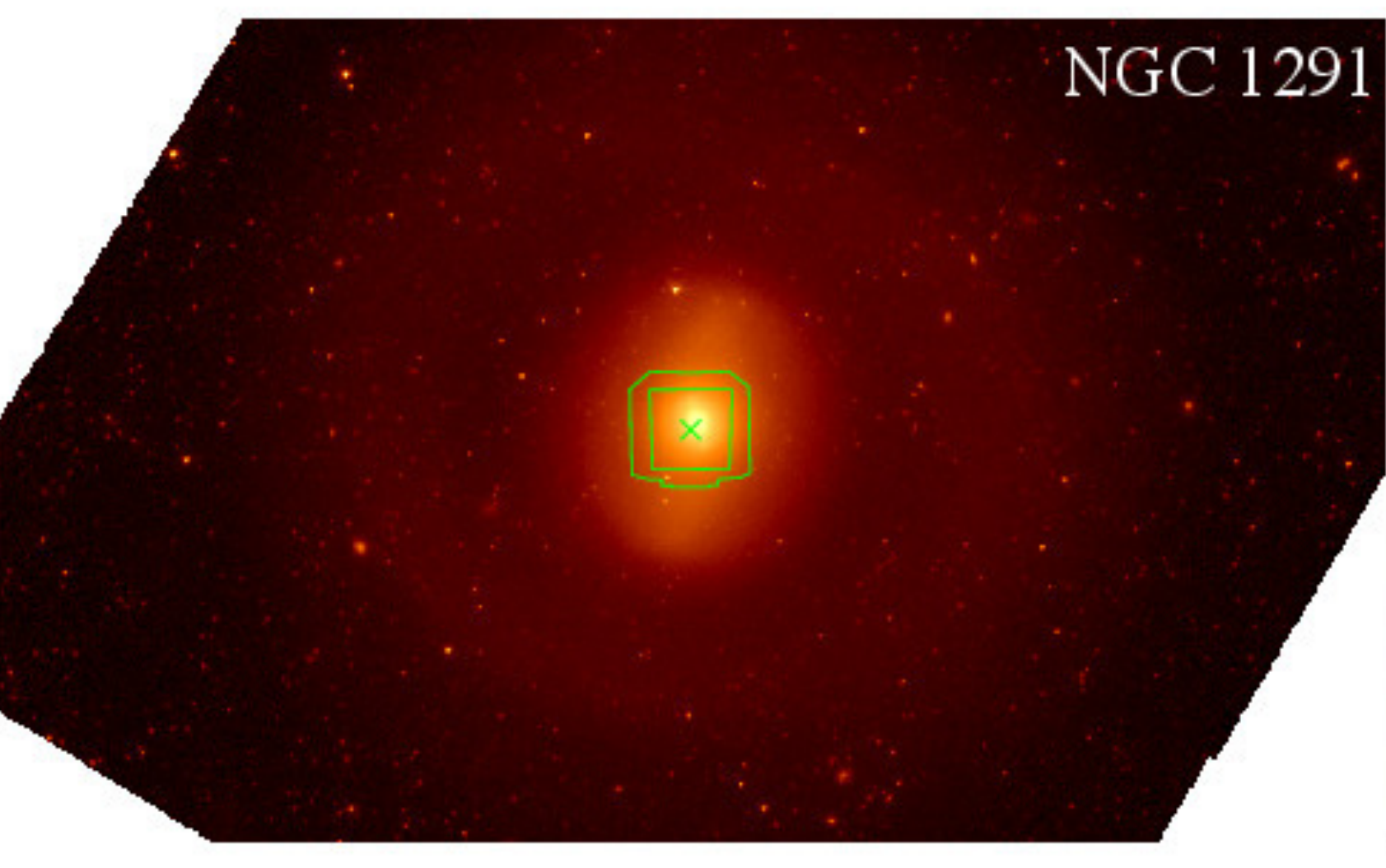}
	\includegraphics[trim=1cm 9.5cm 2.9cm 9.5cm, clip=true, width=0.5\columnwidth]{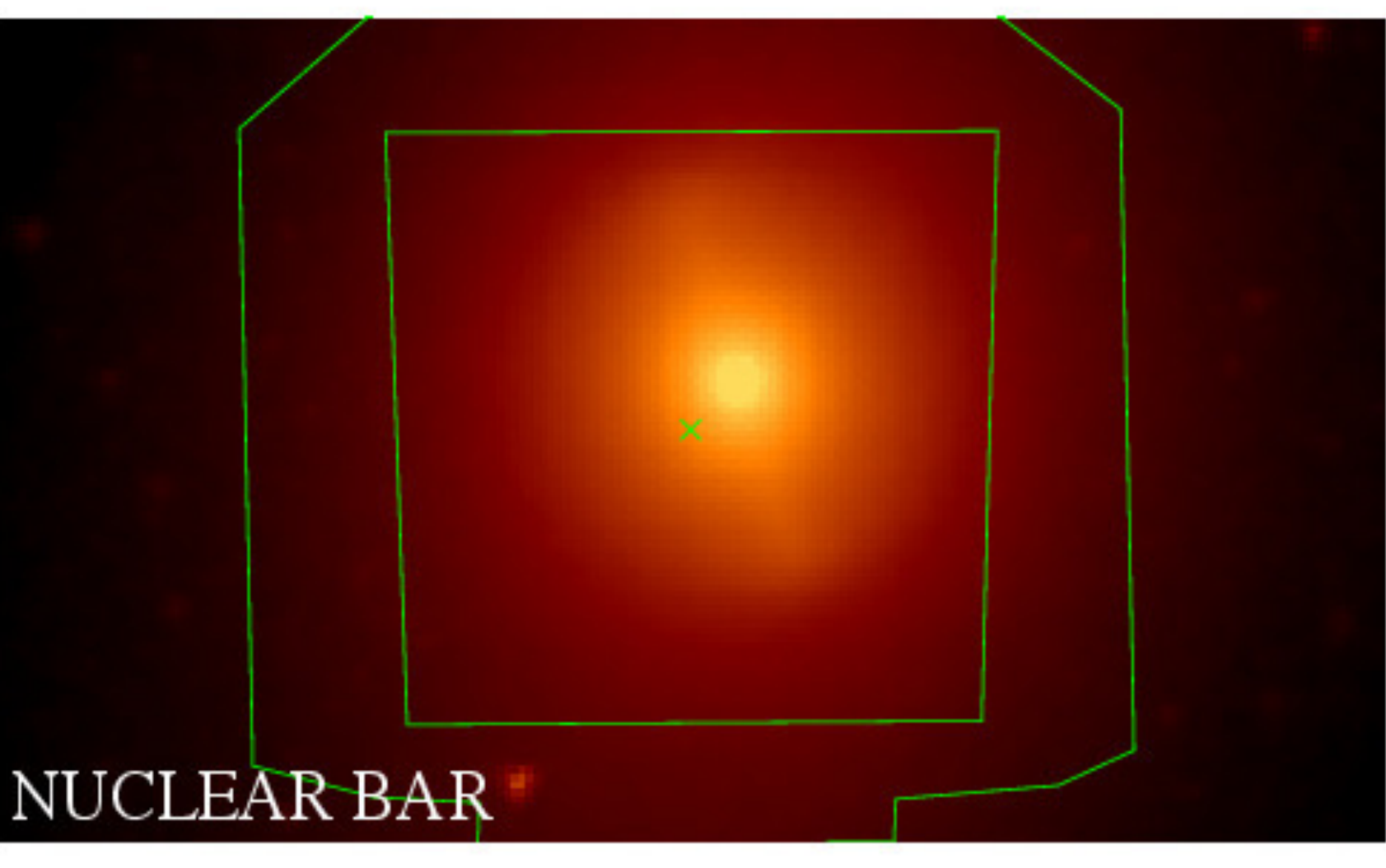}\\
	\includegraphics[trim=1cm 9.5cm 2.9cm 9.5cm, clip=true, width=0.5\columnwidth]{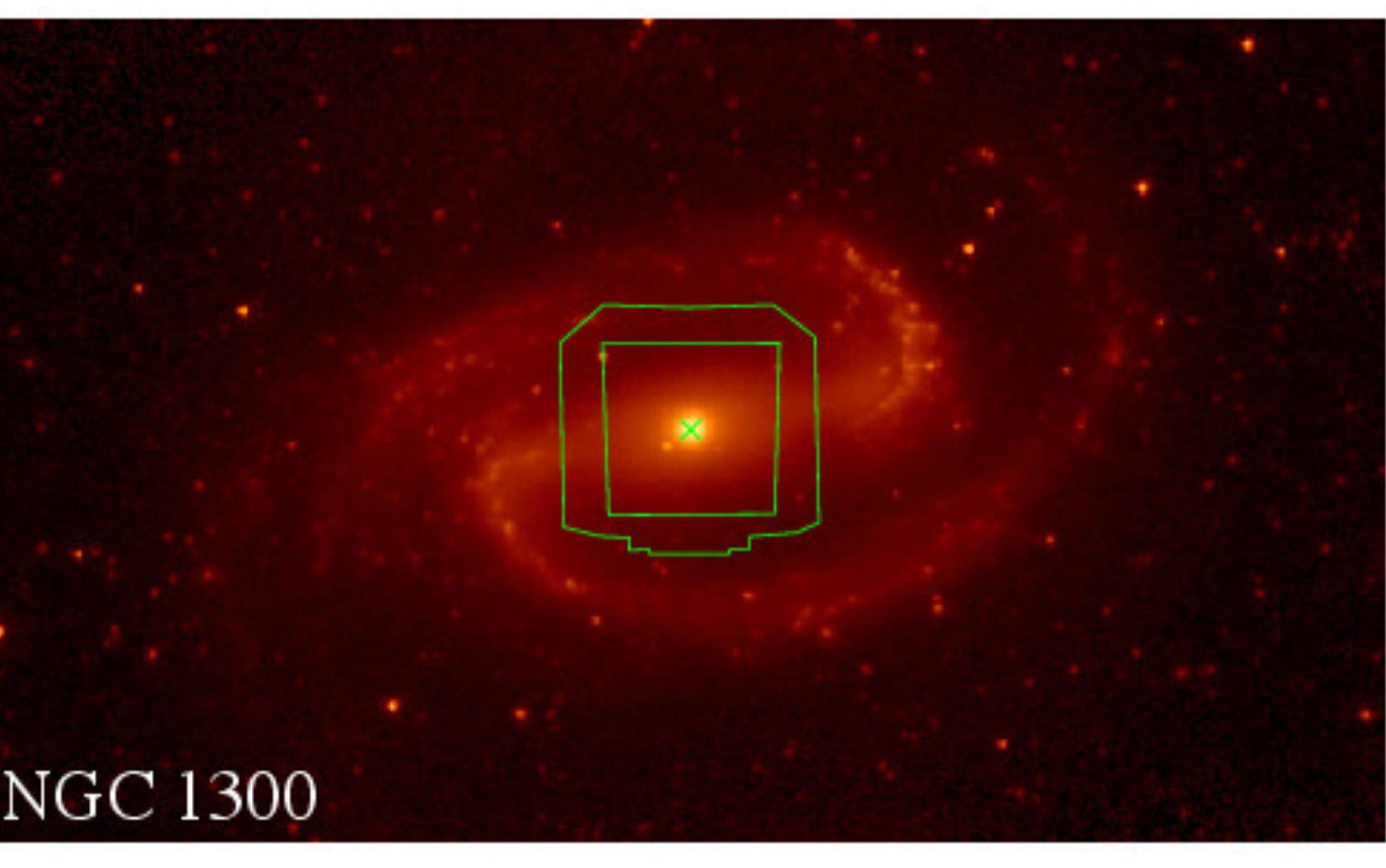}
	\includegraphics[trim=1cm 9.5cm 2.9cm 9.5cm, clip=true, width=0.5\columnwidth]{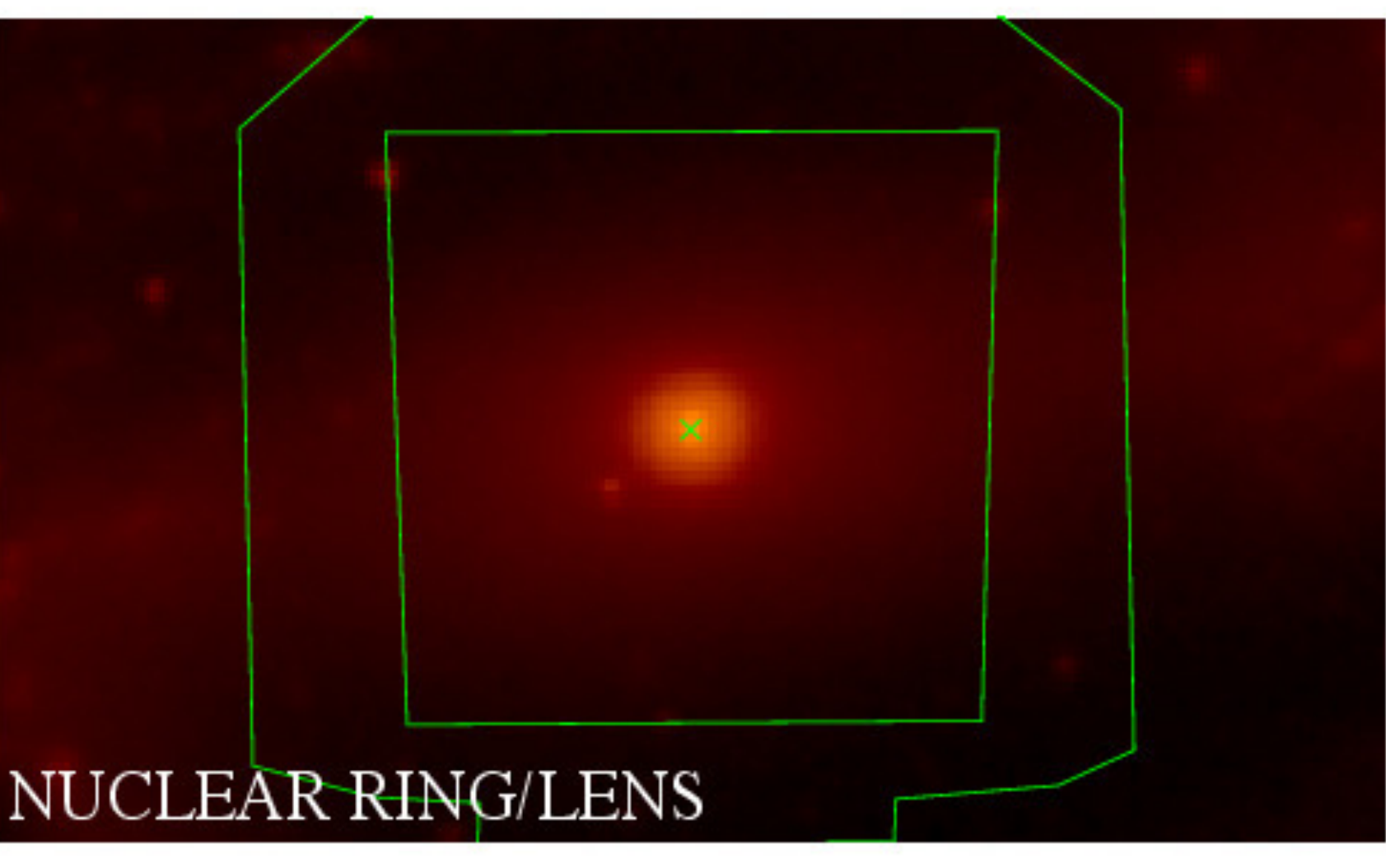}
	\includegraphics[trim=1cm 9.5cm 2.9cm 9.5cm, clip=true, width=0.5\columnwidth]{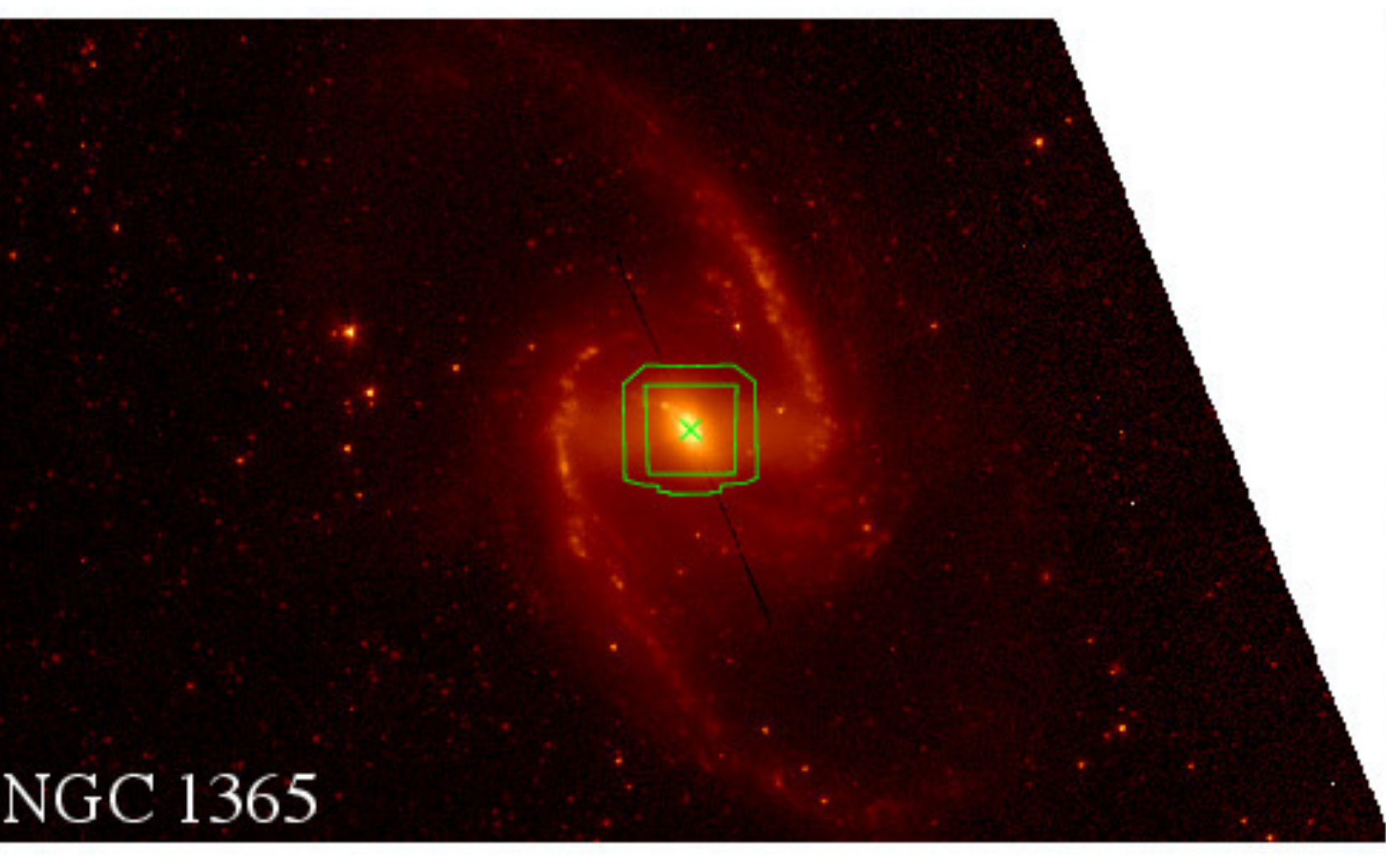}
	\includegraphics[trim=1cm 9.5cm 2.9cm 9.5cm, clip=true, width=0.5\columnwidth]{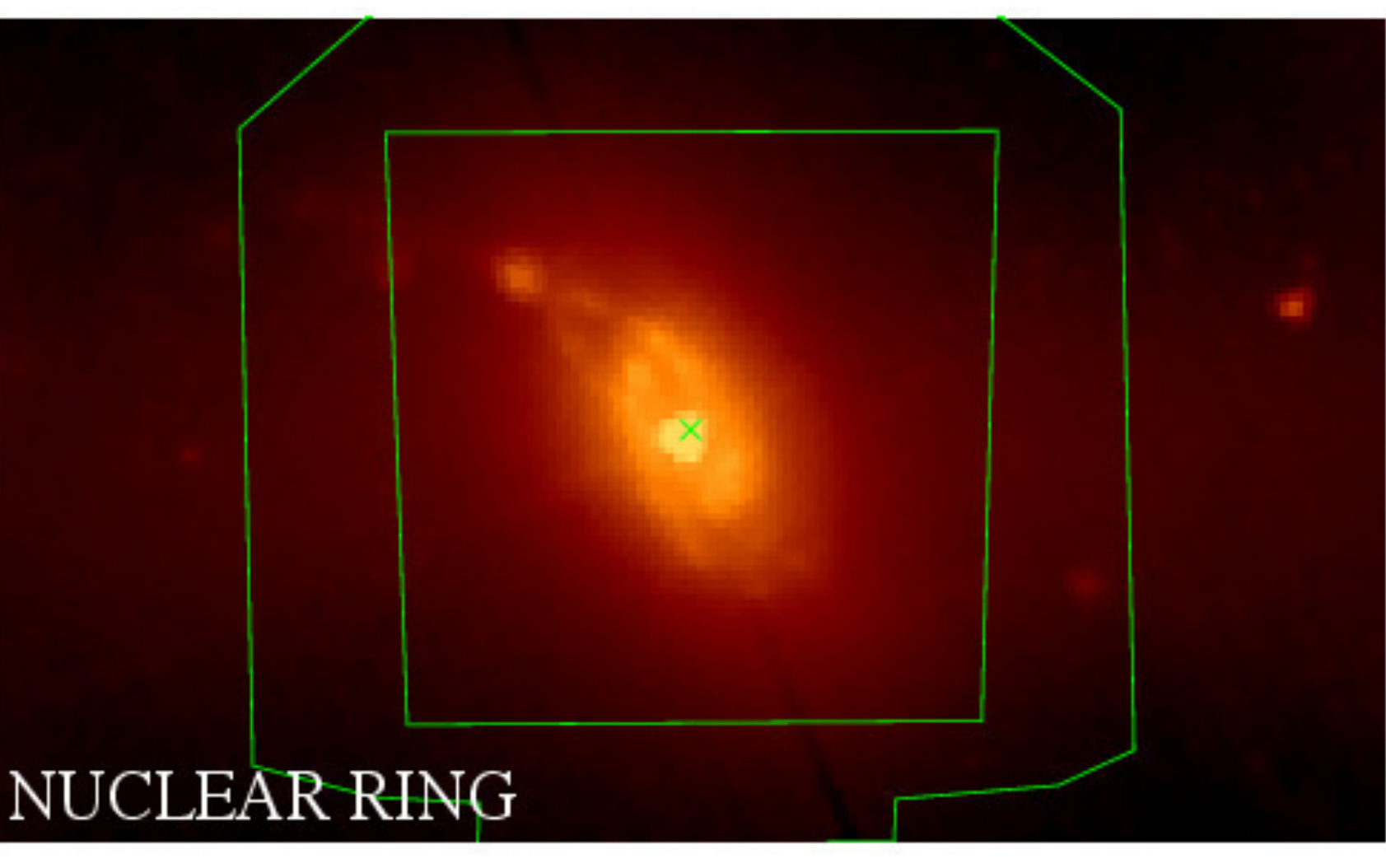}\\
	\includegraphics[trim=1cm 9.5cm 2.9cm 9.5cm, clip=true, width=0.5\columnwidth]{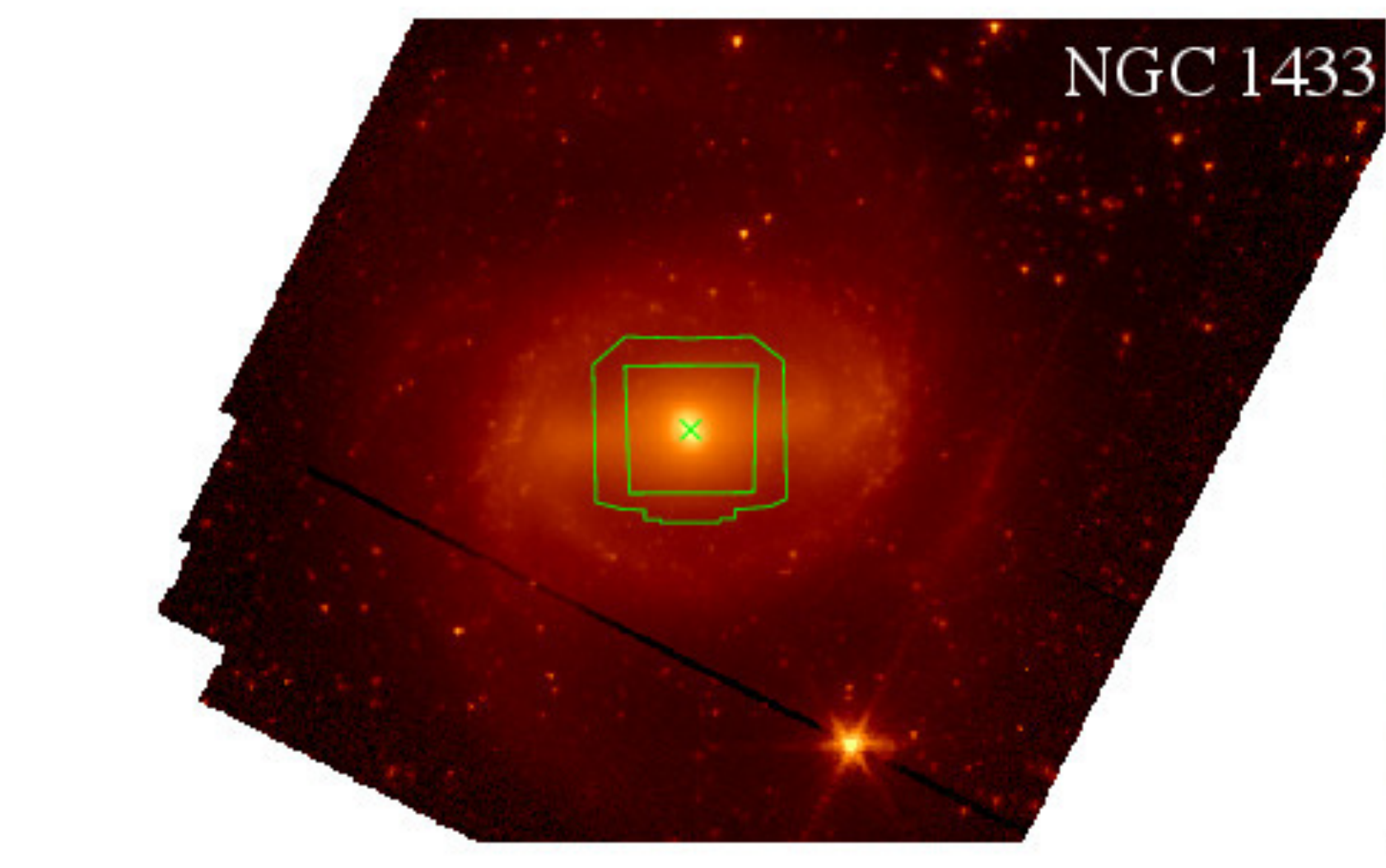}
	\includegraphics[trim=1cm 9.5cm 2.9cm 9.5cm, clip=true, width=0.5\columnwidth]{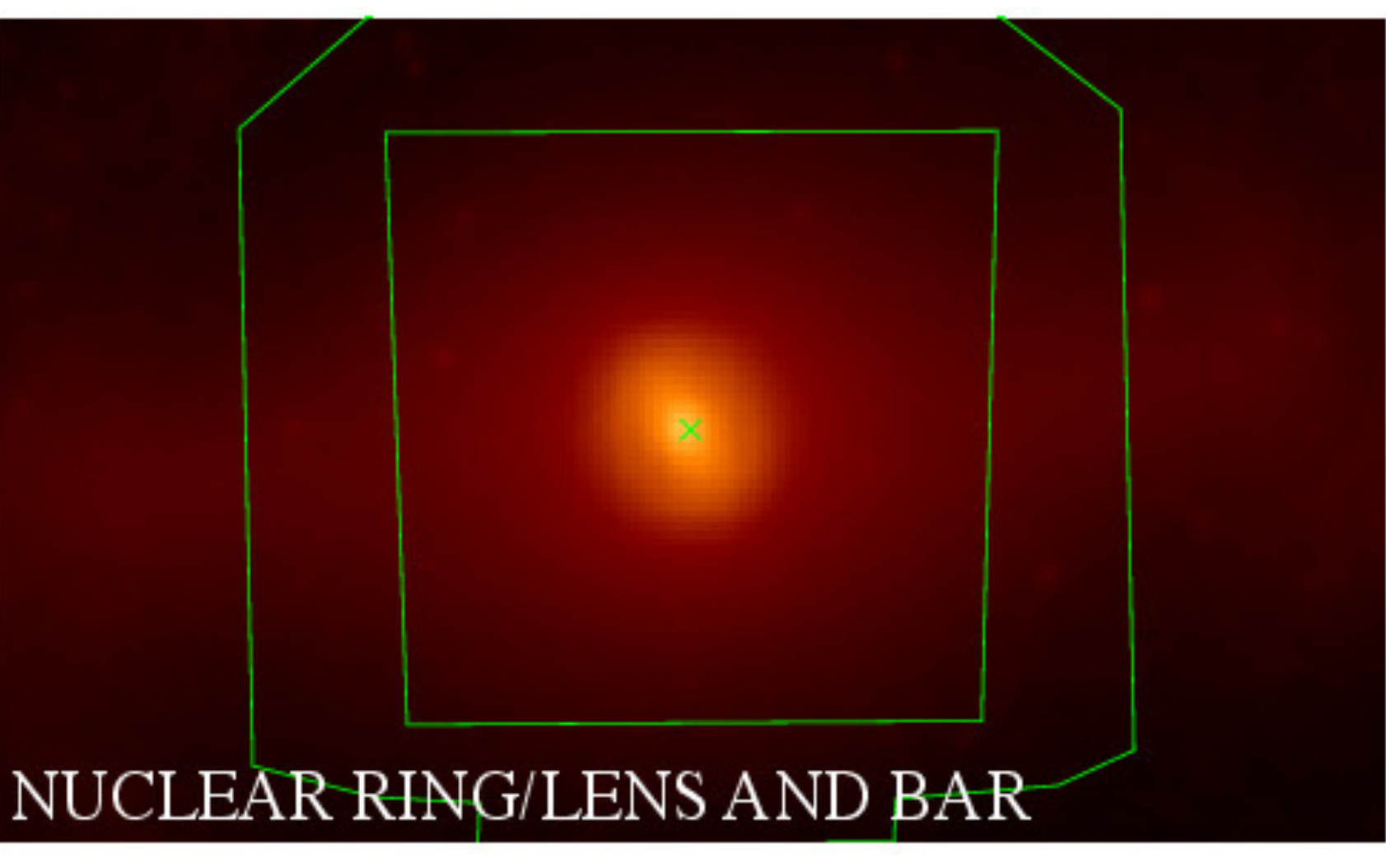}
	\includegraphics[trim=1cm 9.5cm 2.9cm 9.5cm, clip=true, width=0.5\columnwidth]{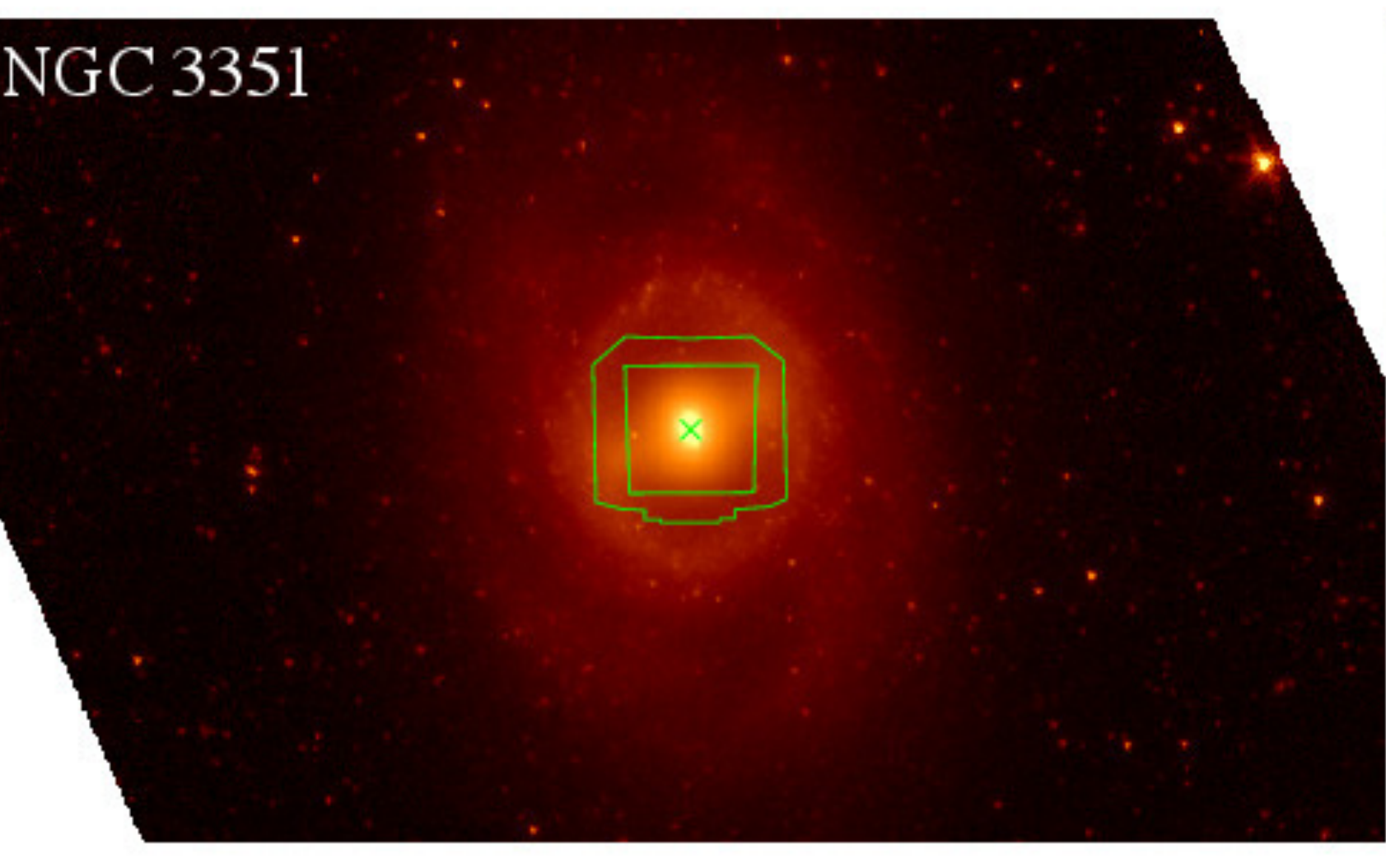}
	\includegraphics[trim=1cm 9.5cm 2.9cm 9.5cm, clip=true, width=0.5\columnwidth]{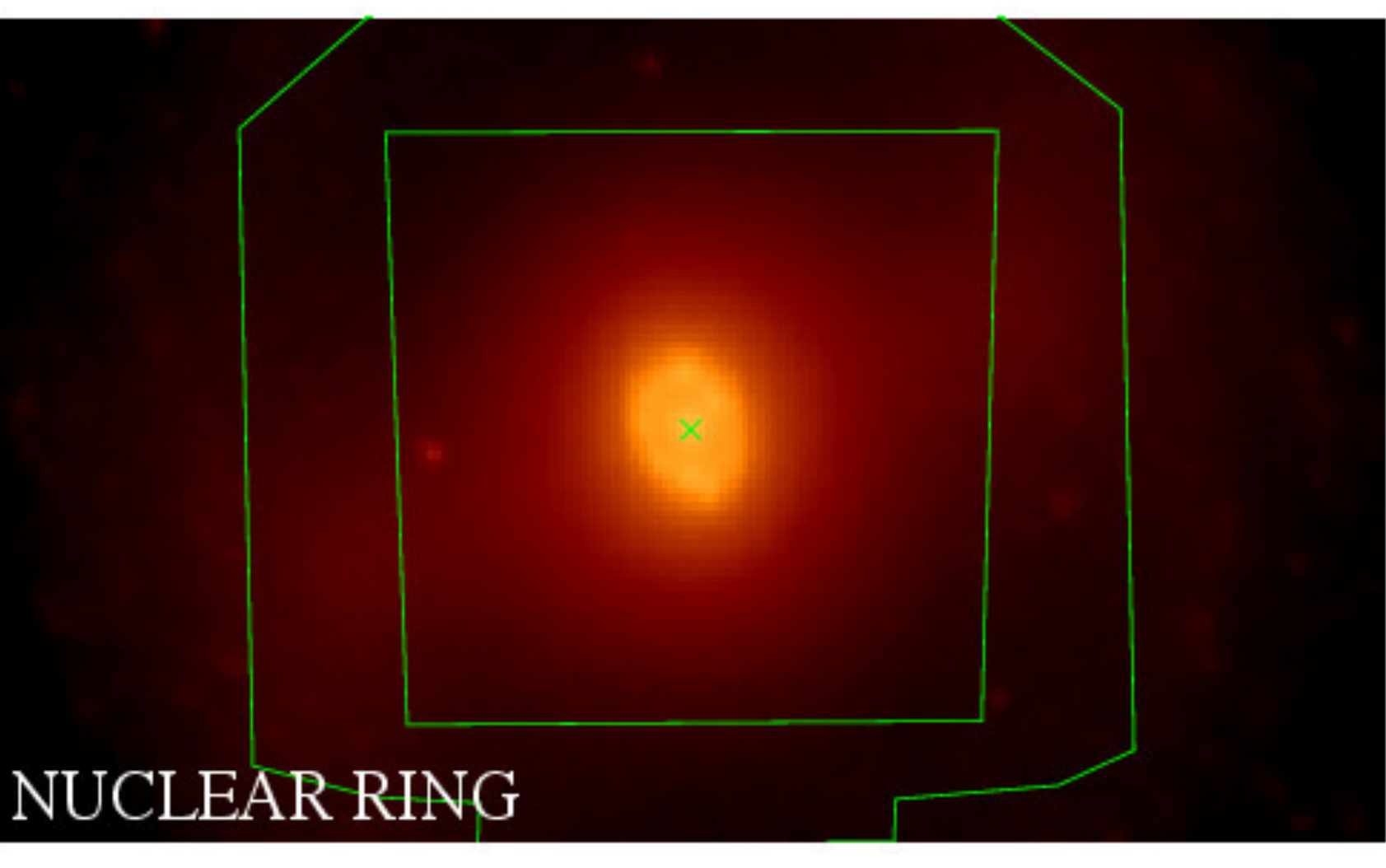}\\
	\includegraphics[trim=1cm 9.5cm 2.9cm 9.5cm, clip=true, width=0.5\columnwidth]{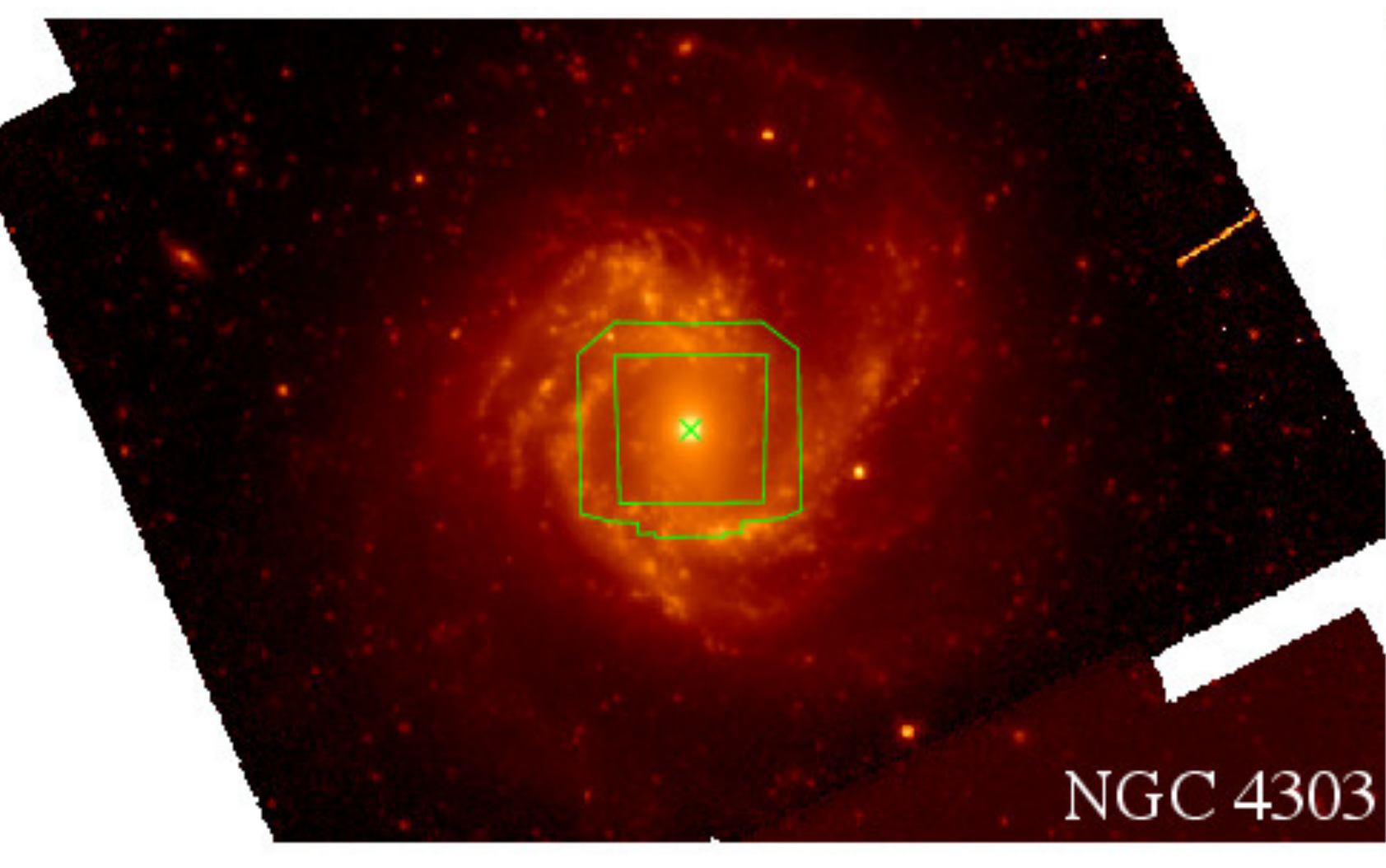}
	\includegraphics[trim=1cm 9.5cm 2.9cm 9.5cm, clip=true, width=0.5\columnwidth]{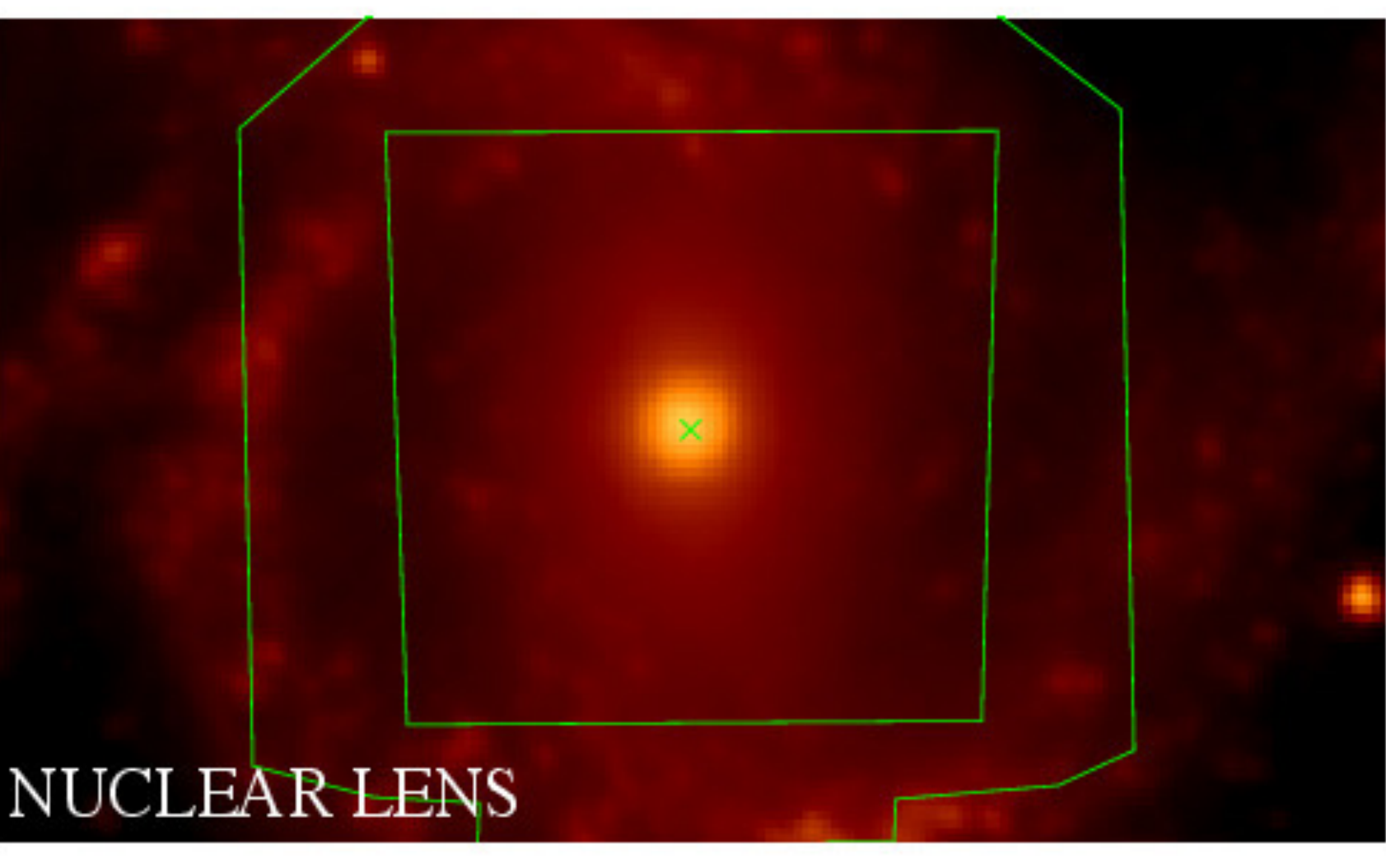}
	\includegraphics[trim=1cm 9.5cm 2.9cm 9.5cm, clip=true, width=0.5\columnwidth]{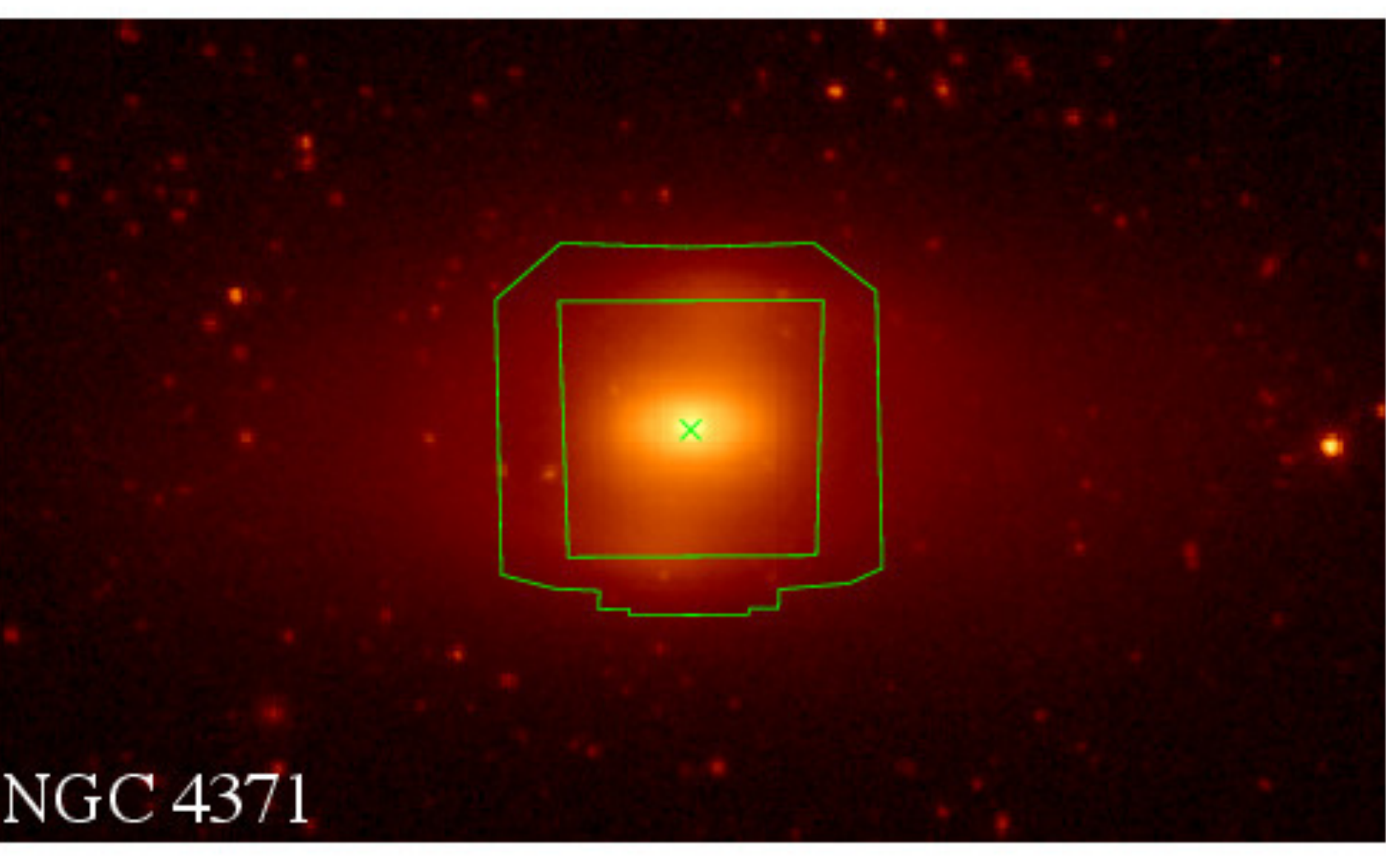}
	\includegraphics[trim=1cm 9.5cm 2.9cm 9.5cm, clip=true, width=0.5\columnwidth]{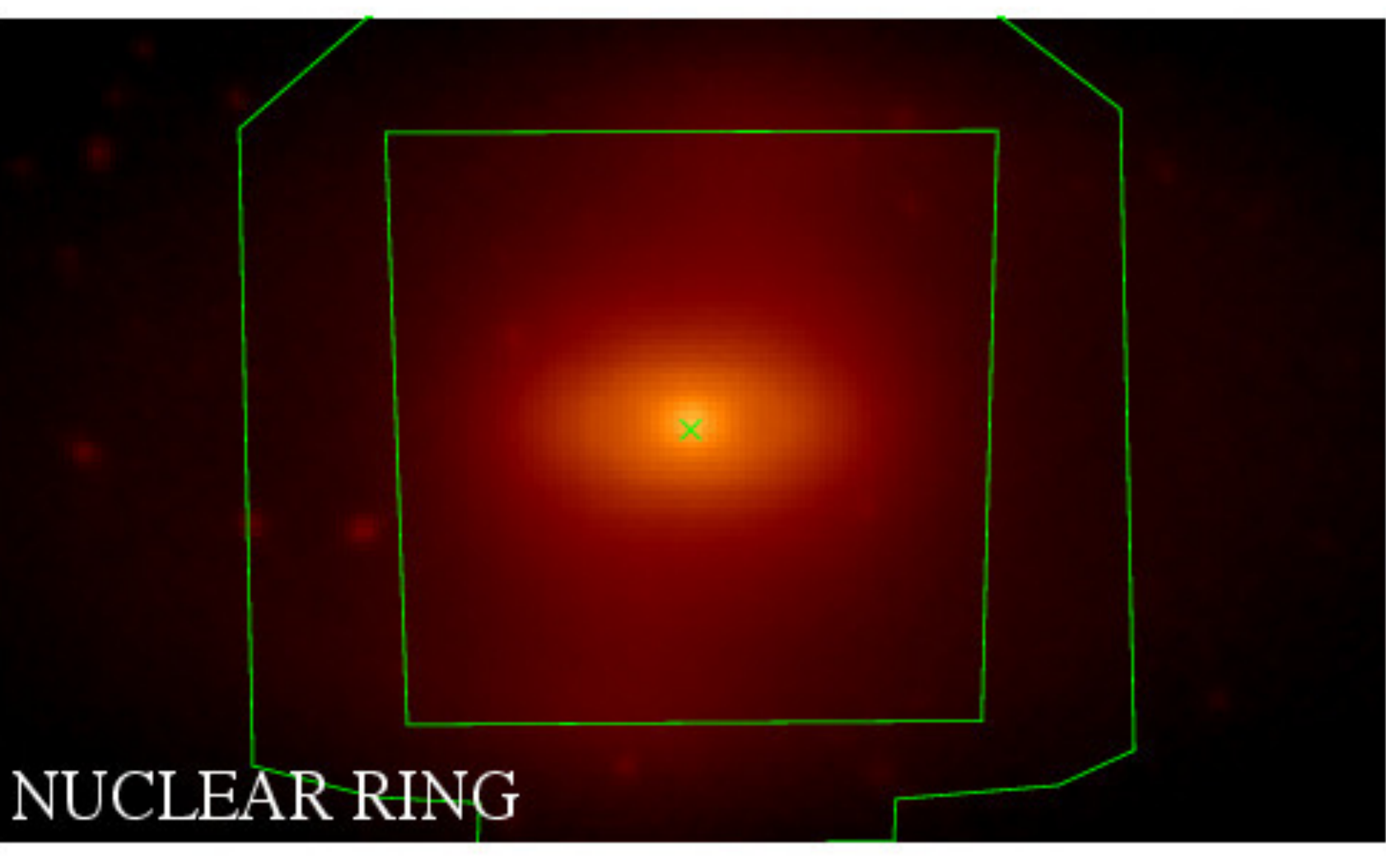}\\
	\includegraphics[trim=1cm 9.5cm 2.9cm 9.5cm, clip=true, width=0.5\columnwidth]{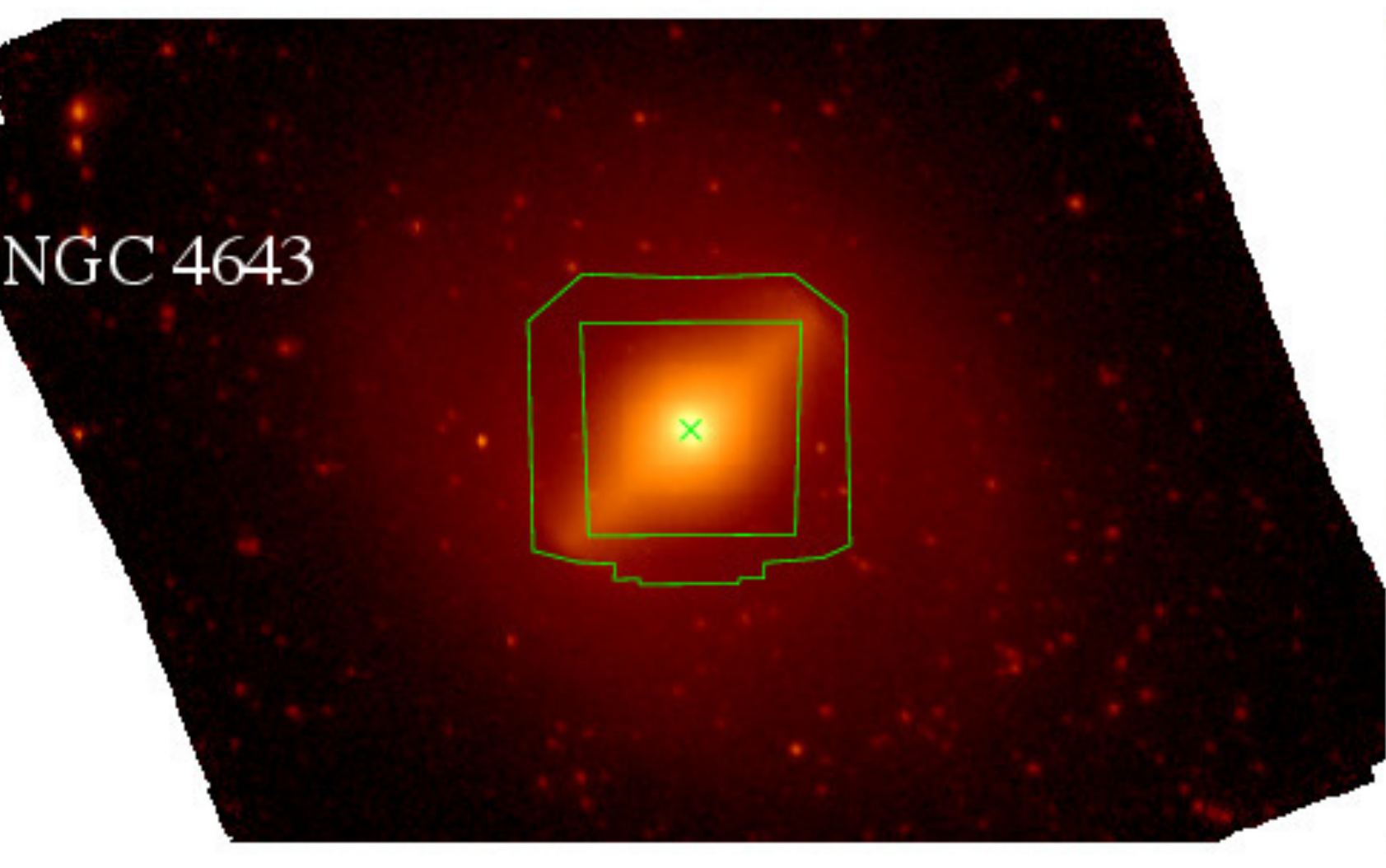}
	\includegraphics[trim=1cm 9.5cm 2.9cm 9.5cm, clip=true, width=0.5\columnwidth]{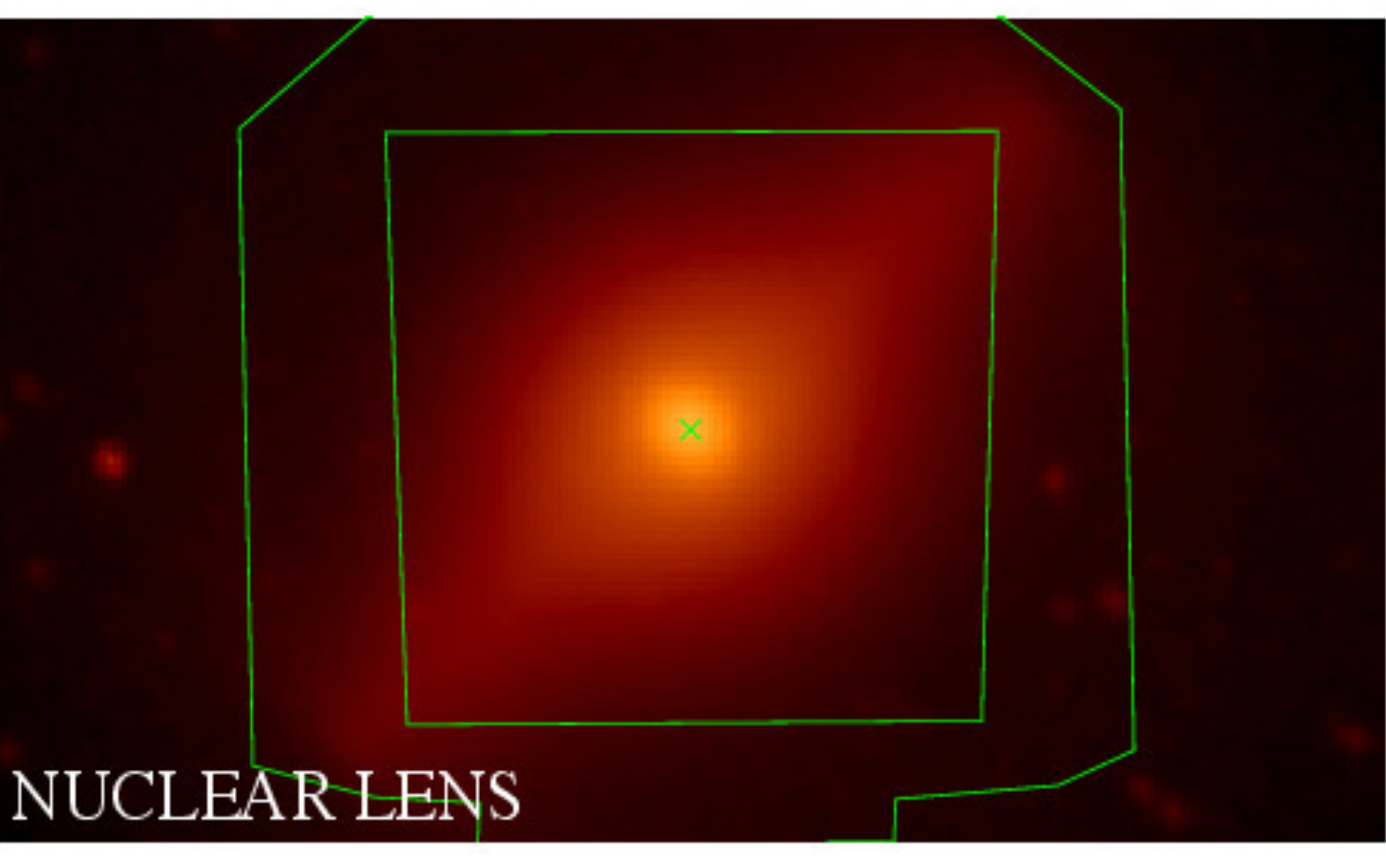}
	\includegraphics[trim=1cm 9.5cm 2.9cm 9.5cm, clip=true, width=0.5\columnwidth]{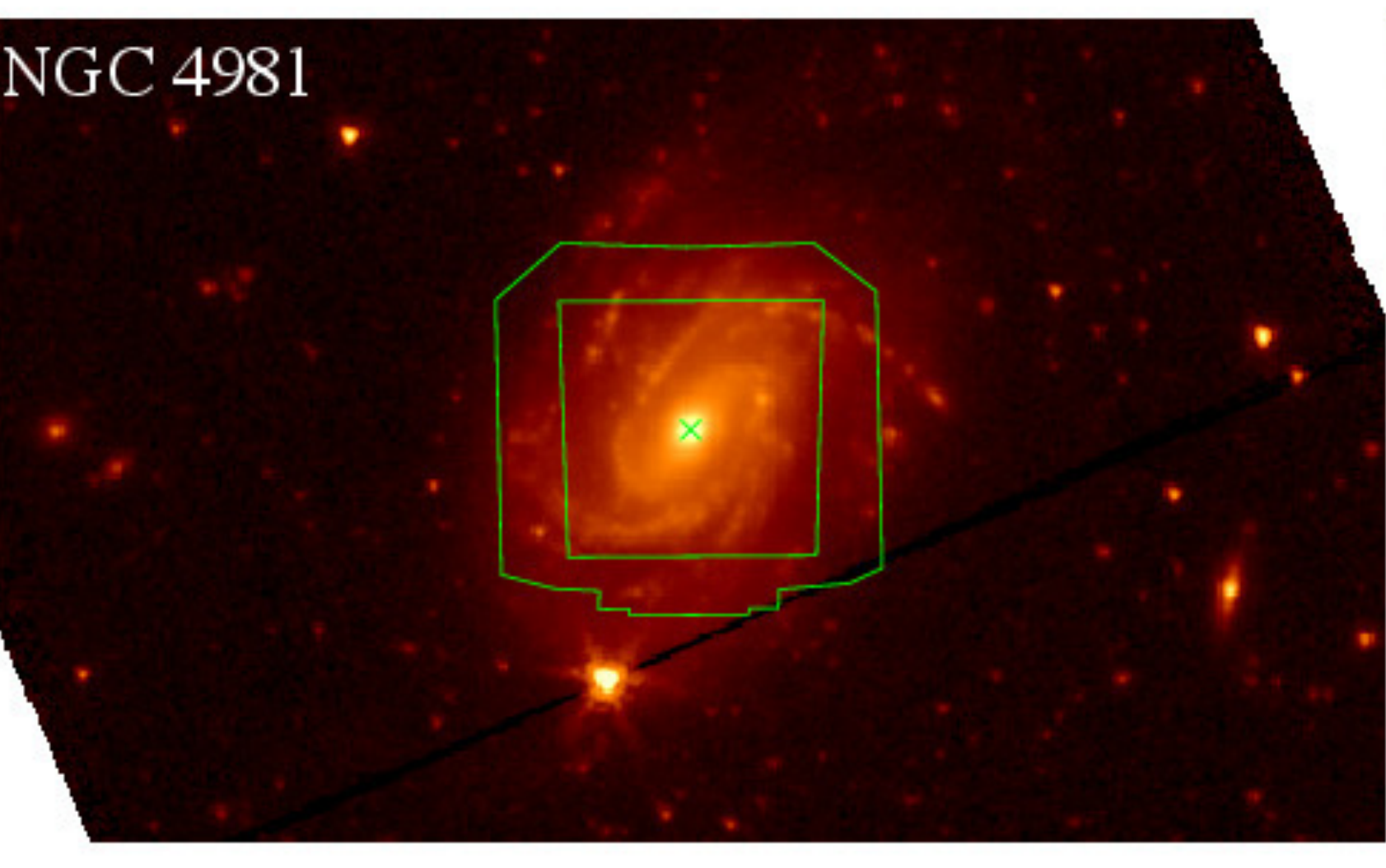}
	\includegraphics[trim=1cm 9.5cm 2.9cm 9.5cm, clip=true, width=0.5\columnwidth]{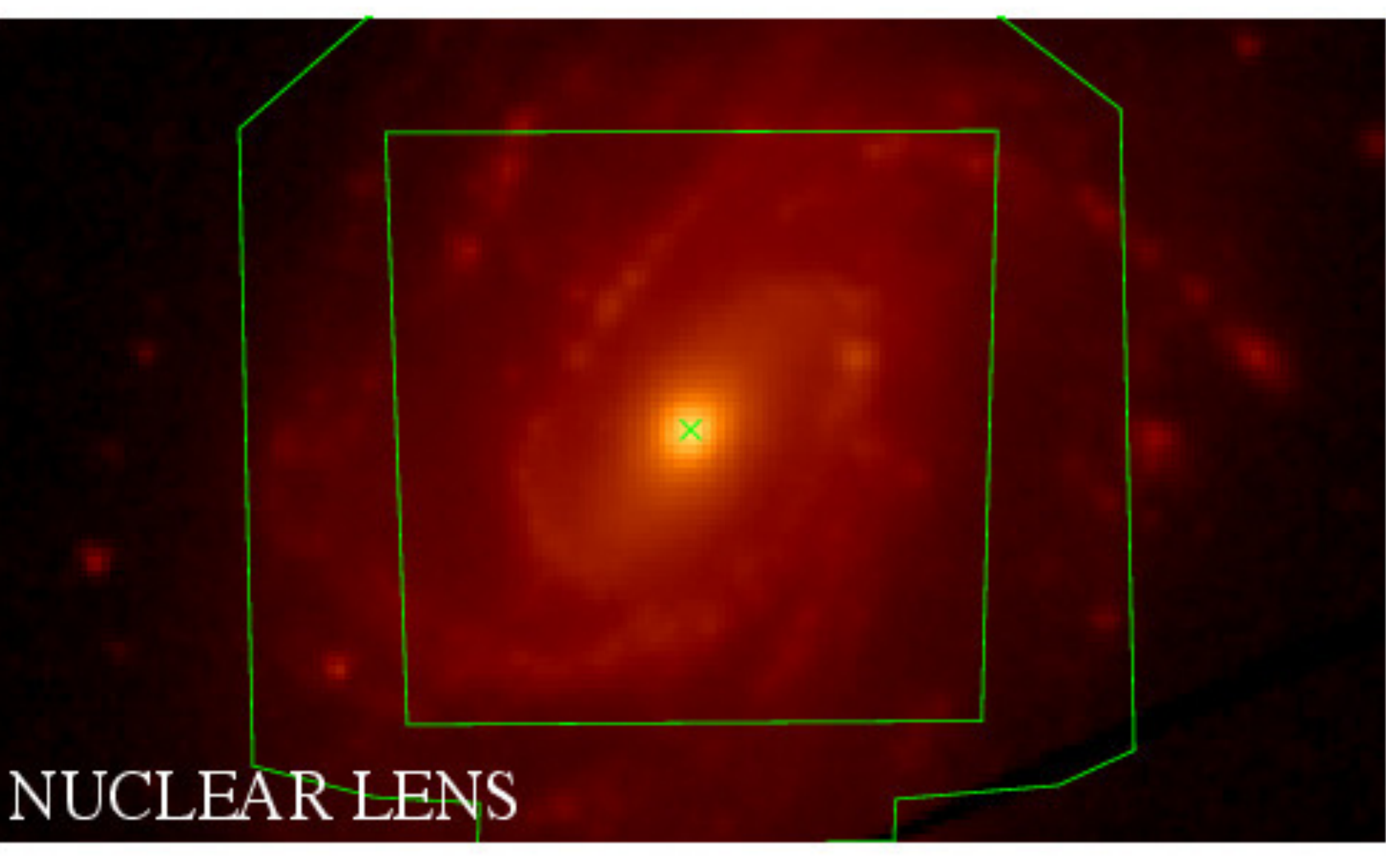}\\
	\includegraphics[trim=1cm 9.5cm 2.9cm 9.5cm, clip=true, width=0.5\columnwidth]{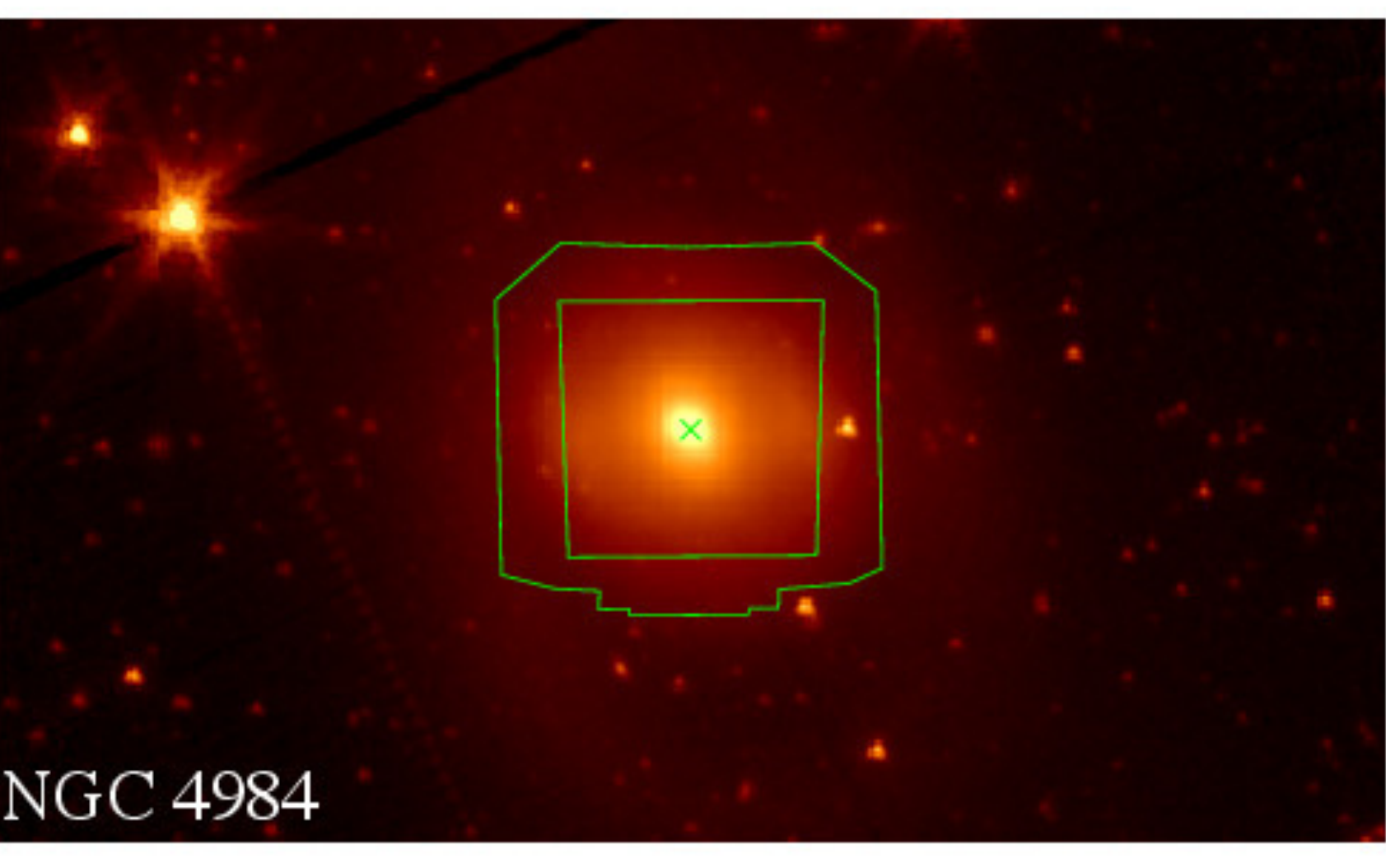}
	\includegraphics[trim=1cm 9.5cm 2.9cm 9.5cm, clip=true, width=0.5\columnwidth]{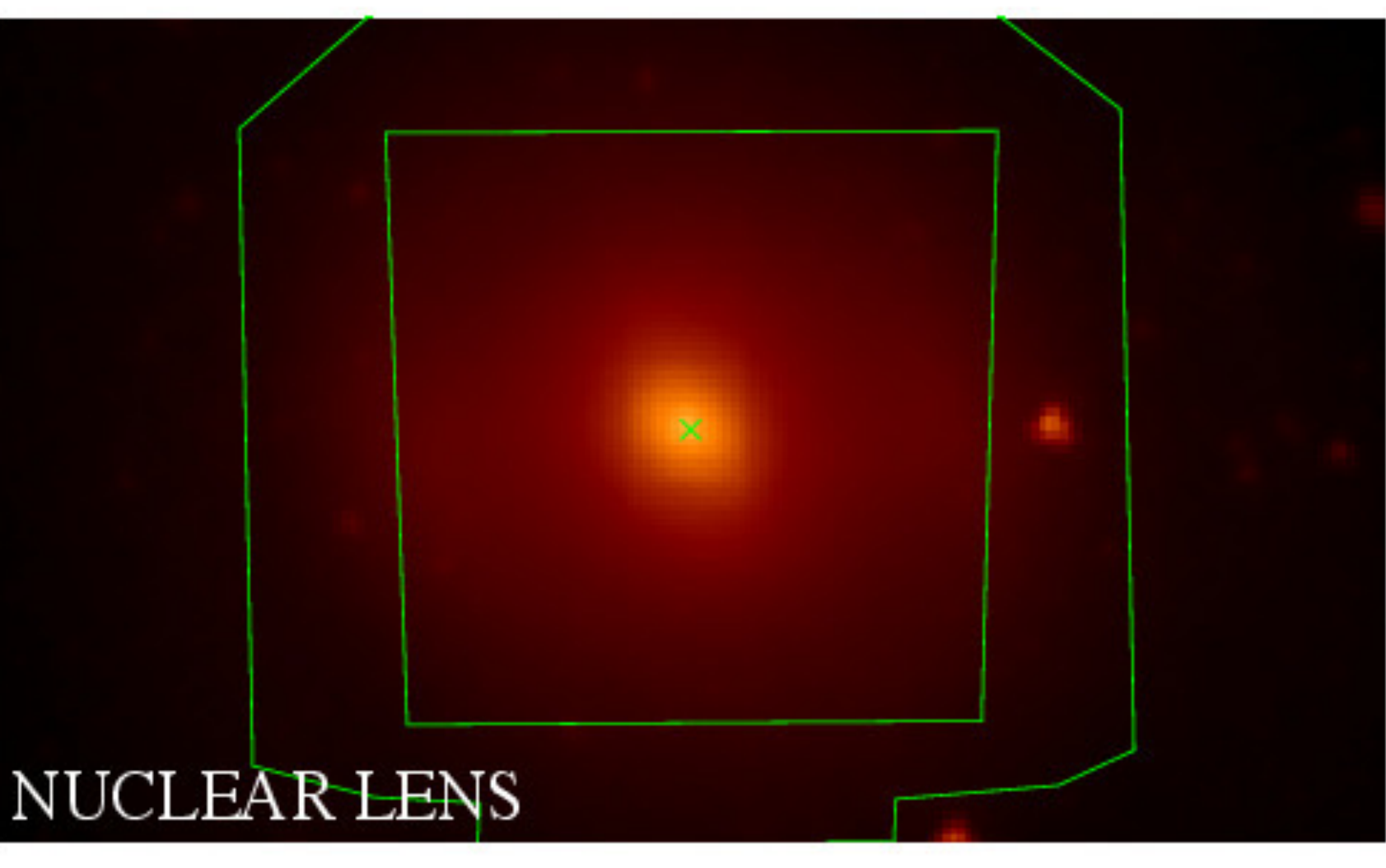}
	\includegraphics[trim=1cm 9.5cm 2.9cm 9.5cm, clip=true, width=0.5\columnwidth]{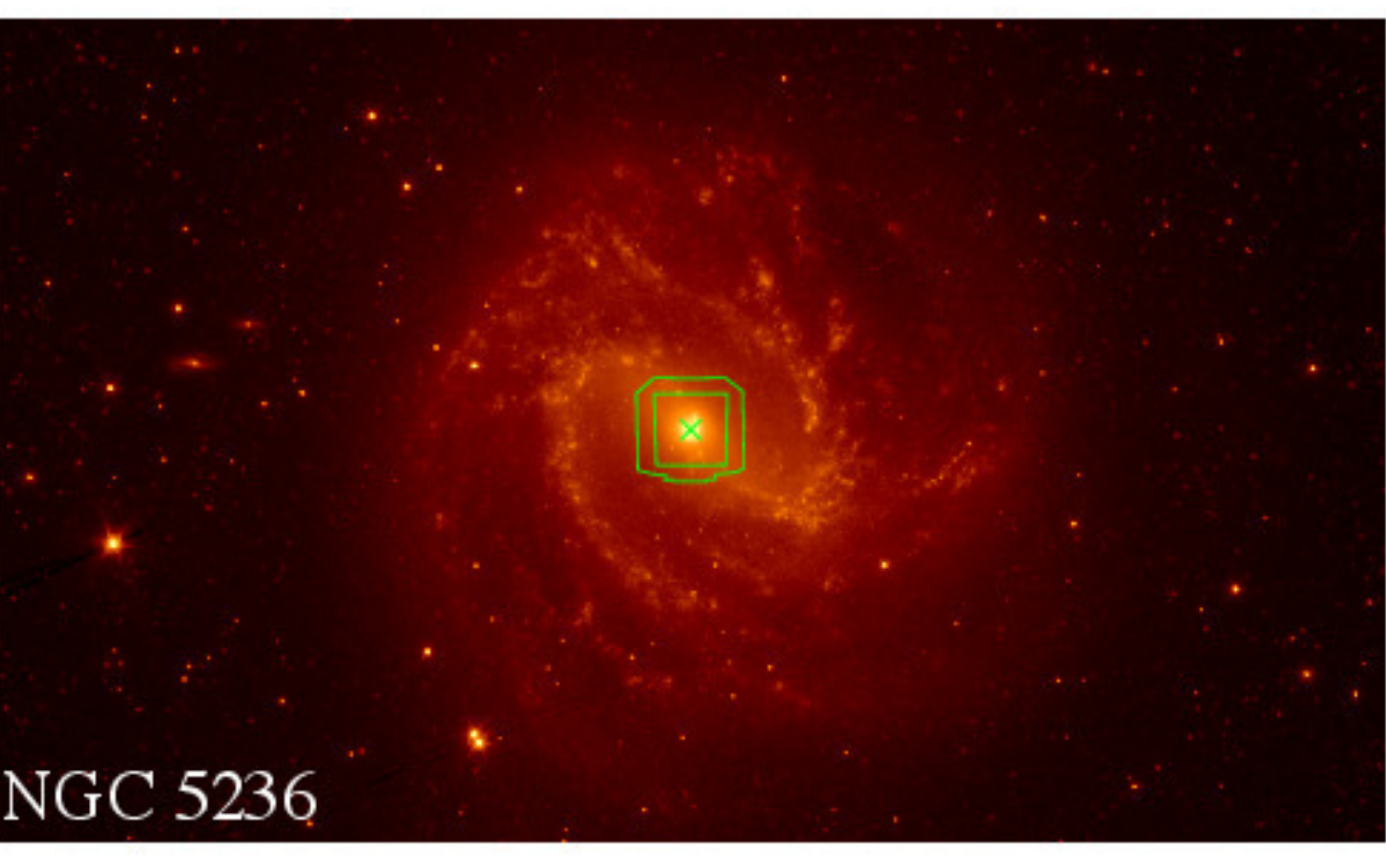}
	\includegraphics[trim=1cm 9.5cm 2.9cm 9.5cm, clip=true, width=0.5\columnwidth]{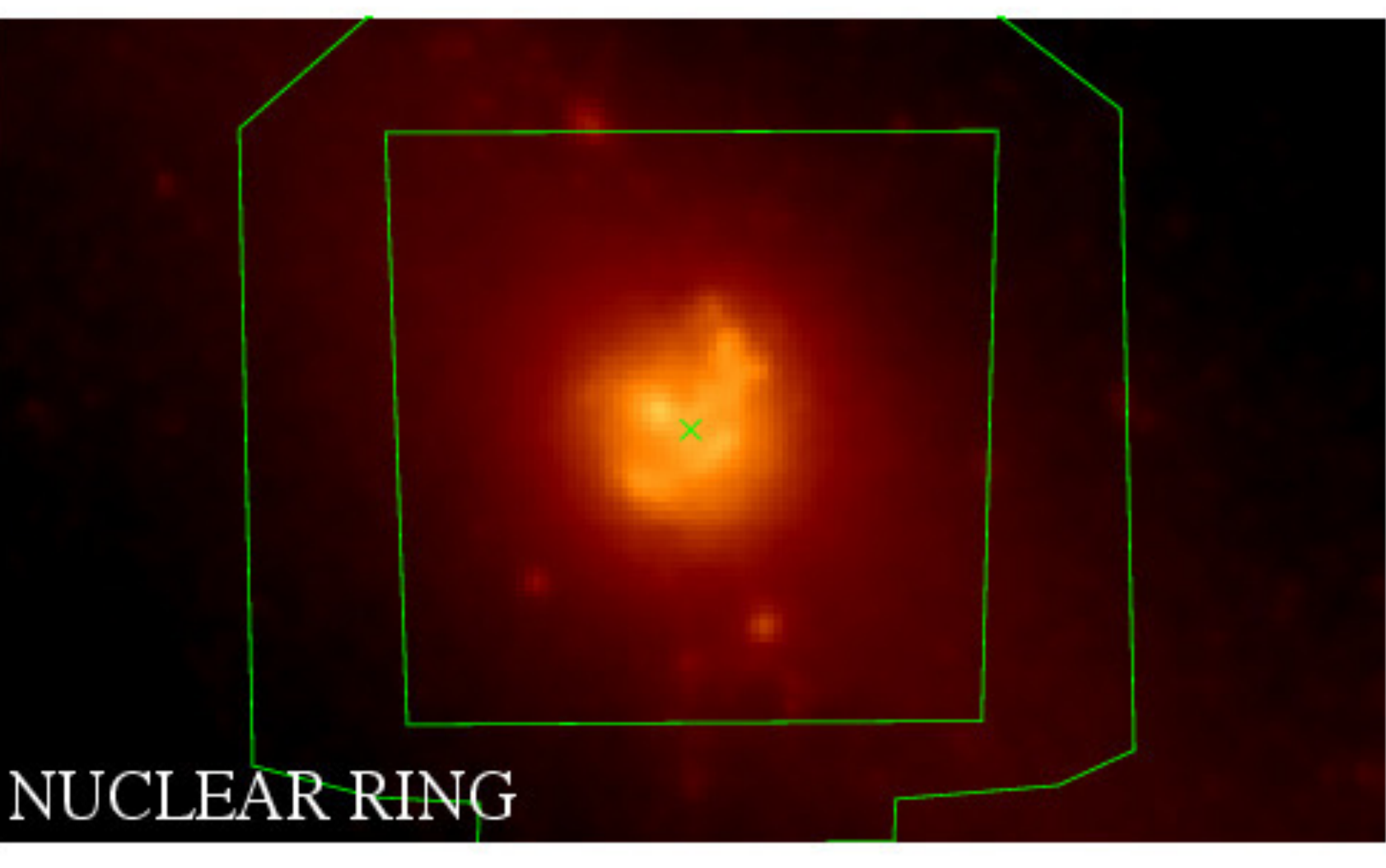}\\
	\includegraphics[trim=1cm 9.5cm 2.9cm 9.5cm, clip=true, width=0.5\columnwidth]{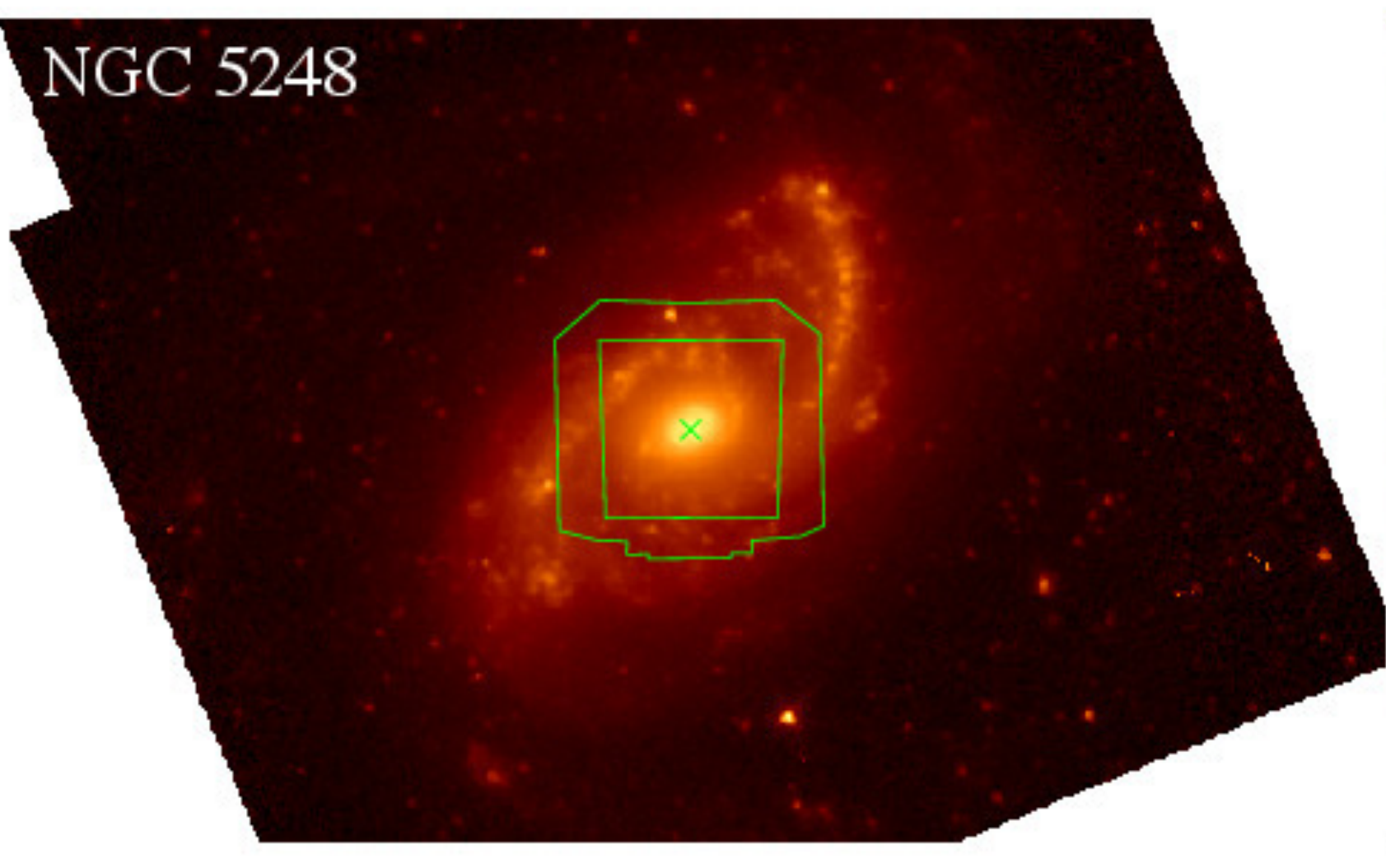}
	\includegraphics[trim=1cm 9.5cm 2.9cm 9.5cm, clip=true, width=0.5\columnwidth]{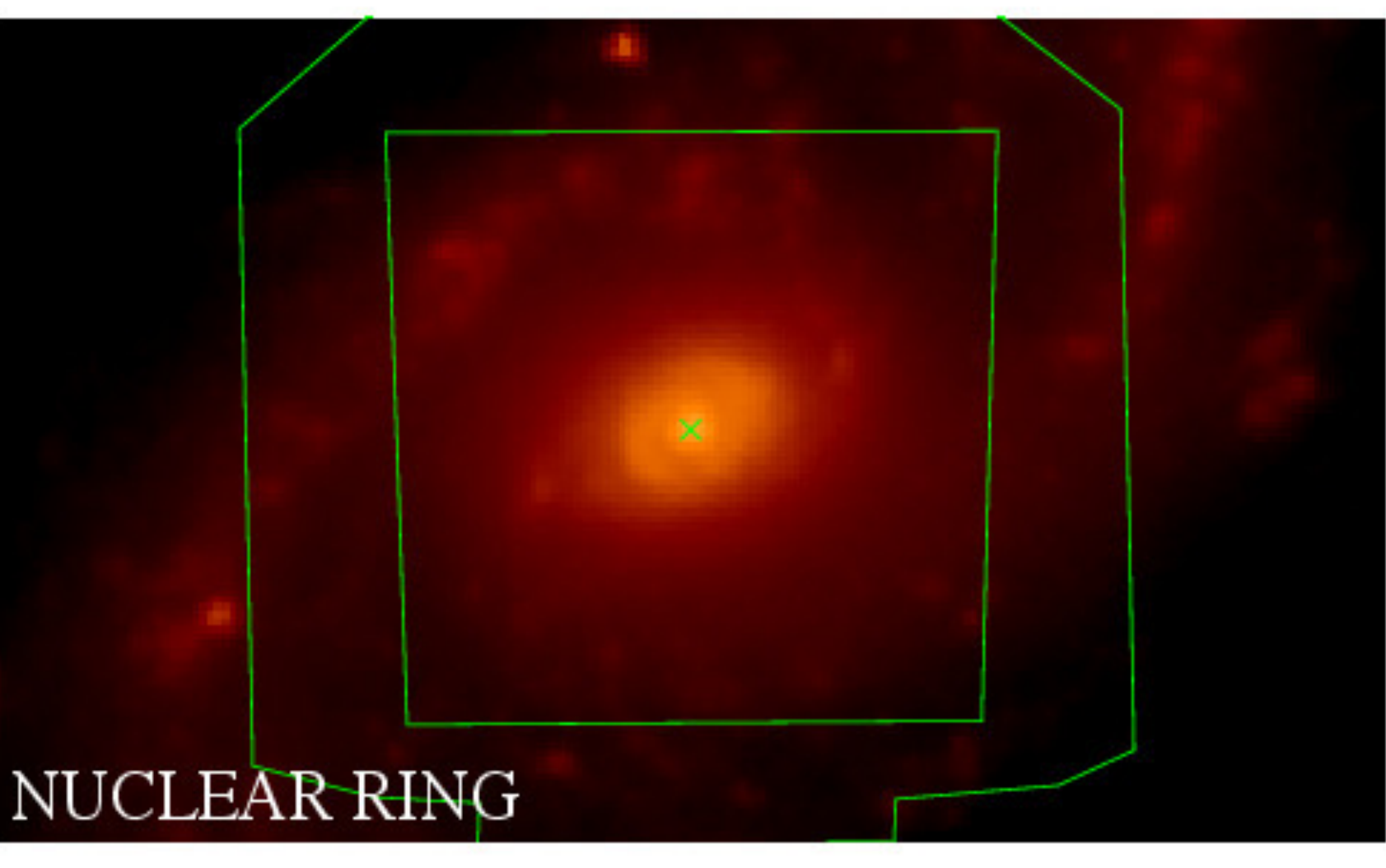}
	\includegraphics[trim=1cm 9.5cm 2.9cm 9.5cm, clip=true, width=0.5\columnwidth]{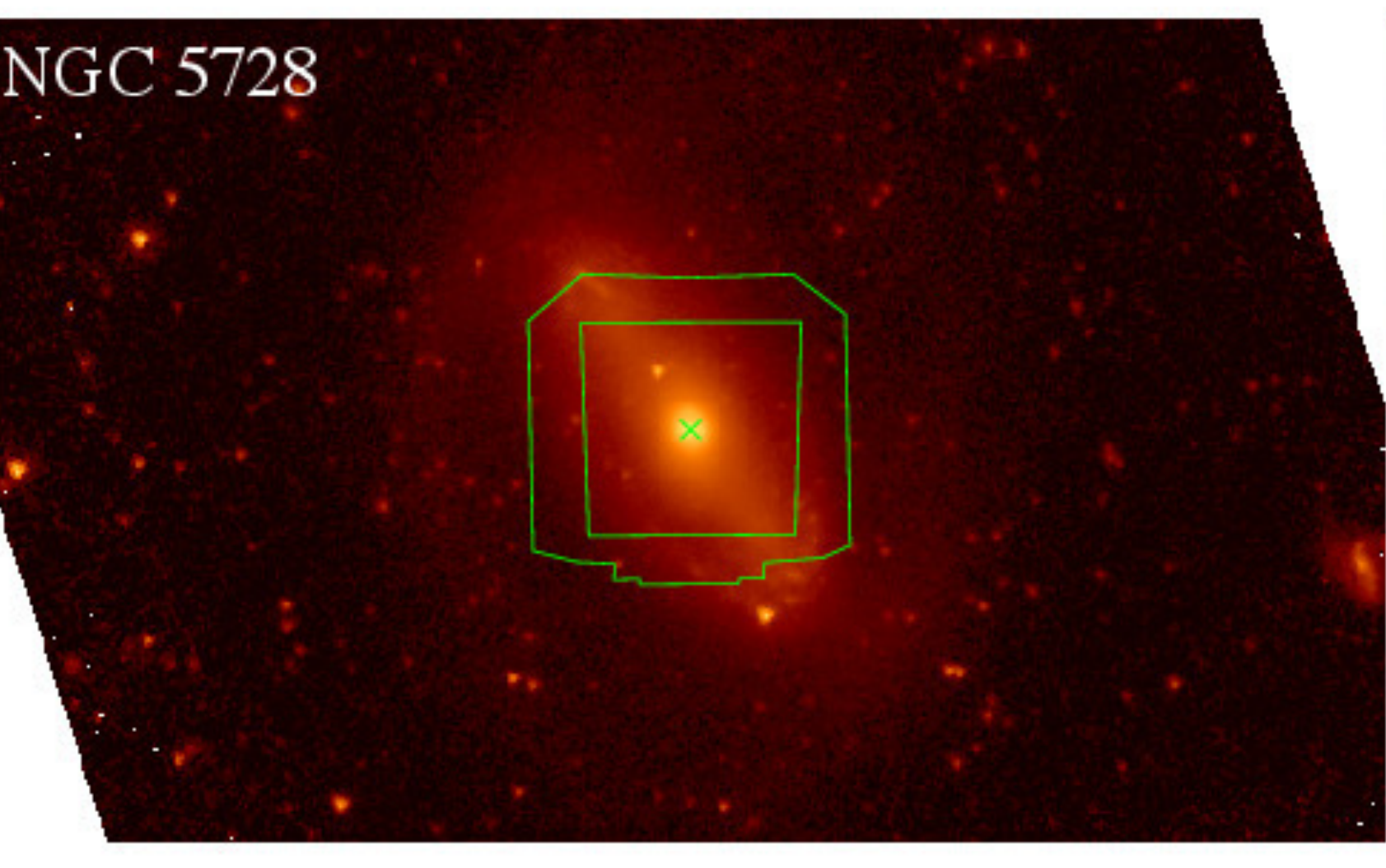}
	\includegraphics[trim=1cm 9.5cm 2.9cm 9.5cm, clip=true, width=0.5\columnwidth]{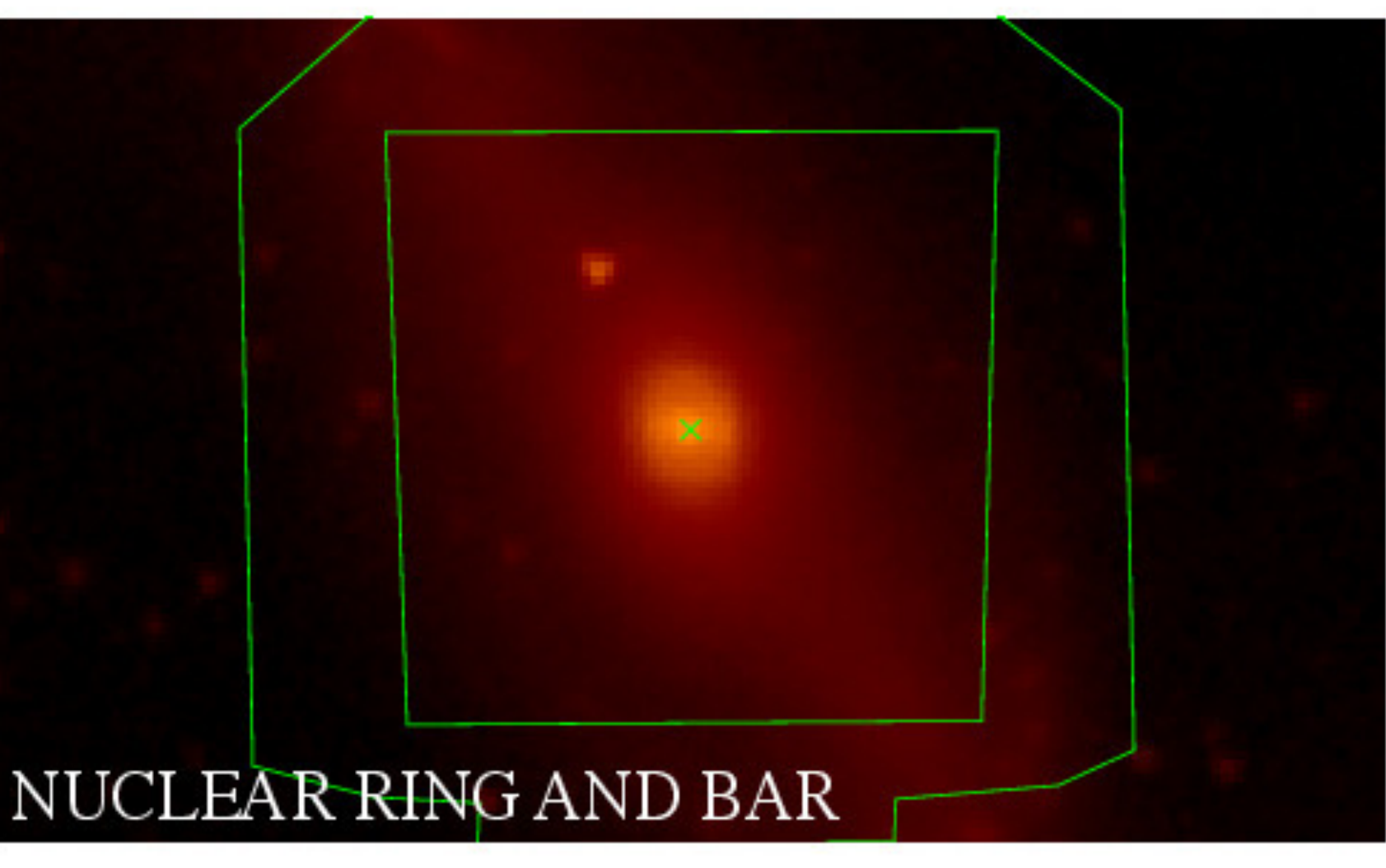}\\
\end{center}
    \caption{S$^4$G images at 3.6\,$\mu$m of all galaxies in the current TIMER sample. For every galaxy we show an image that includes most of the major disc, as well as an image that focuses on the area covered by our MUSE pointing. The MUSE field is represented by the inner trapezoid in green with $\approx1\arcmin$ on a side. The area between that and the outer polygon in green was used by the Slow-Guiding System during the exposures whenever it includes suitable point sources. North is up, east to the left.}
    \label{fig:S4G}
\end{figure*}

\begin{figure*}
\begin{center}
	\includegraphics[trim=1cm 9.5cm 2.9cm 9.5cm, clip=true, width=0.5\columnwidth]{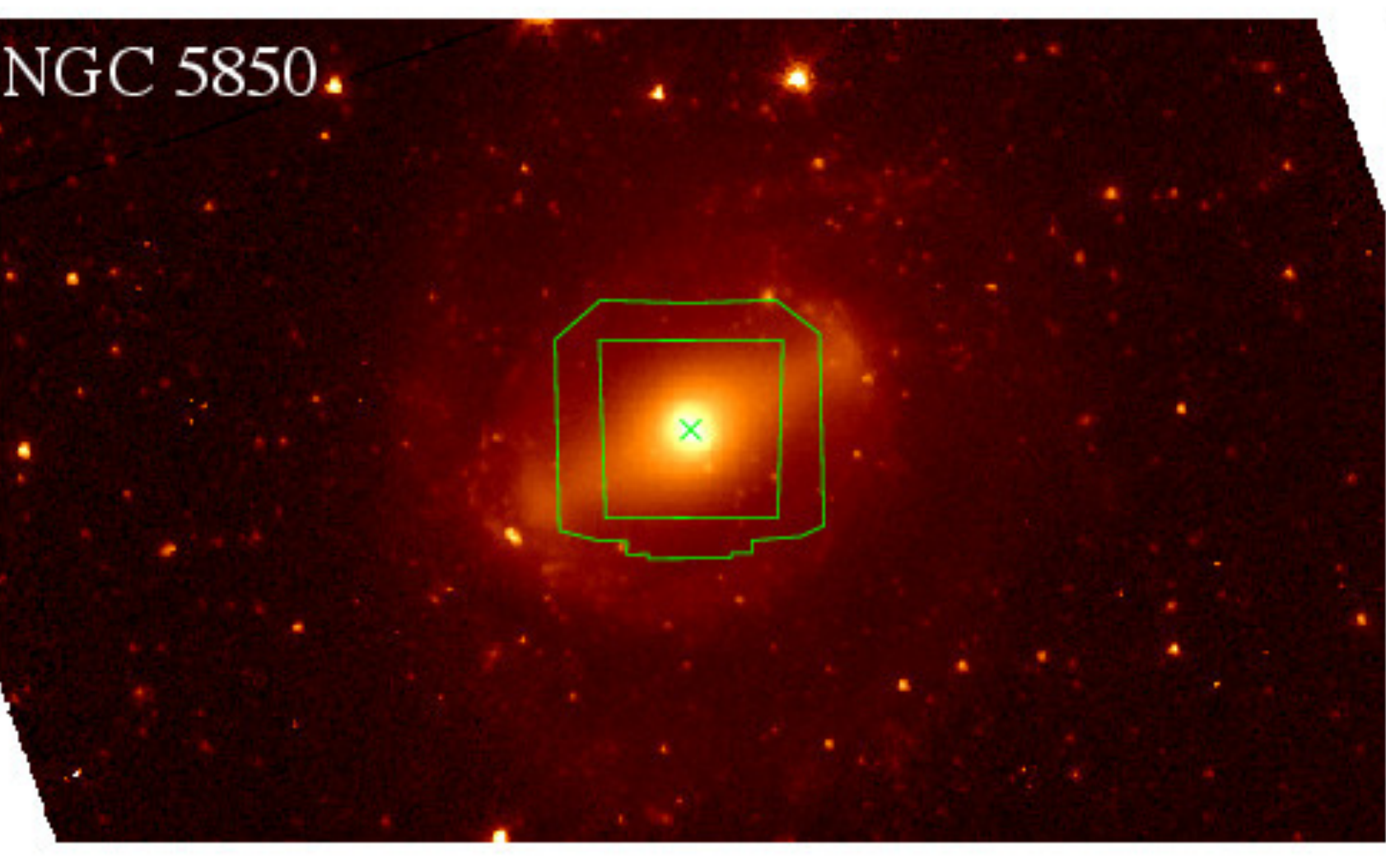}
	\includegraphics[trim=1cm 9.5cm 2.9cm 9.5cm, clip=true, width=0.5\columnwidth]{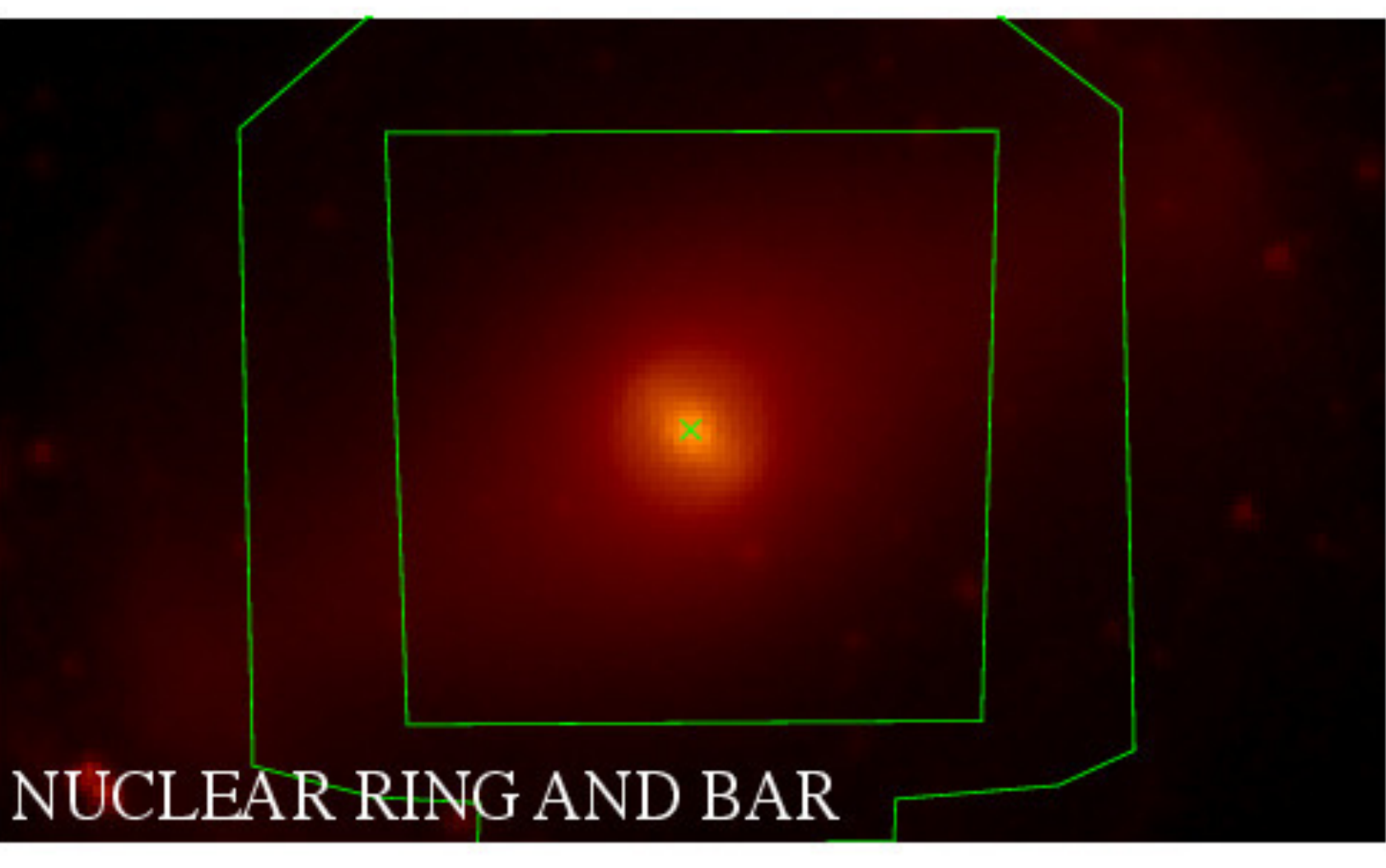}
	\includegraphics[trim=1cm 9.5cm 2.9cm 9.5cm, clip=true, width=0.5\columnwidth]{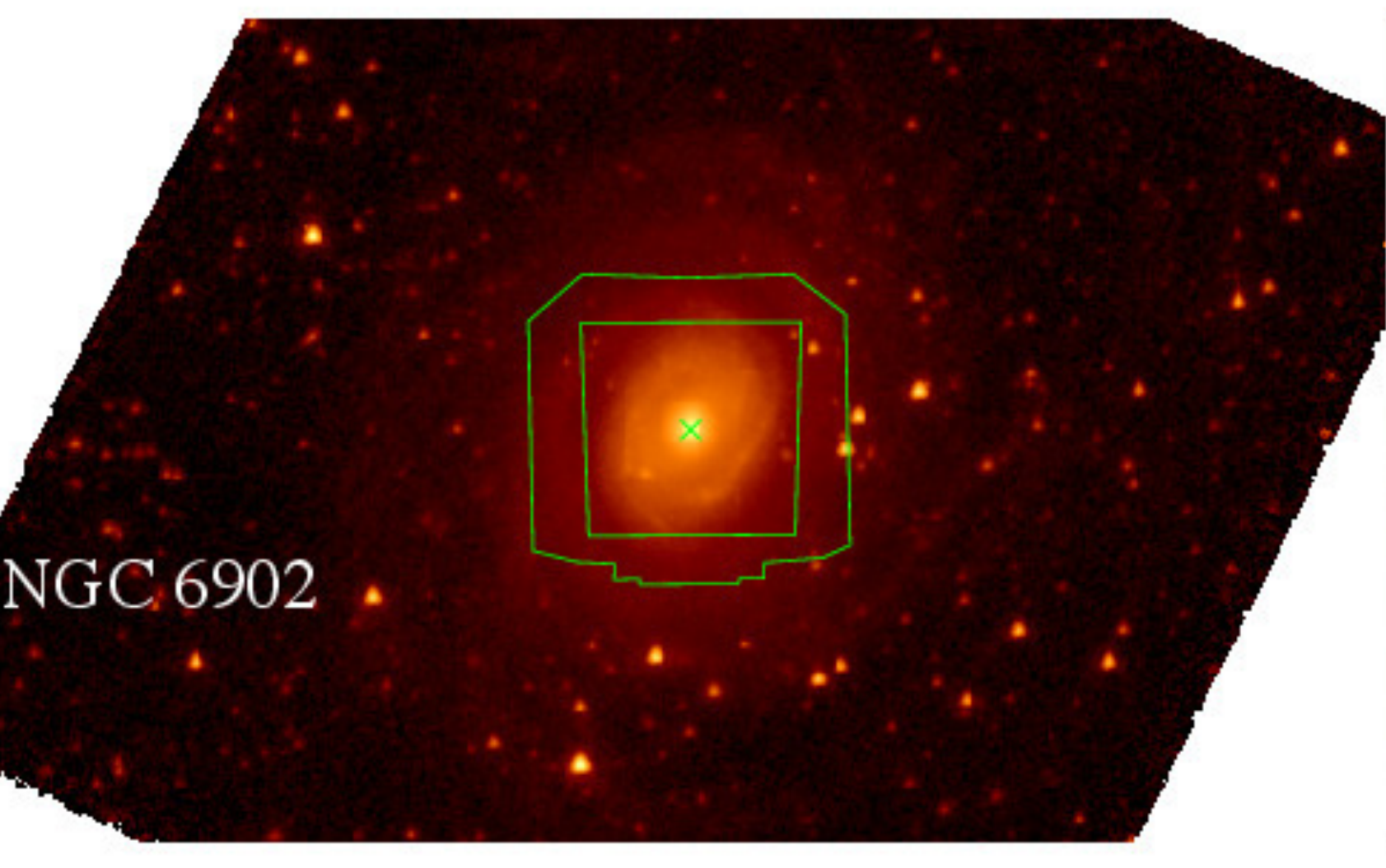}
	\includegraphics[trim=1cm 9.5cm 2.9cm 9.5cm, clip=true, width=0.5\columnwidth]{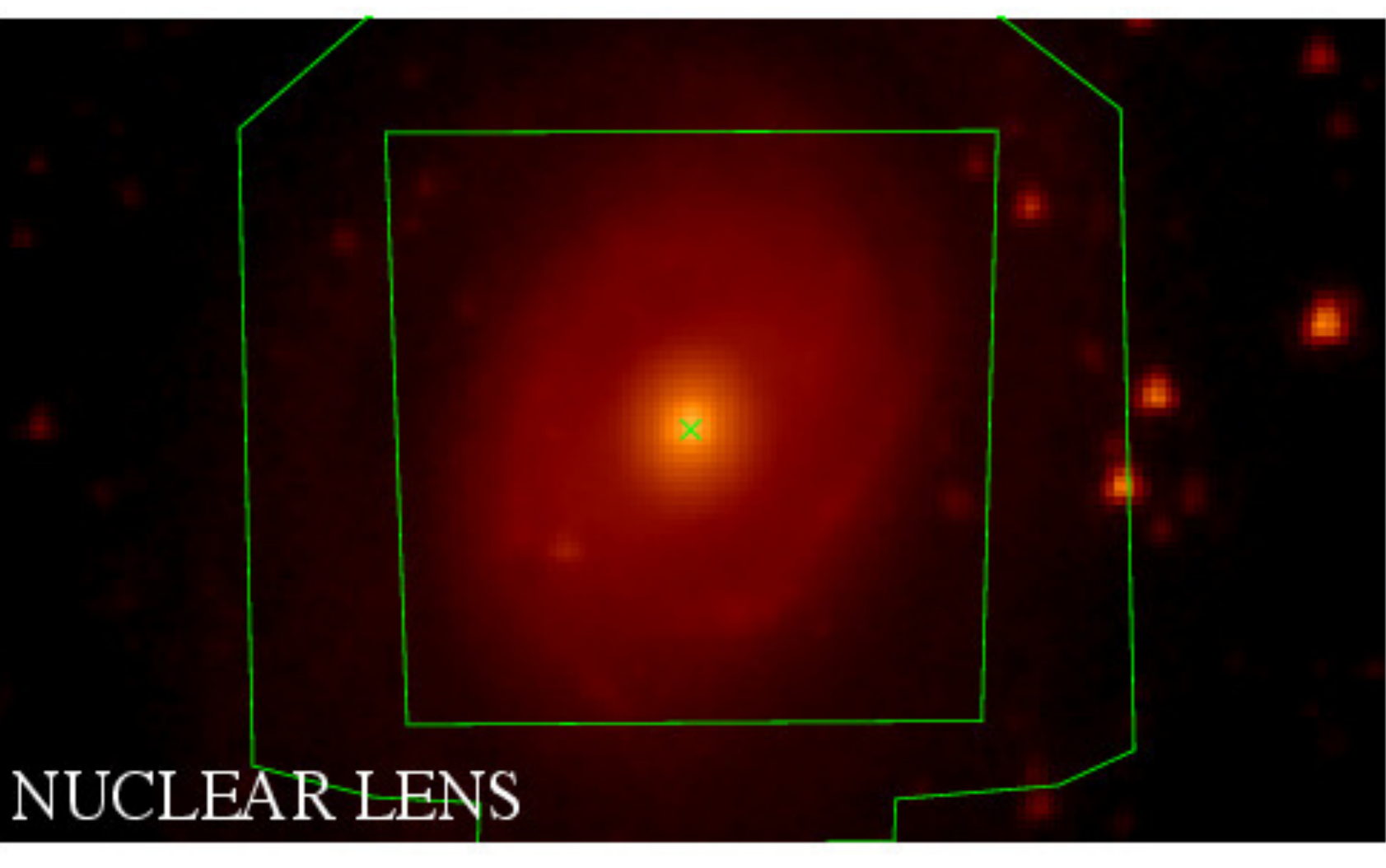}
	\includegraphics[trim=1cm 9.5cm 2.9cm 9.5cm, clip=true, width=0.5\columnwidth]{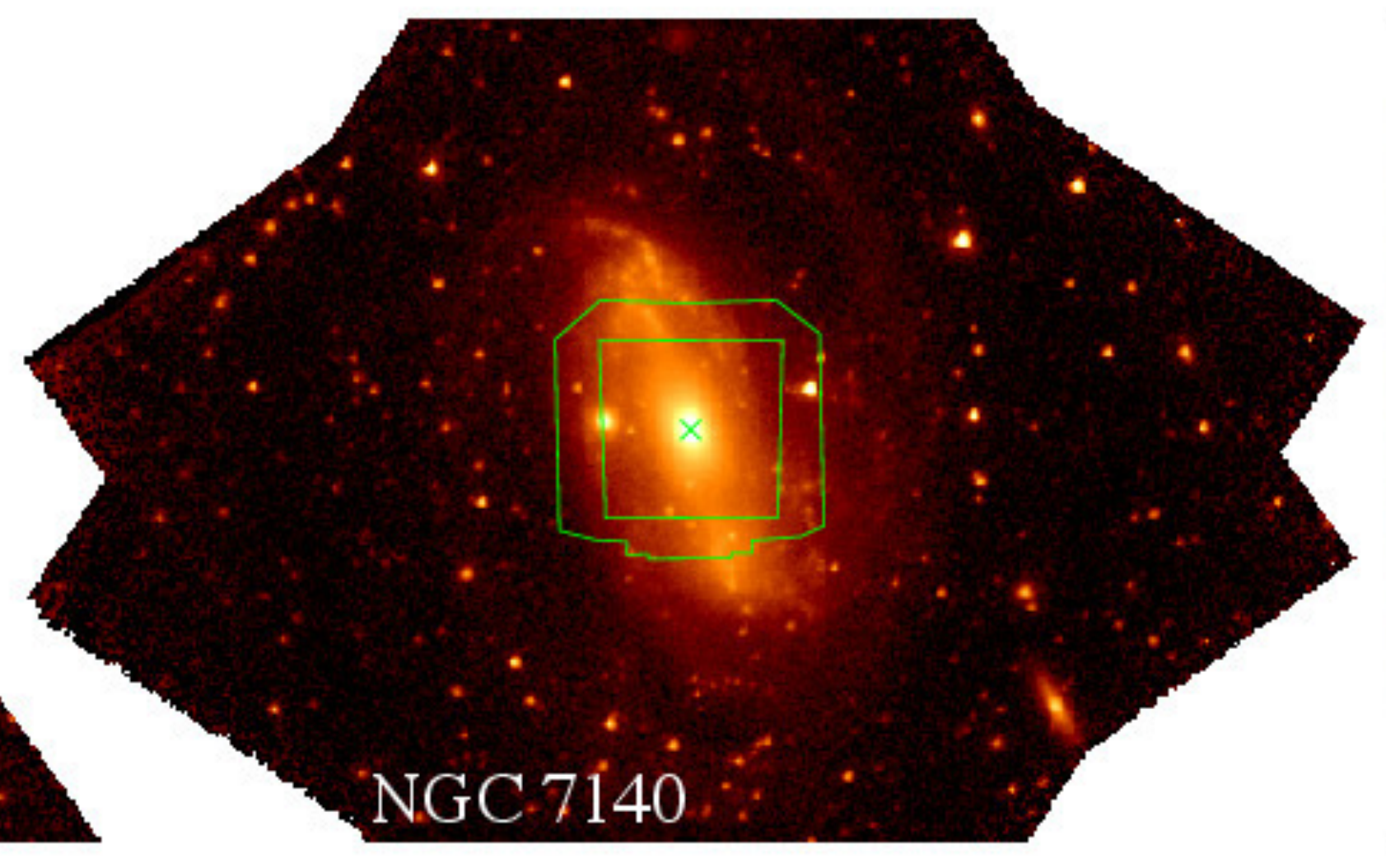}
	\includegraphics[trim=1cm 9.5cm 2.9cm 9.5cm, clip=true, width=0.5\columnwidth]{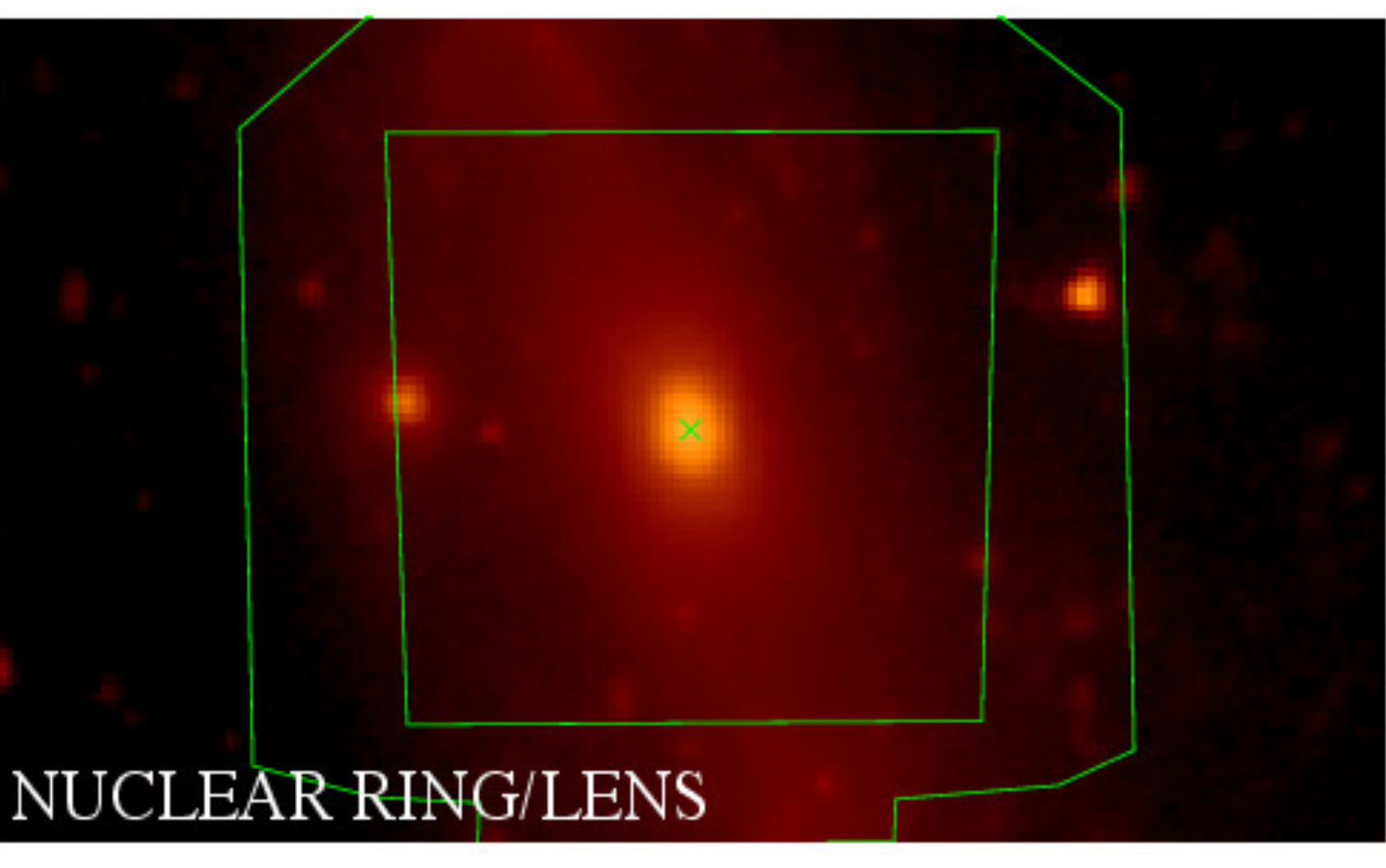}
	\includegraphics[trim=1cm 9.5cm 2.9cm 9.5cm, clip=true, width=0.5\columnwidth]{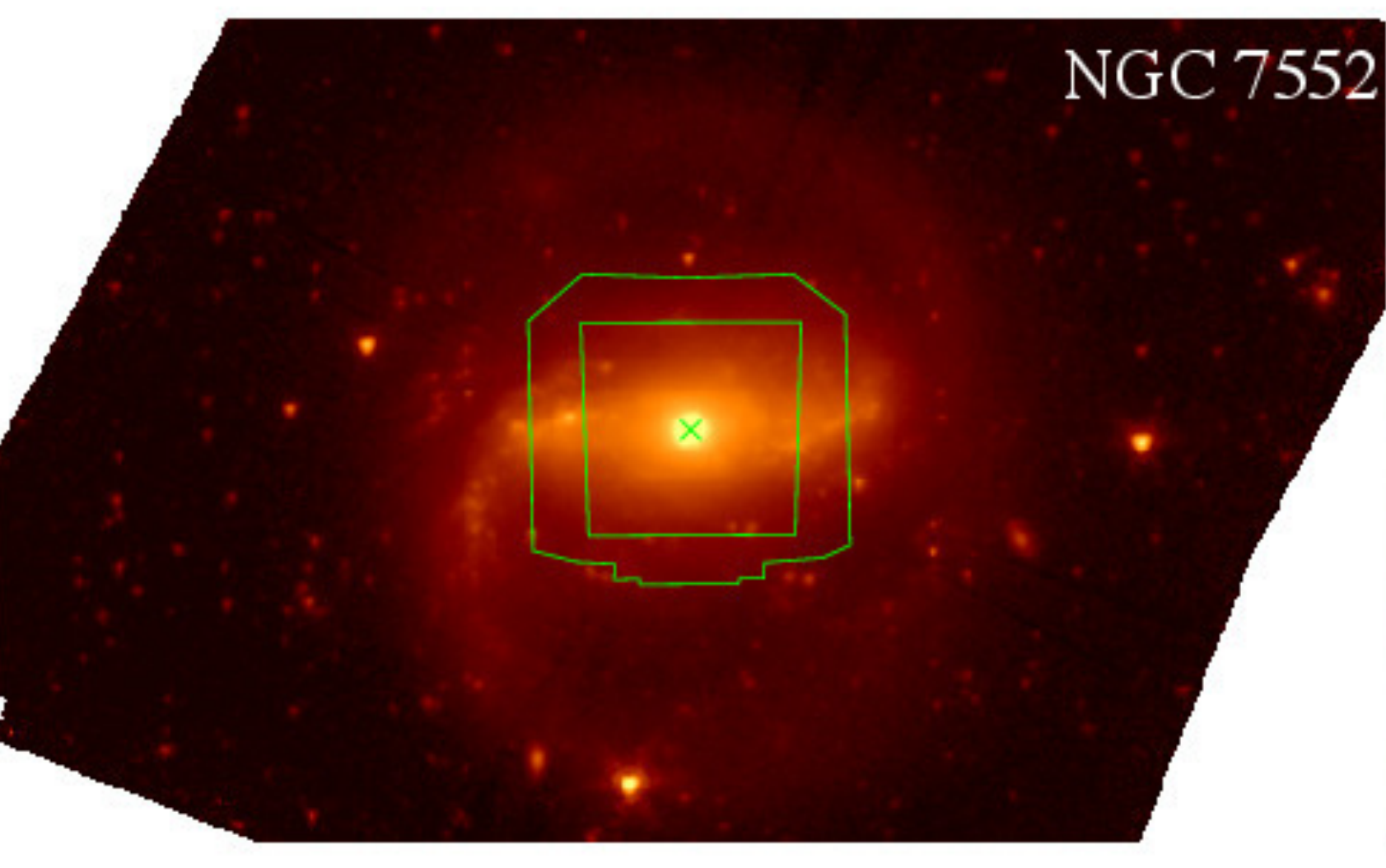}
	\includegraphics[trim=1cm 9.5cm 2.9cm 9.5cm, clip=true, width=0.5\columnwidth]{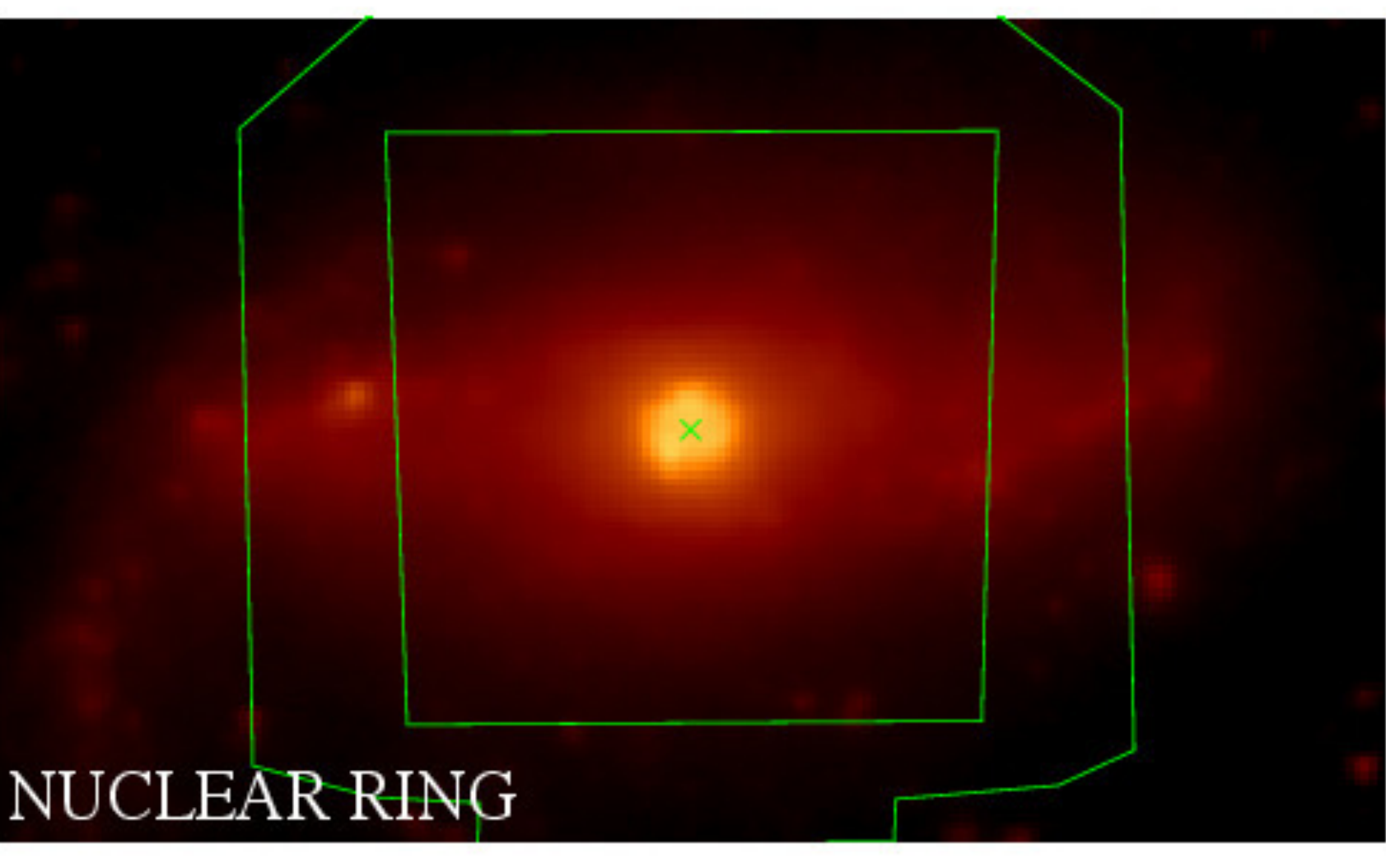}\\
	\includegraphics[trim=1cm 9.5cm 2.9cm 9.5cm, clip=true, width=0.5\columnwidth]{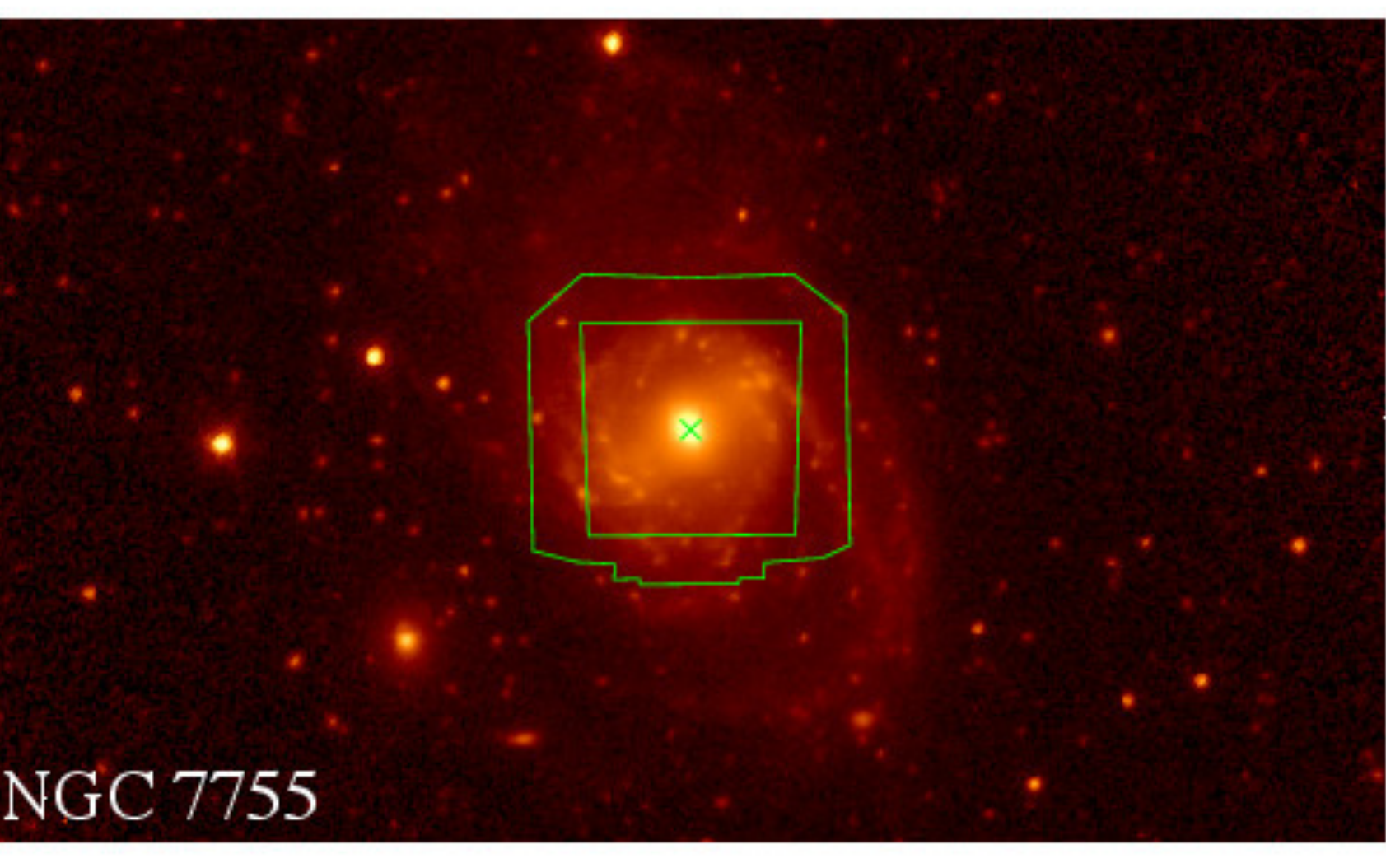}
	\includegraphics[trim=1cm 9.5cm 2.9cm 9.5cm, clip=true, width=0.5\columnwidth]{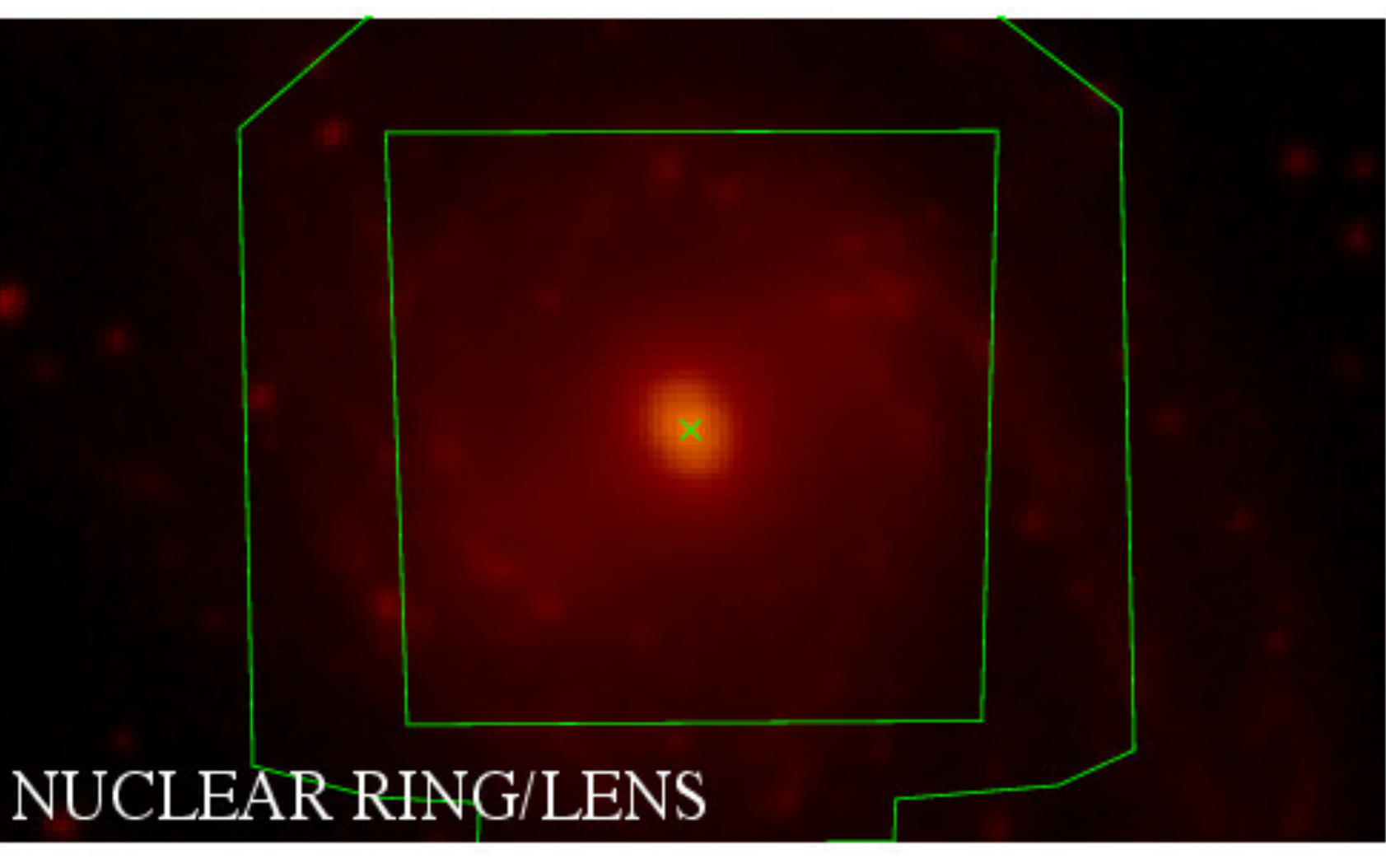}	
\end{center}
    \contcaption{}
\end{figure*}

\begin{table*}
	\centering
	\caption{Some fundamental properties of the full TIMER sample, as presented in \citet{MunSheGil13,MunSheReg15}. Column (1) gives the galaxy designation, and column (2) the morphological classification by \citet{ButSheAth15}. The inclination of the galaxy disc relative to the plane of the sky is given in column (3) and derived using $\cos i = b/a$, where $a$ and $b$ are respectively the semi-major and semi-minor axis of the 25.5\,AB\,mag\,arcsec$^{-2}$ isophote at 3.6\,$\mu$m. Column (4) shows the stellar mass derived within S$^4$G and column (5) the galaxy distance, calculated as the mean of all redshift-independent measurements presented in the NASA Extragalactic Database (NED; \url{http://ned.ipac.caltech.edu/}). The standard deviation of all those measurements is given in column (6). The central stellar velocity dispersion (reduced to a common standard aperture) and the corresponding uncertainty are given in columns (7) and (8), respectively. Those are taken from the Lyon Extragalactic Data Archive (LEDA; \url{http://leda.univ-lyon1.fr/}). Columns (9) and (10) show respectively the maximum rotation velocity corrected for inclination and the corresponding uncertainty, also taken from LEDA. This rotation velocity is calculated from 21-cm (HI) line widths and available rotation curves generally from H$\alpha$ emission. Finally, in column (11) we show the mass of neutral Hydrogen in the galaxy taken from LEDA and calculated from the flux in the 21-cm line.}
	\label{tab:sample}
	\begin{tabular}{llccccccccc}
		\hline
Galaxy & Type & $i$ & $M_\star$ & $d$ & SD($d$) & $\sigma_*$ & err($\sigma_*$) & $v_{\mathrm{rot}}$ & err($v_{\mathrm{rot}}$) & $M_{\mathrm{HI}}$ \\
\omit & \omit & $^\circ$ & 10$^{10}$ M$_\odot$ & Mpc & Mpc & km s$^{-1}$ & km s$^{-1}$ & km s$^{-1}$ & km s$^{-1}$ & $10^{10}$ M$_\odot$ \\
(1) & (2) & (3) & (4) & (5) & (6) & (7) & (8) & (9) & (10) & (11) \\
		\hline
  IC\,1438  & (R$_1$)SAB$_{\mathrm{a}}$(r$'$\underline{l},nl)0/a         & 24 & 3.1  & 33.8 & \omit & \omit & \omit & 154 & 10 & 0.12 \\ 
  NGC\,613  & SB(\underline{r}s,bl,nr)b                                  & 39 & 12.2 & 25.1 & 5.3   & 128   & 19    & 289 & 5  & 0.47 \\ 
  NGC\,1097 & (R$'$)SB(rs,bl,nr)ab pec                                   & 51 & 17.4 & 20.0 & 5.3   & \omit & \omit & 220 & 2  & 0.91 \\ 
  NGC\,1291 & (R)SAB(l,bl,nb)0$^+$                                       & 11 & 5.8  & 8.6  & \omit & 172   & 8     & 42  & 1  & 0.09 \\ 
  NGC\,1300 & (R$'$)SB(s,bl,nrl)b                                        & 26 & 3.8  & 18.0 & 2.8   & 199   & 59    & 144 & 2  & 0.22 \\ 
  NGC\,1365 & (R$'$)SB(r\underline{s},nr)bc                              & 52 & 9.5  & 17.9 & 2.7   & 151   & 20    & 198 & 3  & 0.76 \\ 
  NGC\,1433 & (R$'_1$)SB(r,p,nrl,nb)a                                    & 34 & 2.0  & 10.0 & 2.3   & \omit & \omit & 86  & 2  & 0.07 \\ 
  NGC\,1512 & (R\underline{L})SB(r,bl,nr)a                               & 43 & 2.2  & 12.3 & 1.6   & \omit & \omit & 118 & 1  & 0.34 \\                   
  NGC\,2903 & (R$'$)SB(rs,nr)b                                           & 61 & 4.6  & 9.1  & 1.6   & 101   & 7     & 187 & 4  & 0.30 \\ 
  NGC\,3351 & (R$'$)SB(r,bl,nr)a                                         & 42 & 3.1  & 10.1 & 1.0   & 99    & 20    & 149 & 5  & 0.09 \\ 
  NGC\,4303 & SAB(rs,nl)b\underline{c}                                   & 34 & 7.2  & 16.5 & 10.8  & 109   & 12    & 214 & 7  & 0.45 \\ 
  NGC\,4371 & (L)SB$_{\mathrm{a}}$(r,bl,nr)0$^{0/+}$                     & 59 & 3.2  & 16.8 & 1.6   & 137   & 3     & 162 & 8  & 0.08 \\ 
  NGC\,4394 & (\underline{R}L)SB(rs,bl,nl)0/a                            & 30 & 2.8  & 16.8 & \omit & 138   & 16    & 255 & 11 & 0.05 \\  
  NGC\,4643 & (L)SB(\underline{r}s,bl,nl)0$^{0/+}$                       & 44 & 10.7 & 25.7 & \omit & 163   & 8     & 171 & 7  & 0.03 \\ 
  NGC\,4981 & SA\underline{B}(s,nl)\underline{b}c                        & 54 & 2.8  & 24.7 & 2.3   & \omit & \omit & 163 & 5  & 0.35 \\ 
  NGC\,4984 & (R$'$R)SAB$_{\mathrm{a}}$(l,bl,nl)0/a                      & 53 & 4.9  & 21.3 & \omit & \omit & \omit & 125 & 12 & 0.03 \\ 
  NGC\,5236 & SAB(s,nr)c                                                 & 21 & 10.9 & 7.0  & 4.1   & \omit & \omit & 487 & 21 & 1.95 \\
  NGC\,5248 & (R$'$)SAB(s,nr)bc                                          & 41 & 4.7  & 16.9 & 4.8   & \omit & \omit & 145 & 2  & 0.40 \\
  NGC\,5728 & (R$_1$)SB(\underline{r}$'$l,bl,nr,nb)0/a                    & 44 & 7.1  & 30.6 & 6.4   & 210   & 15    & 208 & 8  & 0.19 \\
  NGC\,5850 & (R$'$)SB(r,bl,nr,nb)\underline{a}b                         & 39 & 6.0  & 23.1 & 7.6   & \omit & \omit & 117 & 3  & 0.11 \\
  NGC\,6902 & (R$'$)S\underline{A}B(\underline{r}s,nl)\underline{a}b     & 37 & 6.4  & 38.5 & 5.0   & \omit & \omit & 234 & 5  & 2.34 \\                                       
  NGC\,7140 & (R$'$)SA\underline{B}$_{\mathrm{x}}$(rs,nrl)a\underline{b} & 51 & 5.1  & 37.4 & \omit & \omit & \omit & 179 & 7  & 1.29 \\
  NGC\,7552 & (R$'_1$)SB(r\underline{s},bl,nr)a                          & 14 & 3.3  & 17.1 & 3.3   & 104   & 20    & 206 & 8  & 0.21 \\  
  NGC\,7755 & (R$'$)SAB(rs,nrl)\underline{b}c                            & 52 & 4.0  & 31.5 & 4.6   & \omit & \omit & 165 & 2  & 0.65 \\   
		\hline
	\end{tabular}
\end{table*}

The parent sample from which the TIMER sample was drawn is that of the Spitzer Survey of Stellar Structure in Galaxies (S$^4$G, \citealt{shereghin10}). The S$^4$G sample is a volume- ($d<40\,\rm{Mpc}$), magnitude- ($m_{B_{\mathrm{corr}}}<15.5$) and size-limited ($D_{25}>1\arcmin$) sample of 2352 galaxies. The galactic latitude is constrained to $|b|>30^\circ$, and the distance determined from neutral Hydrogen redshifts. Each galaxy was observed with the {\it Spitzer Space Telescope} Infrared Array Camera (IRAC) at 3.6 and 4.5\,$\mu$m and mapped to $1.5\times D_{25}$.

When designing the TIMER sample with the project goals in mind, we selected only galaxies with prominent bars, stellar masses above $10^{10}\,{\rm M}_\odot$, inclinations below $\approx60^\circ$, and inner structures built by bars. The prominence of the bar and presence of bar-built inner structures were assessed from the morphological classifications of \citet{ButSheAth15}, who used the S$^4$G images for their work. This means we selected galaxies in the SB family (strongly barred) and with any of the following nuclear varieties: nr (nuclear ring), nr$'$ (nuclear pseudo-ring), ns (nuclear spiral), nl (nuclear lens), nrl (nuclear ring-lens), nb (nuclear bar), nb$_{\rm{a}}$ (nuclear ansae-type bar), nd (nuclear disc) or np (nuclear pattern). Stellar masses and inclinations were taken from the S$^4$G high-level data products \citep{MunSheGil13,MunSheReg15}. An additional constraint in the selection of the sample is that of course all galaxies must be well observable from Paranal, so galaxies with declinations larger than $+25^\circ$ were not included.

After these steps, we visually inspected the S$^4$G images and removed from the TIMER sample galaxies that are overly disturbed morphologically or where projection effects are still too strong, both of which may prevent a straightforward interpretation of the data. During this process we also removed galaxies where the nuclear variety is not clearly distinguished. We also visually inspected all barred galaxies in S$^4$G with the nuclear varieties mentioned above but not classified as strongly barred by Buta et al., {\it i.e.}, in one of the following families: S\underline{A}B, SAB or SA\underline{B}. A number of these galaxies show spectacular nuclear varieties and were therefore included in the sample.

This selection process yielded the final TIMER sample of 24 galaxies. Table \ref{tab:sample} gives some fundamental properties of these galaxies, and Fig.~\ref{fig:S4G} shows their S$^4$G images and the corresponding fields covered by our MUSE observations. As can be seen in Table \ref{tab:sample}, one important characteristic of the TIMER sample is that it spans a wide range of stellar masses, covering nearly an order of magnitude, from $2.0\times10^{10}\,{\rm M}_\odot$ to $1.7\times10^{11}\,{\rm M}_\odot$. This is important because, as mentioned above, one of the major goals of the project is to test the scenario whereby more massive discs settle first. These masses are calculated from the 3.6 and 4.5\,$\mu$m absolute magnitudes derived with the S$^4$G data and are expected to be accurate, with a typical uncertainty of $\sim15\%$ \citep[see \citealt{MunSheReg15} and][for details]{QueMeiSch15}.

Of the 24 galaxies selected, NGC\,4371 was the subject of our MUSE Science Verification programme to demonstrate the TIMER method for estimating the age of the bar and the cosmic epoch of disc settling \citep[see][]{GadSeiSan15}. NGC\,1365 and NGC\,7552 were targeted with MUSE previously, and thus archival data (as described below) suitable to pursue the TIMER goals were already available (from PIs Marconi and Carollo, respectively). We were then left with 21 galaxies to observe. We will next describe the work corresponding to our observations of 18 of those galaxies. The remaining three galaxies (NGC\,1512, NGC\,2903 and NGC\,4394) will be the subject of a future paper, since their observations with MUSE are yet to be executed. The data reduction, derivation of high-level data products and analyses presented elsewhere in this paper correspond to the 21 galaxies for which there is already MUSE data available.

\section{Observations and Data Reduction}
\label{sec:obs}

All observations were carried out with MUSE in ESO Period 97 from March to October 2016, except NGC\,1365 and NGC\,7552, for which we used archival data from ESO programmes 094.B-0321(A), PI: Marconi, and 095.B-0532(A), PI: Carollo, respectively. MUSE covers an almost square $1\arcmin\times1\arcmin$ field of view with a contiguous sampling of $0.2\arcsec\times0.2\arcsec$, which corresponds to a massive dataset of about 90\,000 spectra per pointing. We used the nominal instrument setup that yields a spectral coverage from $4750\,\AA$ to $9350\,\AA$ at a mean resolution of about $2.65\,\AA$ (FWHM) and a sampling of $1.25\,\AA$ per pixel. For most galaxies, the image quality delivered, as measured with the VLT guide star at the $V$ band, was typically around $0.8-0.9\arcsec$ (FWHM), including the two archival galaxies, with the exception of a few galaxies observed at even better conditions and NGC\,1291 observed with an image quality of about $1.1\arcsec$.  At the average distance of the galaxies in our sample $1\arcsec$ corresponds to $\approx100\,\rm{pc}$. Most observations were performed with clear sky transparency but some were affected by the presence of thin cirrus. So in most of our observations the rms of the flux variation was 5 per cent or less, as measured during the night by the VLT Astronomical Site Monitor; except during the observations of NGC\,4303, 4643, 4984, 5236, 6902 and NGC\,7140, when the rms is reported to have reached up to 15 per cent. Nevertheless, the presence of thin cirrus may limit the accuracy of the background subtraction.

As shown in Fig.~\ref{fig:S4G} we targeted the central $1\arcmin\times1\arcmin$ region of all galaxies. The centre of our field for NGC\,1291 is slightly displaced from the galaxy centre (by a few arcseconds) to allow the inclusion of a point source within the Slow-Guiding System area. For each galaxy, the observations were distributed in two 1-hour Observation Blocks (OBs) with a total integration time of 3\,840\,s on source. (For NGC\,1365 the total integration time on source is 4\,000\,s, whereas for NGC\,7552 it is 5\,913\,s.) Because the galaxies fill the MUSE field, we split the exposures in each OB into four exposures of 480\,s each on source, and monitored the sky background by observing a blank sky field for 300\,s following a OBJ-SKY-OBJ-OBJ-SKY-OBJ pattern. To be able to reduce the effects of bad pixels and flat-fielding uncertainties, we applied a small dither of $0.5\arcsec-3\arcsec$ of the field centre after each exposure, and rotated the entire MUSE field by $90^\circ$ such that every galaxy was observed twice at four different position angles.

Some OBs were repeated usually due to changes in the atmospheric conditions. In such cases, when there were exposures from such observations that are nevertheless within our data quality constraints, such exposures were included in the production of the final data cube, and the total integration time on source is thus larger than what is mentioned above. Conversely, due to difficulties in the data reduction (flat-fielding) corresponding to NGC\,1433, the total integration time on source for that galaxy is less than the typical for our observations. In addition, because the image quality between individual exposures of NGC\,613 and NGC\,1300 varies significantly, we obtained for those two galaxies only not only the nominal data cube that combines all suitable exposures, but in addition an alternative data cube that combines only exposures obtained with the best image quality. Using the VLT guide star as above, the image quality in the nominal data cube of those galaxies is $\approx1\arcsec$ whereas it is $\approx0.7-0.8\arcsec$ in the alternative data cubes.

Frames were taken to correct the exposures for bias, to flat-field the exposures, and to perform illumination correction. Wavelength calibration is achieved through a set of different arc lamp frames, and the exposures were flux-calibrated through the observation of a spectrophotometric standard star; the latter was also used to remove telluric features. Finally, the exposures were also finely registered astrometrically, so that the point spread function of the combined cube is similar to that in individual exposures. All these frames were observed as part of the MUSE standard calibration plan, and we applied the calibration frames taken closest in time to the execution of our OBs.

The MUSE pipeline (version 1.6) was used to reduce the dataset \citep{WeiStrUrr12} and we briefly outline the reduction process here. The process is split up into three steps. In the first one, each science frame is calibrated separately to take out instrumental effects. This consists of subtracting the bias level, flat-fielding the images (including illumination correction), extracting the spectra from all slices, and performing the wavelength calibration, after mapping the line spread function. The calibrated science frames are combined into the final data cube in the second step. This includes the process of flux calibration, removal of telluric features, background subtraction adopting a model sky spectrum computed from the sky field, astrometric registration, correction for differential atmospheric refraction, and resampling of the data cube based on the drizzle algorithm \citep[see][]{WeiStrUrr12}. Since the galaxies cover the MUSE field, leaving few or none suitable point sources in the field, often the astrometric registration had to be done manually, in a few cases using only the galaxy centre as reference. To subtract the background from the two consecutive object exposures taken between two sky exposures in each OB, we typically used a sky field obtained after interpolating in time the sky exposures to account for variations in the background and best reproduce the background at the time of the object exposure. Finally, in the last step of our data reduction process, after employing the MUSE pipeline, we have also implemented a method to improve the removal of the sky background from the data cube using principal component analysis. This is similar to the implementation of the ZAP tool, described in \citet{SotLilBac16}. With respect to the two galaxies with archival data, to produce our corresponding data cubes we used only raw data and associated raw calibration frames from the ESO archive, and ran them through our data reduction process as for all other TIMER galaxies. Typically, the SNR per pixel at the central spaxels of our fully-reduced data cubes is approximately 100.

\begin{figure*}
\hskip-9.2cm {\sc colour composites} \hskip1.9cm {\sc colour maps}\\
\begin{center}
	\includegraphics[width=0.5\columnwidth]{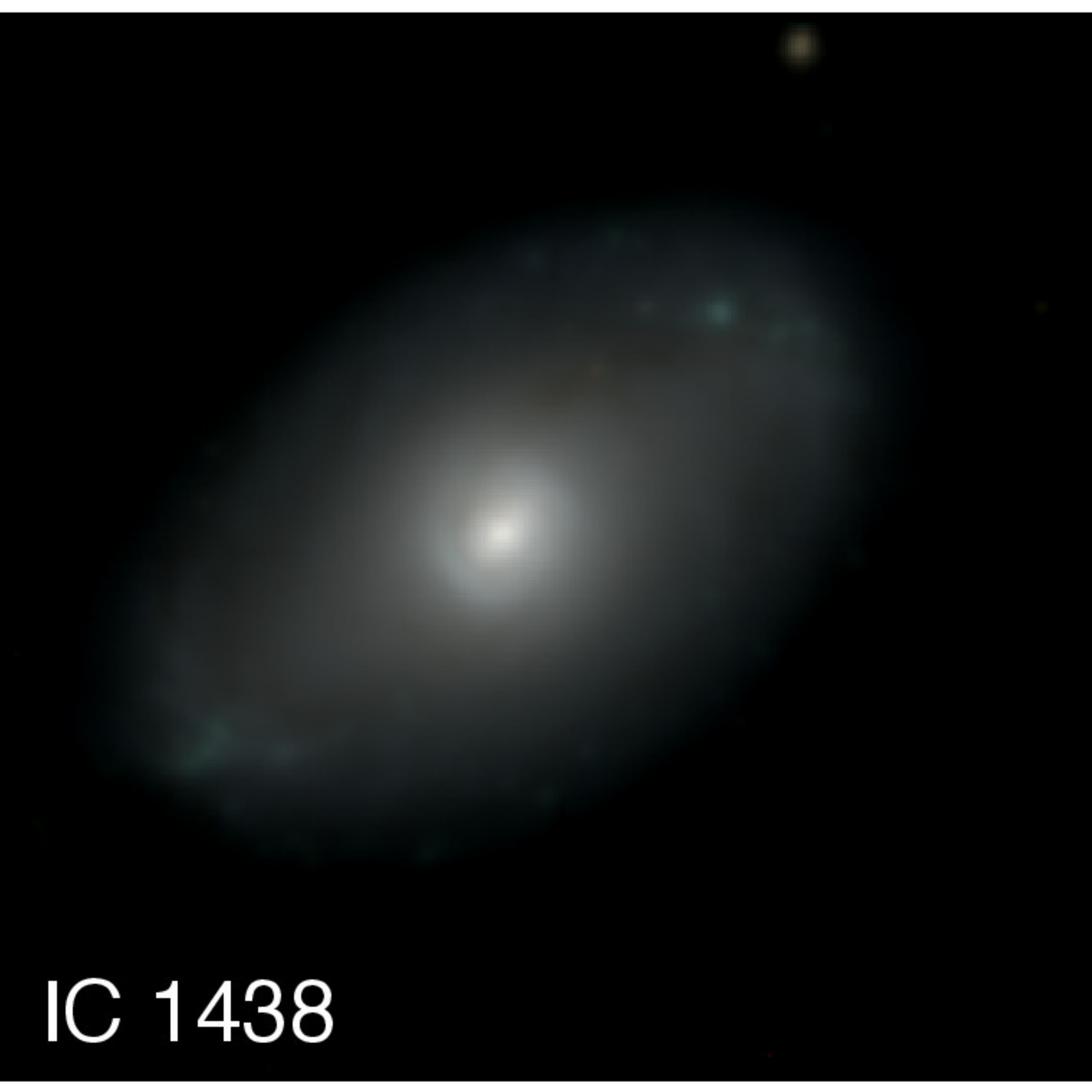}
	\includegraphics[width=0.5\columnwidth]{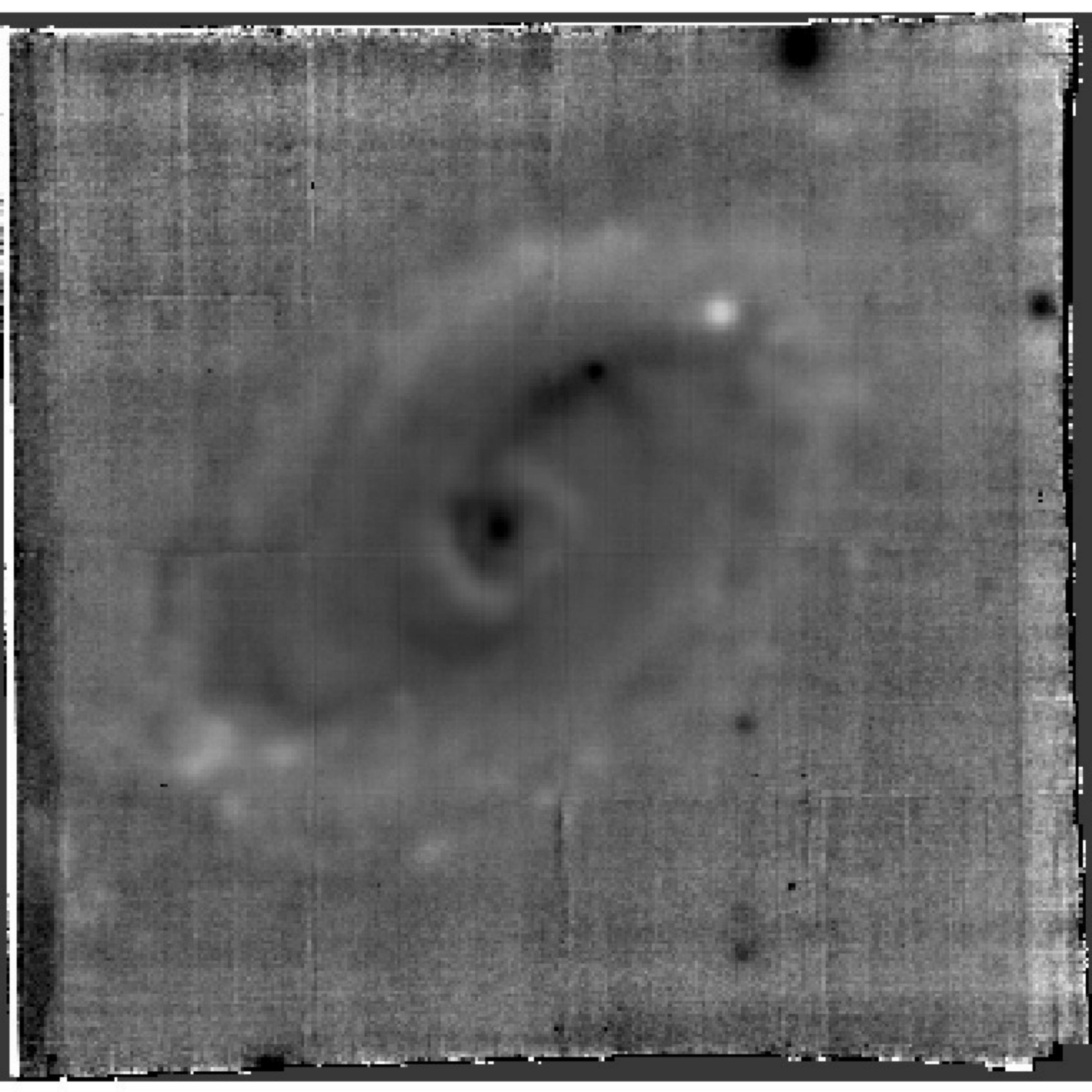}
	\includegraphics[width=0.5\columnwidth]{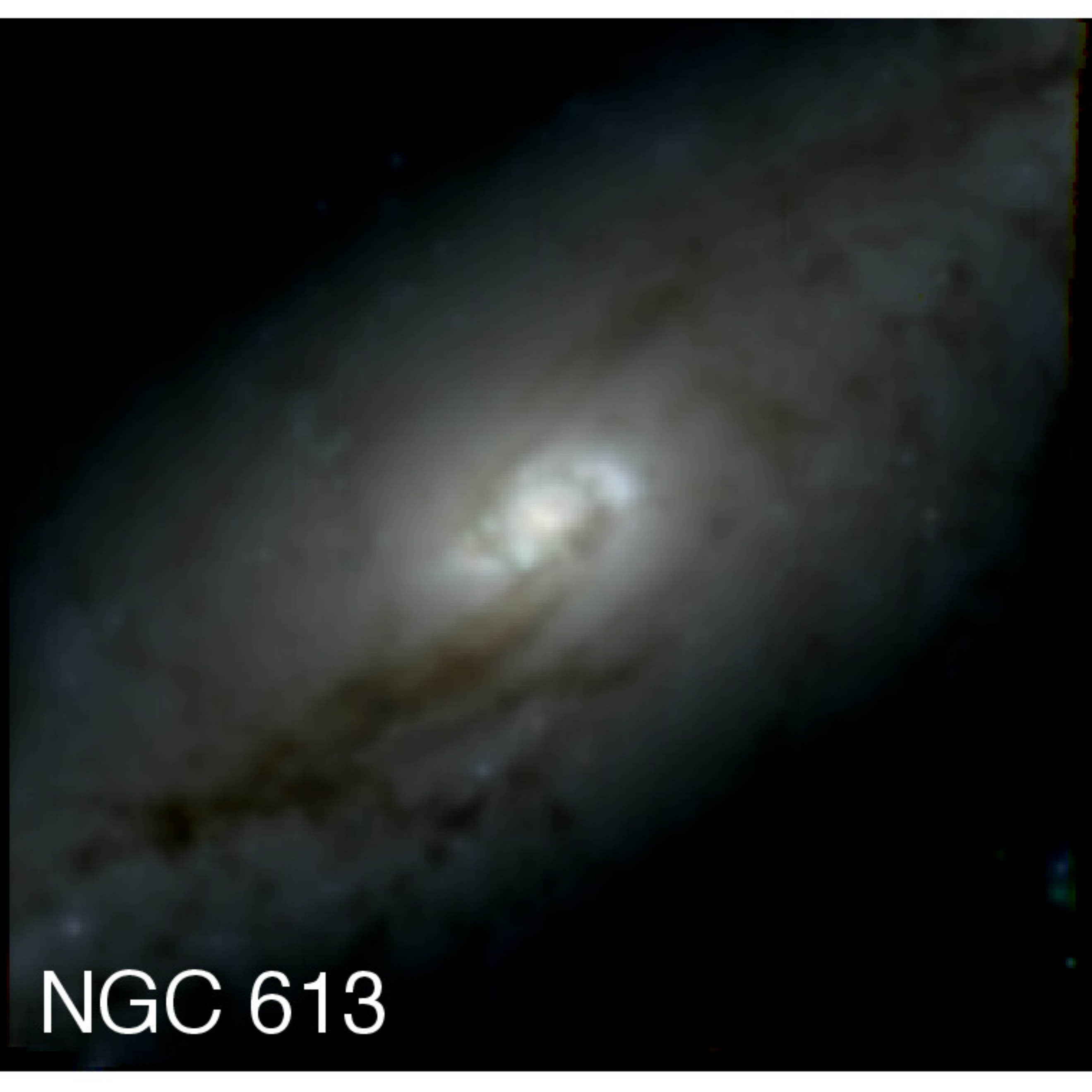}
	\includegraphics[width=0.5\columnwidth]{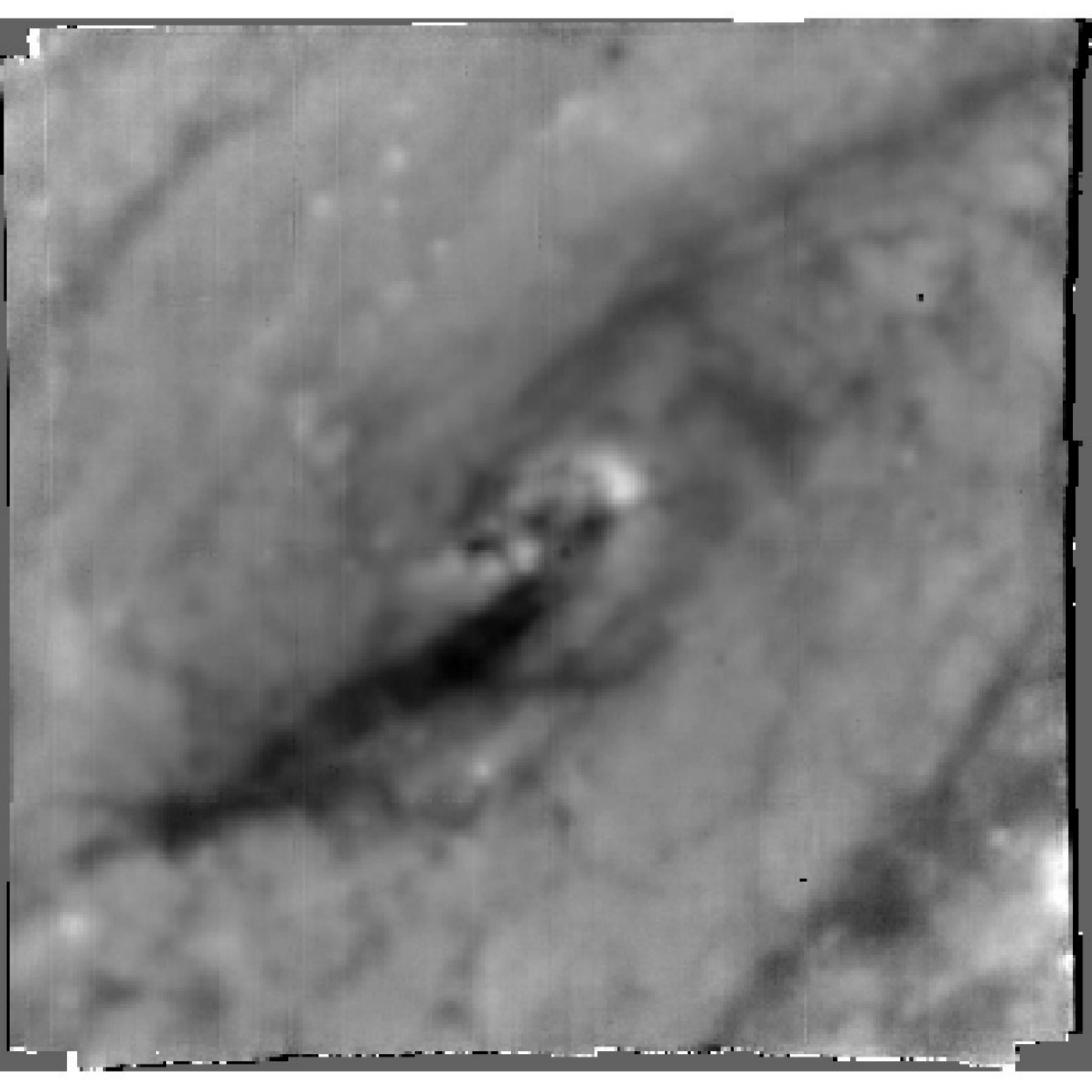}
	\includegraphics[width=0.5\columnwidth]{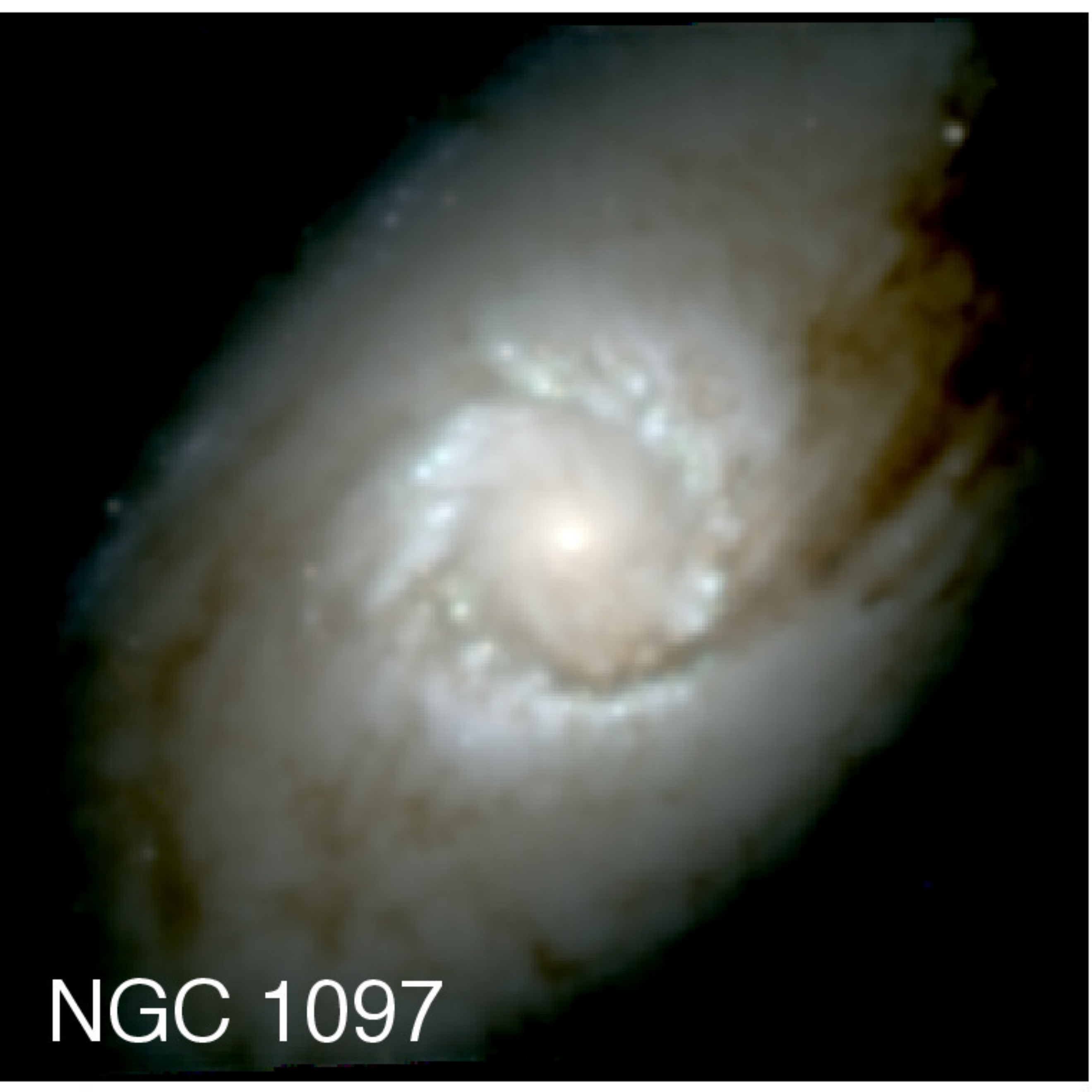}
	\includegraphics[width=0.5\columnwidth]{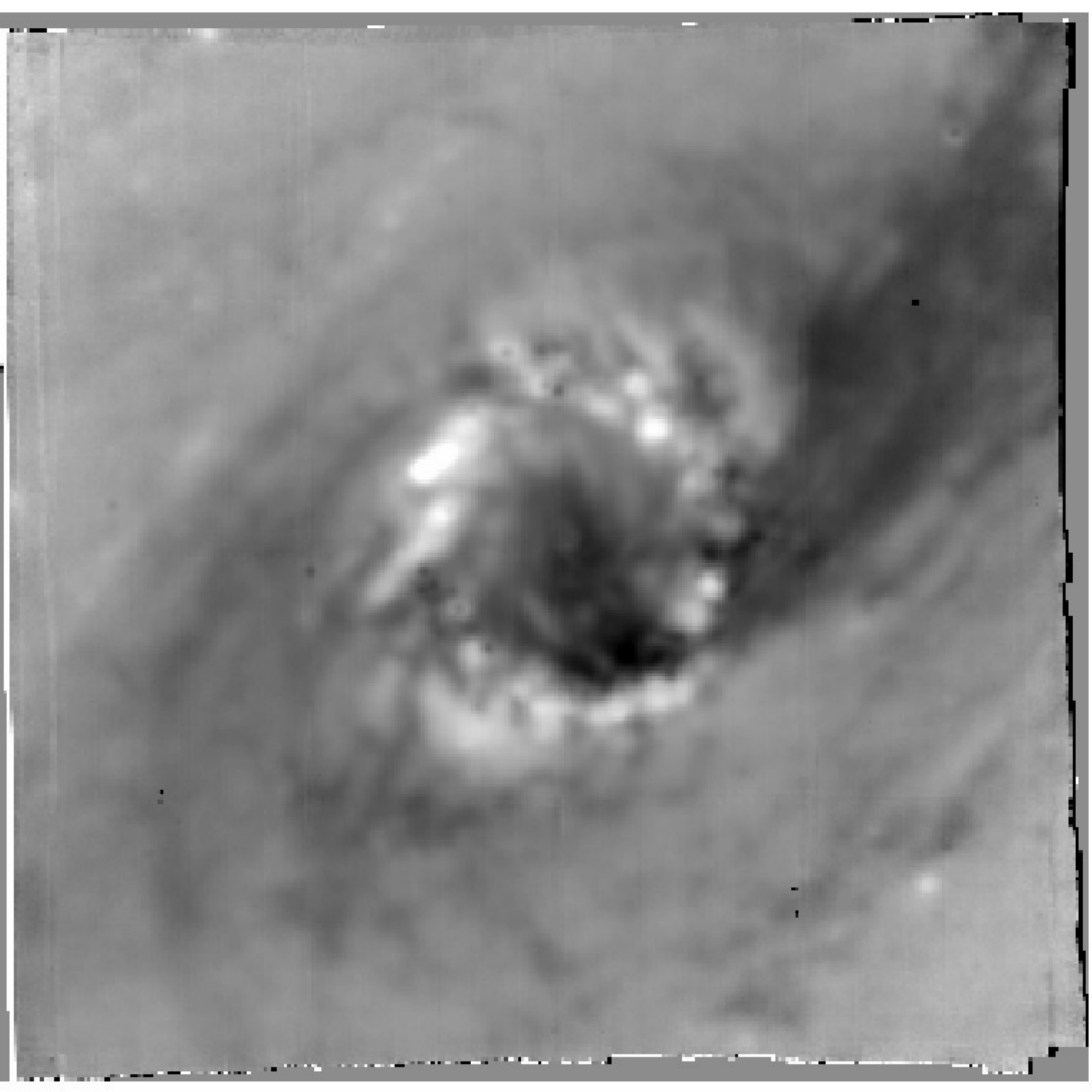}
	\includegraphics[width=0.5\columnwidth]{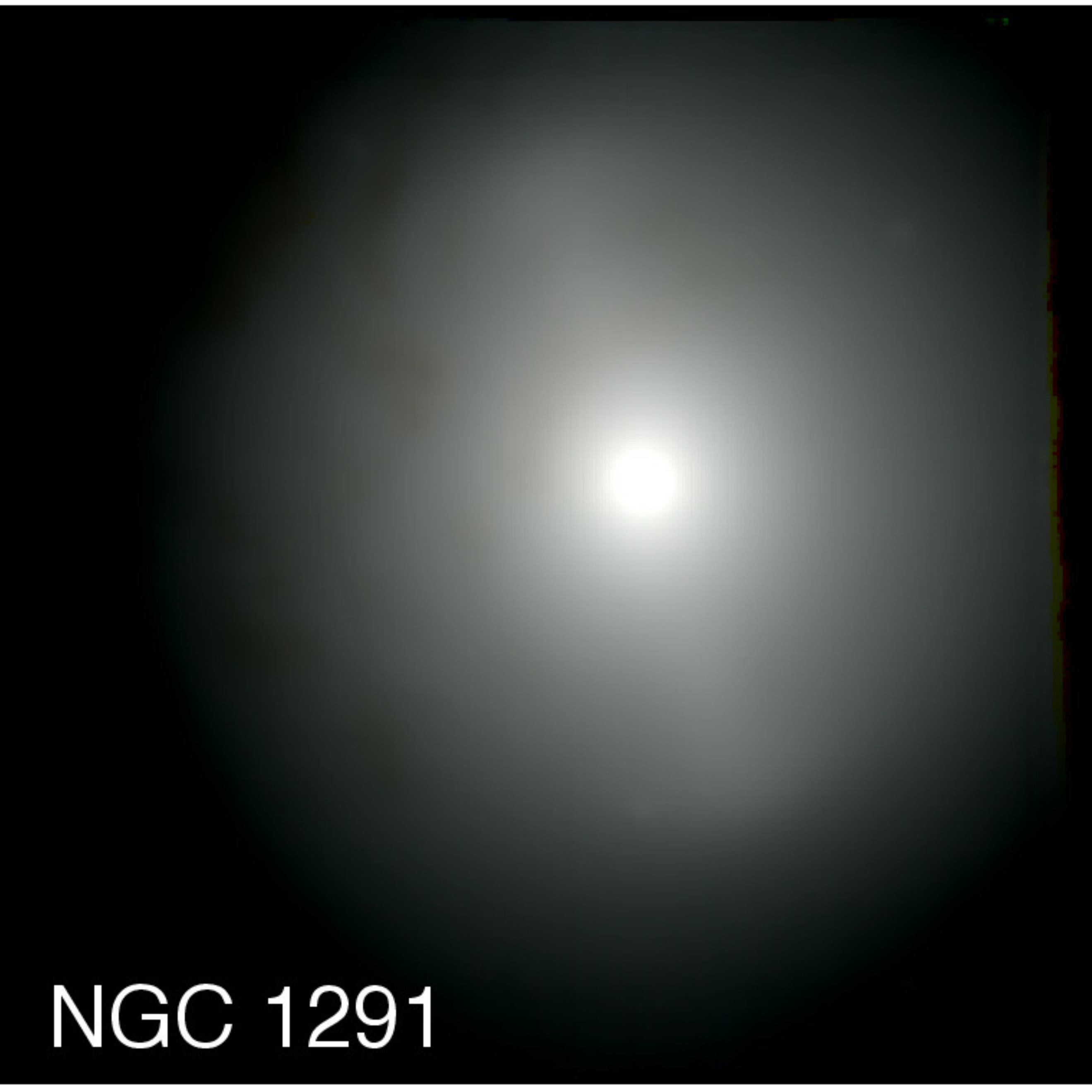}
	\includegraphics[width=0.5\columnwidth]{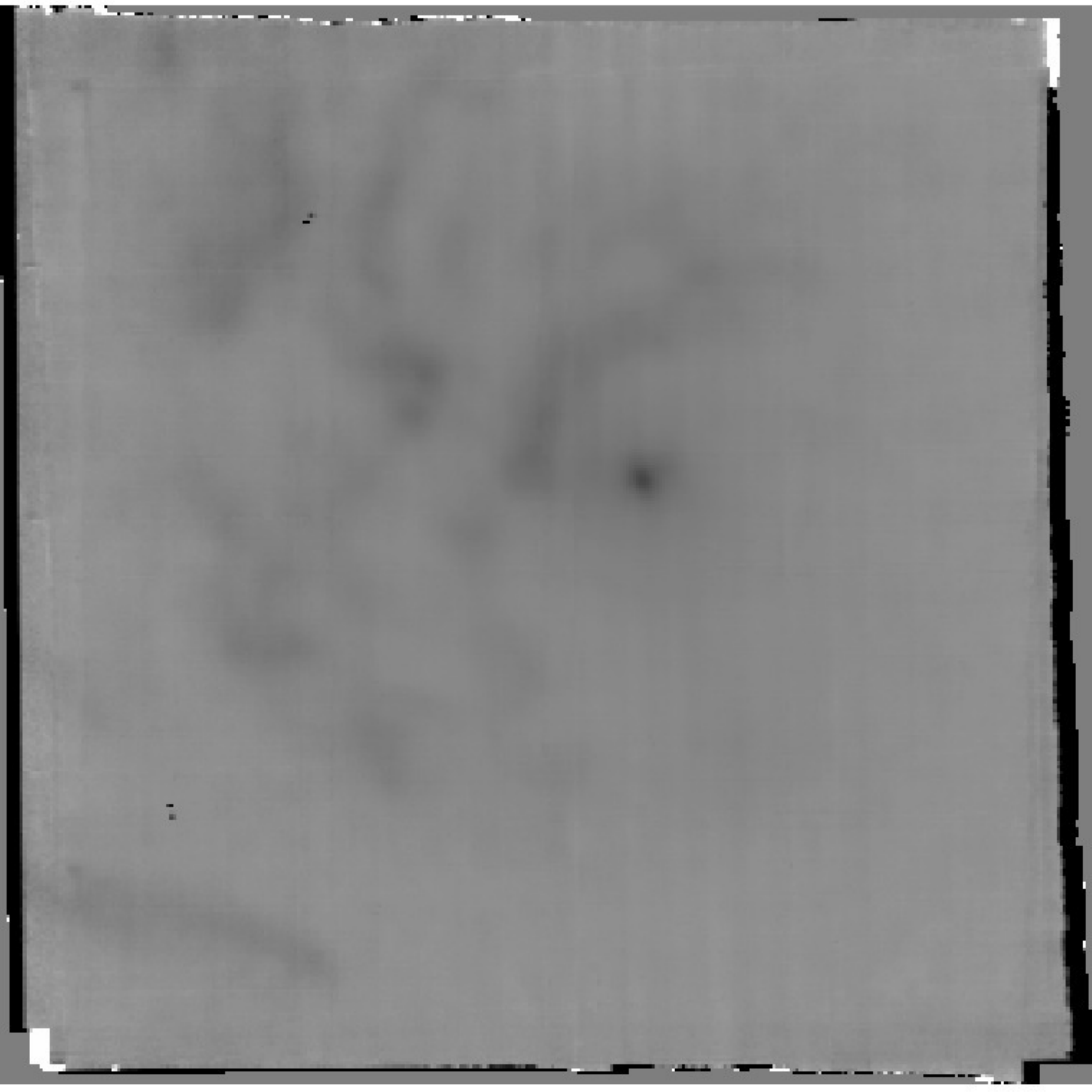}
	\includegraphics[width=0.5\columnwidth]{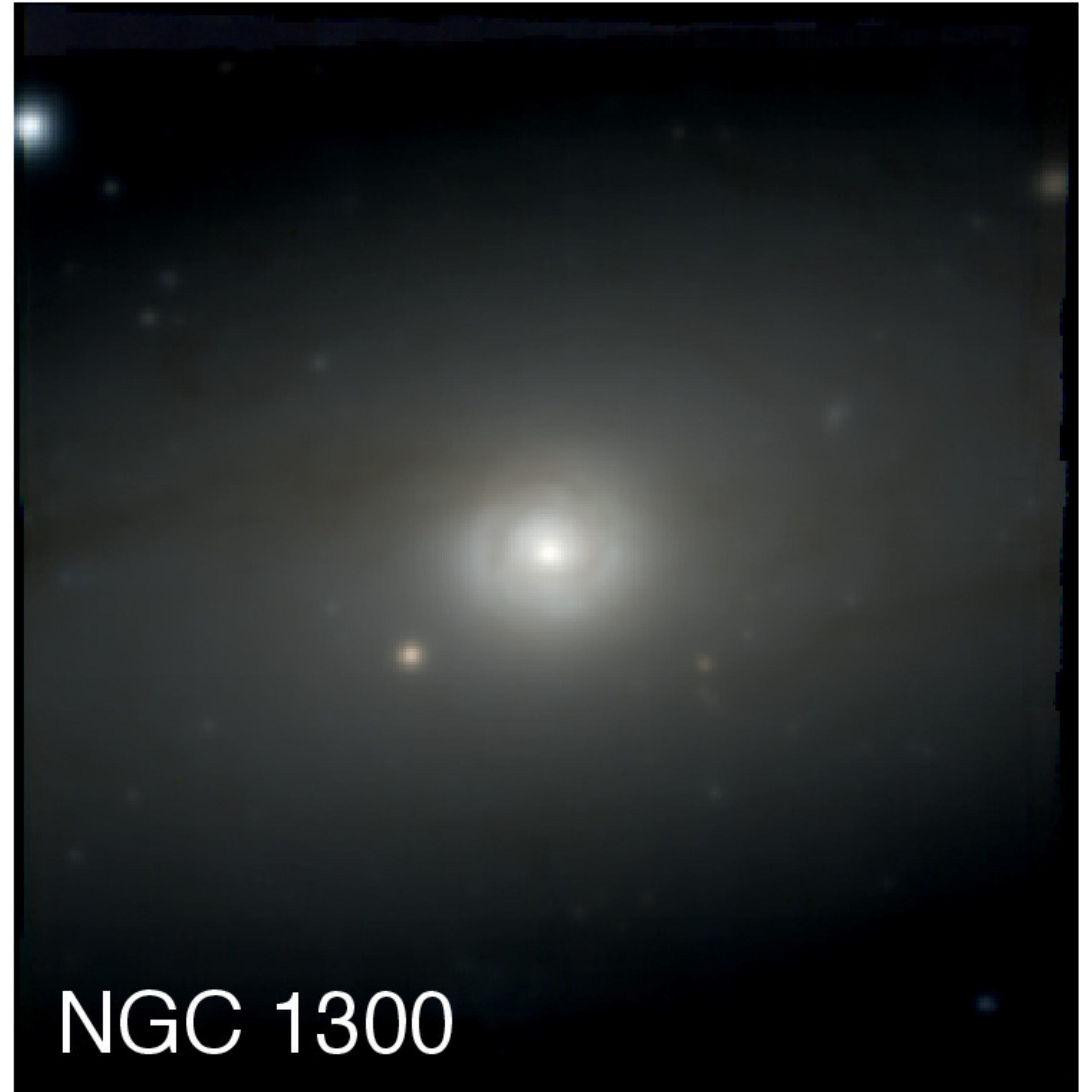}
	\includegraphics[width=0.5\columnwidth]{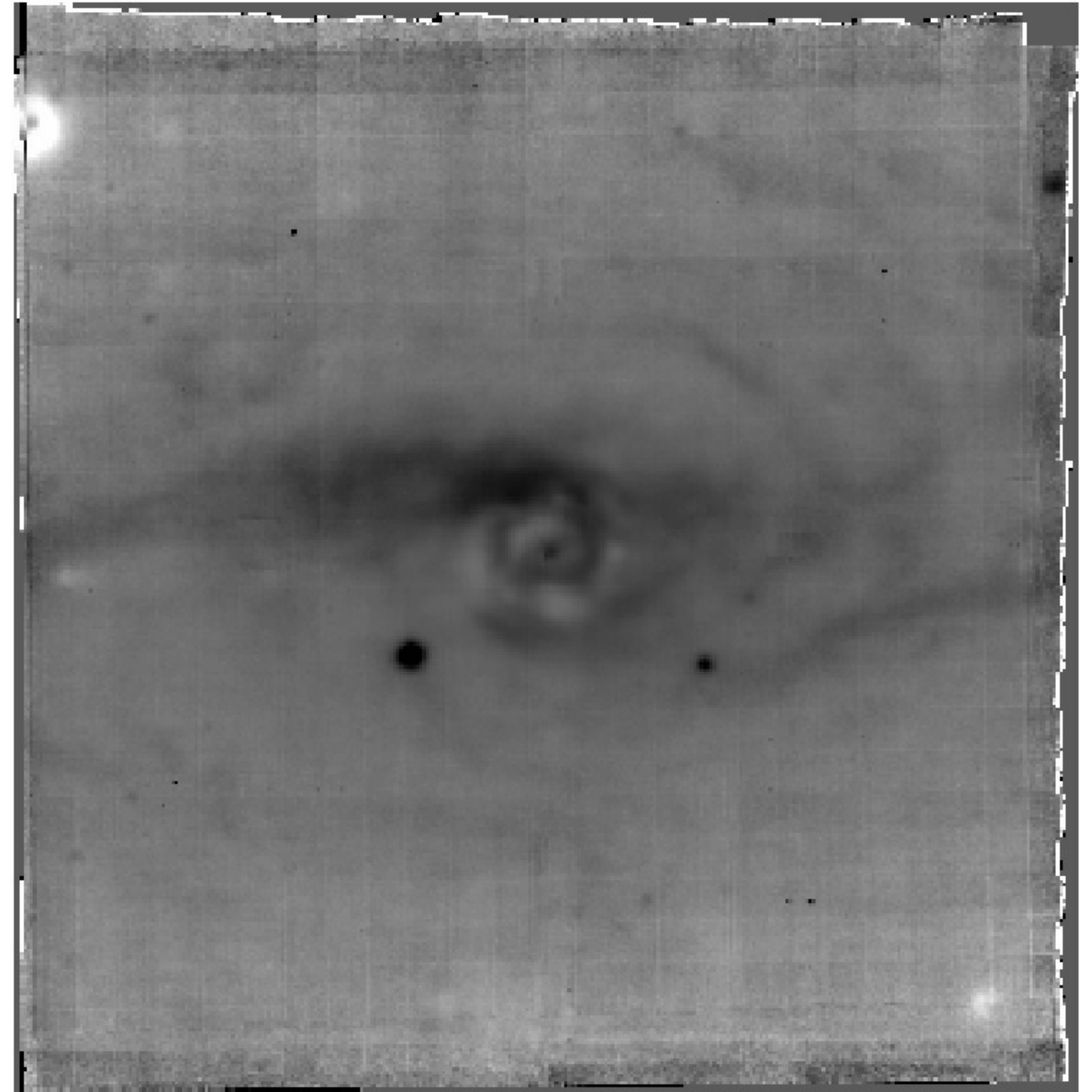}
	\includegraphics[width=0.5\columnwidth]{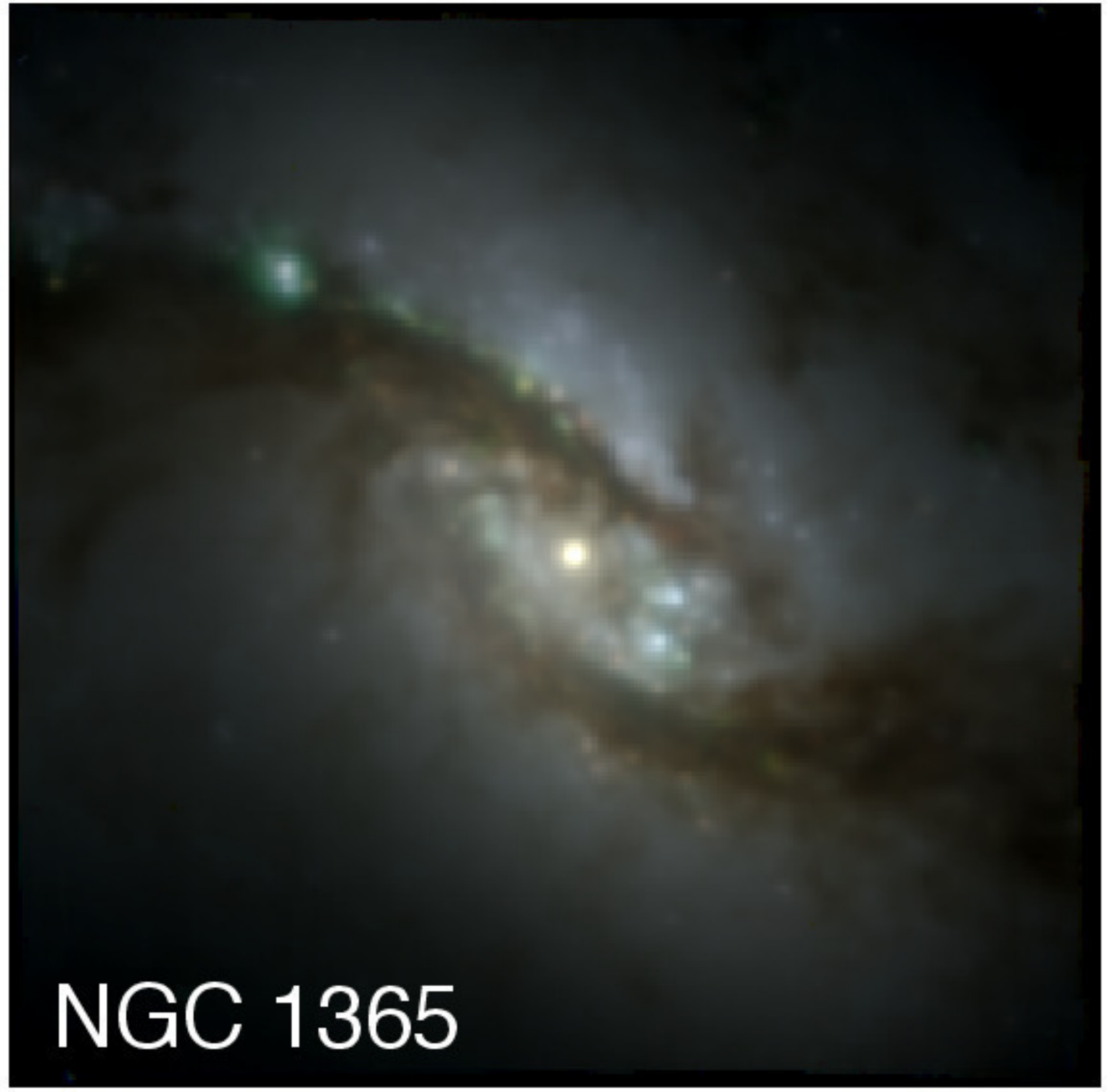}
	\includegraphics[width=0.495\columnwidth]{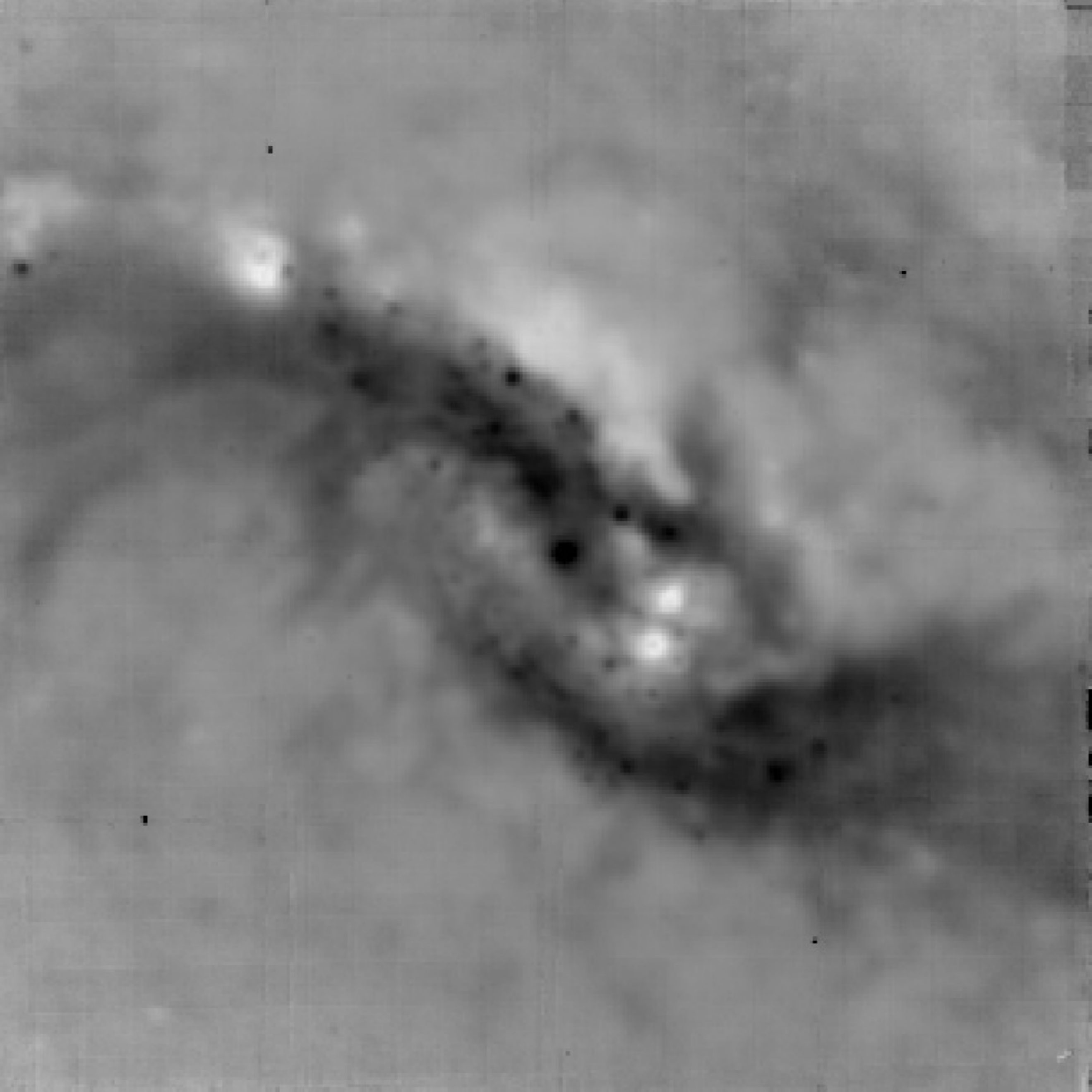}
	\includegraphics[width=0.5\columnwidth]{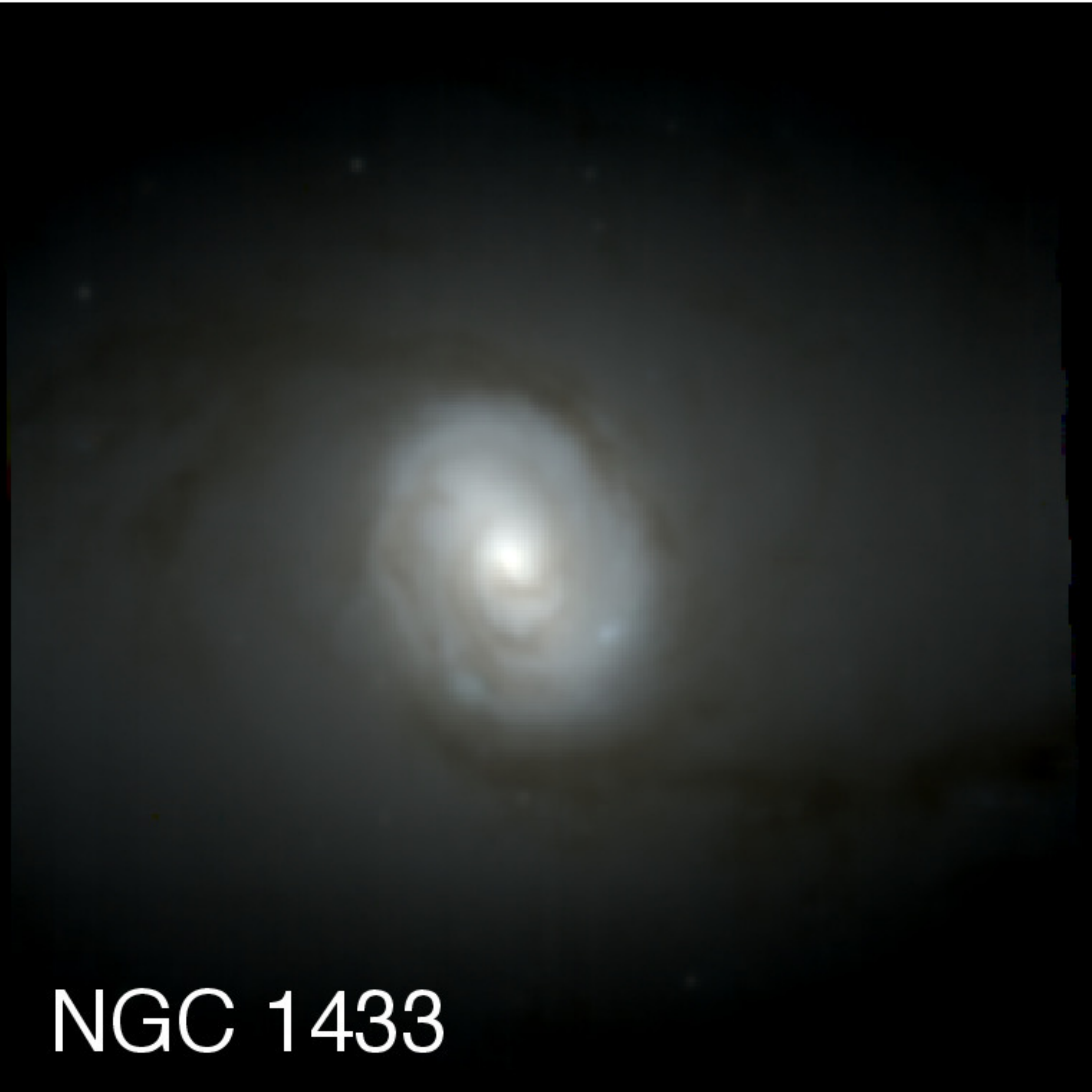}
	\includegraphics[width=0.5\columnwidth]{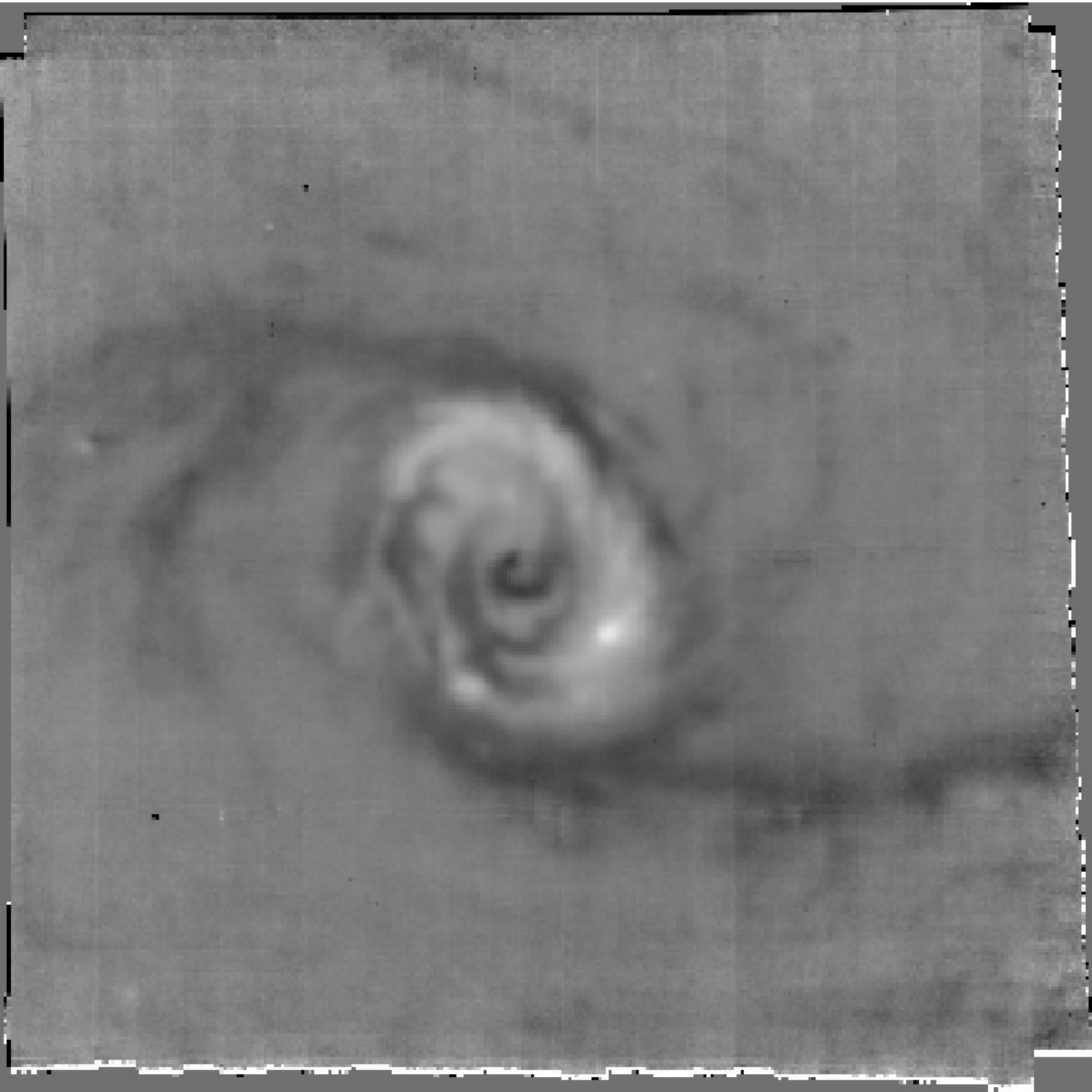}
	\includegraphics[width=0.5\columnwidth]{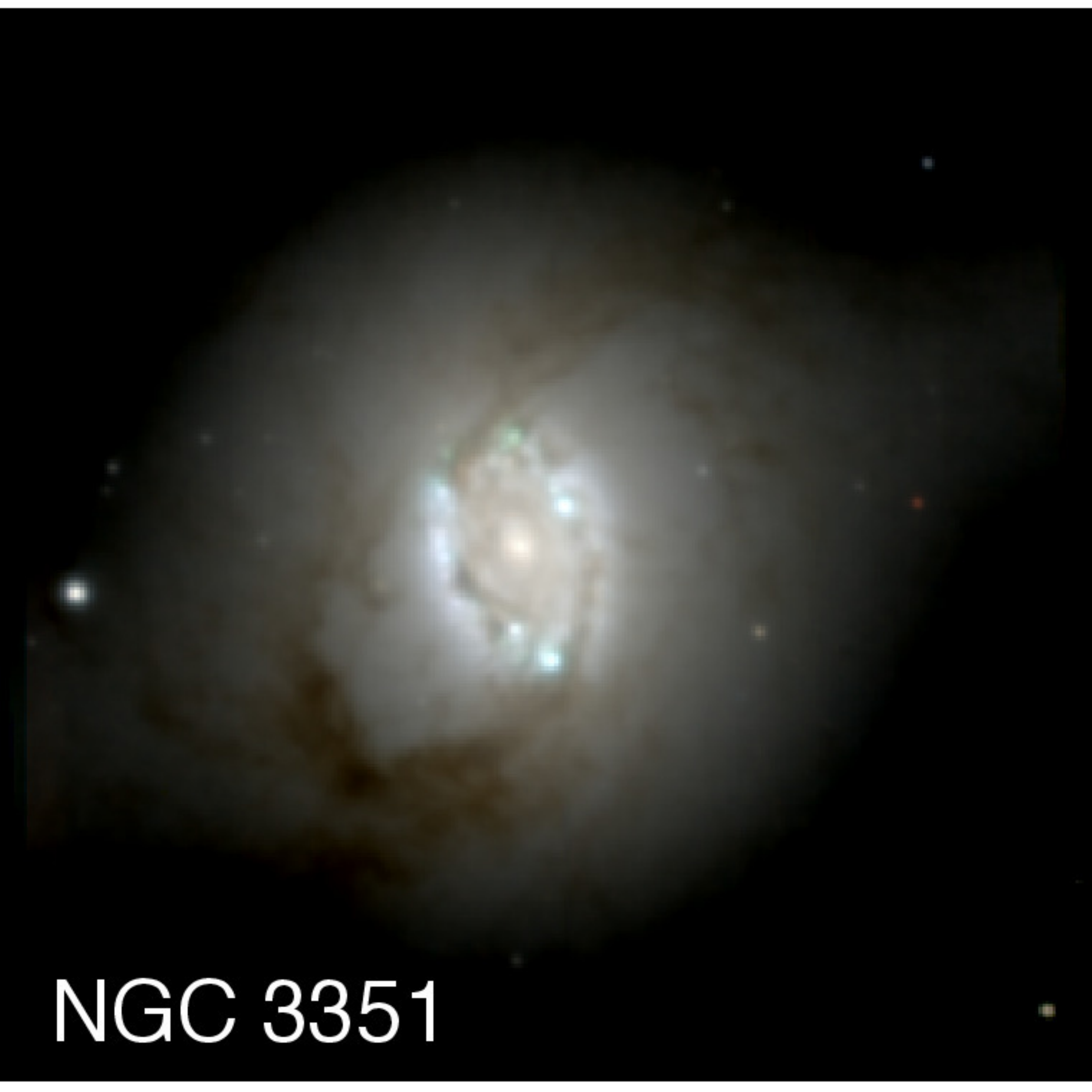}
	\includegraphics[width=0.5\columnwidth]{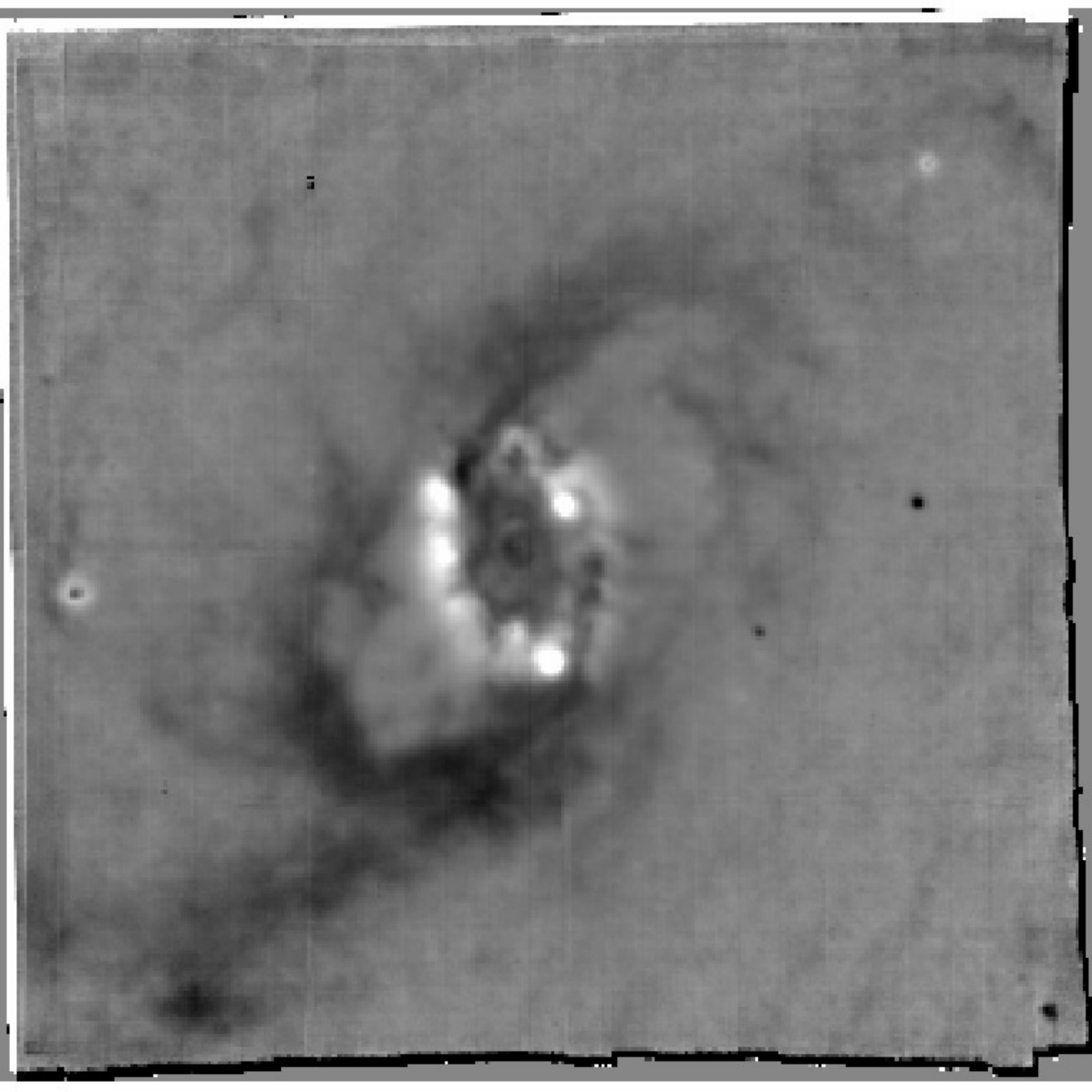}
	\includegraphics[width=0.5\columnwidth]{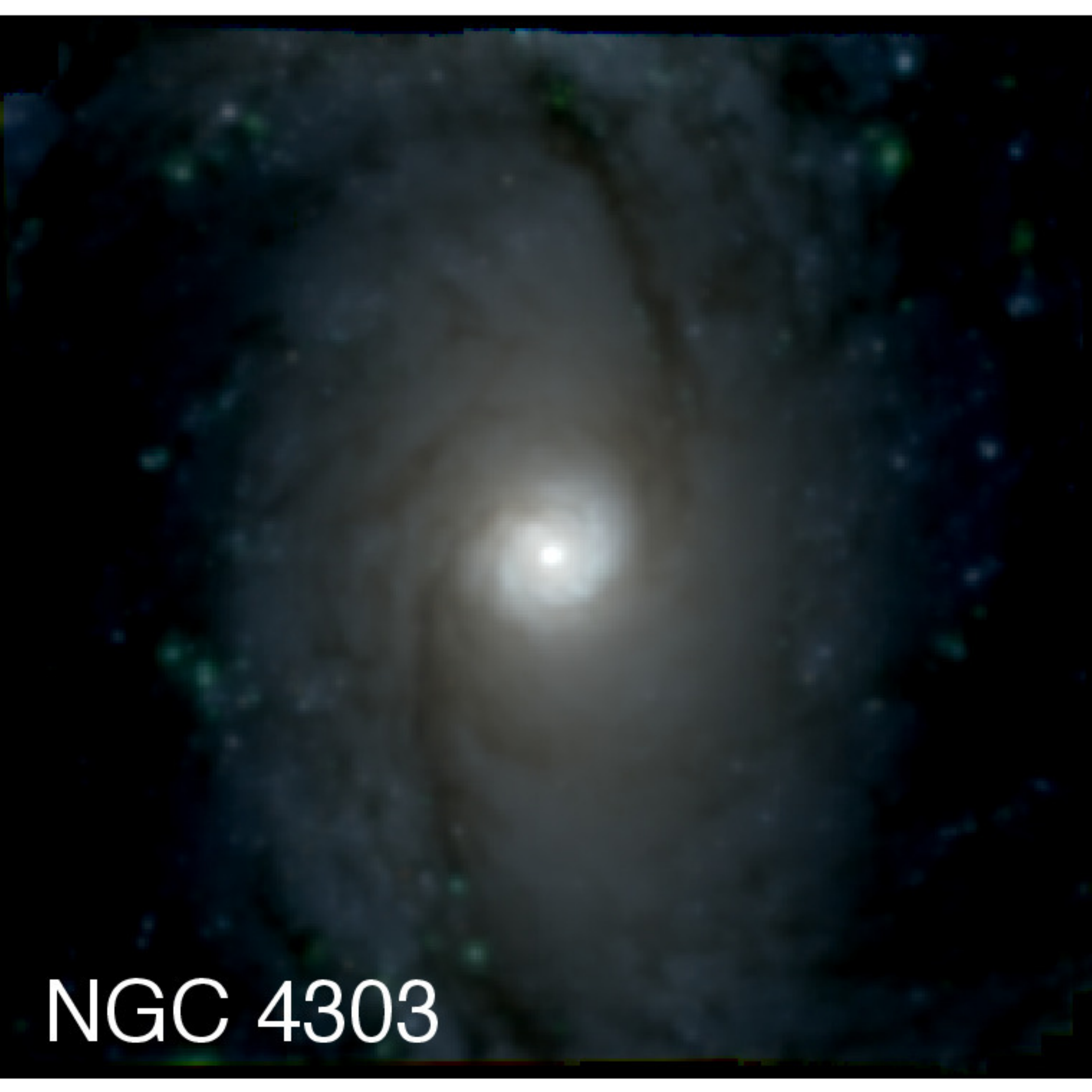}
	\includegraphics[width=0.5\columnwidth]{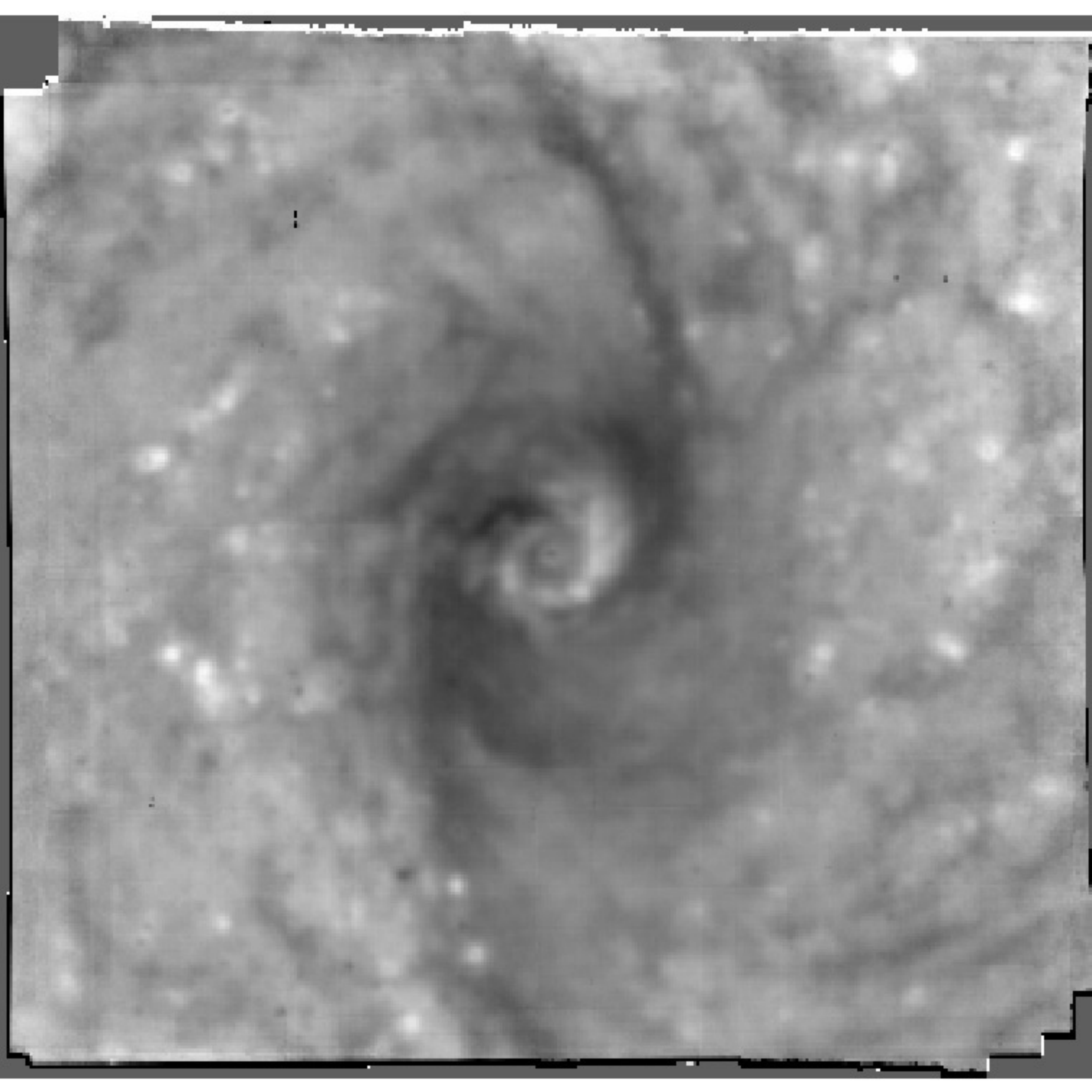}
	\includegraphics[width=0.5\columnwidth]{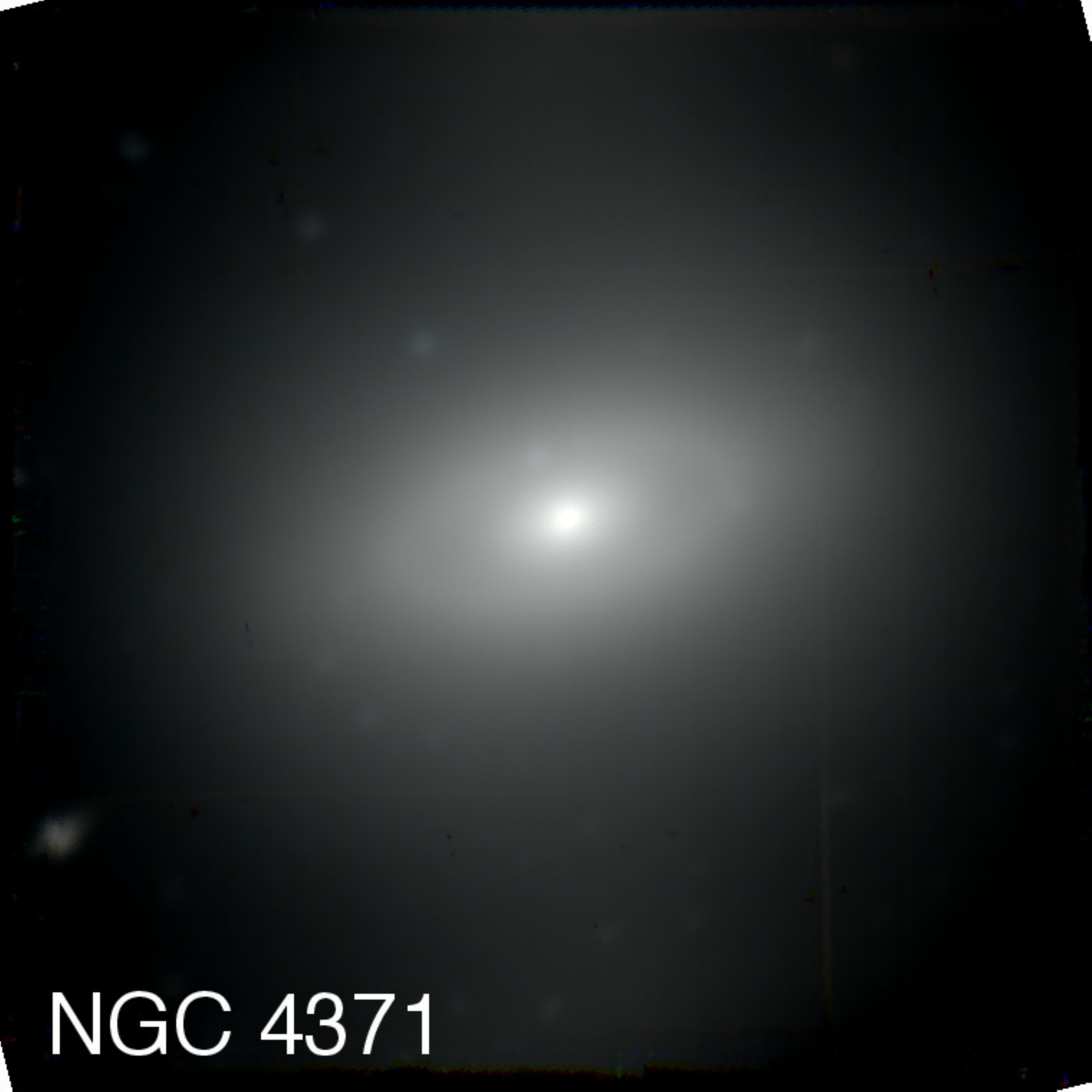}
	\includegraphics[width=0.5\columnwidth]{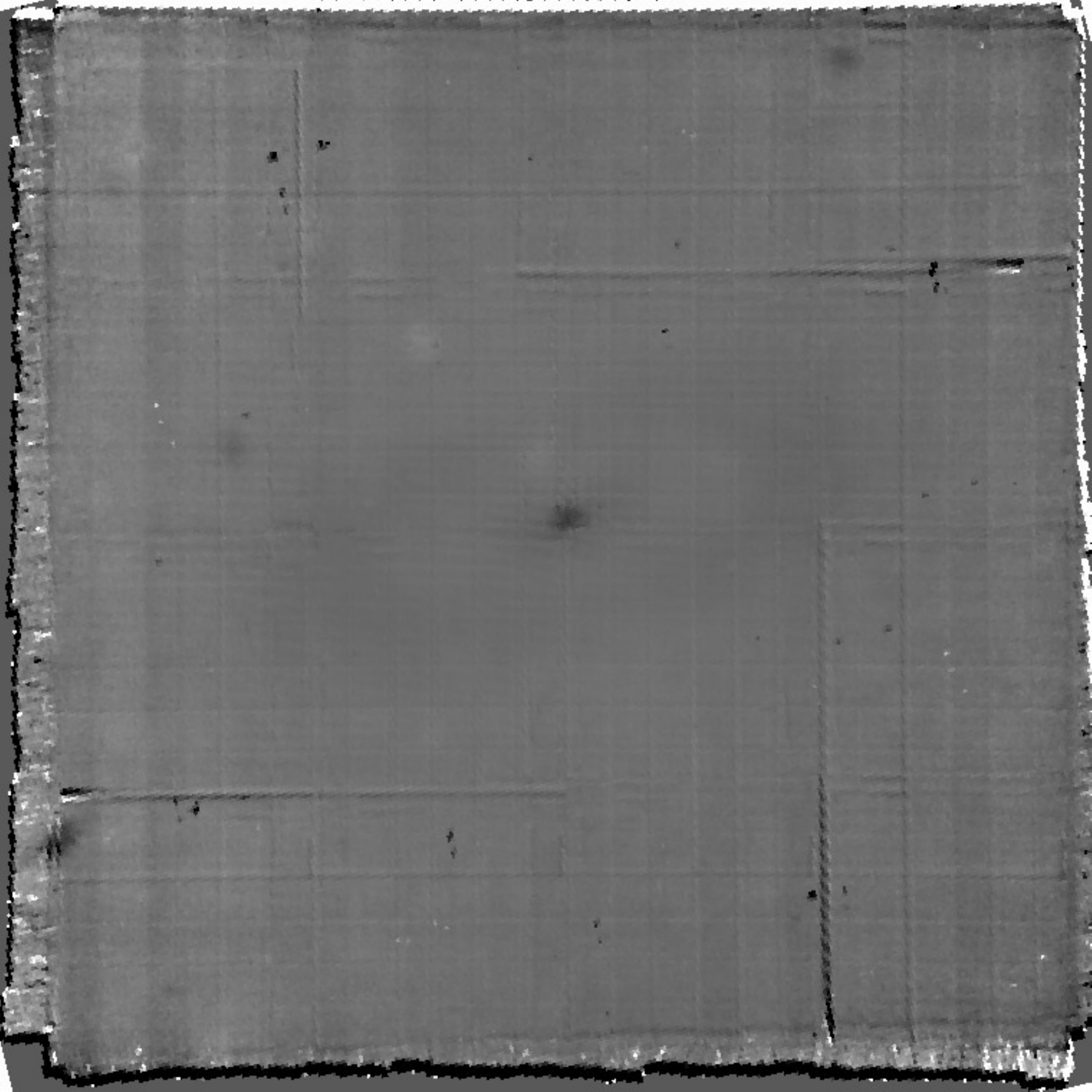}
\end{center}
    \caption{For each galaxy in the current TIMER sample of 21 galaxies we show a colour composite and a colour map built from the TIMER MUSE data cubes, as explained in the text. In the colour maps, redder colours correspond to darker shades.}
    \label{fig:rgb_cm}
\end{figure*}

\begin{figure*}
\begin{center}
	\includegraphics[width=0.5\columnwidth]{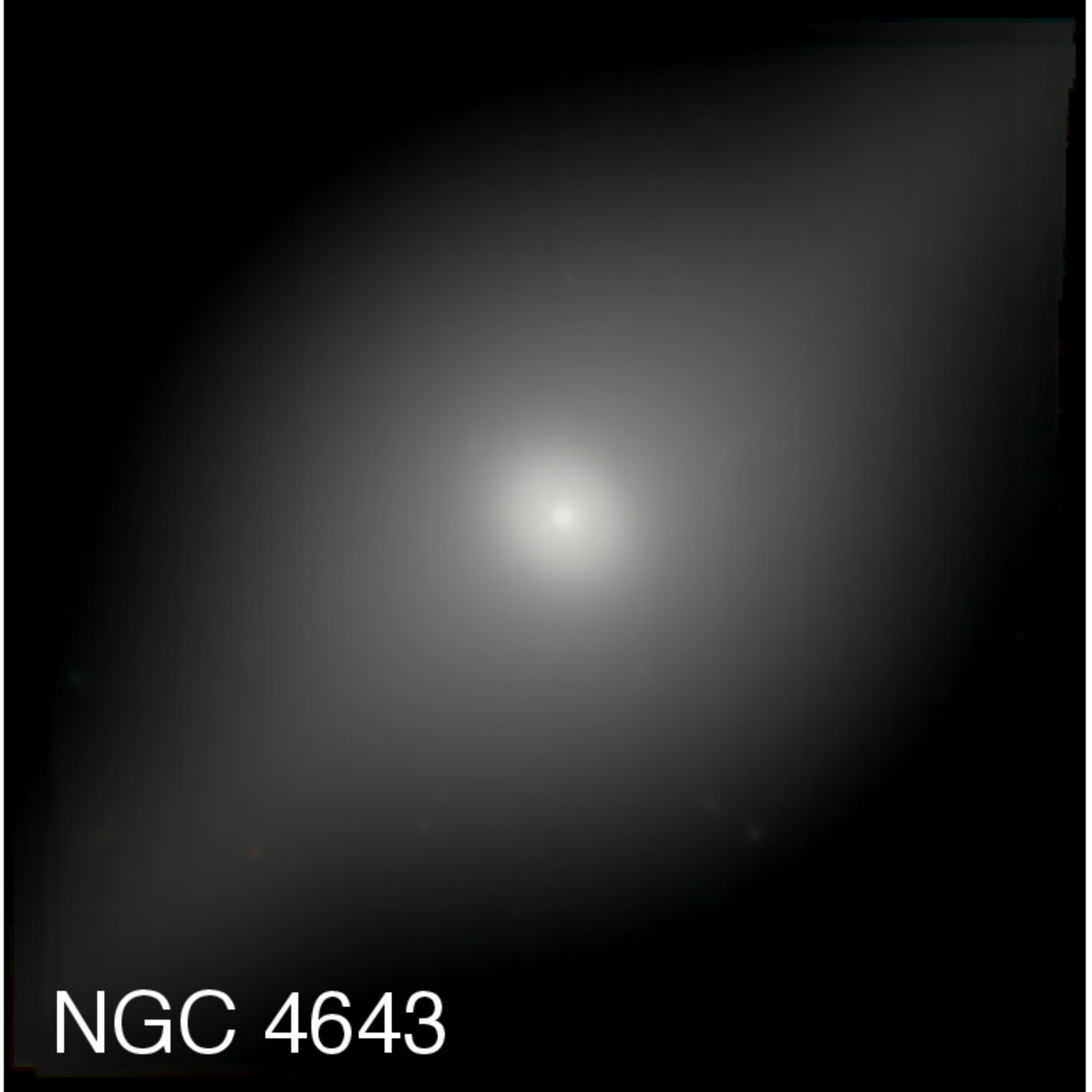}
	\includegraphics[width=0.5\columnwidth]{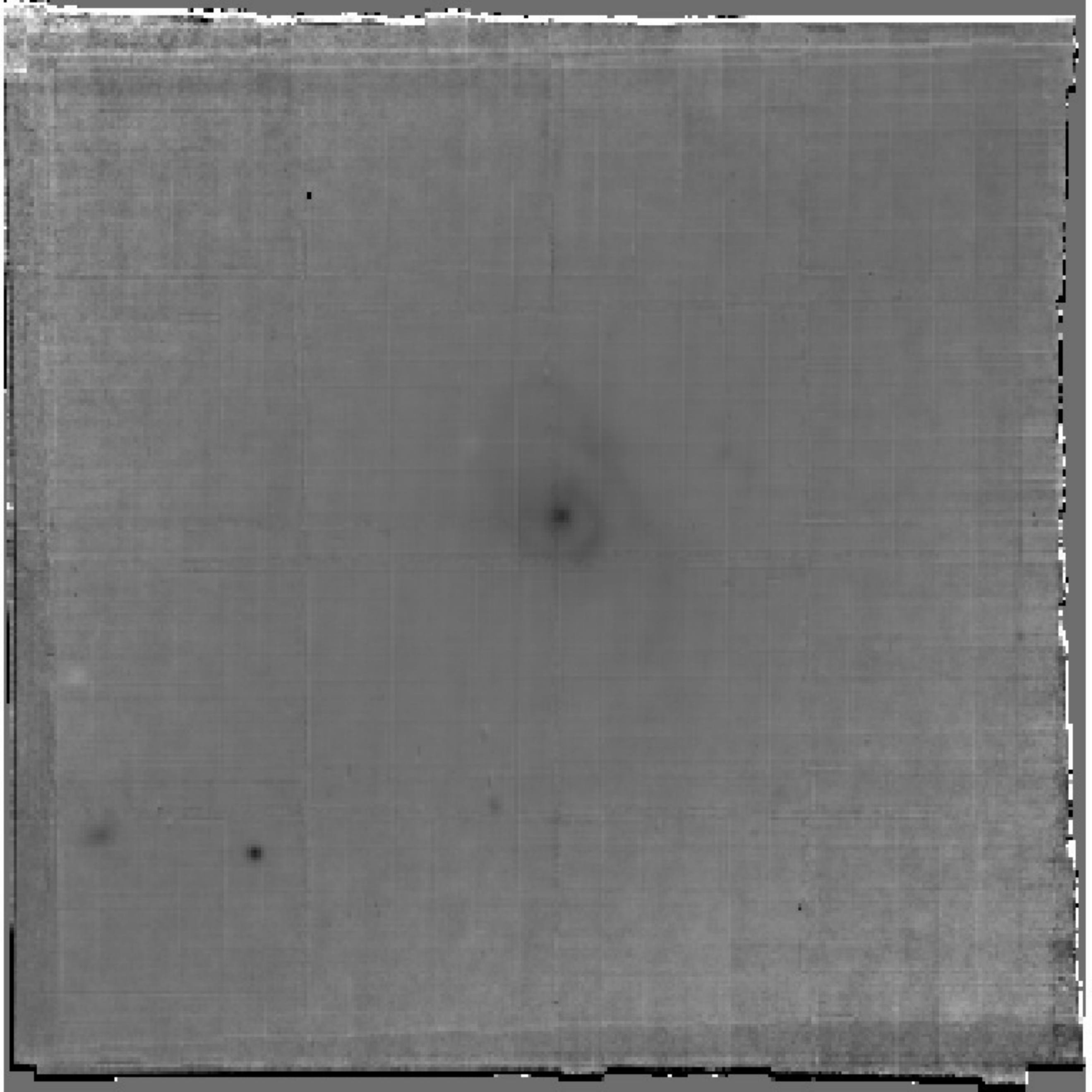}
	\includegraphics[width=0.5\columnwidth]{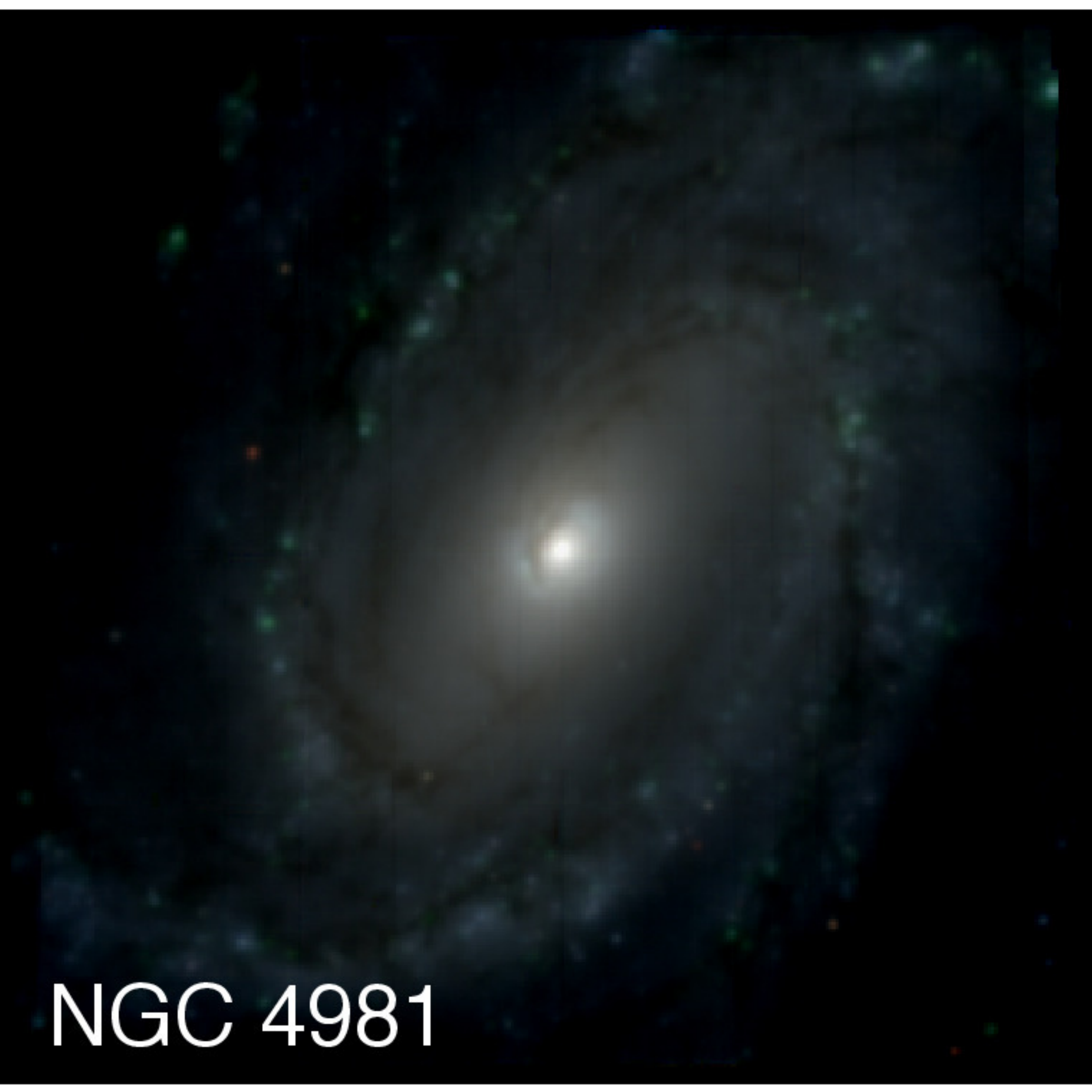}
	\includegraphics[width=0.5\columnwidth]{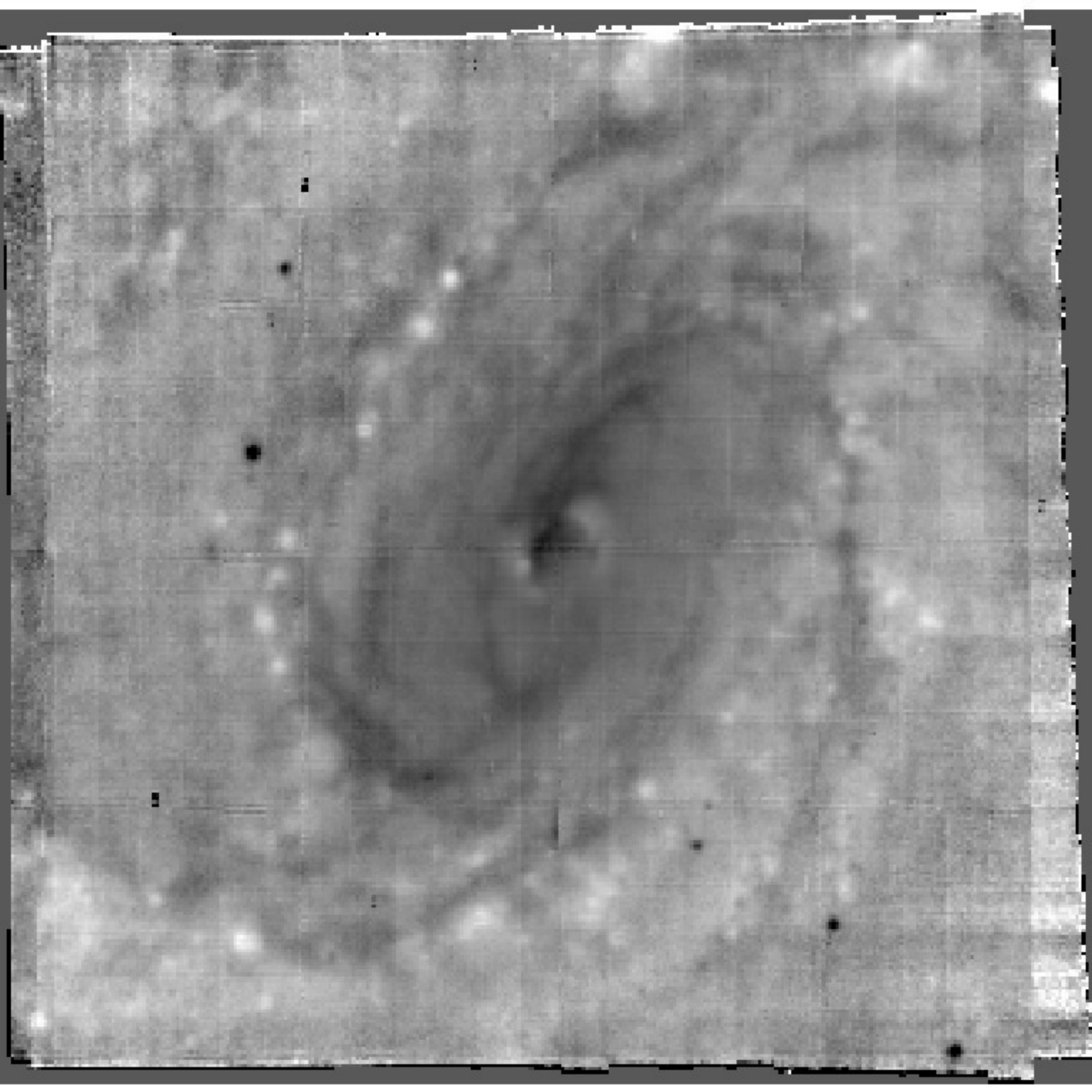}
	\includegraphics[width=0.5\columnwidth]{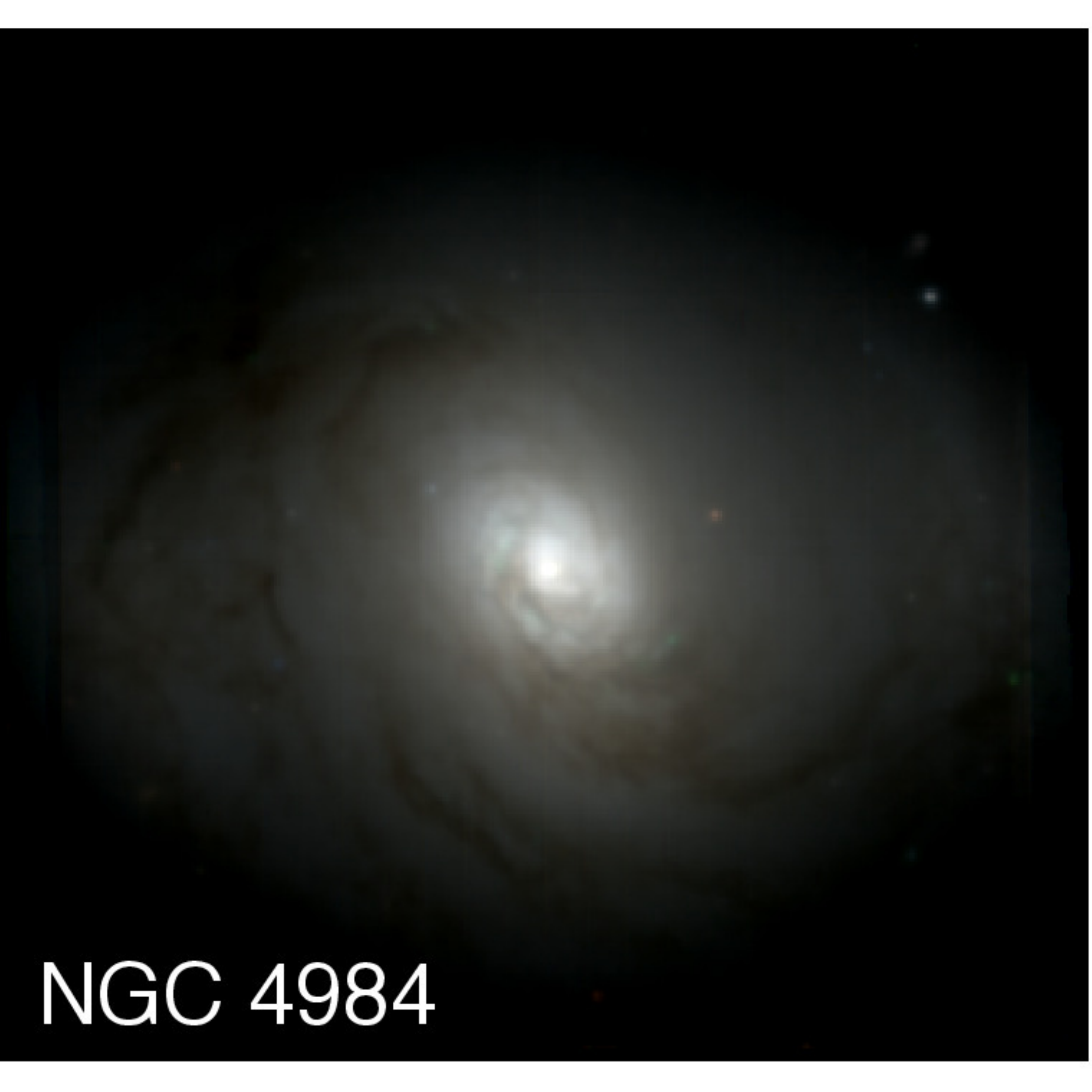}
	\includegraphics[width=0.5\columnwidth]{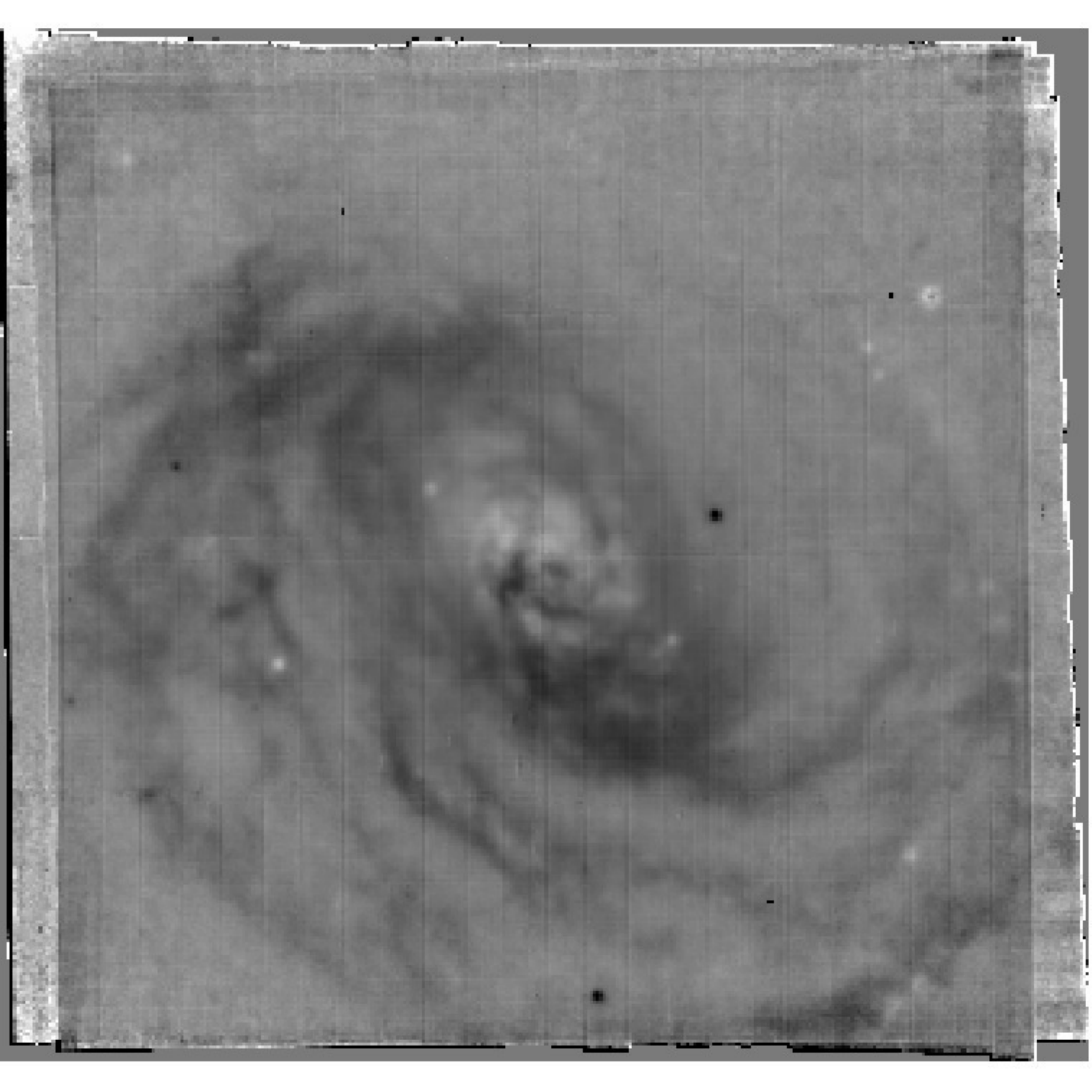}
	\includegraphics[width=0.5\columnwidth]{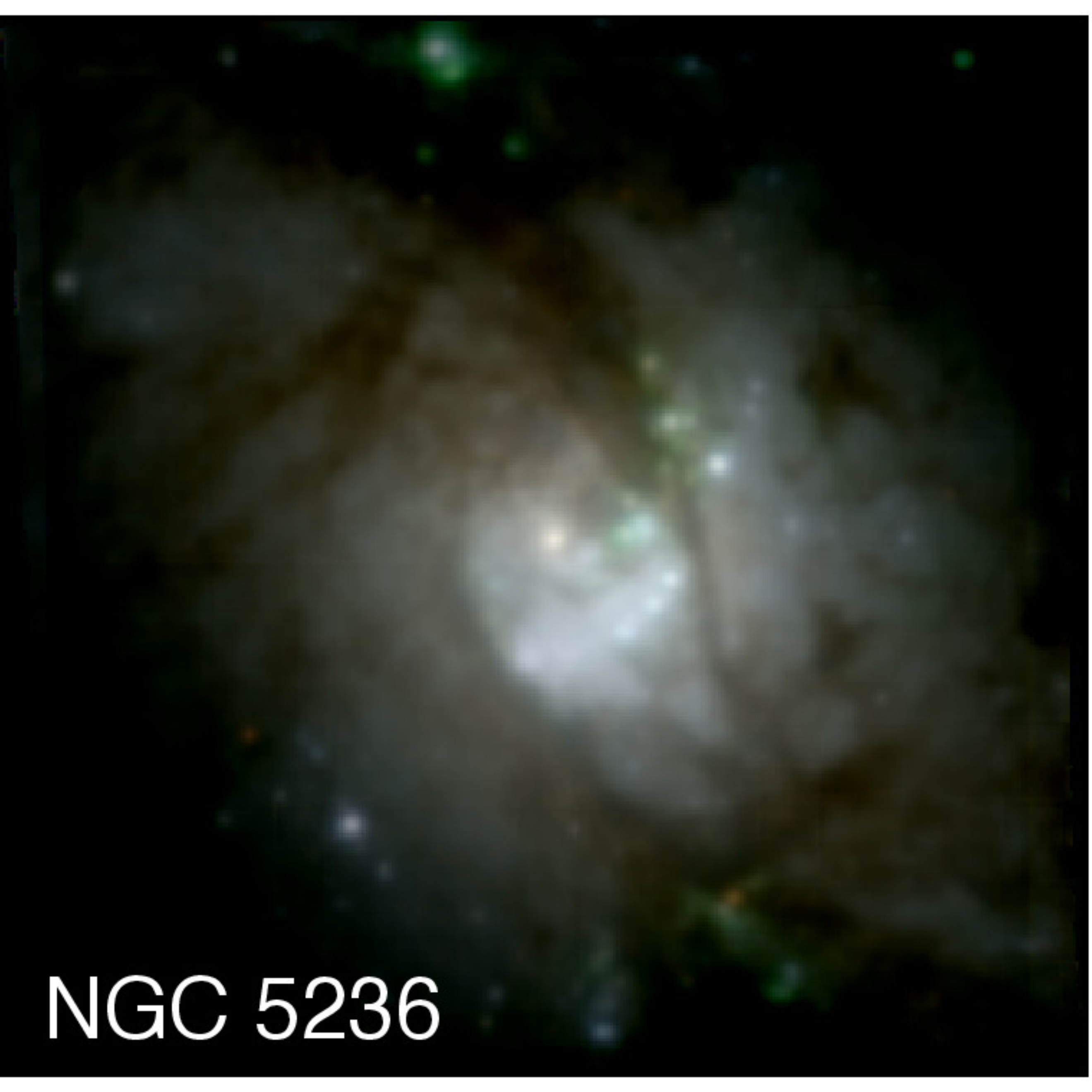}
	\includegraphics[width=0.5\columnwidth]{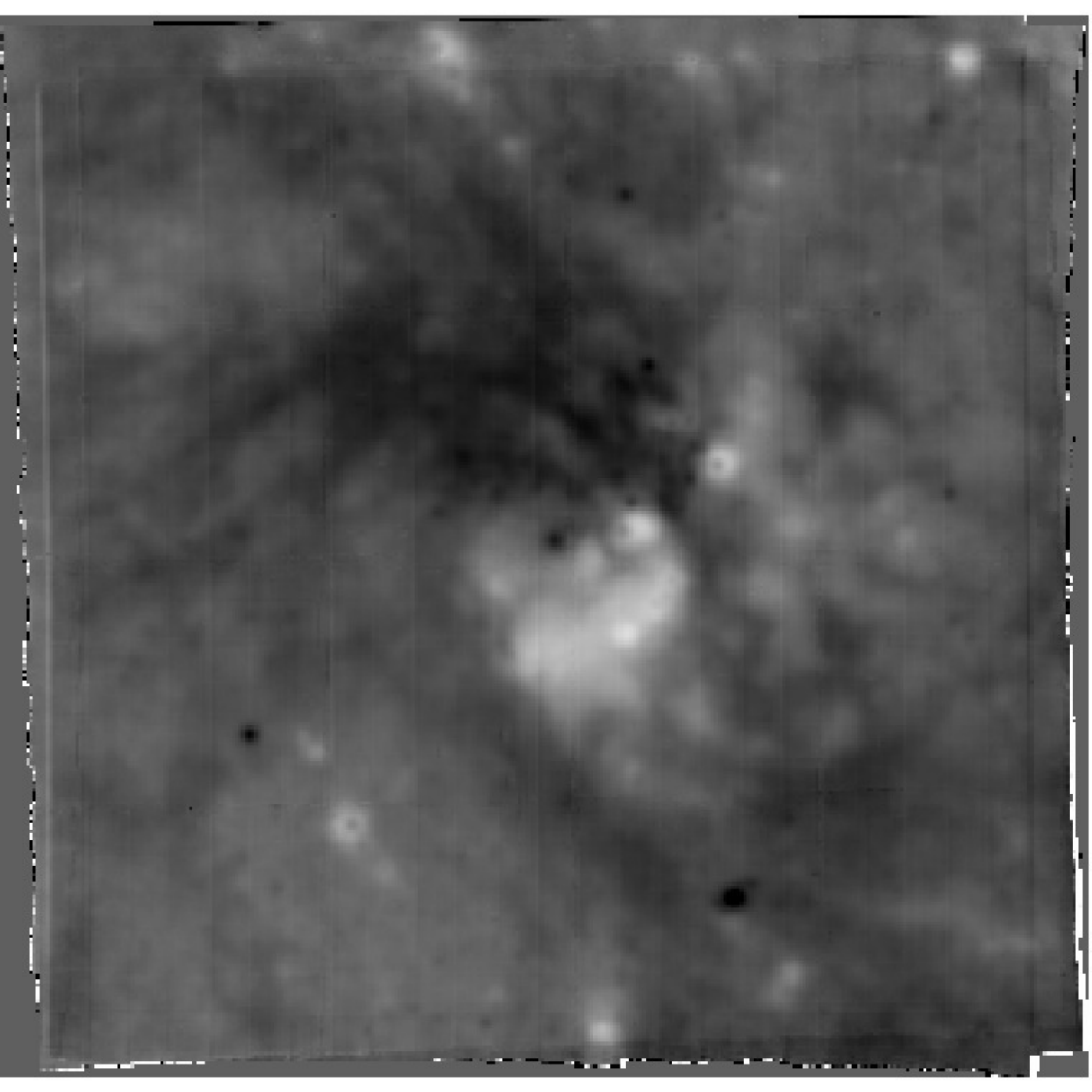}
	\includegraphics[width=0.5\columnwidth]{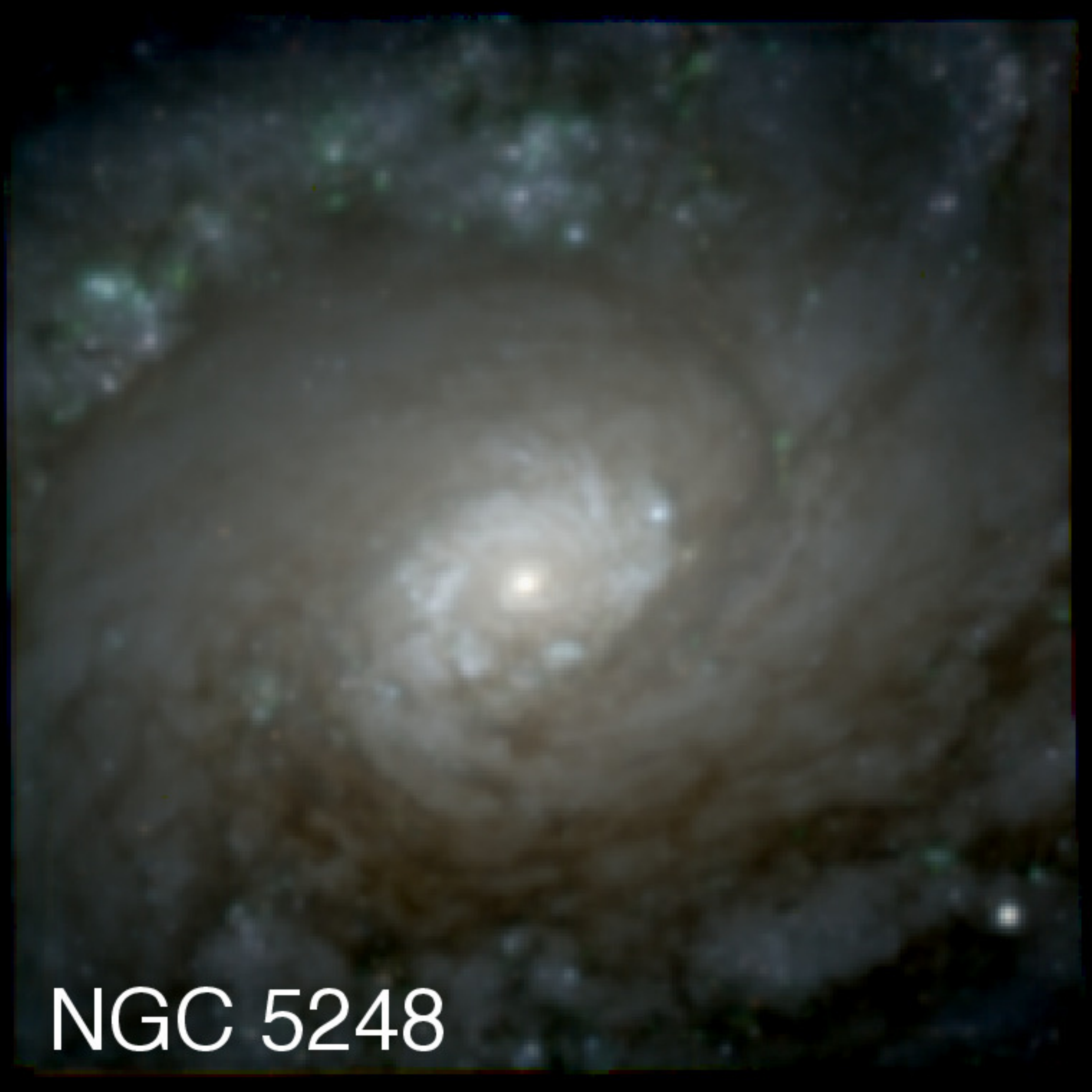}
	\includegraphics[width=0.5\columnwidth]{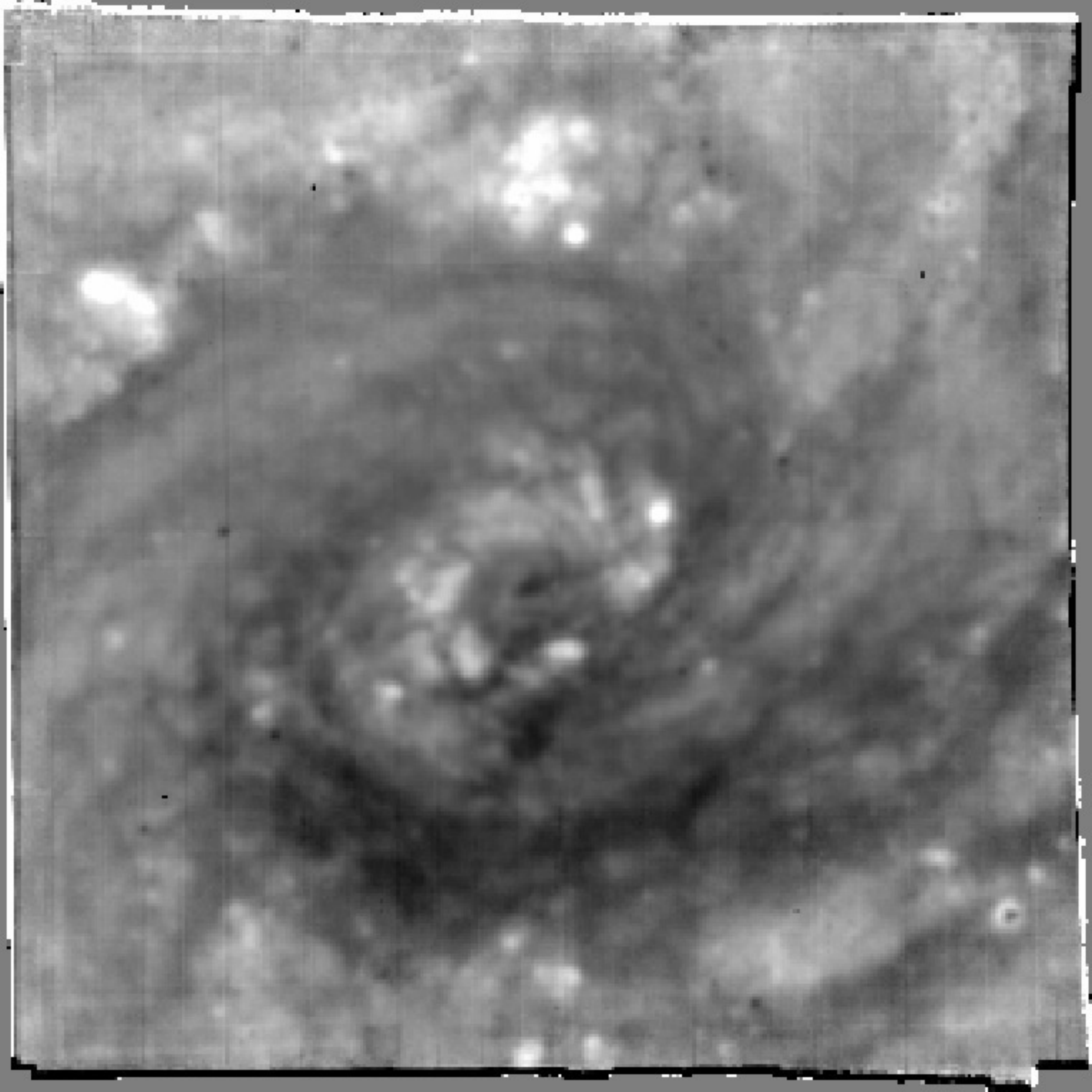}
	\includegraphics[width=0.5\columnwidth]{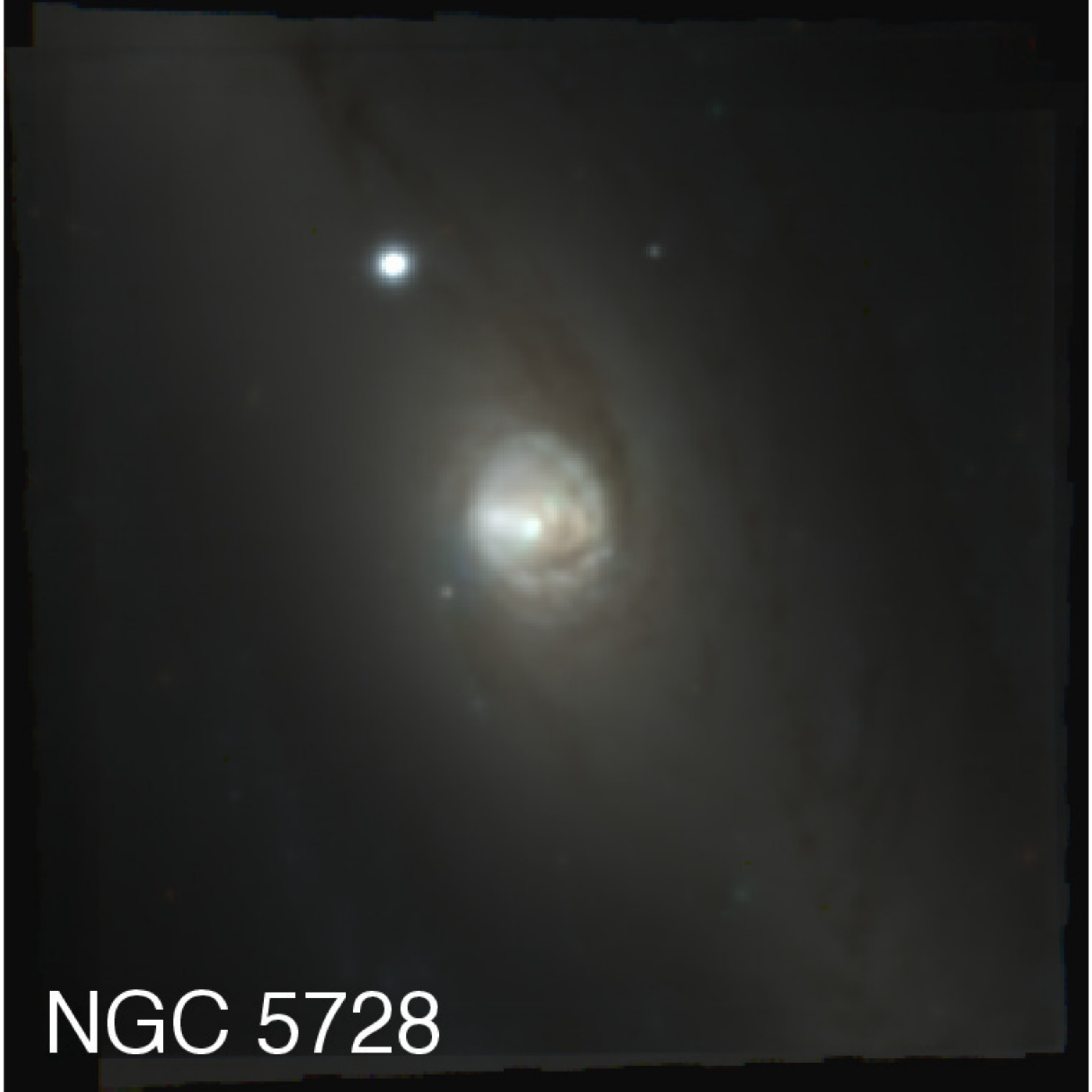}
	\includegraphics[width=0.5\columnwidth]{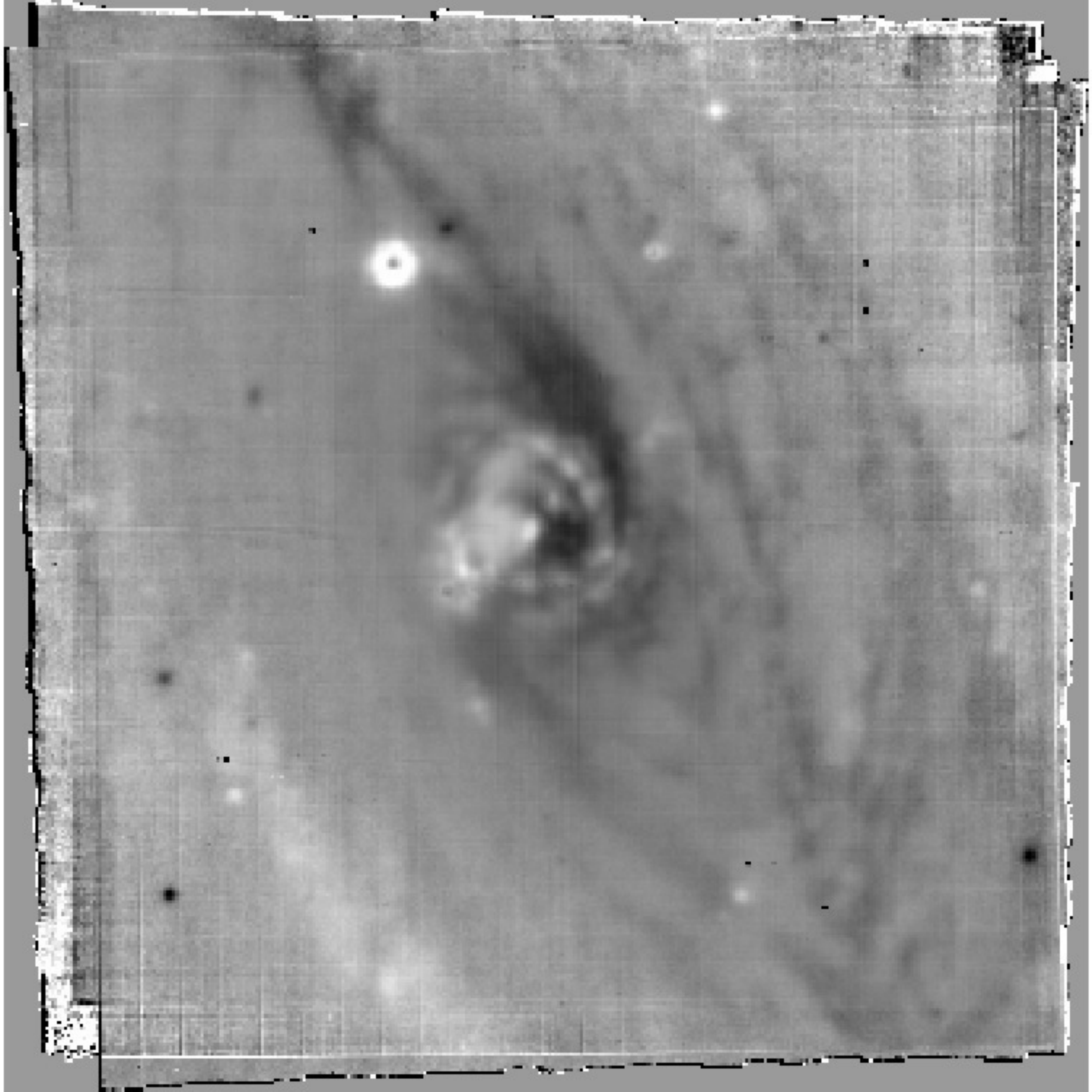}
	\includegraphics[width=0.5\columnwidth]{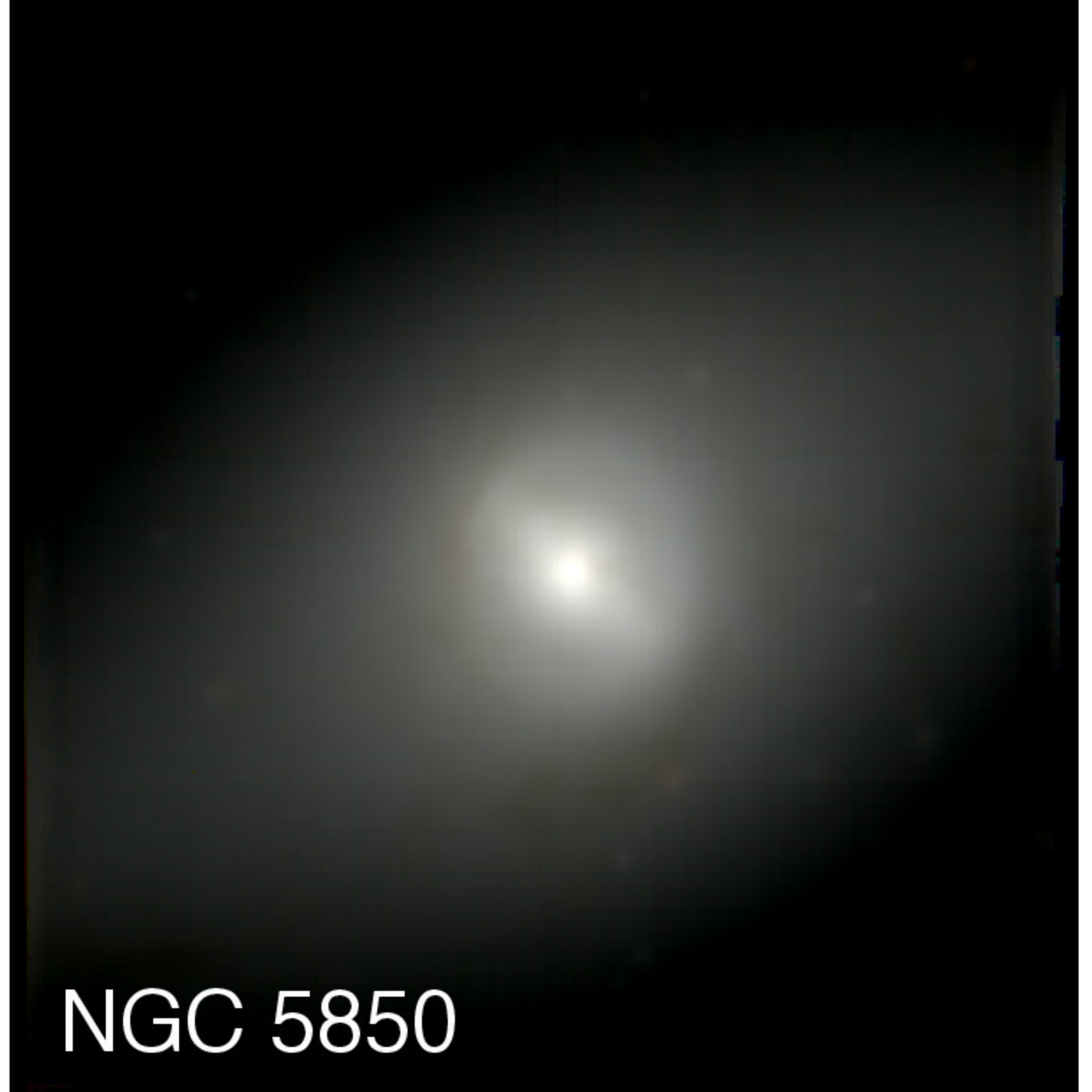}
	\includegraphics[width=0.5\columnwidth]{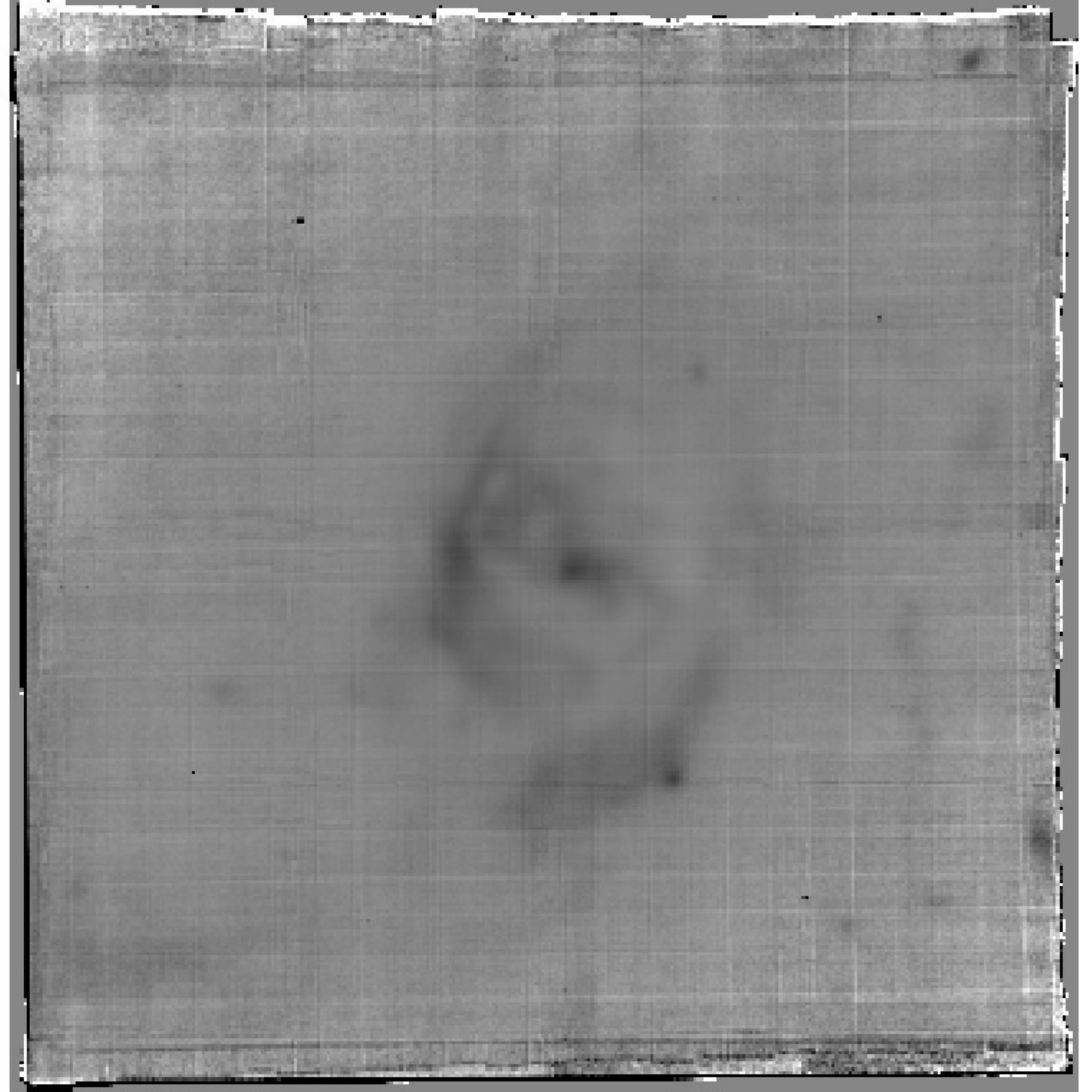}
	\includegraphics[width=0.5\columnwidth]{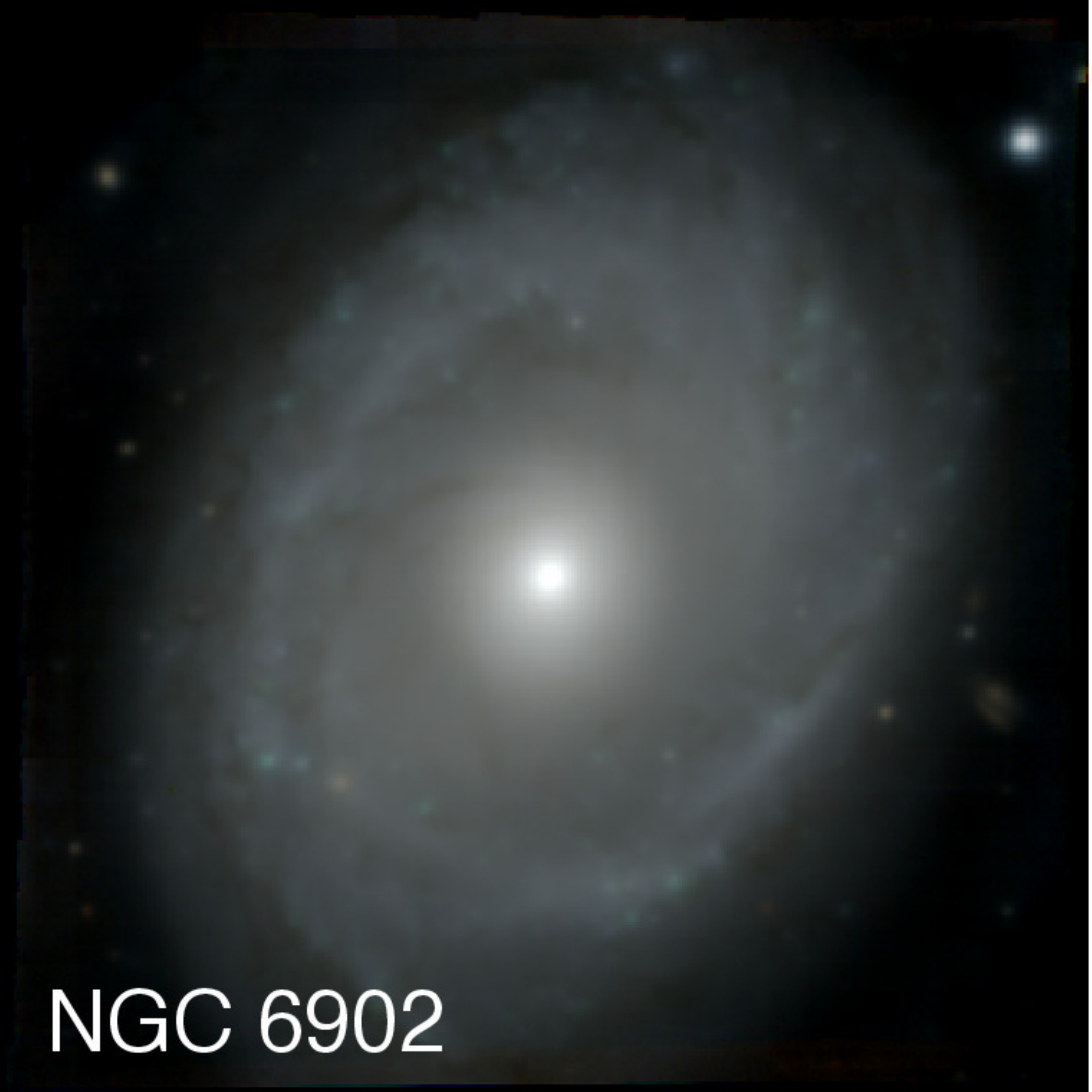}
	\includegraphics[width=0.5\columnwidth]{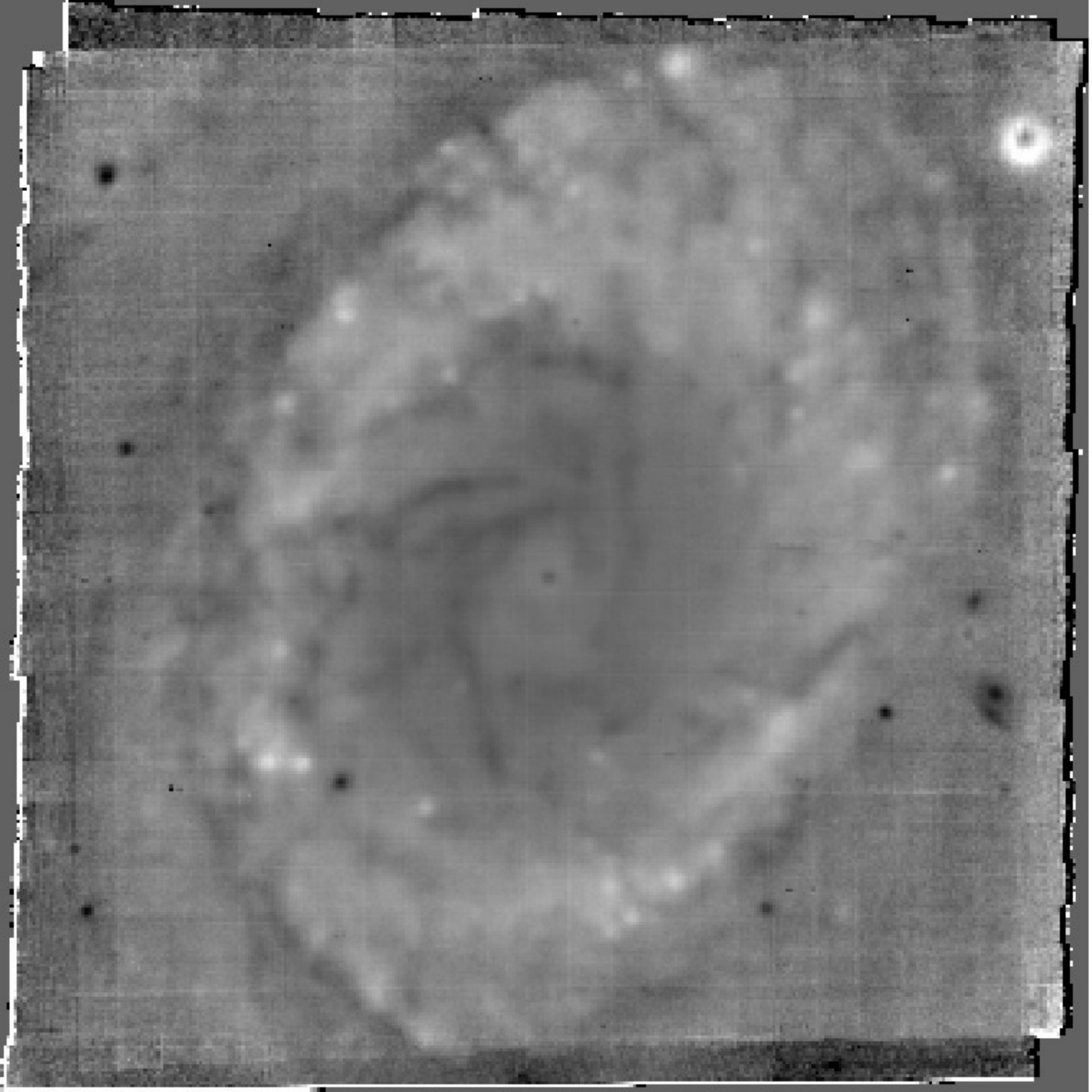}
	\includegraphics[width=0.5\columnwidth]{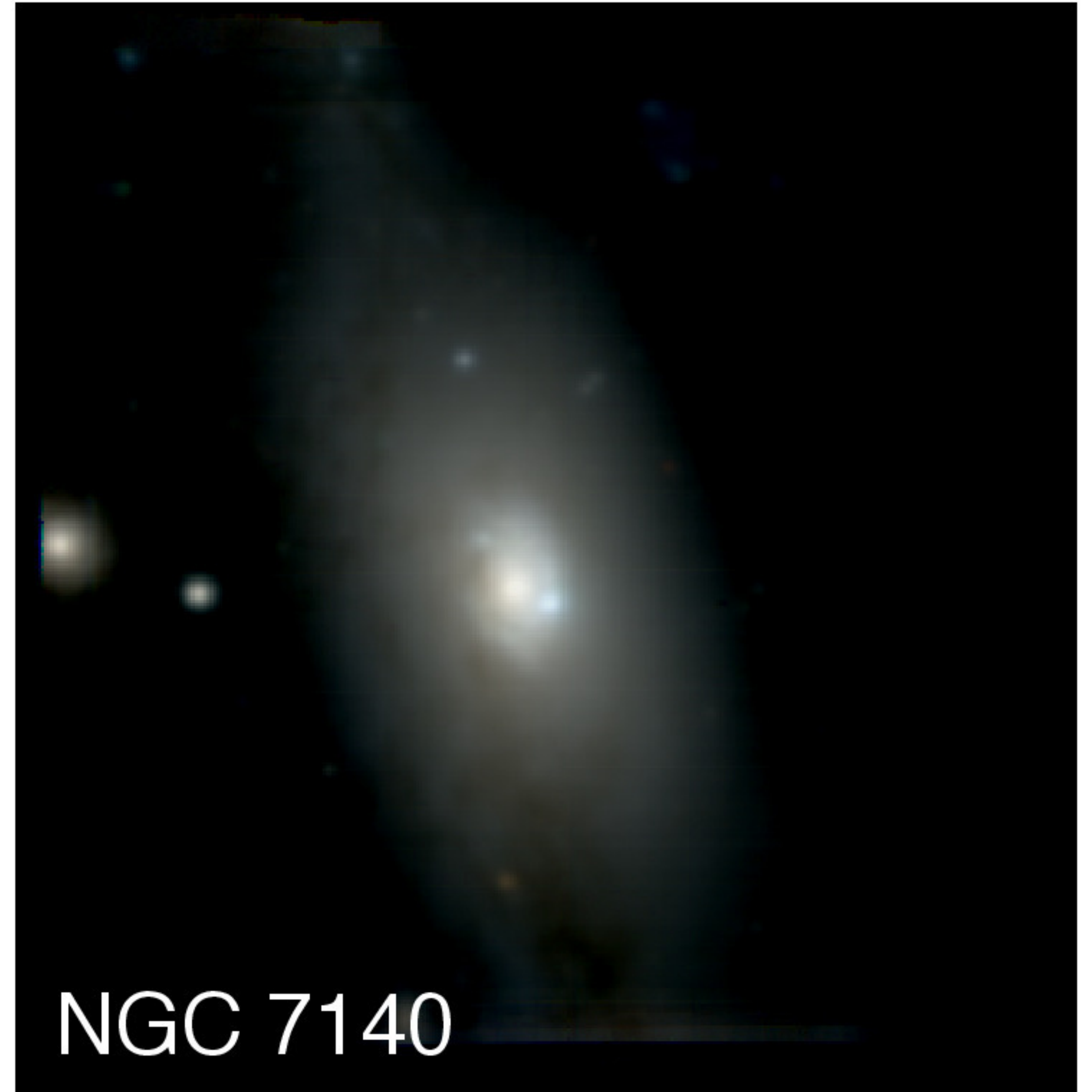}
	\includegraphics[width=0.5\columnwidth]{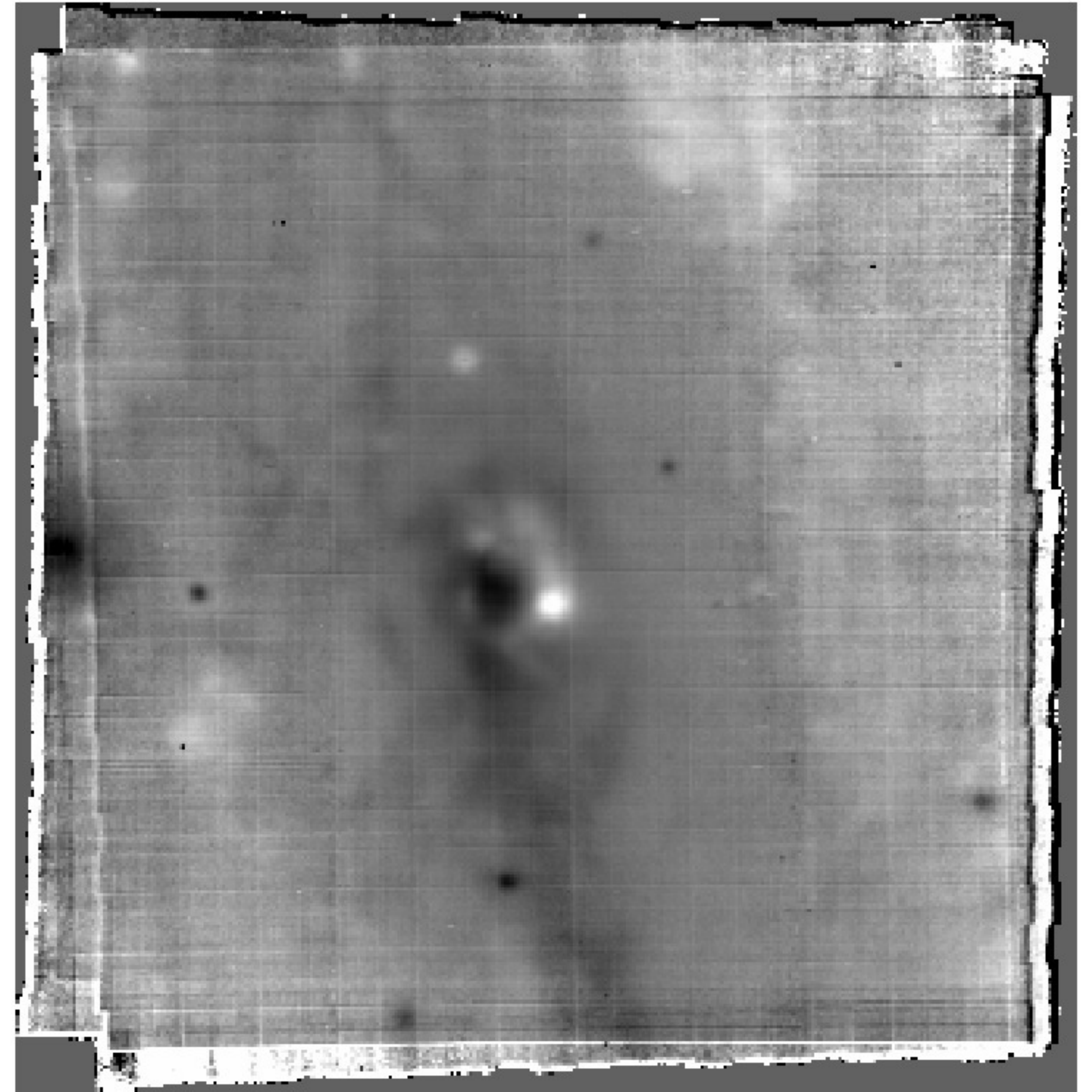}
	\includegraphics[width=0.495\columnwidth]{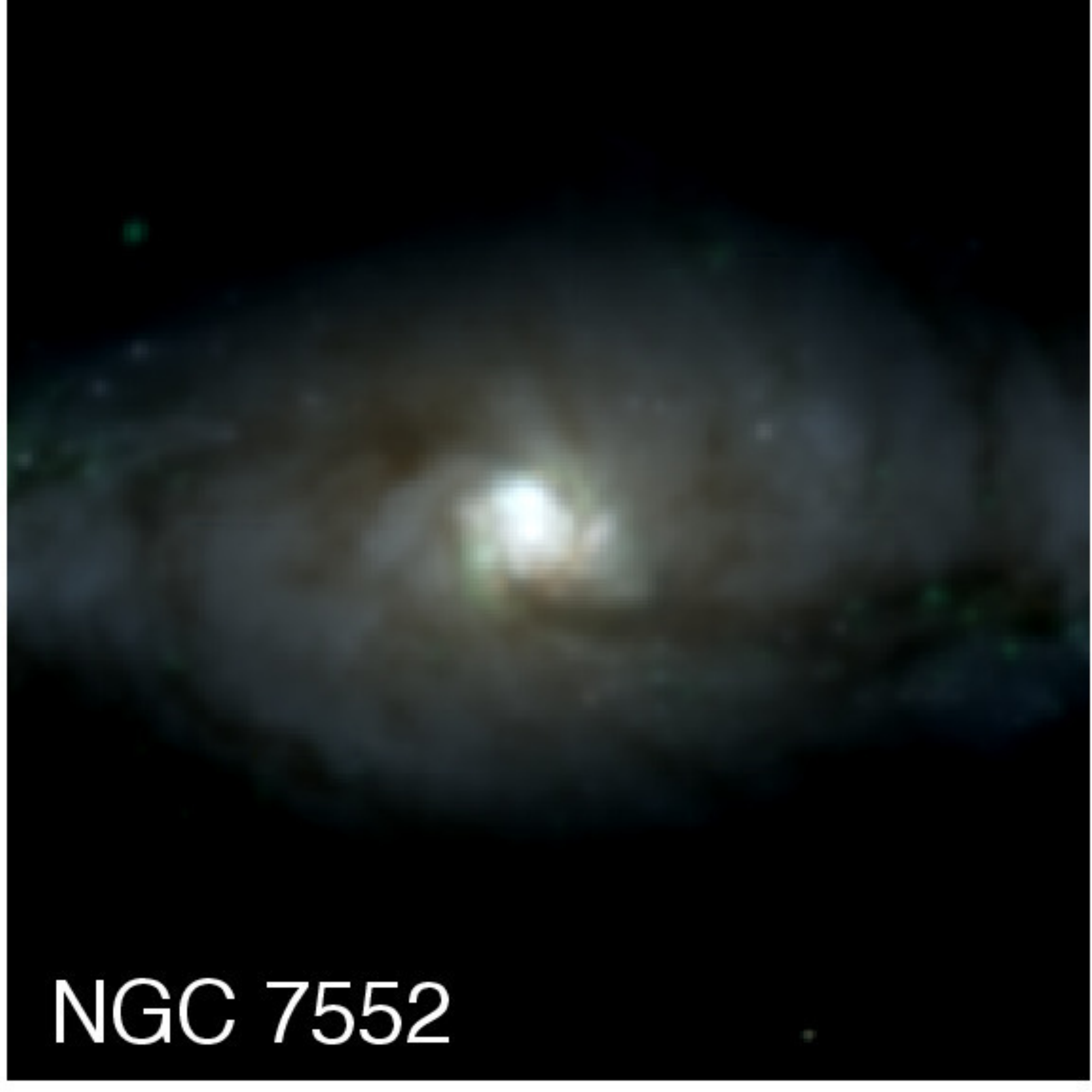}
	\includegraphics[width=0.5\columnwidth]{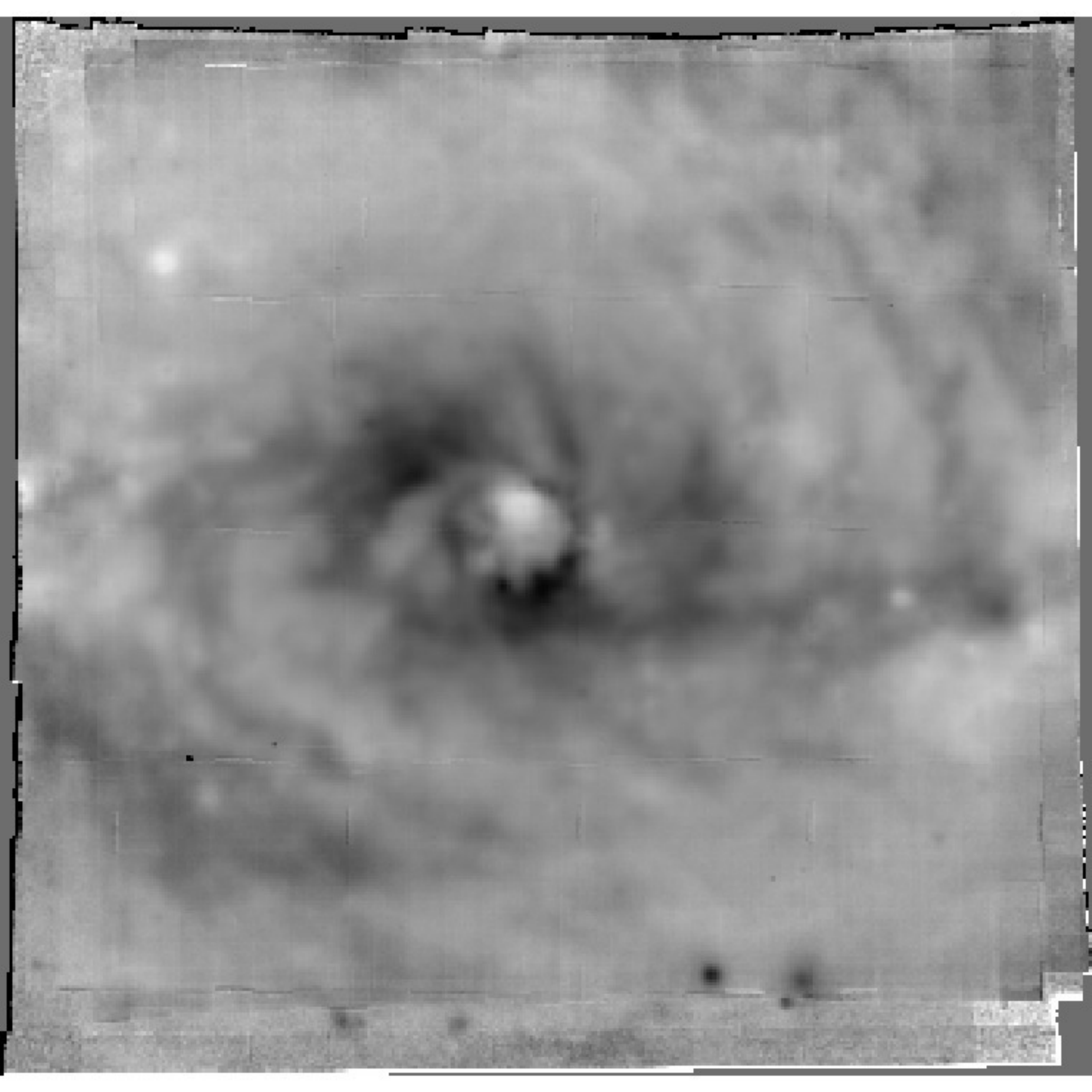}
\end{center}
    \contcaption{}
\end{figure*}

\begin{figure*}
\begin{center}
	\includegraphics[width=0.5\columnwidth]{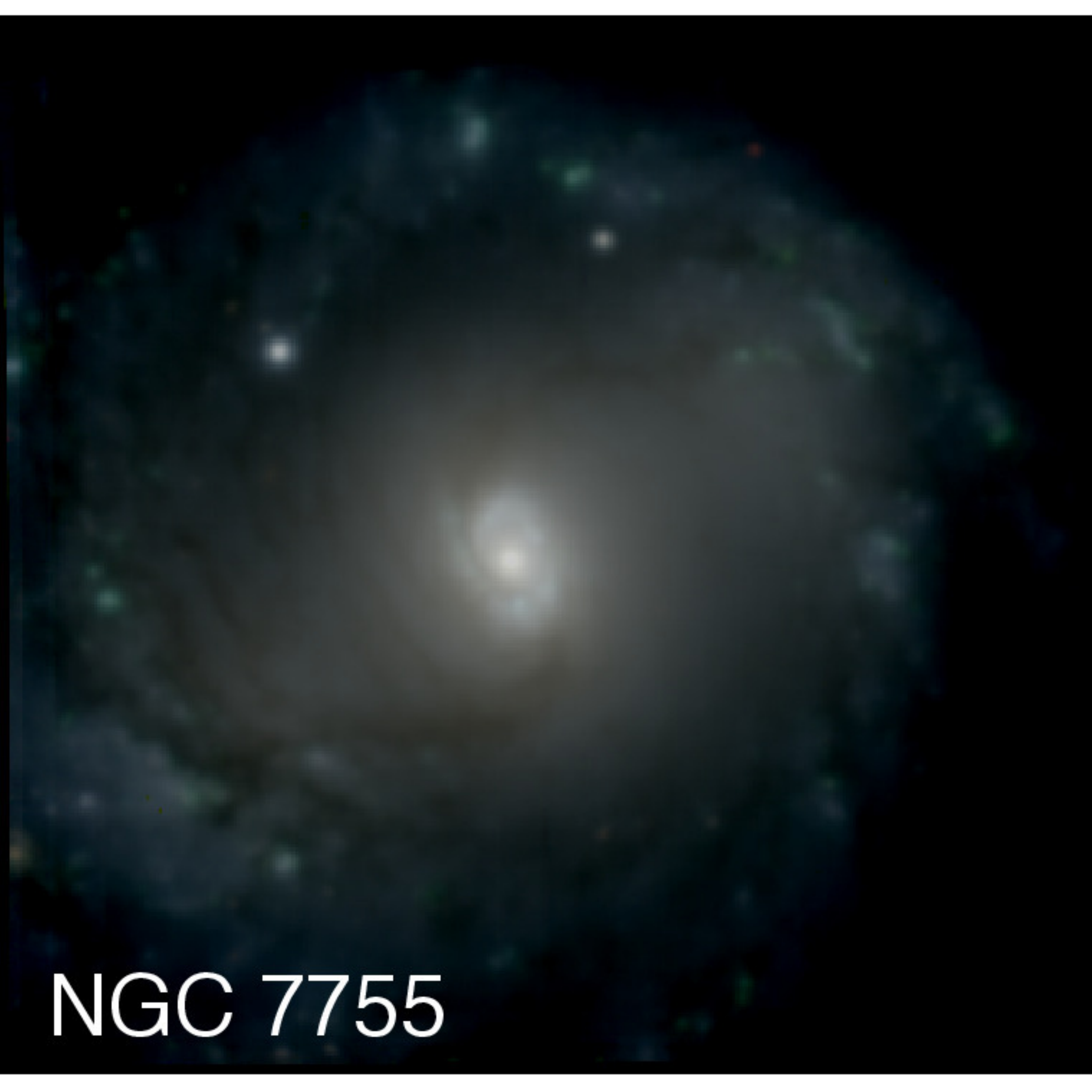}
	\includegraphics[width=0.5\columnwidth]{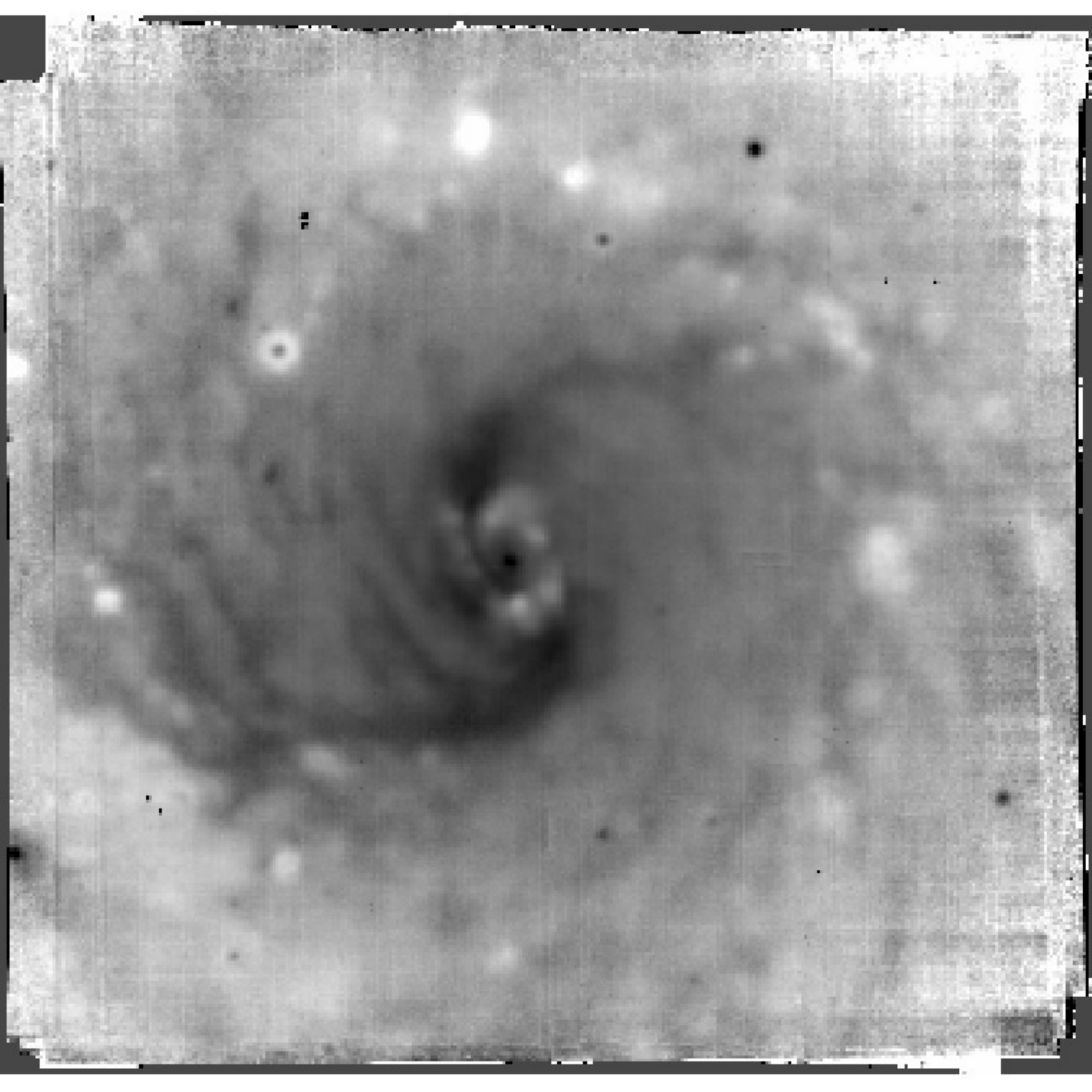}
\end{center}
    \contcaption{}
\end{figure*}

A clear understanding of the area covered by the TIMER fields is crucial to appreciate the extent of information gathered and thus the potential for scientific discovery. For this reason, we show in Fig.~\ref{fig:rgb_cm} colour composites and colour maps built directly from the TIMER data cubes. The colour composites are built using three false-colour images that we call blue, green and red, derived after collapsing the data cubes along the wavelength direction. The blue image covers from the shortest wavelength observed ({\it i.e.}, $\approx4750\,\AA$) until $\approx6000\,\AA$ and thus covers for example the [OIII] lines at $\lambda\lambda4959,5007\,\AA$. The green image covers from $\approx6000$ to $\approx7000\,\AA$, thus covering for example H$\alpha$ at $6563\,\AA$, the [NII] lines at $\lambda\lambda6549,6583\,\AA$ and the [SII] doublet at $\lambda\lambda6717,6731\,\AA$. Finally, the red image covers from $\approx7000$ to $\approx9000\,\AA$. The colour maps are simply built from subtracting the red image from the blue image. We purposely do not use any conventional filter response curve to create these false-colour images -- which would just throw away precious information --, so that instead these images convey all information contained in the MUSE data cubes (with the exception of the range $\approx9000-9300\,\AA$, which we purposely avoid due to the difficult background subtraction in this part of the spectrum).

The colour composites in Fig.~\ref{fig:rgb_cm} also clearly reveal the outstanding image capabilities and spatial sampling of MUSE, at least at these high levels of SNR, when the small-scale variations in the complex flat-fielding of the instrument have no noticeable effect. These colour composites were constructed with the aim at emphasising the central parts of the MUSE fields, and reveal a number of spectacular stellar structures, dust lanes and differences in the stellar population content. See NGC\,1097 and NGC\,3351 for examples of star-forming nuclear rings, NGC\,1433 for a star-forming inner disc, and the complex structure at the centre of NGC\,5236. We also note that, for a number of galaxies, the MUSE field covers the whole (or almost the whole) bar, such as in IC\,1438 and NGC\,4303. The colour maps in Fig.~\ref{fig:rgb_cm}, with a much more restricted dynamical range, reveal in some cases minor artefacts from flat-fielding, but in none of such cases this is a matter of concern considering our scientific objectives. Instead, such maps are extremely revealing. The typical dust lanes along the leading edges of the bar are clearly seen, as well as their entry points towards the nuclear structures they build: spectacular cases include NGC\,1097,  NGC\,1300 and NGC\,5728, where one can appreciate the dust lanes bending once they reach the nuclear structures. Even in the early-type disc galaxy NGC\,1291 the colour map reveals dust patches in the region of the nuclear disc and nuclear bar. The high SNR and detailed spatial sampling achieved with our MUSE exposures is also evident in our high-level data products discussed below.

\section{Stellar Kinematics}
\label{sec:kin}

\subsection{Determination of the line-of-sight velocity distributions}
\label{sec:kin_met}

In this paper we followed closely the steps outlined in \citet{GadSeiSan15}. In short, we used the penalised pixel fitting (pPXF) code developed by \citet{CapEms04} to extract the stellar kinematics, including the Gauss-Hermite higher-order moments \citep[{\it e.g.},][]{vanFra93}. The line-of-sight velocity distribution (LOSVD) was thus parametrised with the mean stellar velocity ($V_*$), stellar velocity dispersion ($\sigma_*$) and the h$_3$ and h$_4$ moments. We made use of the E-MILES model library of single age and single metallicity populations ({\it i.e.}, single stellar populations; SSPs) from \citet{VazCoeCas15} as stellar spectral templates, covering a wide range of ages, metallicities and [$\alpha$/Fe] abundance ratios to avoid template mismatch issues. The models, with a mean spectral resolution of $\approx$\,2.51\,\AA\ \citep{FalSanVaz11}, were broadened to the spectral resolution of the MUSE data ($\sim2.65\,\AA$) before the fitting process \citep[see, {\it e.g.},][]{KraWeiUrr15}. A non-negative linear combination of the SSP models convolved with the LOSVD was fitted to each individual spectrum. The best-fitting parameters were determined by $\chi^2$ minimisation in pixel space. We restricted our analysis to the rest-frame wavelength range between 4750 and 5500\,\AA\ for the minimisation, after checking that including the whole spectral range produces no noticeable differences in the derived kinematics. This restriction avoids adding effects from the masking of many emission lines at the longest wavelengths and complications from the spectral resolution dependence on wavelength. In the restricted range only a few potential emission lines were masked during the fitting procedure (H$\beta$, [OIII], [NI]). Additionally, a multiplicative low-order Legendre polynomial was included in the fit to account for small differences in the continuum shape between the galaxy spectra and the input library of synthetic models. For the derivation of stellar population properties (see Sect. \ref{sec:stelpop} below) the whole spectral range of the MUSE data was employed.

To guarantee a reliable kinematics extraction, the data were spatially binned using the Voronoi binning scheme by \citet{CapCop03}. Before binning, we selected only spaxels within the isophote level where the average SNR is larger than three. This cut ensures the removal of low-quality spaxels, which could introduce undesired systematic effects in our data at low surface brightness regimes. The data cubes were spatially binned to achieve an approximately constant SNR of 40 per pixel. Note however that the high quality of the data means that a large fraction of the original spaxels remains un-binned. The Voronoi-binned spectra were used not only in the derivation of the stellar kinematics but also in the derivation of the stellar population properties (see Sect. \ref{sec:stelpop} below).

\subsection{Spatial distributions of $V_*$, $\sigma_*$, h$_3$ and h$_4$}
\label{sec:kin_kin}

\begin{figure*}
\begin{center}
	\includegraphics[trim=1cm 0.5cm 1cm 1cm, clip=true, width=1.04\columnwidth]{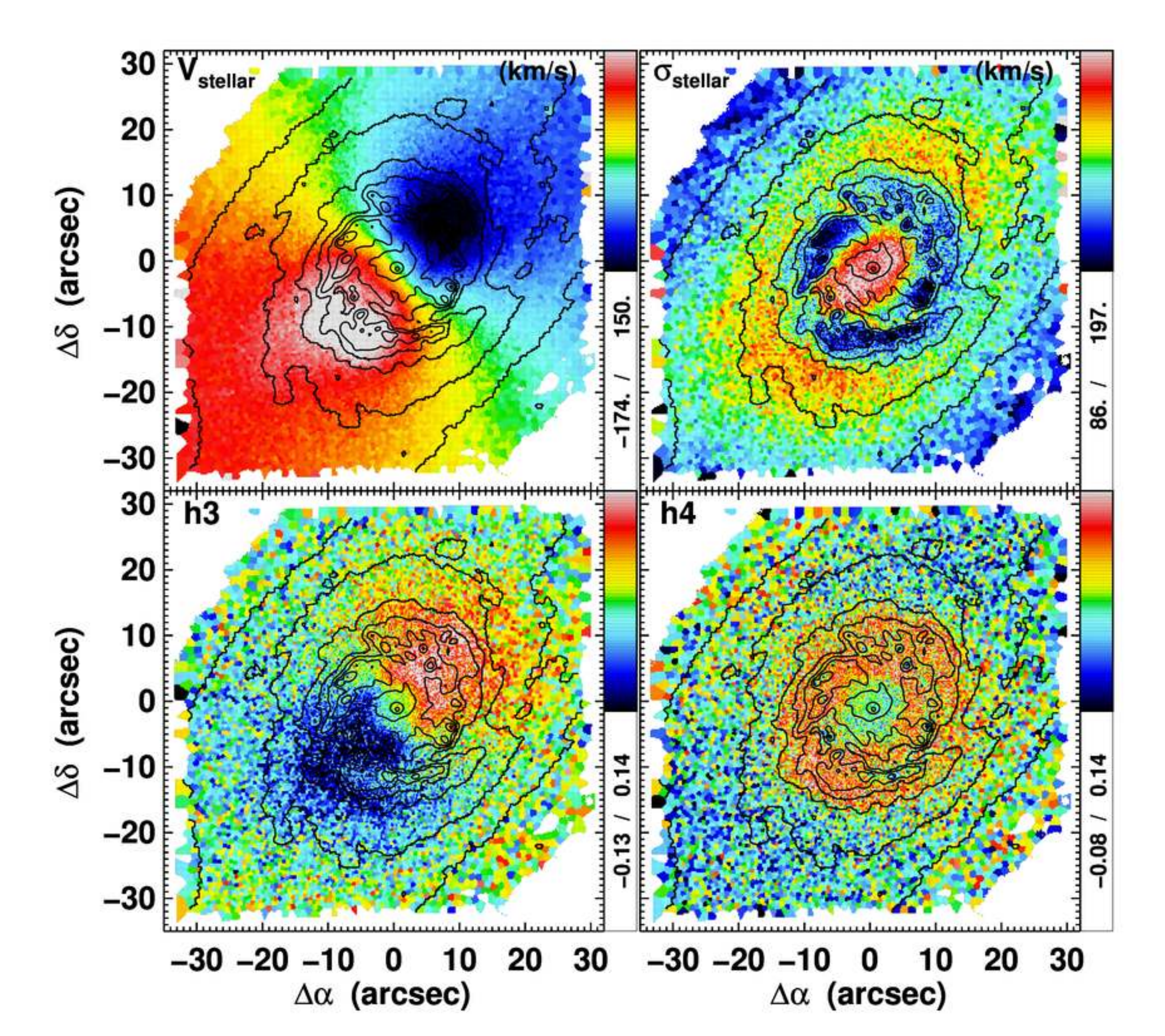}
	\includegraphics[trim=1cm 0.5cm 1cm 1cm, clip=true, width=1.04\columnwidth]{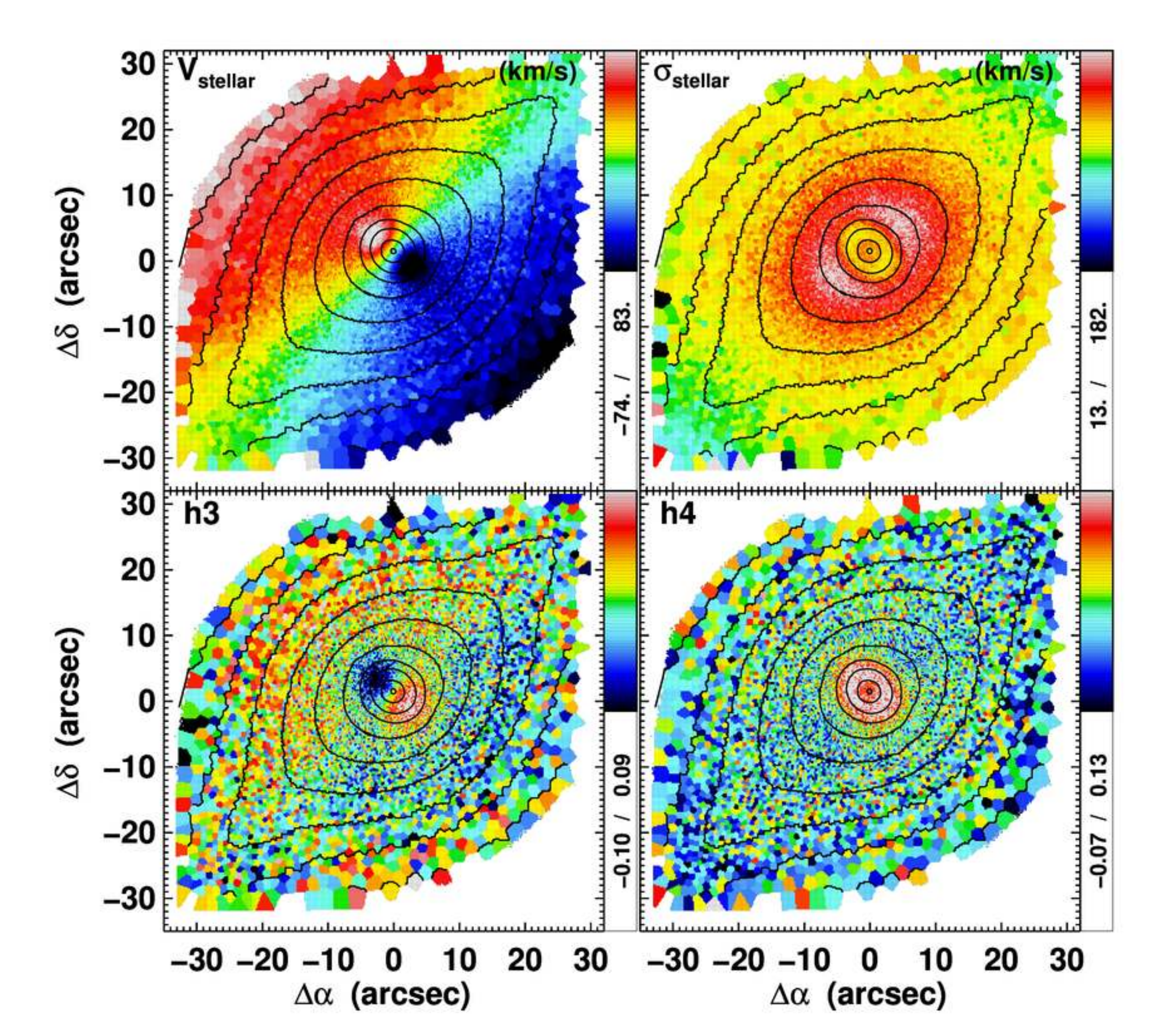}
\end{center}
    \caption{Radial velocity, velocity dispersion, h$_3$ and h$_4$ maps for the stellar component in NGC\,1097 (left) and NGC\,4643 (right), as indicated. The colour bars on the side of each panel indicate the plotted range of the parameter measured. For radial velocity and velocity dispersion these are given in km\,s$^{-1}$. The isophotes shown are derived from the MUSE data cube reconstructed intensities and are equally spaced in steps of about 0.5 magnitudes. At the distances shown in Table \ref{tab:sample}, $1\arcsec$ corresponds to $\approx100\,\rm{pc}$ for NGC\,1097 and $\approx125\,\rm{pc}$ for NGC\,4643. North is up, east to the left.}
    \label{fig:kin}
\end{figure*}

To illustrate the quality and potential of the TIMER dataset, in this and in the next sections we will present part of our high-level data products for two galaxies. For results concerning stellar dynamics and stellar populations, we chose NGC\,1097 and NGC\,4643 as examples of the different galaxy properties covered in the TIMER sample. The former shows a conspicuous star-forming nuclear ring, whereas the latter has currently a quiescent central region. In addition, our MUSE field of NGC\,1097 shows regions with strong dust absorption, whereas our MUSE field of NGC\,4643 shows a much less conspicuous presence of dust (see Fig.~\ref{fig:rgb_cm}).

Stellar kinematics maps derived for NGC\,1097 and NGC\,4643 are shown in Fig.~\ref{fig:kin}. The richness in spatial detail is immediately evident, even with Voronoi bins with a minimum SNR of 40. For example, while the conspicuous rapidly-rotating inner disc in NGC\,1097 is evident in the radial velocity map, the much more subtle inner disc in NGC\,4643 is clearly distinct also as rapidly rotating. Much detail and differences between those two inner discs are seen in the velocity dispersion maps. That of NGC\,4643 shows a uniform distribution of $\sigma_*$ with lower values than the immediate surroundings. In contrast, the inner disc of NGC\,1097 interestingly shows an inner region ($\approx10\arcsec$ from the centre) with low $\sigma_*$ and an outer region ($\approx15\arcsec$ from the centre) of more elevated $\sigma_*$ values. The centre of NGC\,1097 shows even higher $\sigma_*$ values. There is a clear structural difference between these two inner discs that may be associated to the differences seen in their $\sigma_*$ maps, namely the fact that NGC\,1097 shows strong nuclear spiral arms. Two regions of elevated $\sigma_*$ values are seen {\em outside} the inner disc of NGC\,4643, along the major axis of the bar (which is oriented diagonally across the field, from the bottom-left corner to the top-right corner), and may indicate the presence of a box/peanut \citep[see discussion in][for the case of NGC\,4371]{GadSeiSan15}.

However, both inner discs present an anti-correlation between $V_*$ and h$_3$, and this holds even in the regions with elevated $\sigma_*$ values of the inner disc in NGC\,1097. This anti-correlation is a signature of near-circular orbits. Conversely, in the bar-dominated region of NGC\,4643 one clearly sees a {\em correlation} between $V_*$ and h$_3$, which is a signature of orbits with high eccentricity, such as the x1 orbits that are the backbone of bars. The h$_4$ maps of both galaxies also show similar properties for the inner discs, {\it i.e.}, both structures are clearly regions of elevated h$_4$ values (although the centre of NGC\,1097 shows lower values of h$_4$, as opposed to the centre of NGC\,4643). High values of h$_4$ suggest the superposition of structures with different LOSVDs \citep[see, {\it e.g.},][and references therein]{BenSagGer94}. Interestingly, the outer regions of the inner disc of NGC\,1097, where $\sigma_*$ is high, show intermediate values of h$_4$. Further discussion on these maps, combined with additional high-level data products, will be presented in Sect. \ref{sec:sci}.

\section{Stellar Ages, Metallicities and Star Formation Histories}
\label{sec:stelpop}

\subsection{Extraction of stellar population parameters}
\label{sec:stelpop_met}

\begin{figure*}
\centering
\hspace{0.4cm}
\includegraphics[width=\columnwidth]{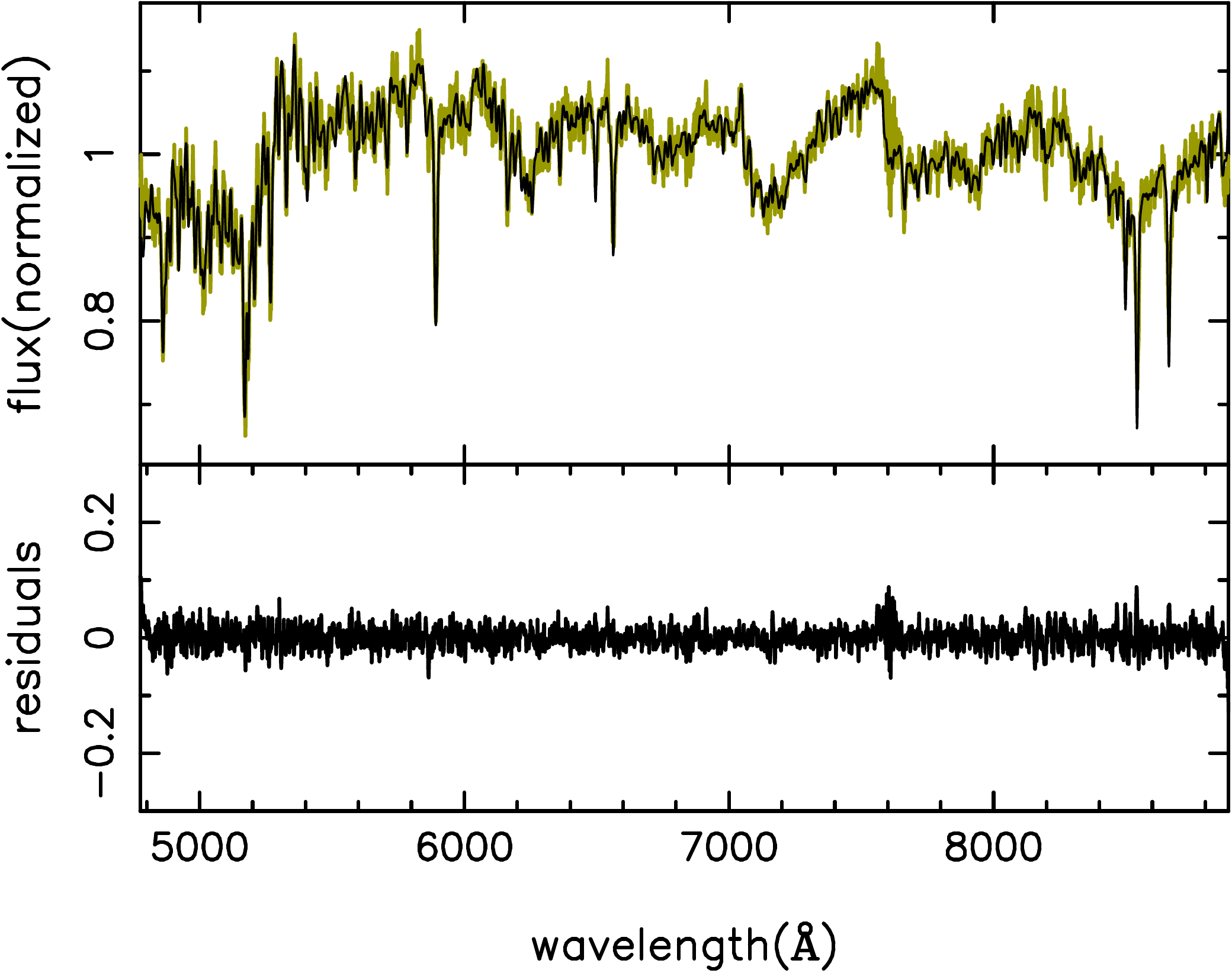}\hskip0.3cm
\includegraphics[width=0.99\columnwidth]{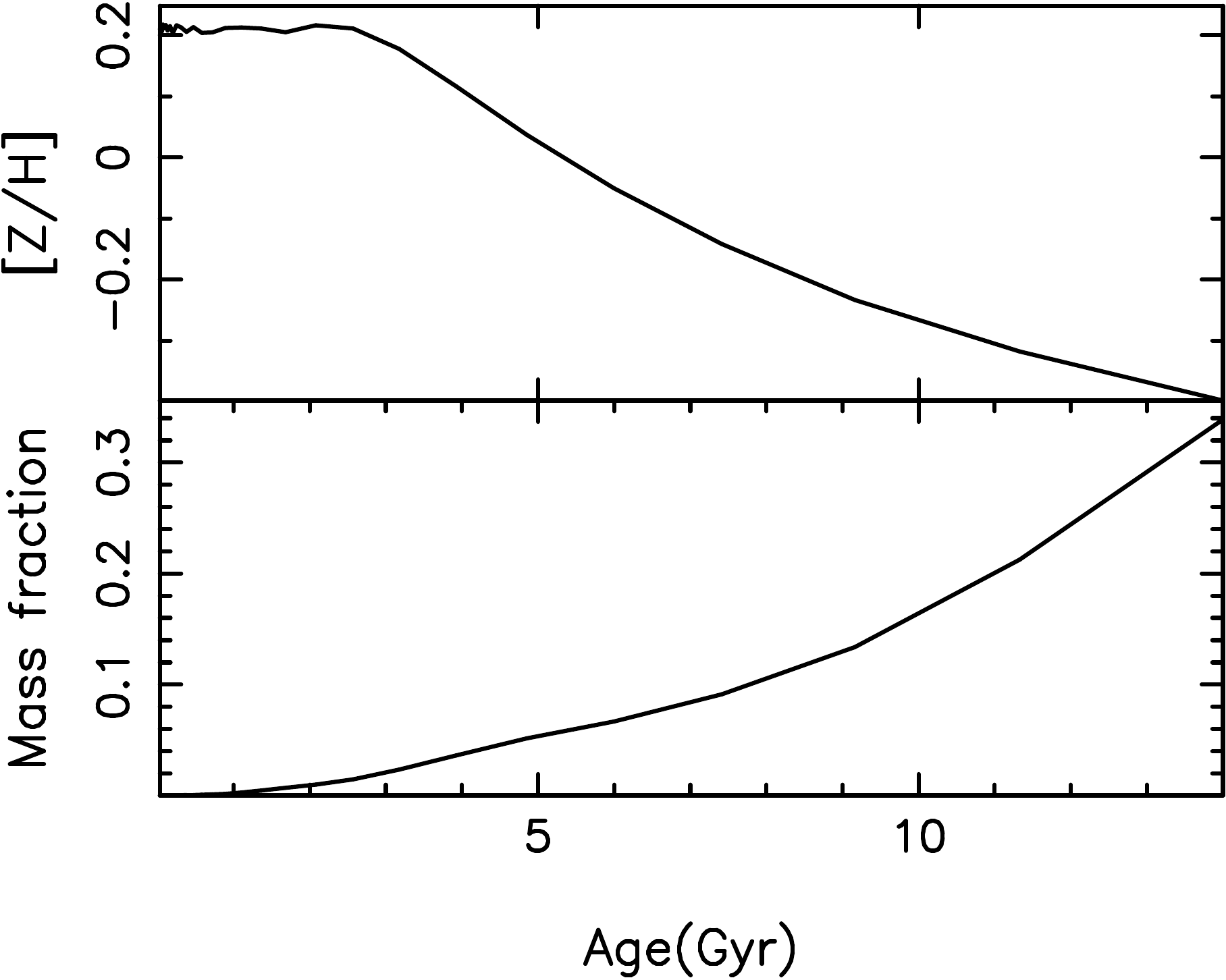}
\caption{Left: observed (green) and fitted (black) spectrum from one of the Voronoi bins of NGC\,4643 (with SNR $=40$; upper panel) and residuals from the fit (bottom panel). Right: stellar age distribution (in mass fraction; bottom panel) and age-metallicity relation (upper panel) for the same spectrum.}
\label{fig:steckmap_fit}
\end{figure*}

While stellar absorption features are dominant in the wavelength range under study, the typical spectra from our sample of galaxies contain significant amounts of nebular emission. This emission has to be removed for a meaningful stellar population study. We used the code {\tt GANDALF} \citep{SarFalDav06,FalBacBur06} for this purpose. It searches iteratively for the emission-line velocities and velocity dispersions, while linearly solving at each step for the emission-line amplitudes and the optimal combination of the stellar templates. The stellar kinematics is held fixed during the fitting process to minimise degeneracies. With {\tt GANDALF} we thus produced models of the nebular emission for the spectra from each of the Voronoi bins described above. The nebular emission was thus removed so that the spectra could be further processed.






With the emission lines removed, we derived stellar ages, metallicities and star formation histories using the code {\tt STECKMAP} \cite[STEllar Content and Kinematics via Maximum a Posteriori;][]{OcvPicLan06a,OcvPicLan06b}, along with the E-MILES stellar population models \citep{VazCoeCas15}. The models cover a wavelength range between 3645 and 9469\,\AA, allowing us to use the full spectral range of MUSE. We employed the BASTI isochrones \citep{PieCasSal04,PieCasSal06,PieCasSal09,PieCasSal13} with a range of stellar ages spanning the interval $0.03-14.0\,\rm{Gyr}$ and metallicities Z from 0.0001 to 0.05, equivalent to a [Z/H] ranging from -2.3 to 0.4; in addition, we assumed a \citet{Kro01} Initial Mass Function (IMF). {\tt STECKMAP} has been used to derive stellar population parameters with other datasets, and the results compared with those from other methods, including results from colour-magnitude diagrams of resolved stellar populations, showing excellent agreement \citep[see, {\it e.g.},][]{SanOcvGib11,RuiPerGal15}.

{\tt STECKMAP} projects the observed spectrum onto a temporal sequence of single stellar population models to determine the linear combination of these models that fits the observed spectrum best. The weights of the various components of this linear combination indicate the stellar population content. The reconstruction of the stellar age distribution and age-metallicity relation by maximising the likelihood term only is a discrete, ill-conditioned problem ({\it i.e.}, small variations in the initial conditions can lead to large variations in the final solution). It thus requires regularisation so that the solutions obtained are physically meaningful. In practise, this is achieved in {\tt STECKMAP} by adding a penalisation function to the function to be minimised, that is:
\[
Q_{\mu} = \chi^2 [s({\rm{age}}, Z)] + P_{\mu}({\rm{age}},Z),
\]
where $s$ represents the synthetic spectrum for a single stellar population of a given age and metallicity ($Z$), $P_{\mu}$ is the penalising function, and $\mu \equiv (\mu_X,\mu_Z)$ represents the smoothness parameters for age ($\mu_X$) and metallicity ($\mu_Z$). The penalising function has small values for solutions with smooth variations of age and $Z$, and large values for solutions with overly irregular (jagged) variations of age and $Z$. The smoothness parameters for both the age and the metallicity ($\mu_X$ and $\mu_Z$, respectively) need to be set as a compromise between not over-interpreting the solution ({\it i.e.}, with small values for $\mu$) and not losing information contained in it (with large values). Appropriate values for the smoothness parameters depend on the quality of the data, in particular the SNR and spectral resolution. To determine the appropriate values for this study, we thus run {\tt STECKMAP} on artificial spectra with similar SNR and resolution as our TIMER spectra, and verify for which values of the smoothness parameters the code reproduces best the known stellar population parameters of the artificial spectra.

The continuum is normalised by multiplying the model by a smooth non-parametric transmission curve. This curve has 30 nodes spread uniformly along the wavelength range, and is obtained by spline-interpolating between the nodes. The latter are treated as additional parameters and adjusted during the minimisation procedure.

Before running {\tt STECKMAP} on the spectrum from each Voronoi bin, we corrected resolution differences across the wavelength range by convolving our MUSE spectra, as well as the template spectra from the E-MILES library, with a Gaussian broadening function with a wavelength-dependent width. We run {\tt STECKMAP} with the stellar kinematics held fixed to avoid possible degeneracies between the stellar velocity dispersion and metallicity \citep[{\it e.g.},][]{SanOcvGib11}. Figure~\ref{fig:steckmap_fit} illustrates a typical fit together with the obtained solutions for the mass fraction as a function of stellar age and the age-metallicity relation.

In Appendix \ref{sec:app_err}, we present a detailed analysis on the uncertainties in the derivation of mean stellar ages and metallicities from the star formation histories produced with {\tt STECKMAP} with our spectra. Using 5\,000 TIMER spectra, we show that the choice for input and smoothness parameters in {\tt STECKMAP} has no effect on our results. In addition, we show how the uncertainty in mean stellar age and metallicity depends on age and metallicity for luminosity-weighted and mass-weighted measurements. Typical values for these uncertainties are $0.5-1$ Gyr for age, and $0.005-0.010$ for metallicity ($Z$).

\subsection{Spatial and temporal distributions of stellar age and metallicity}
\label{sec:stelpop_div}

\begin{figure*}
\begin{center}
	\includegraphics[trim=1cm 0.5cm 1cm 0.2cm, clip=true, width=1.04\columnwidth]{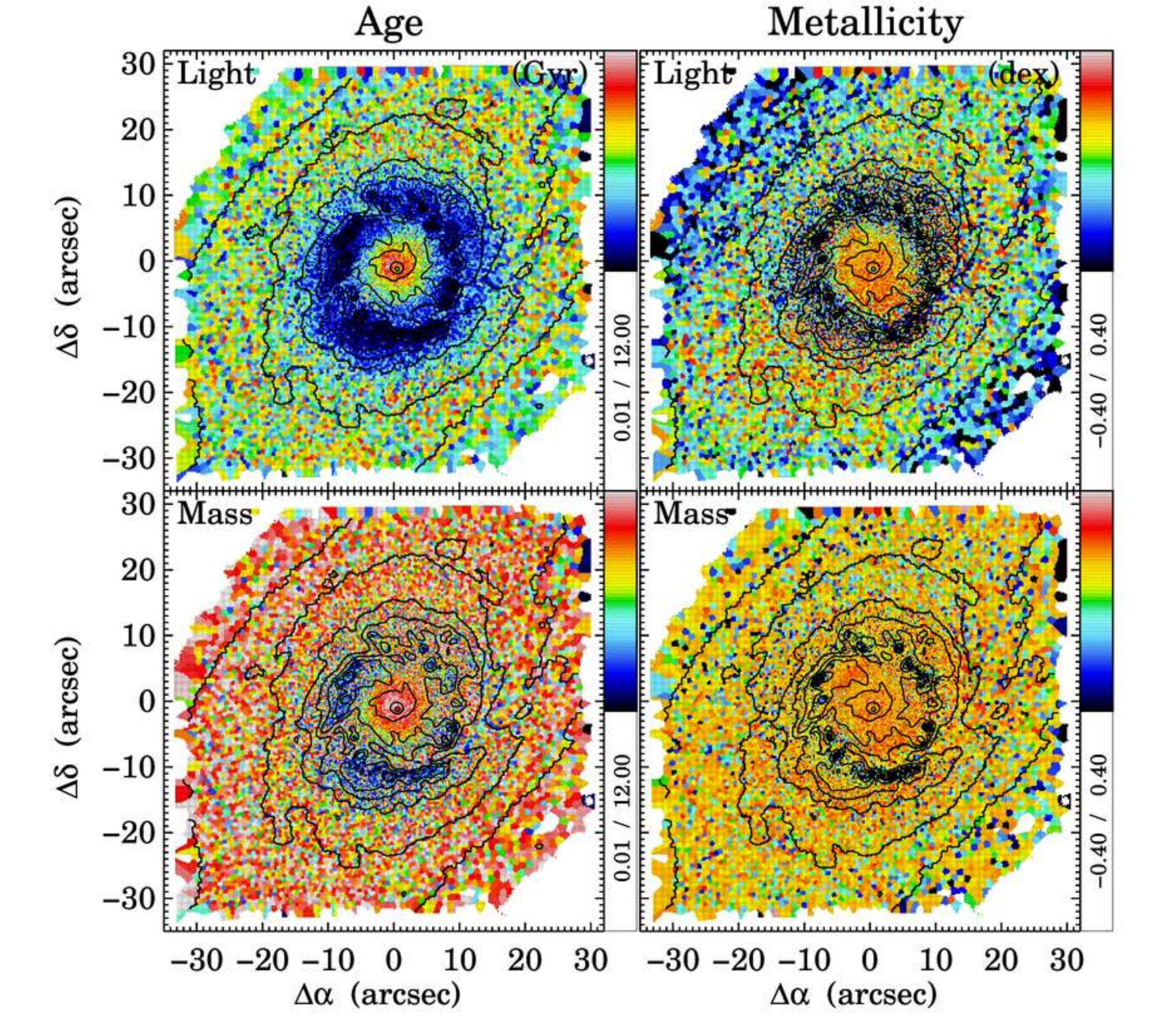}
	\includegraphics[trim=1cm 0.5cm 1cm 0.2cm, clip=true, width=1.04\columnwidth]{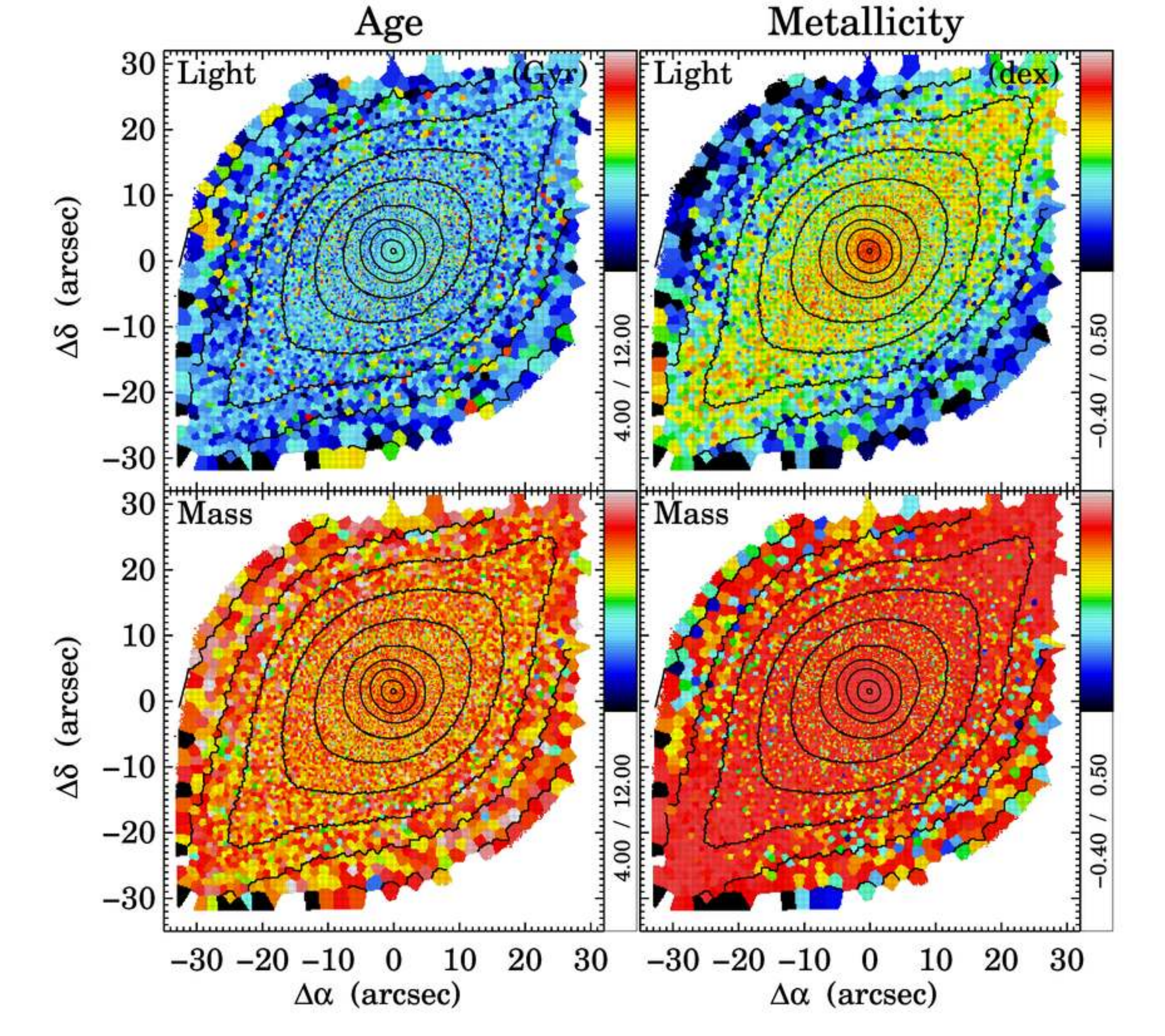}
\end{center}
    \caption{Maps for NGC\,1097 (left) and NGC\,4643 (right) of mean stellar age and mean stellar metallicity weighted by both luminosity and mass, as indicated. The age maps have units of 1\,Gyr, and metallicities are given in the spectroscopic notation (logarithmic scale normalised to the solar value). The isophotes shown are derived from the MUSE cube reconstructed intensities and are equally spaced in steps of about 0.5 magnitudes. At the distances shown in Table \ref{tab:sample}, $1\arcsec$ corresponds to $\approx100\,\rm{pc}$ for NGC\,1097 and $\approx125\,\rm{pc}$ for NGC\,4643. North is up, east to the left.}
    \label{fig:agemet}
\end{figure*}

\begin{figure*}
\begin{center}
	\includegraphics[width=0.51\columnwidth]{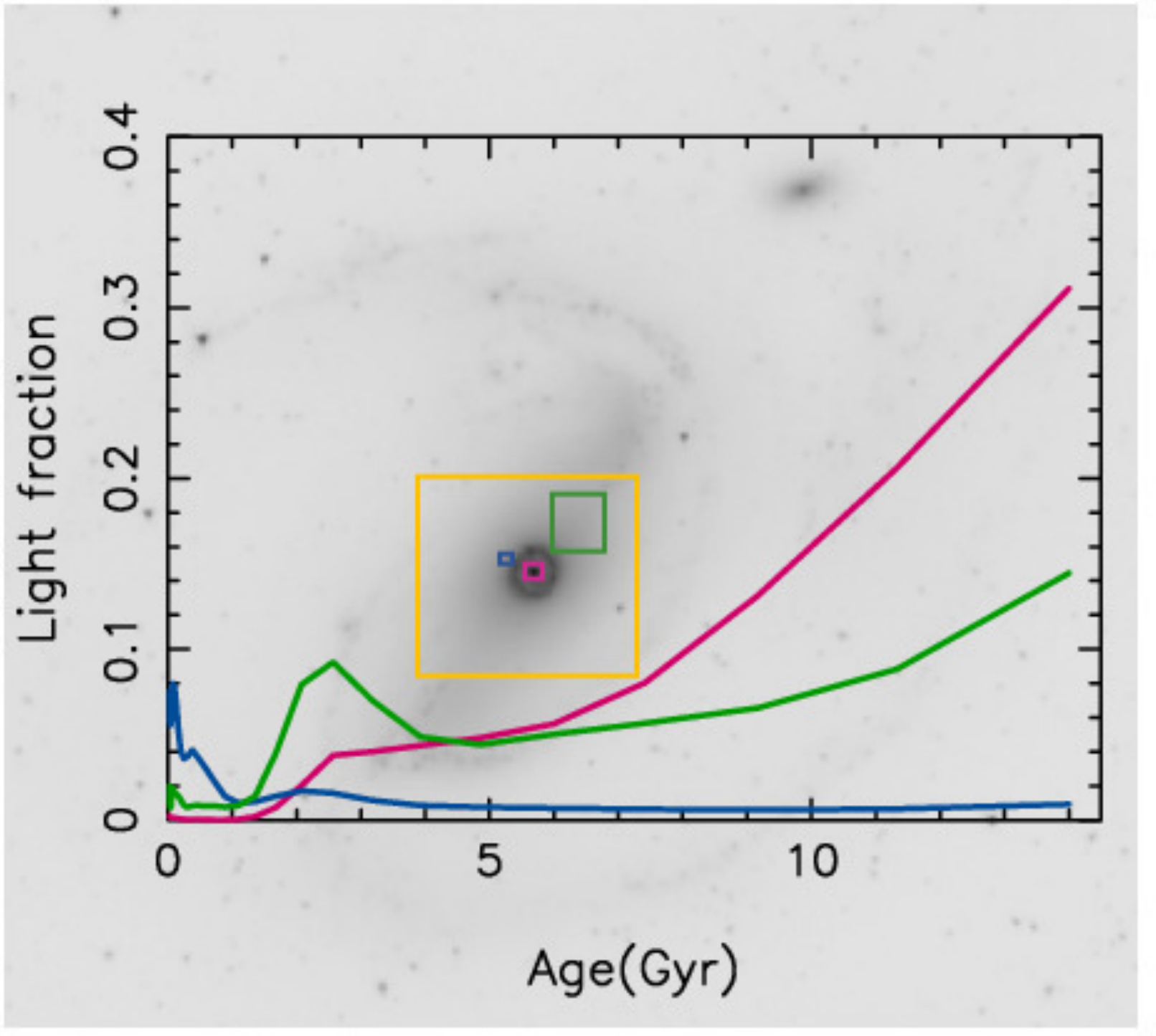}
	\includegraphics[width=0.51\columnwidth]{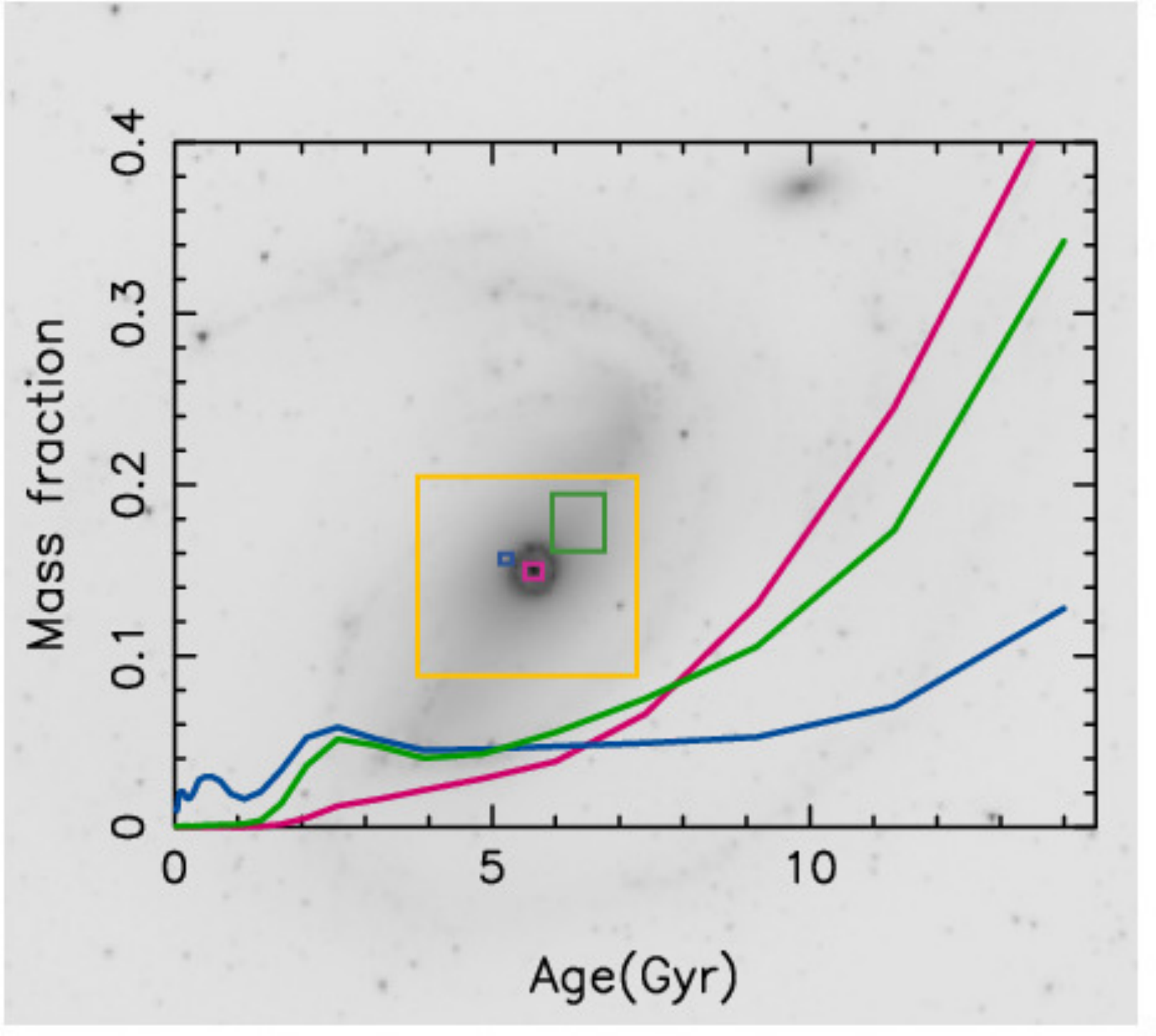}
	\includegraphics[width=0.51\columnwidth]{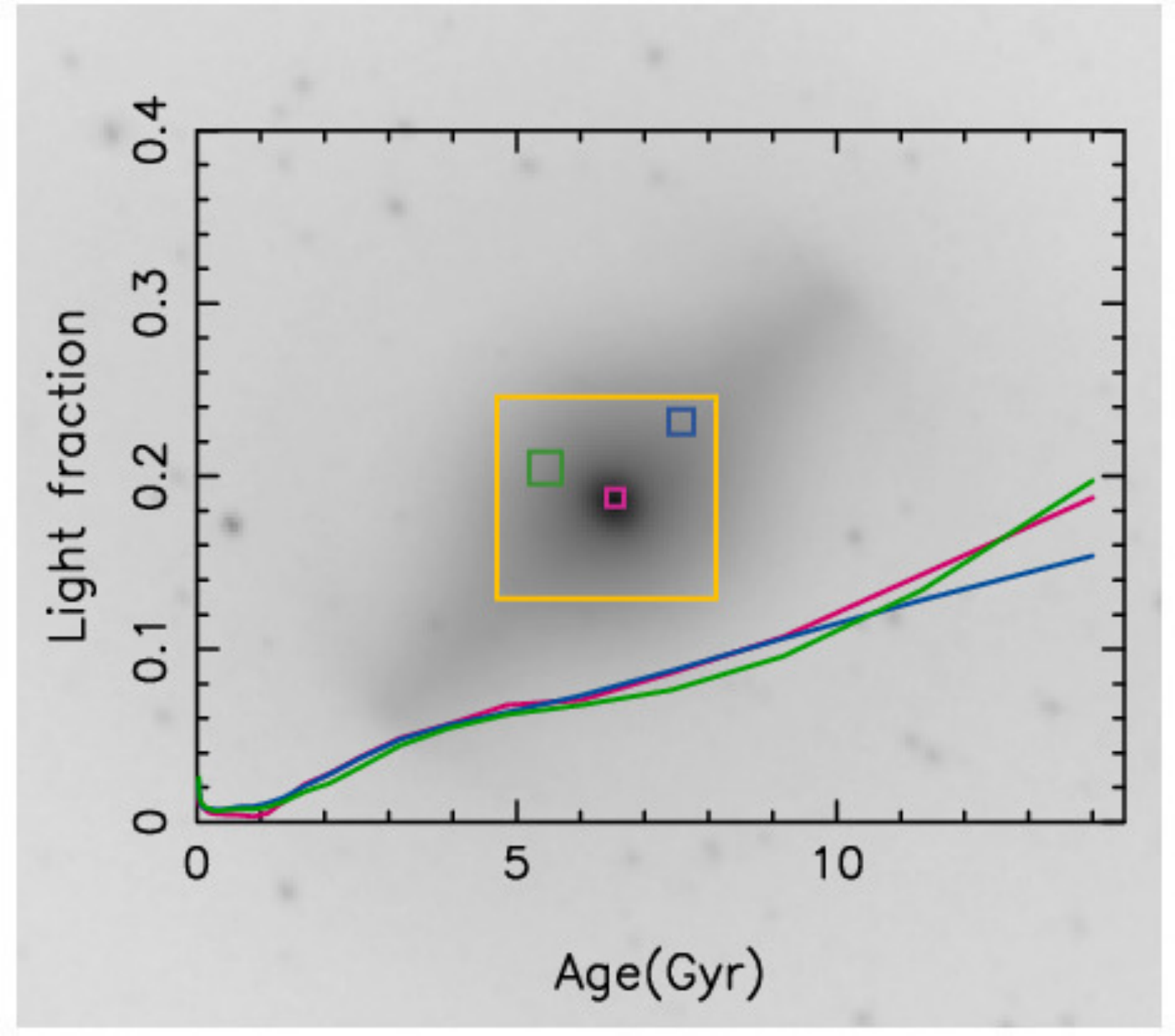}
	\includegraphics[width=0.51\columnwidth]{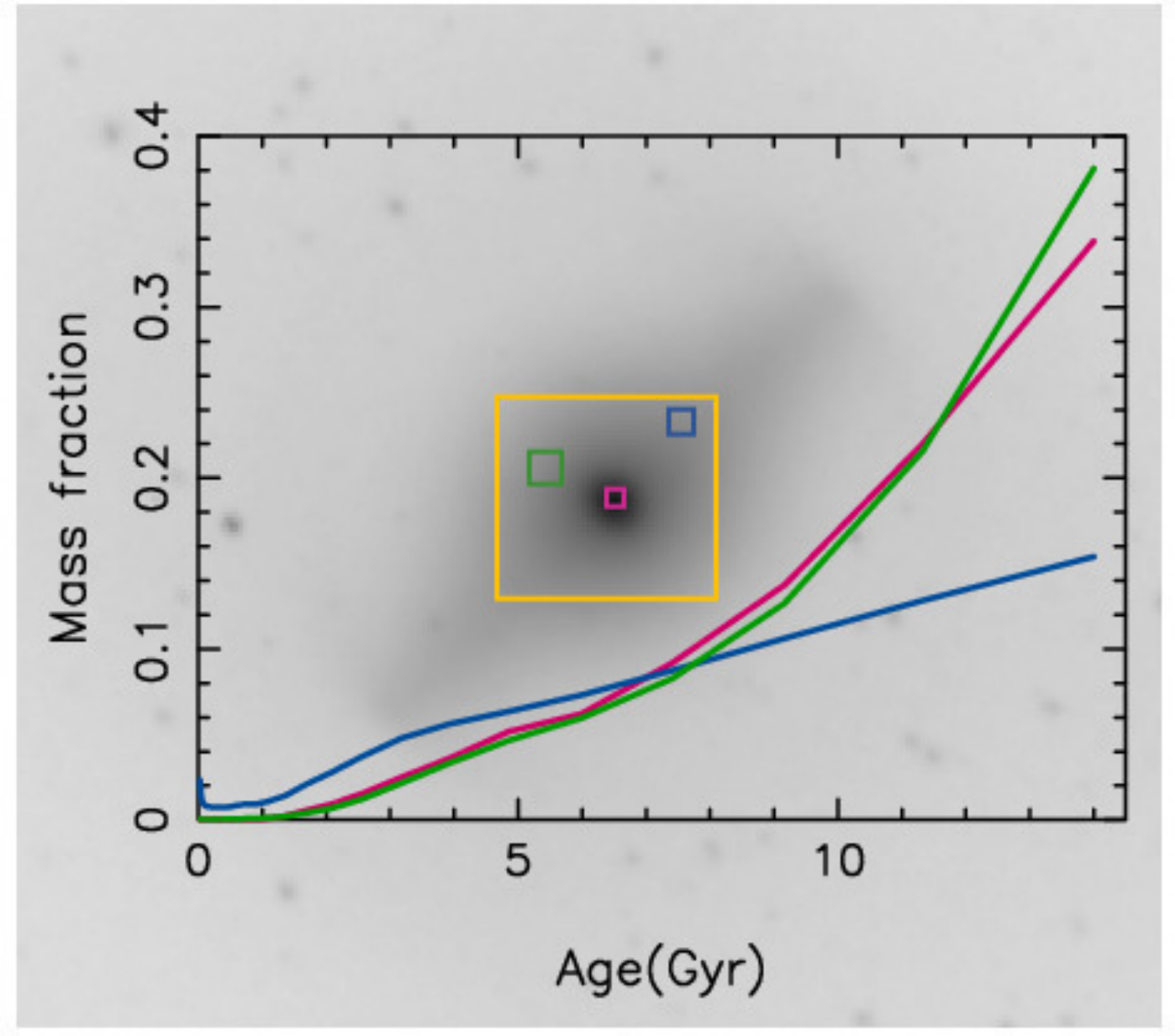}
	\includegraphics[width=\columnwidth]{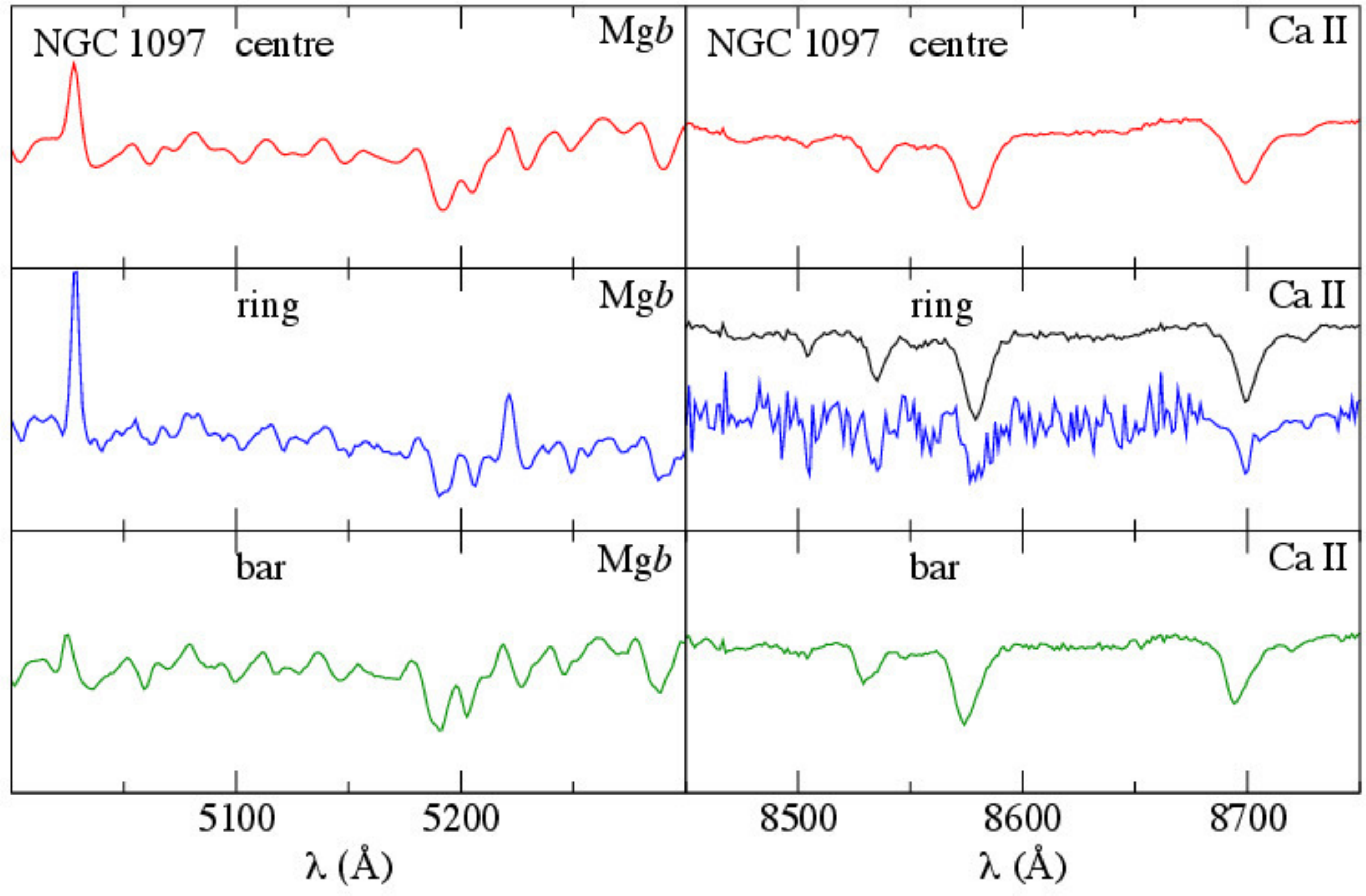}\hskip0.37cm\includegraphics[width=\columnwidth]{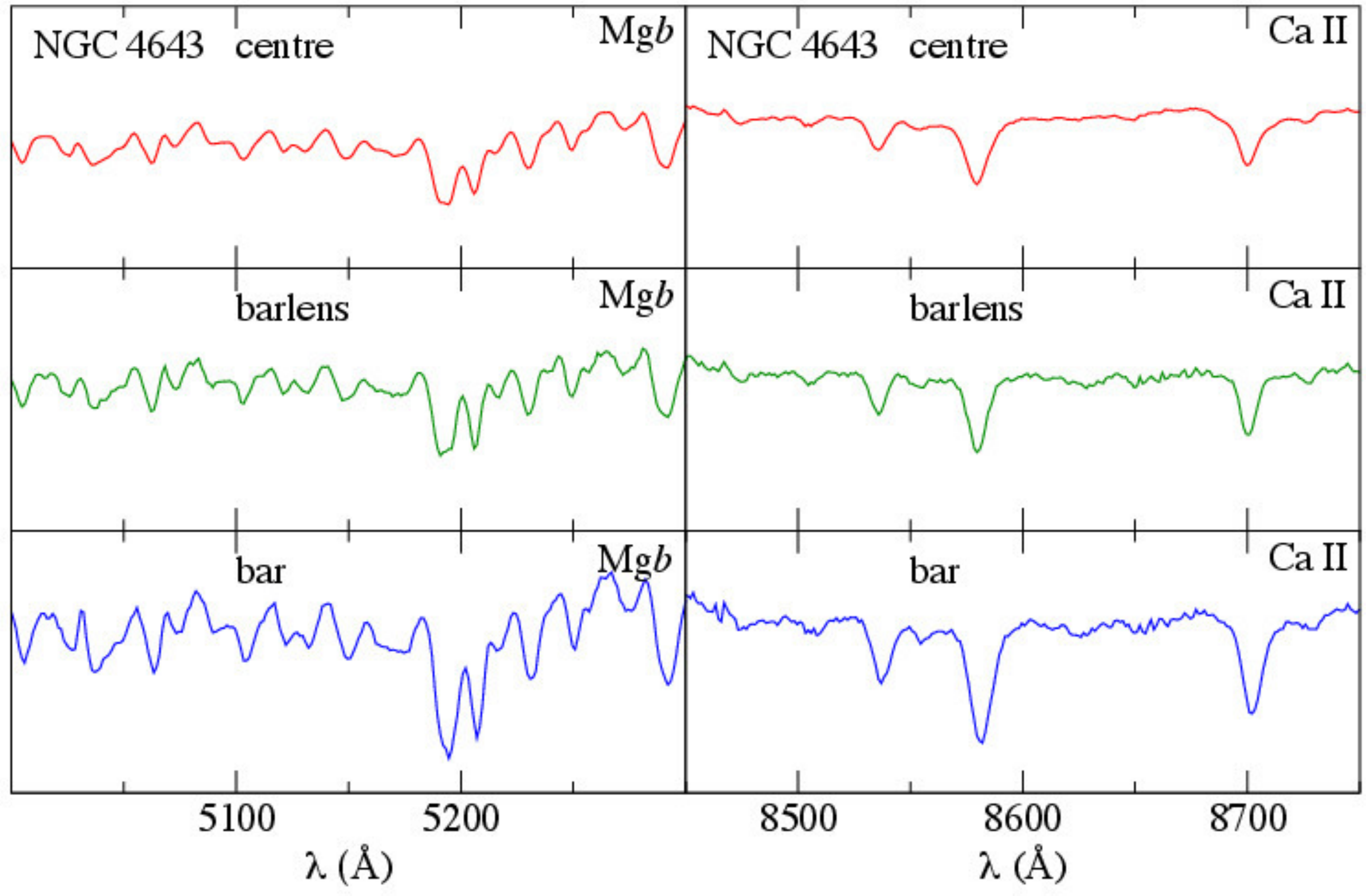}
\end{center}
    \caption{\textit{Top row:} Star formation histories in different regions of NGC\,1097 (left panels) and NGC\,4643 (right panels) weighted by both luminosity and mass, as indicated. These are built from the combined spectra within each region. In each panel, our MUSE field is the yellow square overlaid on a S$^4$G 3.6$\,\mu$m image. Selected regions of the galaxies are shown within the different coloured rectangles, and their corresponding star formation histories are described by the lines with the same colour. These star formation histories correspond to the fraction of the stellar light (left) or mass (right) within each region as a function of stellar age. \textit{Bottom panels:} For both galaxies we show the combined MUSE spectra from the three different regions (with the same colour coding) in the spectral regions around the Mgb feature and the CaII triplet, as indicated. Note that the CaII triplet region in the spectrum from the nuclear ring in NGC\,1097 is heavily affected by Paschen lines, since this region coincides with a spot with very intense star formation and elevated dust content \citep[see][]{CenCarGor01}. The spectrum shown in black in the corresponding panel is taken from a region in the immediate vicinity and with the same size, and it does not show this effect.}
    \label{fig:sfhs}
\end{figure*}

For each Voronoi bin across the entire dataset we obtained a solution such as the one in Fig.~\ref{fig:steckmap_fit}, which therefore allows us to derive the spatial distribution of mean stellar ages, metallicities and star formation histories for all TIMER galaxies. In Fig.~\ref{fig:agemet} we show maps of the mean stellar age and metallicity (weighted by both luminosity and mass) for NGC\,1097 and NGC\,4643.

NGC\,1097 shows a very conspicuous star-forming ring with very young bright stars. Even in the mass-weighted map in Fig.~\ref{fig:agemet} the nuclear ring is distinctively younger than its surroundings as well as the very centre of the galaxy. Interestingly, the region surrounding the nuclear ring has a relatively smooth age distribution, but there is a clear offset in age between the luminosity-weighted and mass-weighted maps. While the typical mean stellar age in this region in the luminosity-weighted map is around 6\,Gyr, that in the mass-weighted map is around 10\,Gyr, showing how indeed low-mass old stars dominate the mass budget in this region of the galaxy.

The maps of NGC\,4643 tell a different story. There are no spatial bins with a mean stellar age below 3\,Gyr. In addition, the nuclear structure built by the bar is hardly discernible, which indicates that the central structures in this galaxy are coeval.
The metallicity maps show that the bar is clearly more metal-rich than the surrounding region, and that the very centre is, as for NGC\,1097, the most metal-rich stellar component.

Figure~\ref{fig:sfhs} shows luminosity-weighted and mass-weighted star formation histories for different regions in our data cubes of NGC\,1097 and NGC\,4643. The data show that the nuclear ring in NGC\,1097 had two additional star-forming bursts apart from the one that is ongoing: one $\sim0.5$\,Gyr ago and another $\sim2.5$\,Gyr ago. This may indicate the times of different events of external gas accretion that led the bar to bring gas to the central region. Another possibility to explain these two peaks in star formation without invoking accretion of external gas is to consider the growth of the bar. Numerical simulations suggest that bars grow longer and stronger with time \citep[see, {\it e.g.},][]{AthMis02,AthMacRod13}, which is however challenging to establish observationally \citep[but see][]{Gad11}. An increasing bar length implies that an increasingly larger fraction of the disc is affected by the torques that remove angular momentum from the gas, thus enhancing the gas inflow through the bar. One would expect that this process is continuous though, so it is not clear if it can be the cause of the more recent starbursts in the nuclear ring. Interestingly, the bar region also shows a star-forming burst $\sim2.5$\,Gyr ago. Nevertheless, a fraction of the stars in the region dominated by the nuclear ring was formed over 10\,Gyr. It remains to be shown whether these stars are part of the nuclear ring or part of the underlying disc or bar. Consistent with the analysis above, Fig.~\ref{fig:sfhs} shows that most stars in the centre were formed over 10\,Gyr ago.

Figure~\ref{fig:sfhs} also shows that, consistently with the analysis above, the stellar structures in the central region of NGC\,4643 are coeval, at least in the luminosity-weighted star formation histories. Most of the stars were formed over 10\,Gyr ago. The mass-weighted star formation histories show fewer old stars in the bar region than in the other two regions included in these analyses. We discuss these results further in Sect. \ref{sec:sci}.

\section{Emission Lines}
\label{sec:emli}

\subsection{Derivation of physical parameters}
\label{sec:emli_met}

\begin{figure*}
\includegraphics[width=0.333\textwidth]{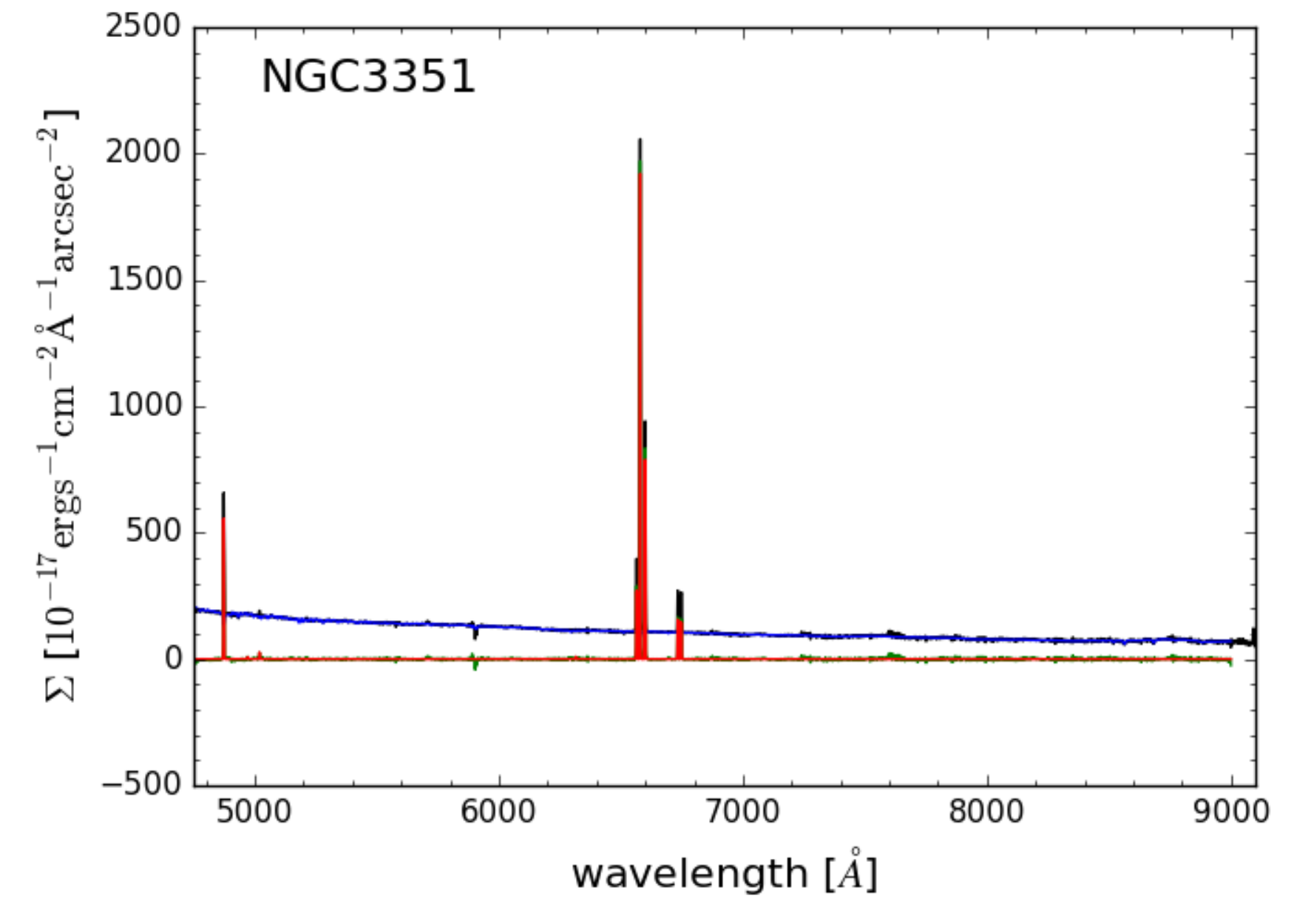}\includegraphics[width=0.333\textwidth]{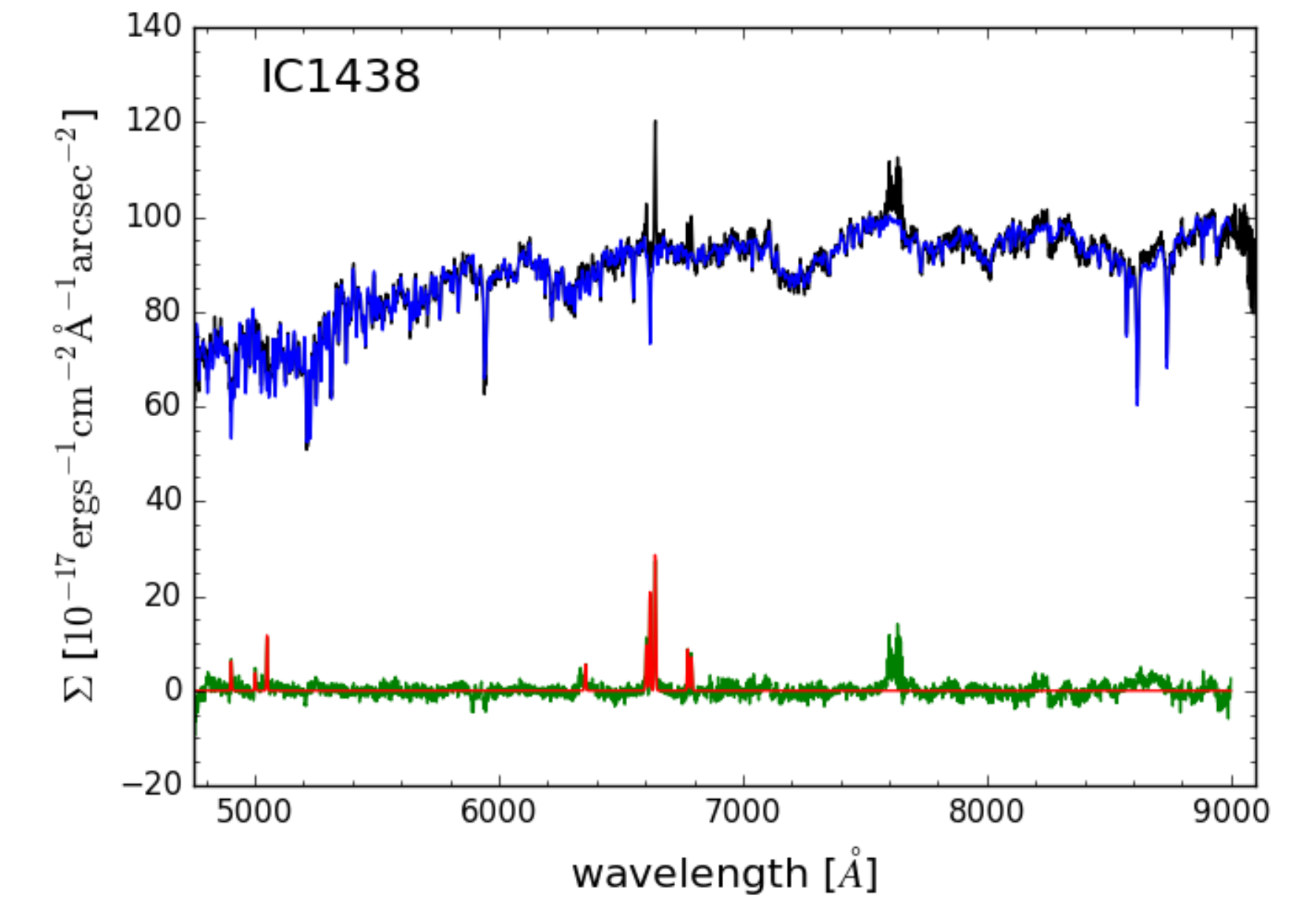}\includegraphics[width=0.333\textwidth]{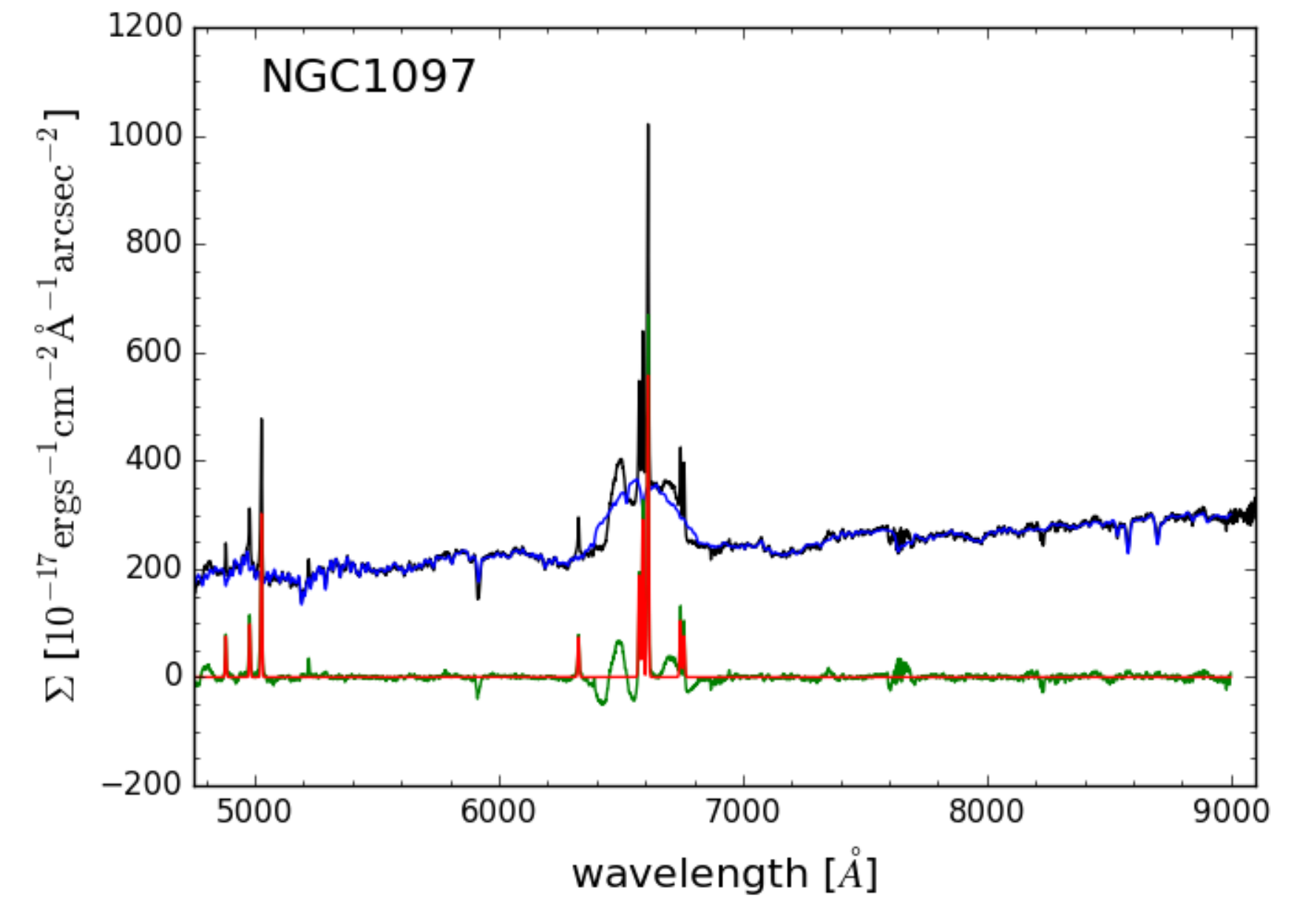}
\caption{Examples of best-fit results from the {\tt PyParadise} spectral modelling for three TIMER galaxies. In each panel we show the observed spectrum (black line), the stellar spectrum model (blue line), the residual after subtracting from the observed spectrum the stellar spectrum only (green line), and the emission line model (red line).}
\label{fig:example_fit}
\end{figure*}

To extract accurate emission line information, we need to subtract the combined stellar population spectrum to recover emission line fluxes even if trapped in prominent absorption lines, such as H$\beta$. While we used {\tt GANDALF} to model and remove emission lines in order to derive stellar population properties from our spectra (see Sect. \ref{sec:stelpop} above), in this work we opted to use {\tt PyParadise} for our study on the emission line properties. {\tt PyParadise} is an extended python version of Paradise \citep[see][]{WalCoeGal15} and iteratively combines the stellar spectrum fit and emission line measurements. One of the benefits of using {\tt PyParadise} is that template mismatches in the fitting of the stellar spectra are propagated in the derivation of the uncertainties in the emission line measurements. In the following we briefly outline our procedure.

The goal is to map the emission line properties on the native spatial sampling that MUSE provides ({\it i.e.}, 0.2\arcsec -- typically 20\,pc for the TIMER galaxies), but spatial binning needs to be applied to retrieve a robust measure of the stellar kinematics. This is because the signal in the emission lines is stronger than that in the stellar spectra typically by about an order of magnitude. In addition, the spatial distribution of the emission line flux is characterised by much more fine structure than that of the stellar light, so spatially binning the emission line signal leads to the loss of important spatial information on the distribution of emission line flux and kinematics. On the other hand, uncertainties in the {\em stellar} spectra are propagated to the derived emission line properties, which further motivates our spatial binning in the measurement of the stellar kinematics.

Therefore, in the first step of our procedure, the approach we followed is to estimate the stellar kinematics from large Voronoi bins and use the kinematic maps on a spaxel-by-spaxel basis to find the matching stellar spectral model that is then subtracted before fitting the emission lines. For designing the Voronoi bins we required a SNR of $\sim40$ per pixel in the continuum, and estimated the stellar velocity $V_*$ and velocity dispersion $\sigma_*$ separately in each bin. The higher-order moments h$_3$ and h$_4$ were not taken into account to speed up the process. {\tt PyParadise} estimates the stellar kinematics in an iterative way\footnote{We do not use the stellar kinematics parameters as obtained above with pPXF to keep the derivation of the emission line properties entirely self-consistent, but we do not expect significant differences.}. Firstly, {\tt PyParadise} normalises the continuum in the observed spectrum -- and in the template library \citep[here we used the Indo-U.S. Library of Coud\'e Feed Stellar Spectra;][]{ValGupRos04} -- with a running average 100 pixels wide, after linearly interpolating regions with emission lines or residuals of sky lines. After that, one template spectrum is taken from the template library and the best fitting kinematics are estimated using an MCMC algorithm. The entire template library is then convolved with the best fitting kinematics kernel, and the best-fitting non-negative linear combination of templates is found, after masking emission line regions. Finally, the best-fitting linear combination of spectra is used again for another kinematics fit and this iterative process is stopped after four repetitions.

In the second step of our procedure, we used {\tt PyParadise} to find the best non-negative linear combination of template spectra per spaxel, after convolving the template library with the kinematics kernel valid for the Voronoi bin to which the spaxel belongs. The resulting best-fit stellar spectrum is then subtracted from the original spectrum. We are now able to fit all the major emission lines, namely, H$\beta$, [OIII]\,$\lambda\lambda4960,5007$, [OI]\,$\lambda6300$, [NII]\,$\lambda\lambda6548,6583$, H$\alpha$ and [SII]\,$\lambda\lambda6717,6730$, with Gaussian functions that are all coupled in line width and radial velocity. Forcing the kinematics to match in this way we ensure that line ratios are computed for physically connected regions. When deemed clearly necessary after inspecting the fits and corresponding residuals, we modelled all lines with a second or even third kinematics component (also modelled as Gaussian functions) in specific regions of the galaxy. Additionally, to reduce the number of free parameters, we fixed forbidden doublet lines to their expected theoretical values, {\it i.e.}, [OIII] $\lambda 5007$/[OIII] $\lambda 4960=3$ and [NII] $\lambda 6584$/[NII] $\lambda 6548=3$, following \citet{StoZei00}. To estimate the uncertainties on all emission line parameters we repeated the measurements 30 times using the following Monte-Carlo approach. We first varied each pixel in the original spectrum within their corresponding variance, and then refitted the stellar spectrum at each spaxel using only 80\% of the spectra of the stellar template library, {\it i.e.}, a randomly chosen subset, before refitting the emission lines. This ensures that the estimated uncertainties in the emission line parameters account for stellar template mismatches.

Figure~\ref{fig:example_fit} shows {\tt PyParadise} stellar and emission line fits for the brightest spaxel of three TIMER galaxies. These examples illustrate the variety of stellar spectra and emission line properties we deal with in TIMER, ranging from starburst galaxies (left panel), old stellar populations with weak emission lines (middle panel) and AGN spectra that are a mix of stellar and AGN emission (right panel). Concerning the latter, the broad lines naturally affect the continuum fit, but we can still recover reasonable emission line measurements due the continuum normalisation of {\tt PyParadise}. Two of the TIMER galaxies are AGN with broad emission lines (NGC\,1097 and NGC\,1365), and for these galaxies the broad lines in the central spectra need to be flagged manually and processed interactively.

\subsection{Classification of galaxy nuclei}
\label{sec:emli_agn}

\begin{figure*}
\includegraphics[width=\textwidth]{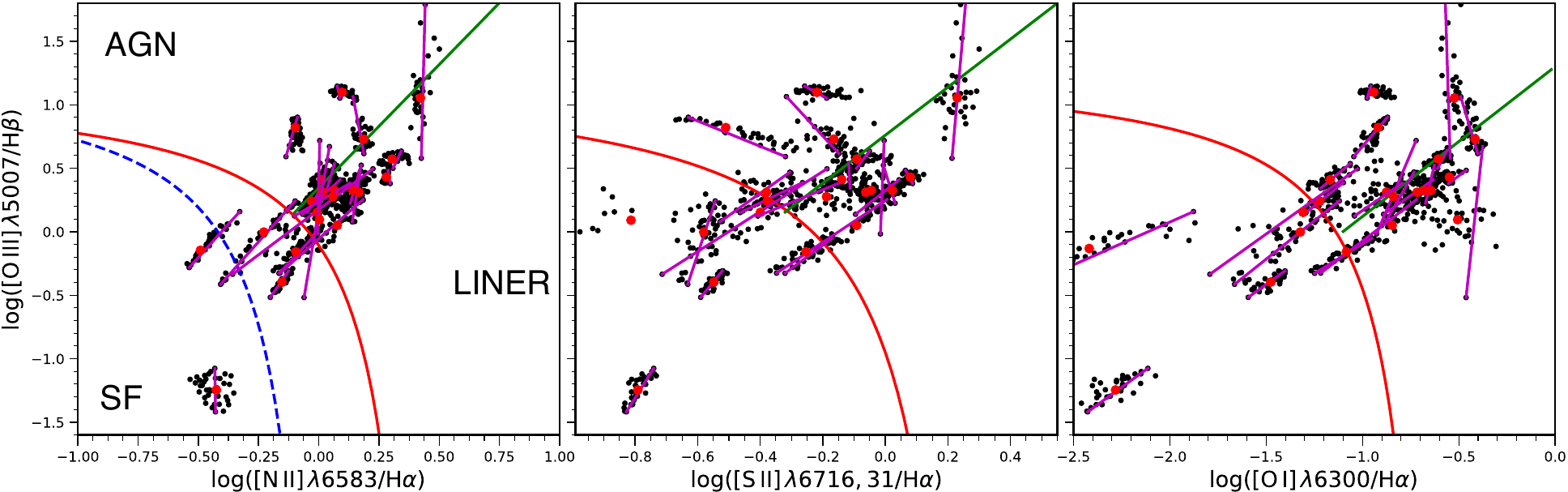}
\caption{Standard emission line diagnostics (BPT diagrams) inferred for the central $6\times6$ spaxels of each TIMER galaxy. The dashed blue line in the left panel is the empirical line from \citet{KauHecTre03} to separate star-forming galaxies from AGN. The red solid curves are likewise the theoretical extreme starburst lines from \citet{KewDopSut01}, and the diagonal, green solid lines are from \citet{KewGroKau06} and separate Seyfert galaxies from LINERs. The magenta line connects the highest and lowest [OIII]/H$\beta$ line ratios for each system to highlight the spread within this small aperture. Furthermore, we highlight the line ratios integrated over the entire aperture with filled red circles. All different nuclei, from star-forming to LINERs and Seyferts, are present in the sample, with a strong mix of ionisation sources even within physical scales of only about a hundred parsecs.}
\label{fig:BPT_nucleus}
\end{figure*}

\begin{table*}
\caption{Emission-line classifications of the integrated spectra within the central $1.2\arcsec$ (or typically $\sim120\,\rm{pc}$). Two categories are displayed when the individual spaxels lie in a region crossing the demarcation lines (\textit{cf.} Fig. \ref{fig:BPT_nucleus}).}
\begin{tabular}{cccc}\hline\hline
Galaxy & [NII]/H$\alpha$ vs. [OIII]/H$\beta$ & [SII]/H$\alpha$ vs. [OIII]/H$\beta$ & [OI]/H$\alpha$ vs. [OIII]/H$\beta$\\\hline
IC\,1438  & LINER & LINER & LINER\\
NGC\,613 & SF/LINER & SF/LINER & SF/LINER\\
NGC\,1097$^{\rm a}$ & LINER    & LINER/SEY & LINER/SEY\\
NGC\,1291 & LINER    & LINER     & LINER\\
NGC\,1300 & LINER    & LINER     & LINER\\
NGC\,1365$^{\rm a}$ & SF       & SF        & SF\\
NGC\,1433 & LINER/SEY & LINER/SEY & SEY\\
NGC\,3351 & SF       & SF        & SF\\
NGC\,4303 & LINER/SEY & SF/SEY   & SF/SEY \\
NGC\,4371 & SF/LINER  & SF       & LINER    \\
NGC\,4643 & LINER & SF/SEY   & LINER/SEY\\
NGC\,4981 & LINER/SEY & LINER   & LINER/SEY \\
NGC\,4984 & SEY   & SEY  & SEY\\
NGC\,5236 & SF    & SF/LINER & SF/LINER\\
NGC\,5248 & SF/LINER & SF/LINER/SEY & SF/SEY\\
NGC\,5728 & SEY   & SEY  & SEY\\
NGC\,5850 & LINER/SEY & LINER/SEY   & SEY\\
NGC\,6902 & LINER/SEY & SEY & LINER/SEY   \\
NGC\,7140 & SF/SEY & SF & SF/SEY\\
NGC\,7552 & SF     & SF & SF\\
NGC\,7755 & LINER  & LINER & LINER/SEY\\\hline
\end{tabular}\\
$^{\rm a}$ NGC\,1097 and NGC\,1365 are AGN with broad emission lines. The classification in this table corresponds to the narrow lines only.
\label{tab:BPT_nucleus}
\end{table*}

A first application of our emission line measurements is the classification of the galaxy nuclei based on standard emission line diagnostic diagrams. These have been first introduced by \citet{BalPhiTer81}, and are now commonly known as BPT diagrams \citep{OstPog85,CidStaMat11}. Significant variations in the emission line ratios have been identified to originate from different ionisation sources and ionisation parameters that are modulated by the gas abundances. Various demarcation lines have been introduced \citep[{\it e.g.},][]{KewDopSut01,KewGroKau06,StaTenRod07}, which we use to assign a primary ionisation source: star formation, LINER-like, and AGN (Seyfert). From the MUSE data, we use to this aim the central $6\times6$ spaxels (corresponding to $1\farcs2\times1\farcs2$) around the light-weighted centre of the galaxies. In Fig.~\ref{fig:BPT_nucleus} we show the line ratios thus derived integrated over the central area, as well as in each individual spaxel. The magenta lines indicate the major variations within the central region.

In Table \ref{tab:BPT_nucleus} we show for each galaxy the emission-line classification from the integrated spectra within the central $6\times6$ spaxels, using the standard emission line diagnostic diagrams as in Fig. \ref{fig:BPT_nucleus}. In the standard [NII] diagram we use the demarcation line from \citet{KewDopSut01} to separate star-forming galaxies. Note that there is variation in the classification depending on which diagnostic diagram is used, but based on the line ratios we can classify five nuclei as AGN, five nuclei as star-forming, and eleven nuclei as LINER-like. This reflects the great diversity in the properties of the nuclei of the TIMER galaxies. Of course, the physical conditions will further vary across the galaxy, as we show below, but the characterisation of the nuclei is important for splitting the sample in bins of activity and discussing different feedback effects on the stars and gas in the individual systems.

\subsection{Spatially-resolved emission line analysis and physical parameters}
\label{sec:emli_phys}

\begin{figure*}
\begin{center}
\includegraphics[width=0.33\textwidth]{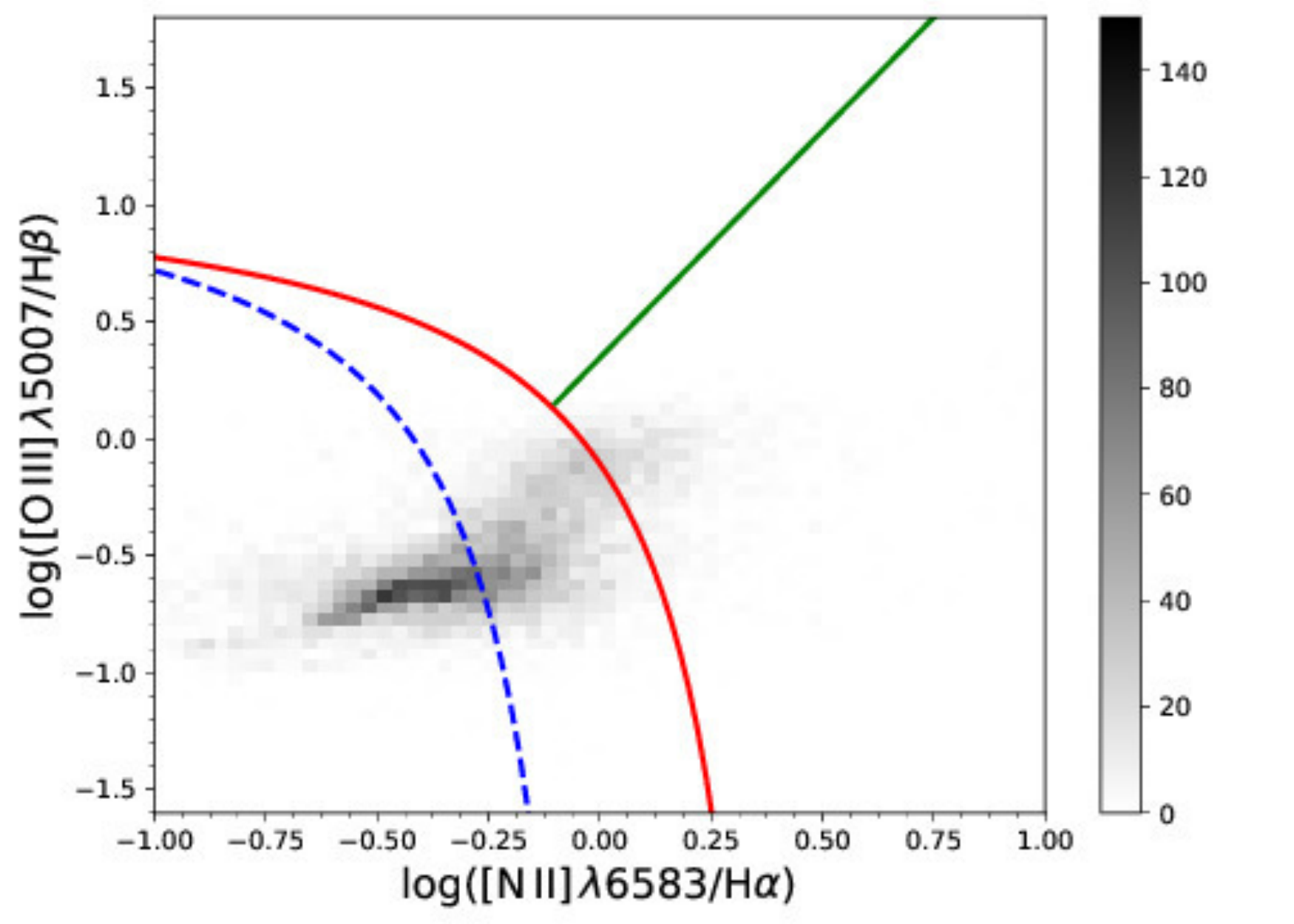}\includegraphics[width=0.33\textwidth]{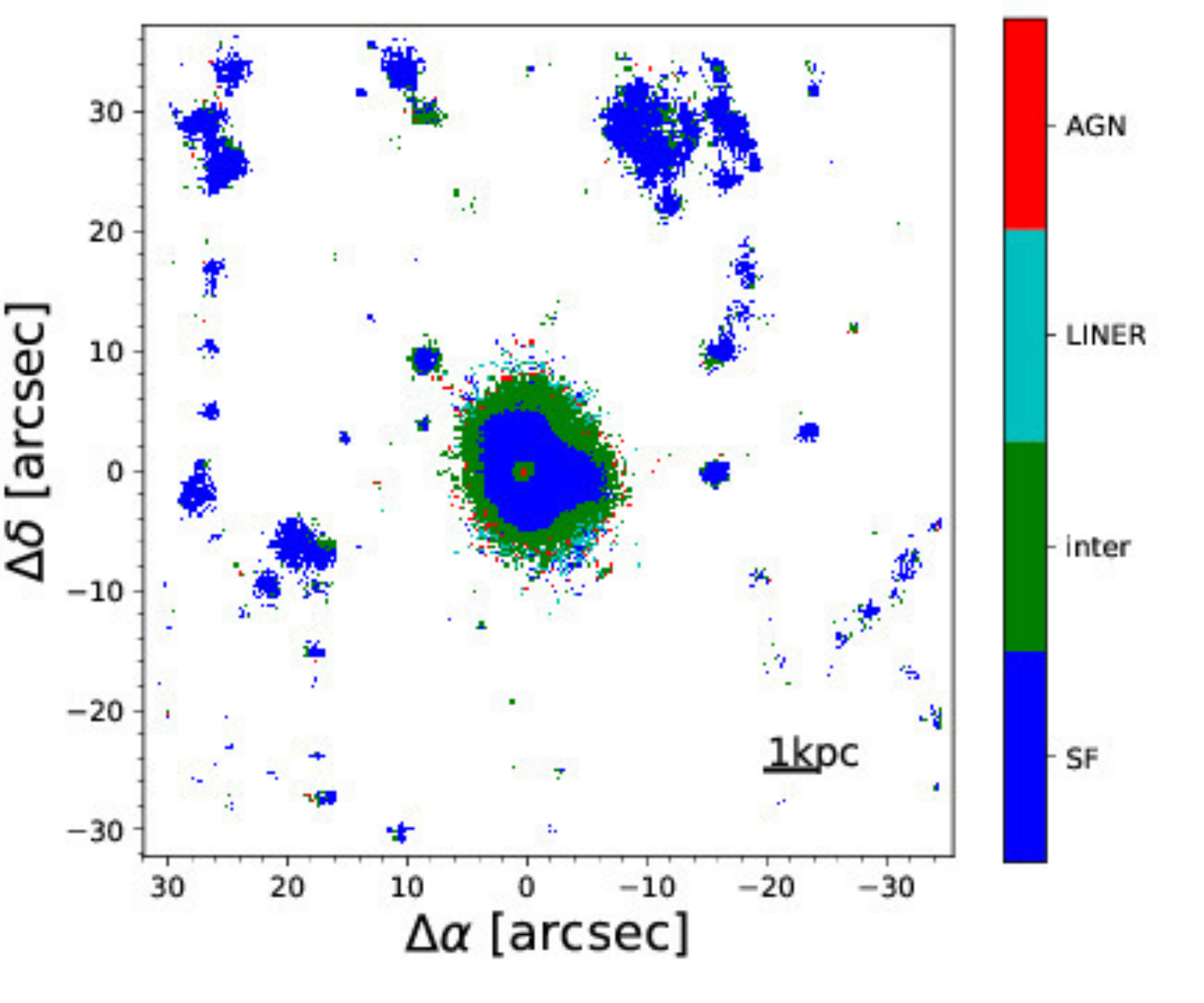}\includegraphics[width=0.33\textwidth]{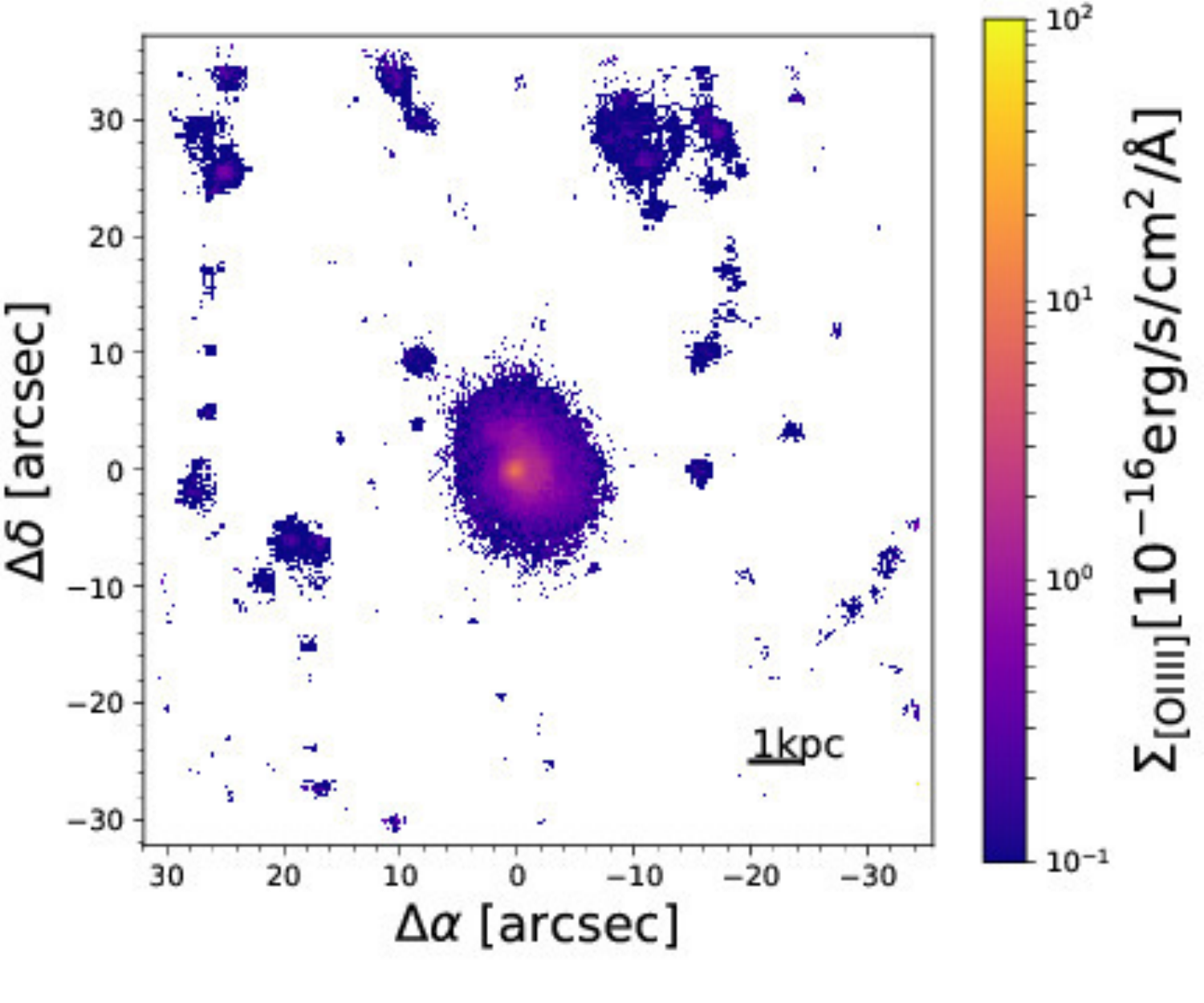}\\
\includegraphics[width=0.33\textwidth]{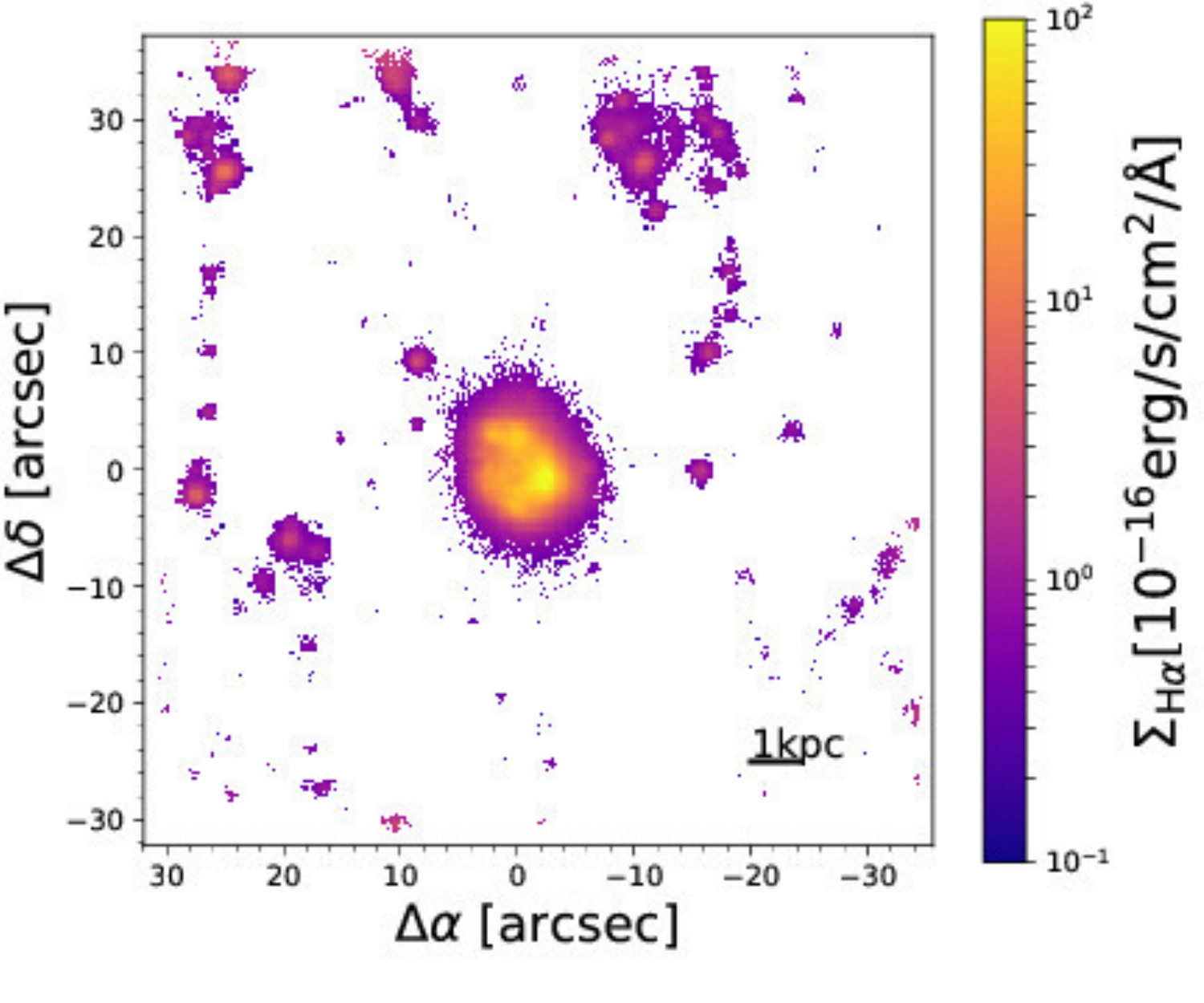}\includegraphics[width=0.33\textwidth]{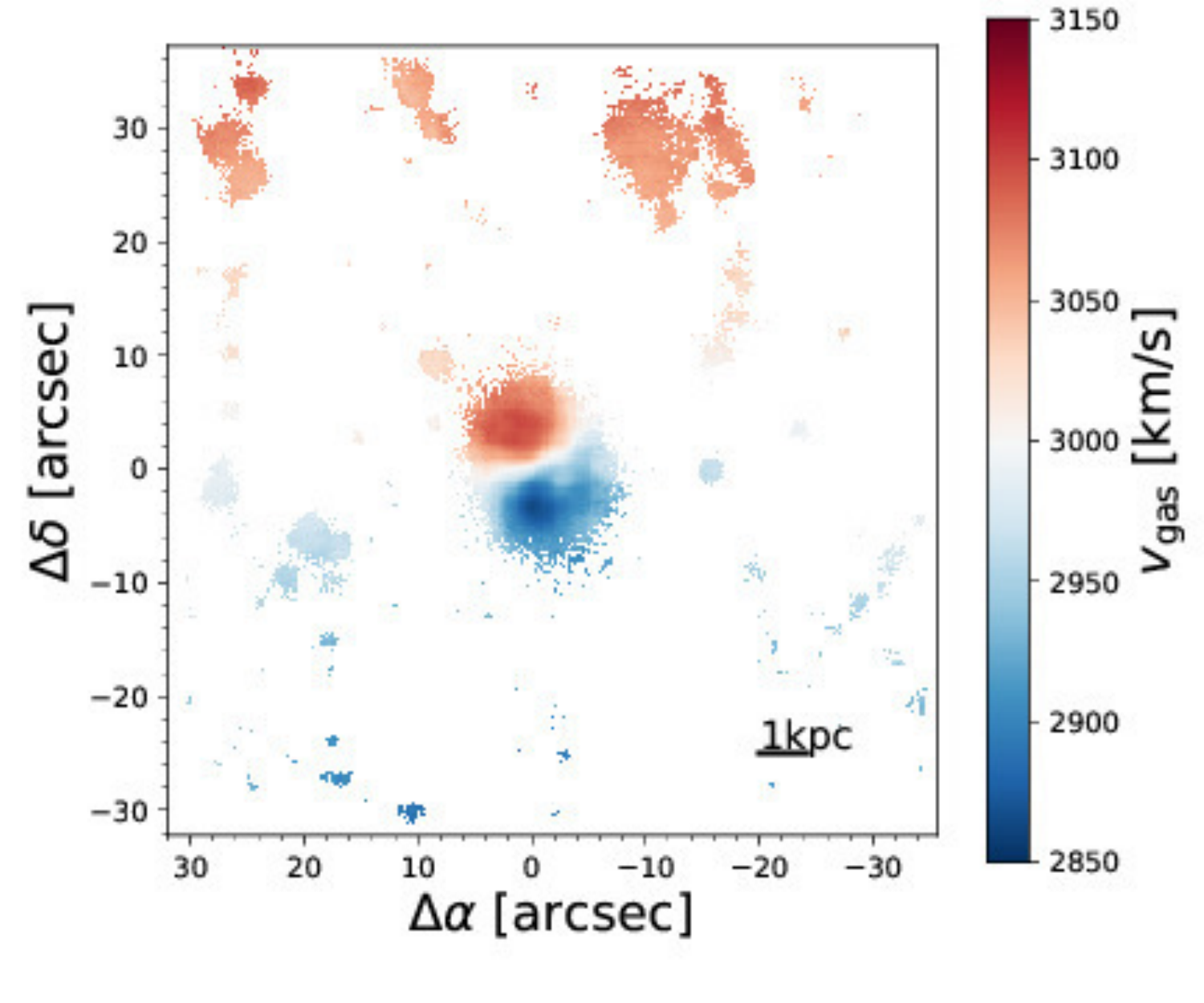}\includegraphics[width=0.33\textwidth]{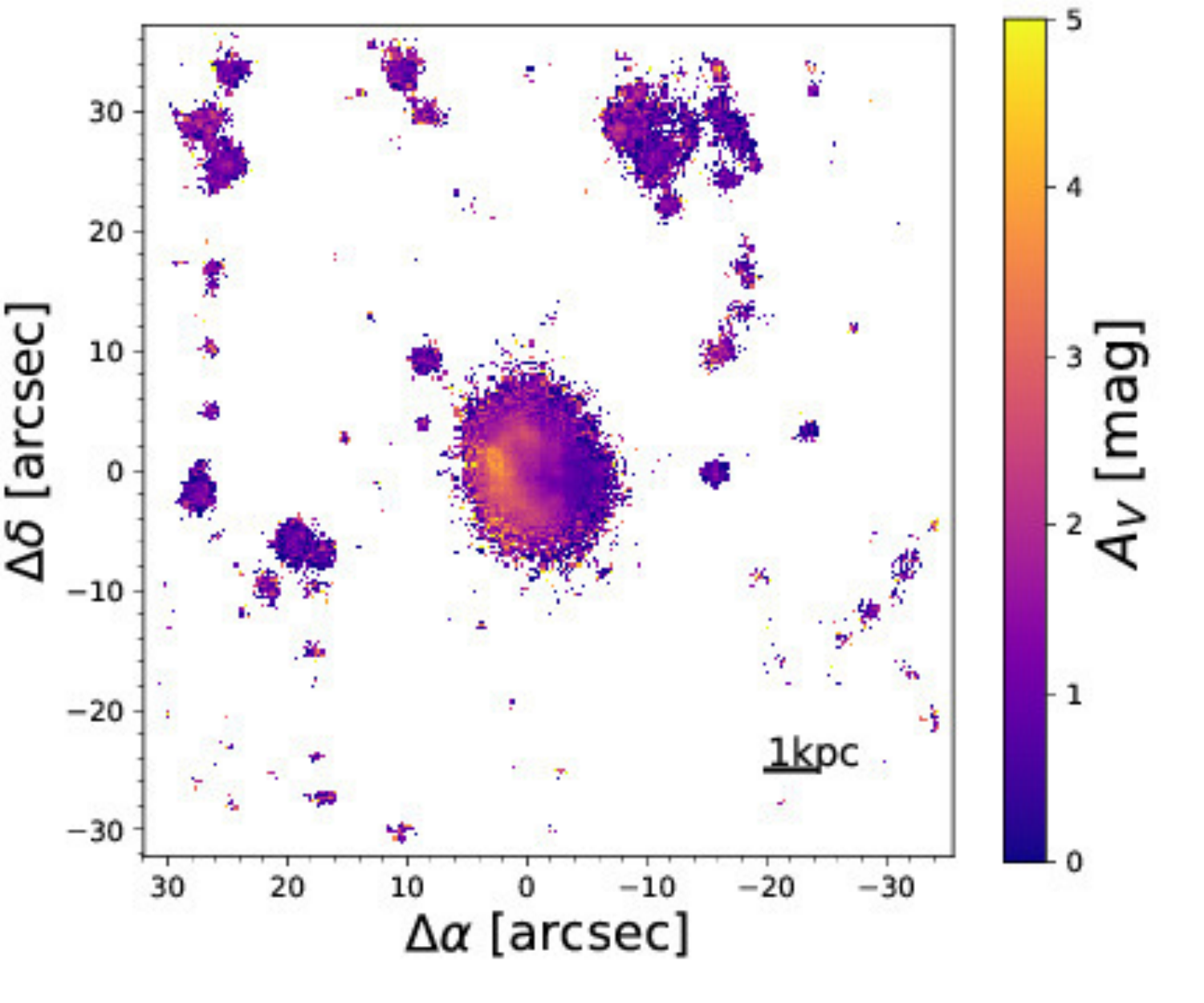}\\
\includegraphics[width=0.33\textwidth]{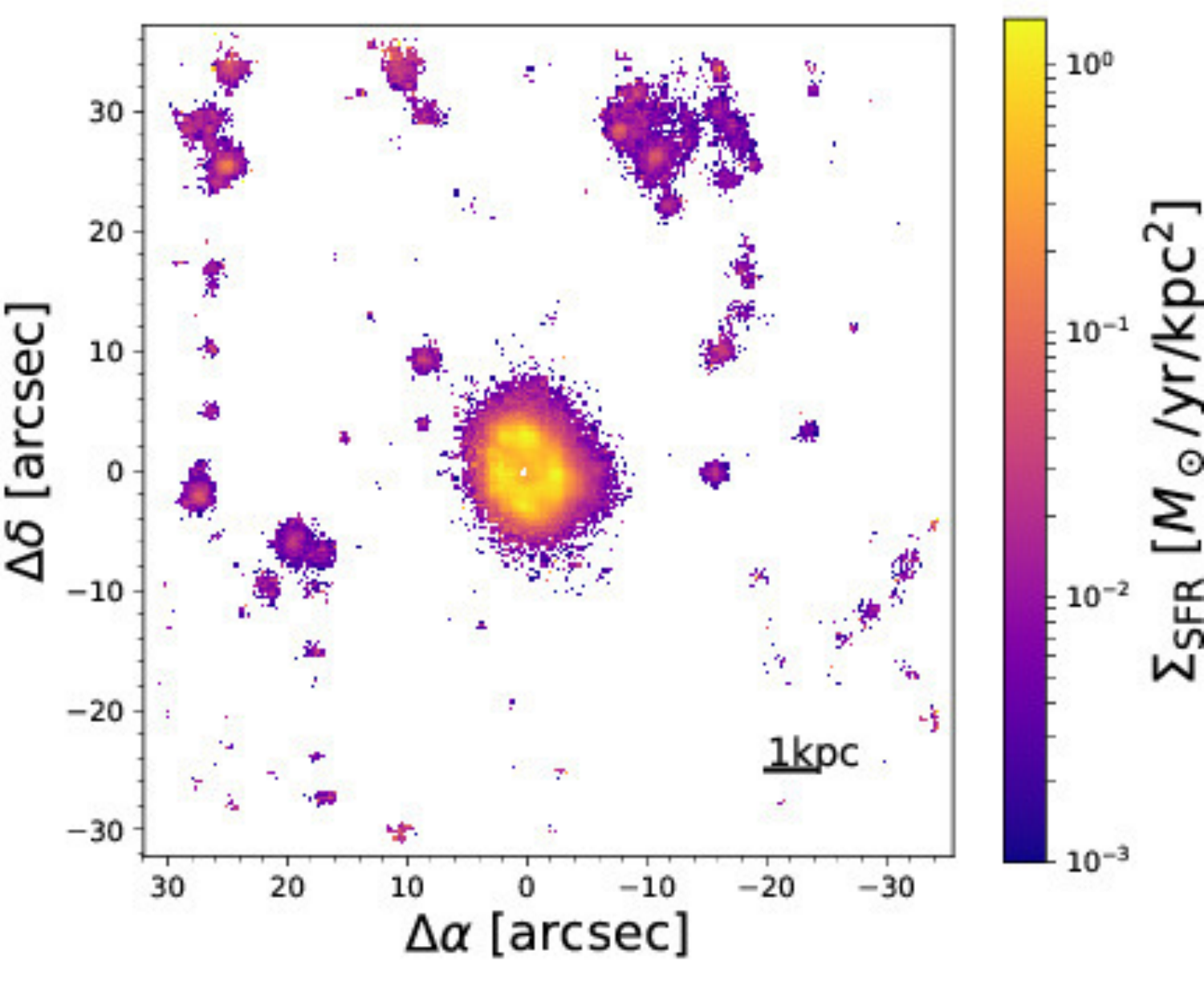}\includegraphics[width=0.33\textwidth]{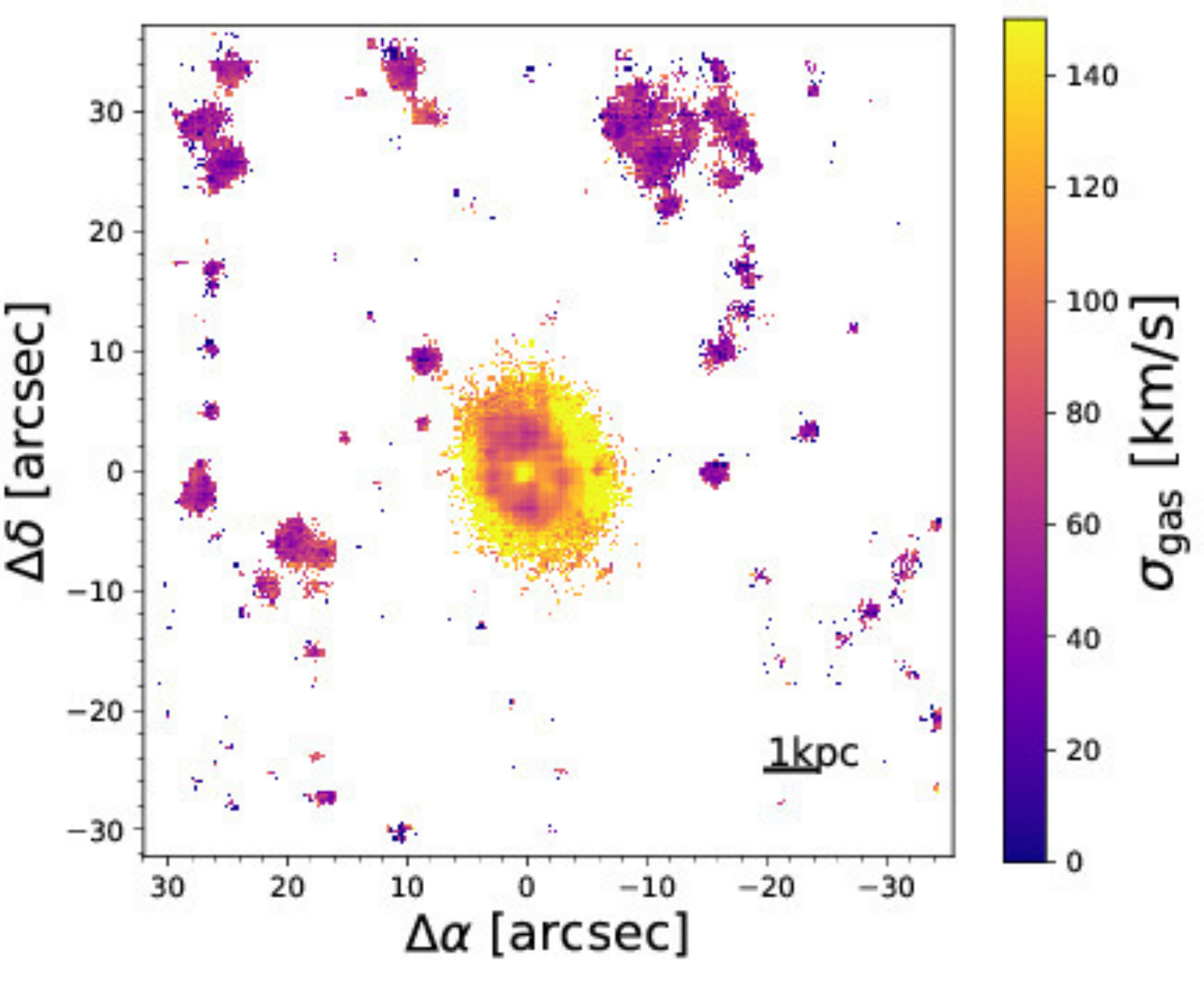}
\includegraphics[width=0.33\textwidth]{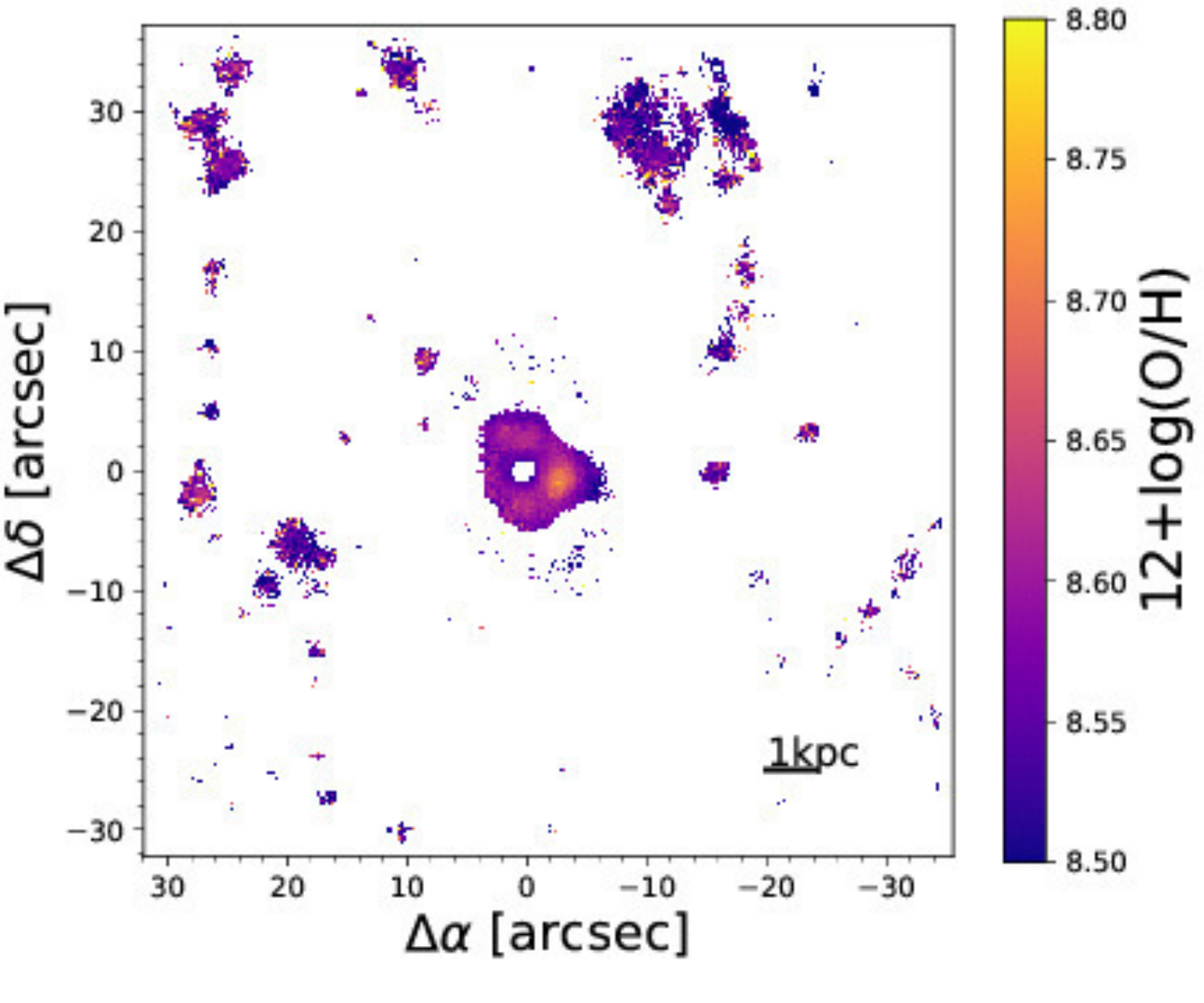}
\end{center}
 \caption{Results from the emission-line analysis for NGC\,7140. \textit{Upper panels:} From left to right we show the classical BPT diagram, a map delineating regions with the different line-excitation mechanisms, and the [OIII] surface brightness map. \textit{Middle panels:} From left to right we show the H$\alpha$ surface brightness map, the gas radial velocity field, and a map of the line-of-sight $V$-band attenuation estimated from the H$\alpha$/H$\beta$ decrement. \textit{Lower panels:} From left to right we show the star formation rate surface density map, the gas velocity dispersion field, and the [O/H] gas-phase metallicity map. An horizontal bar indicates the physical scale in every panel. North is up, east to the left.}
\label{fig:maps_NGC7140}
\end{figure*}

\begin{figure*}
\begin{center}
\includegraphics[width=0.33\textwidth]{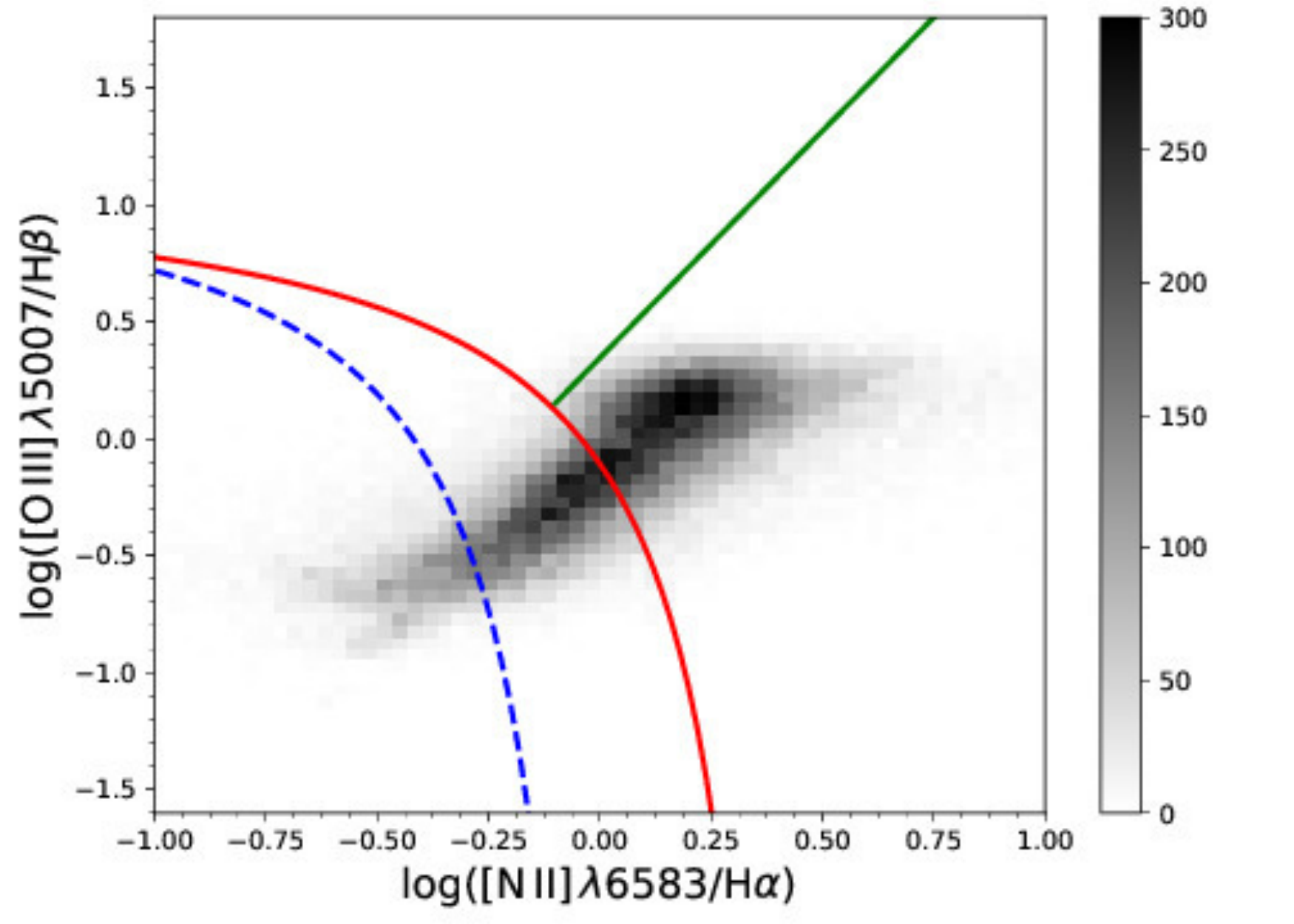}\includegraphics[width=0.33\textwidth]{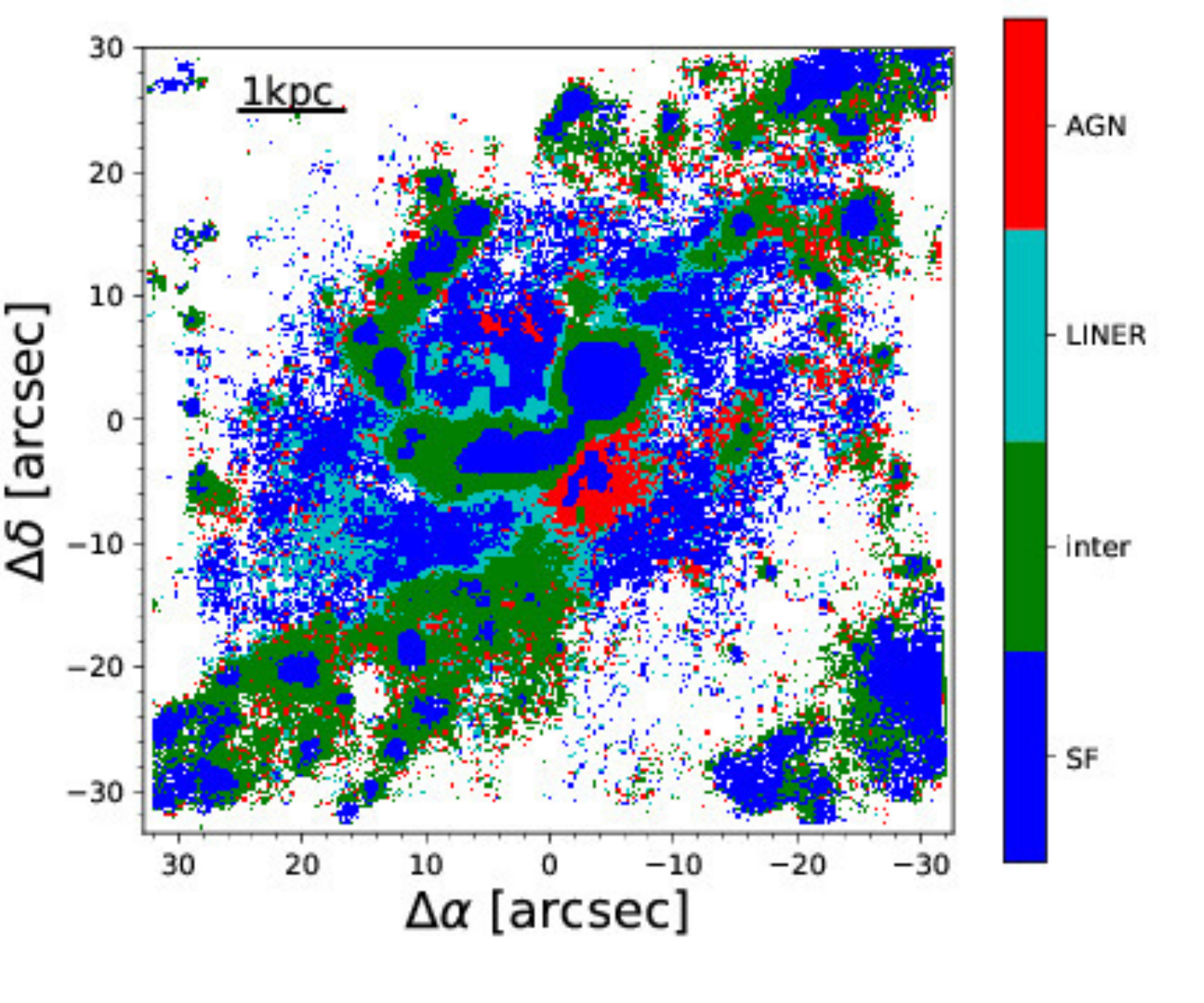}\includegraphics[width=0.33\textwidth]{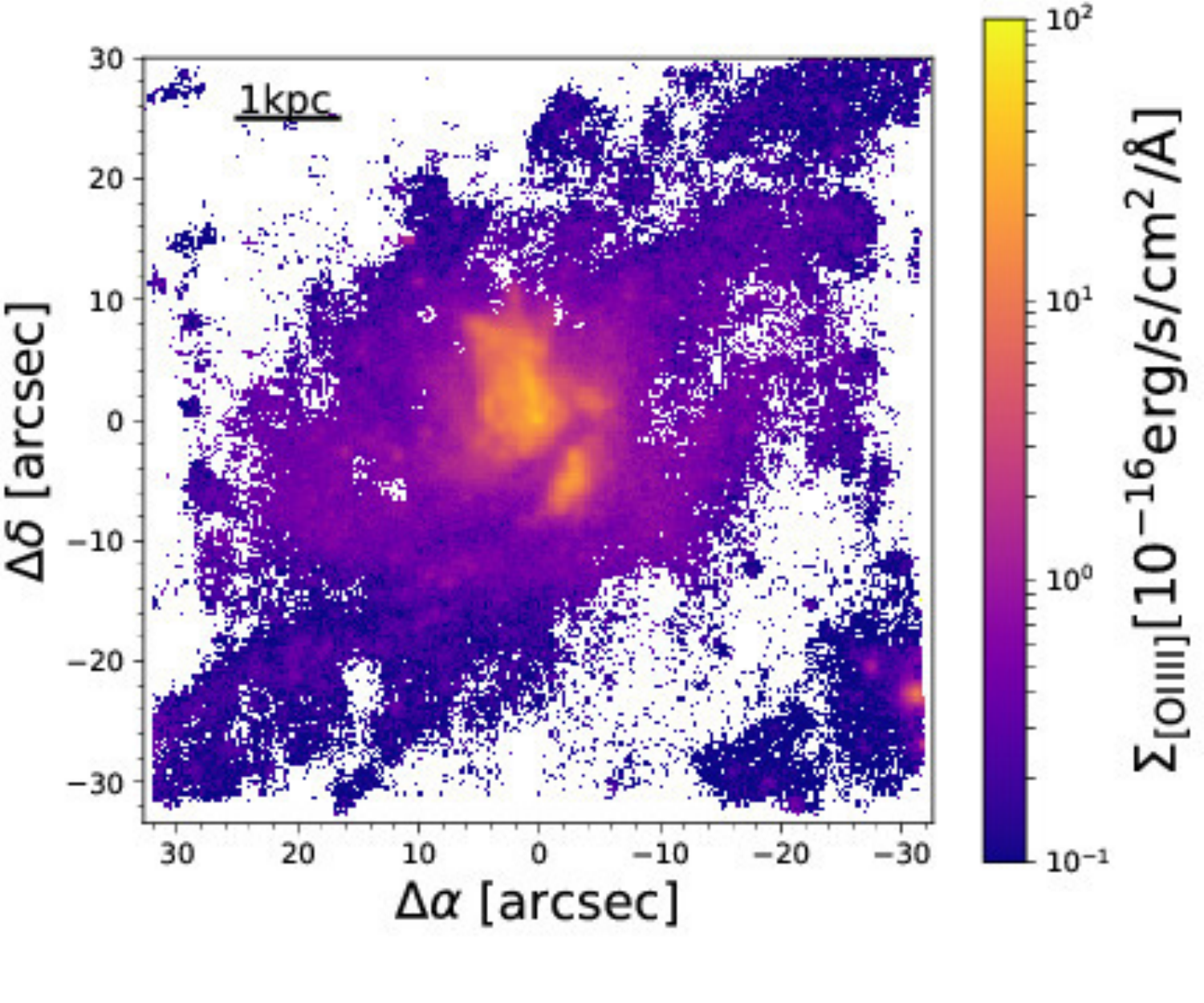}\\
\includegraphics[width=0.33\textwidth]{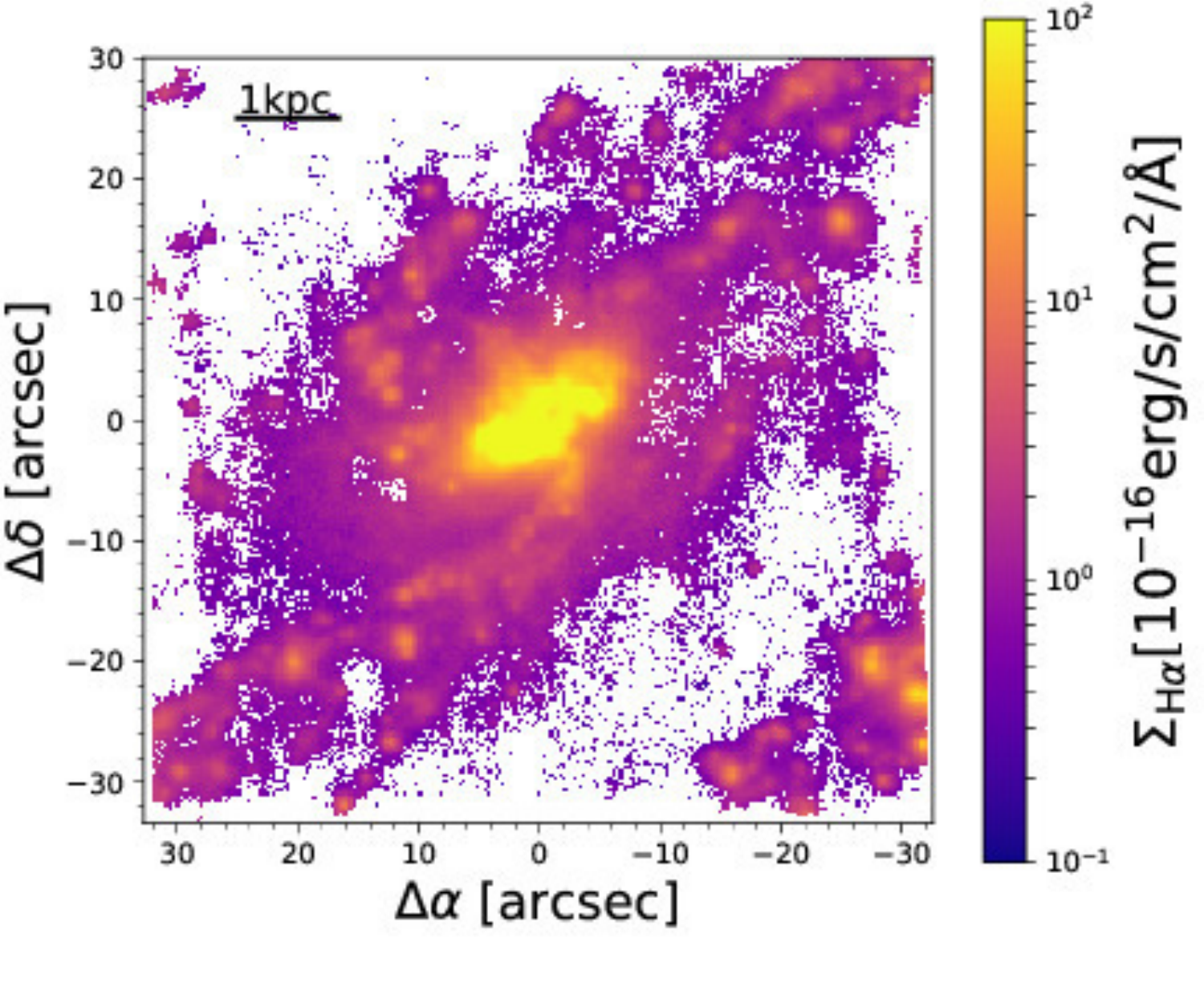}\includegraphics[width=0.33\textwidth]{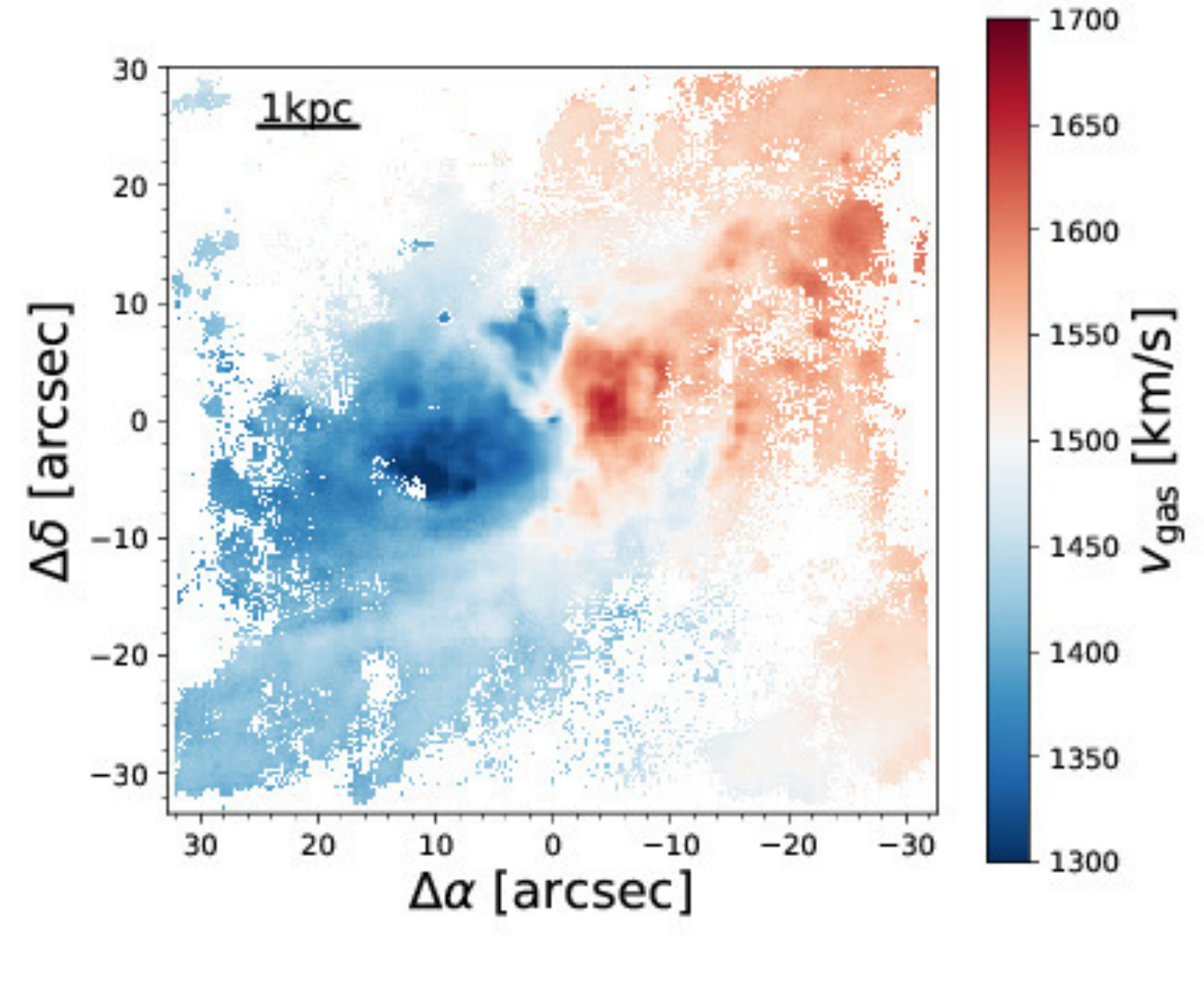}\includegraphics[width=0.33\textwidth]{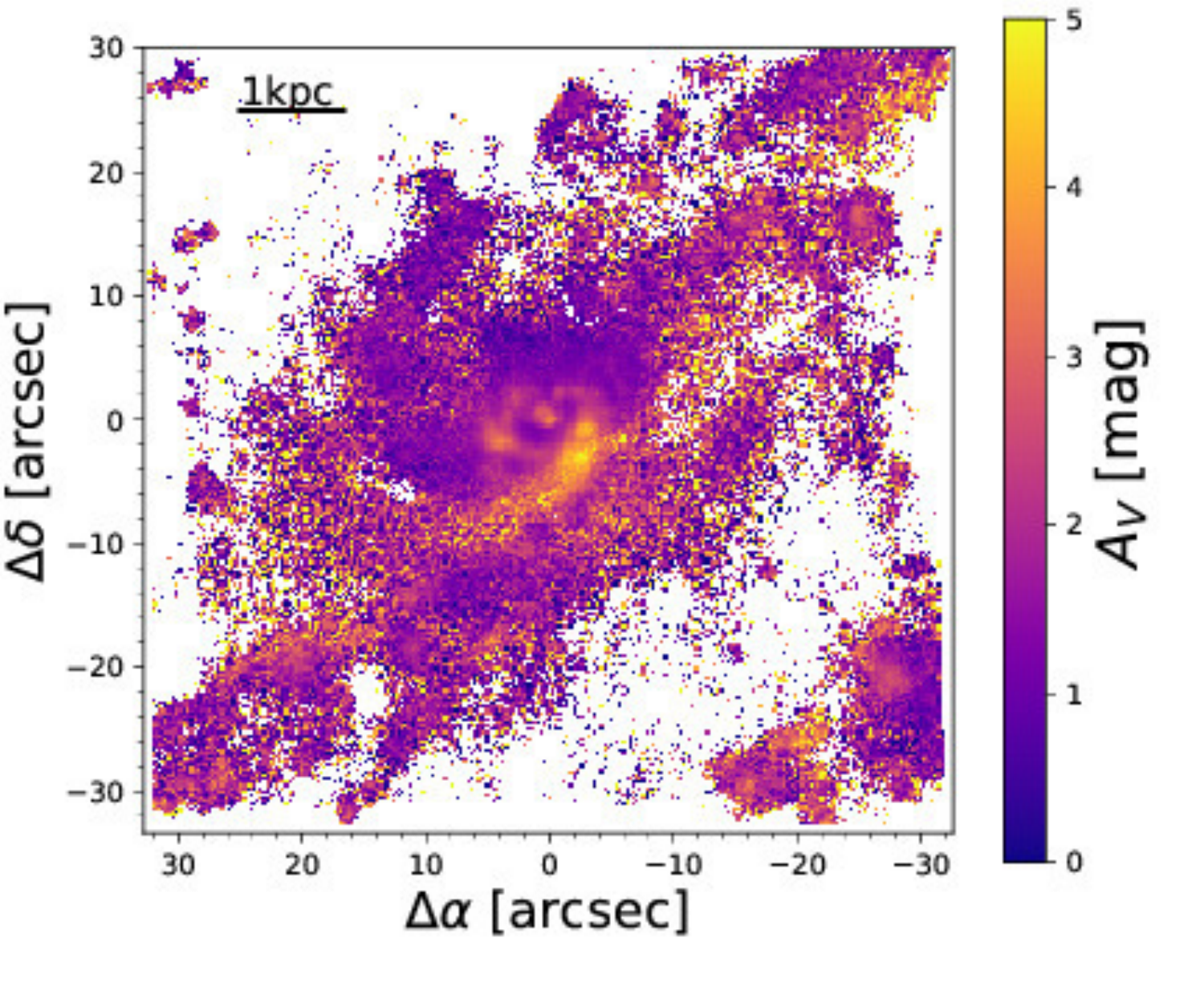}\\
\includegraphics[width=0.33\textwidth]{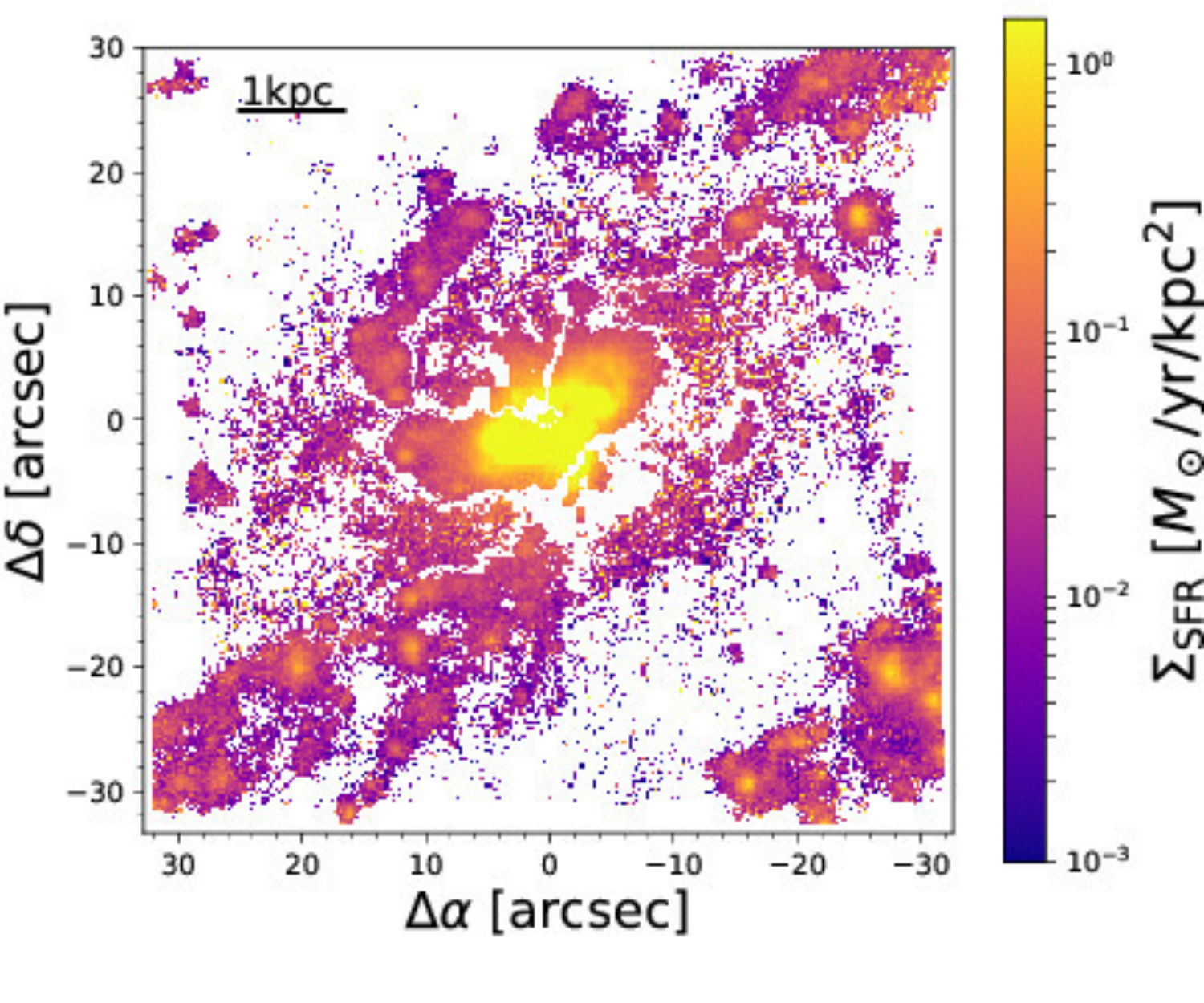}\includegraphics[width=0.33\textwidth]{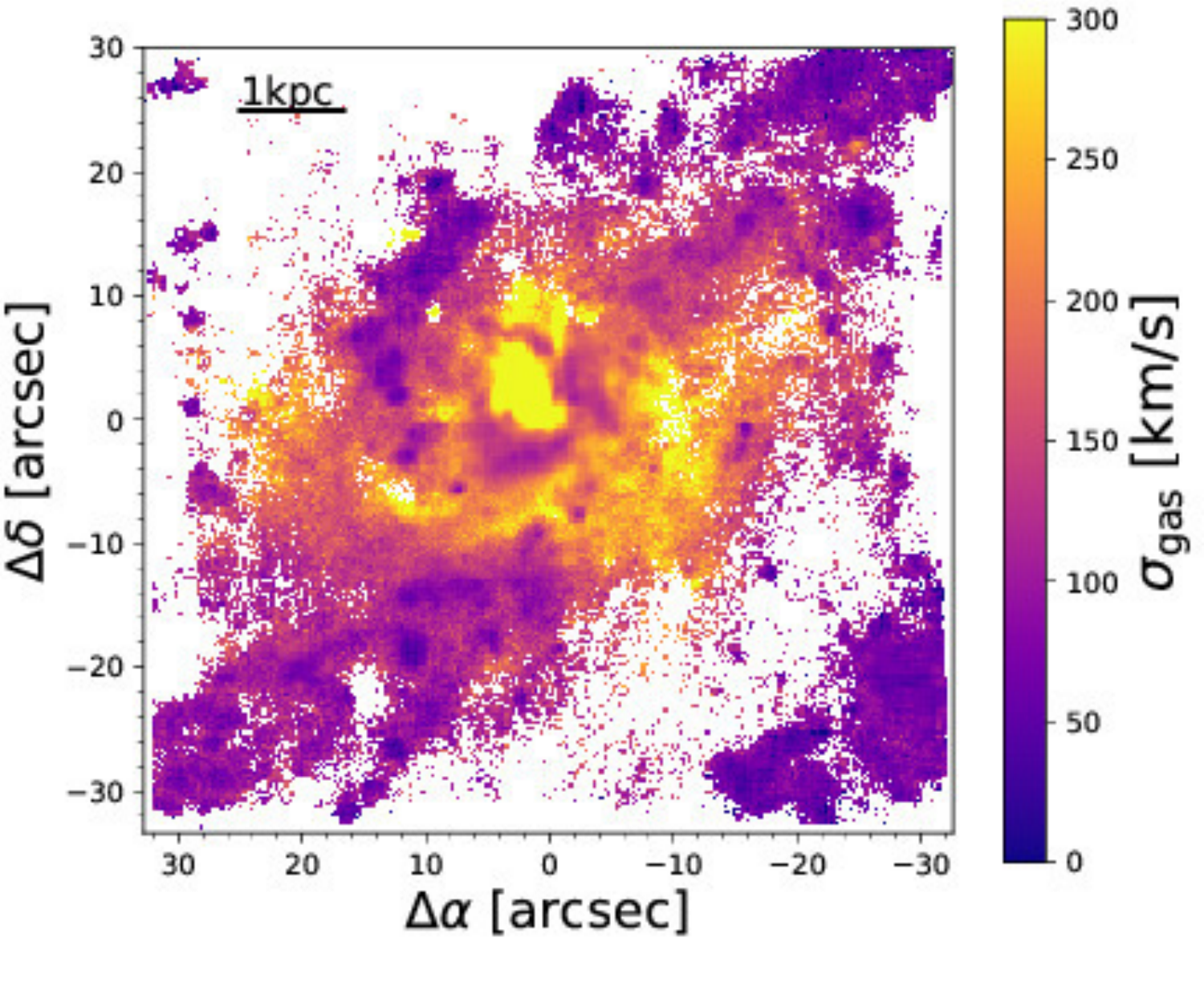}
\includegraphics[width=0.33\textwidth]{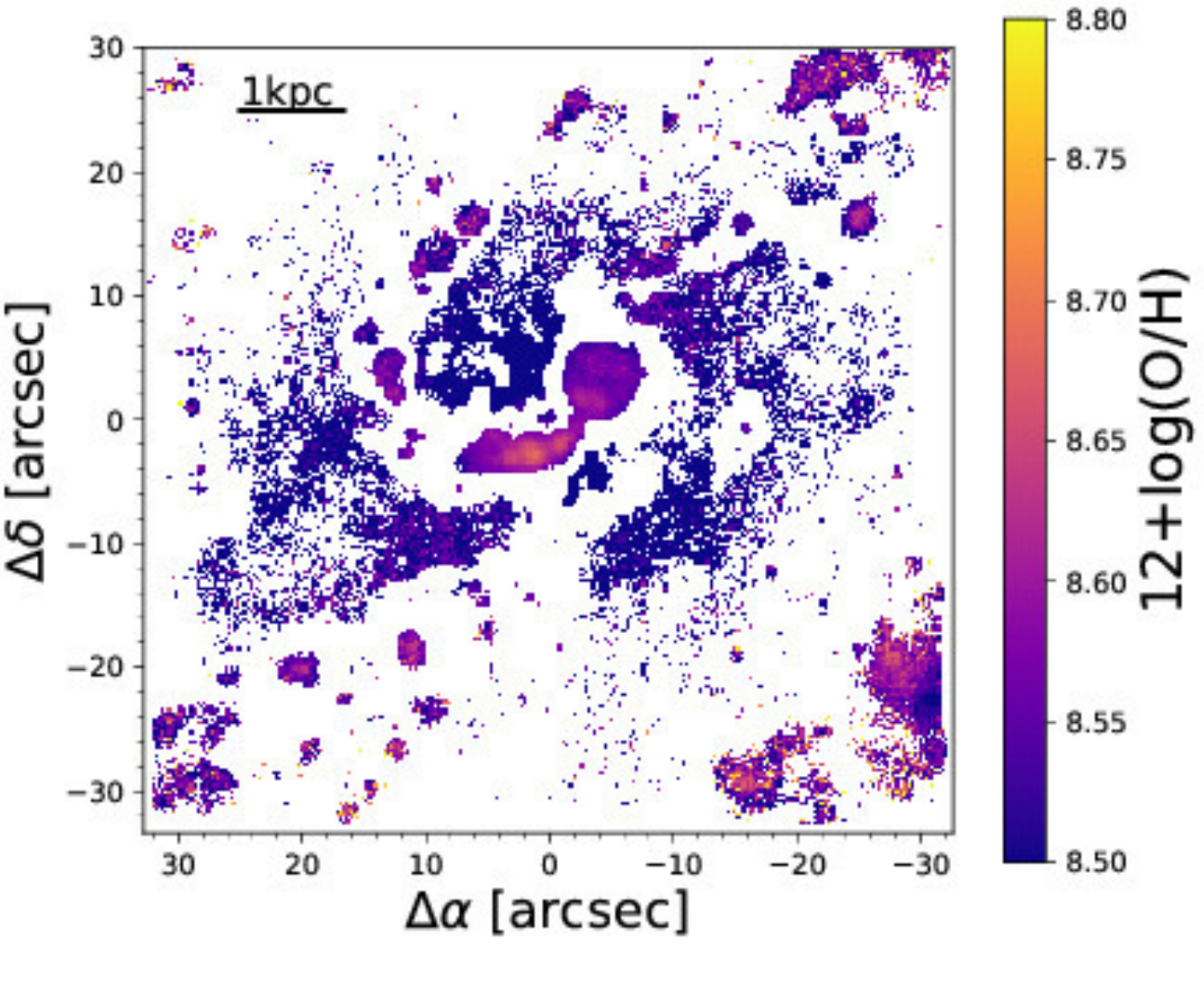}
\end{center}
\caption{Same as Fig.~\ref{fig:maps_NGC7140} but for NGC\,613.}
\label{fig:maps_NGC0613}
\end{figure*}

We now expand our analysis to a full 2D characterisation of the emission line fluxes and kinematics and derived physical properties. In particular, similar to the characterisation of the nuclei, we characterise the ionisation conditions through the classification into AGN, LINER-like and star formation emission per spaxel, to create an excitation map now across our entire MUSE fields. We also produce a map of the strength of the dust attenuation using the H$\alpha$/H$\beta$ Balmer decrement, assuming Case B recombination, a temperature of 10\,000\,K, an electron density of 100\,cm$^{-3}$, and a standard Milky-Way-like extinction curve. Furthermore, combining the excitation map with the dust extinction map allows us to create maps of the star formation rate (SFR) surface density and of the [O/H] gas-phase metallicity of star-forming regions only -- {\it i.e.}, uncontaminated by emission from other line-excitation mechanisms -- and corrected from the effects of dust extinction. In the derivation of the gas-phase metallicity we employed the O3N2 calibration from \citet{MarRosSan13}.

In Figs.~\ref{fig:maps_NGC7140} and \ref{fig:maps_NGC0613}, we show the results of this analysis for NGC\,7140 and NGC\,613 as examples that reveal very different characteristics. While the emission line properties of NGC\,7140 mainly point to star-forming regions in a rotating disc near the centre and in various places across the galaxy, excitation is increasing even to the Seyfert level towards the centre, which suggests the presence of an AGN. NGC\,613 appears significantly more complex, with a mix of star-forming regions, shocks and a well-defined AGN ionisation cone in the [OIII] surface brightness distribution. While the global gas kinematics field follows a disc rotation pattern, there are regions which greatly differ from it, indicating significant non-gravitational motion, such as that from outflows due to stellar and/or AGN feedback. While dust extinction is prominent only at the central region of NGC\,7140, in NGC\,613 dust extinction is important across a much larger region, and a dust lane along one of the leading edges of the bar is clearly recovered.

\section{Complexity and diversity in bar-built nuclear structures}
\label{sec:sci}

\begin{figure*}
\begin{center}
	\includegraphics[width=2\columnwidth]{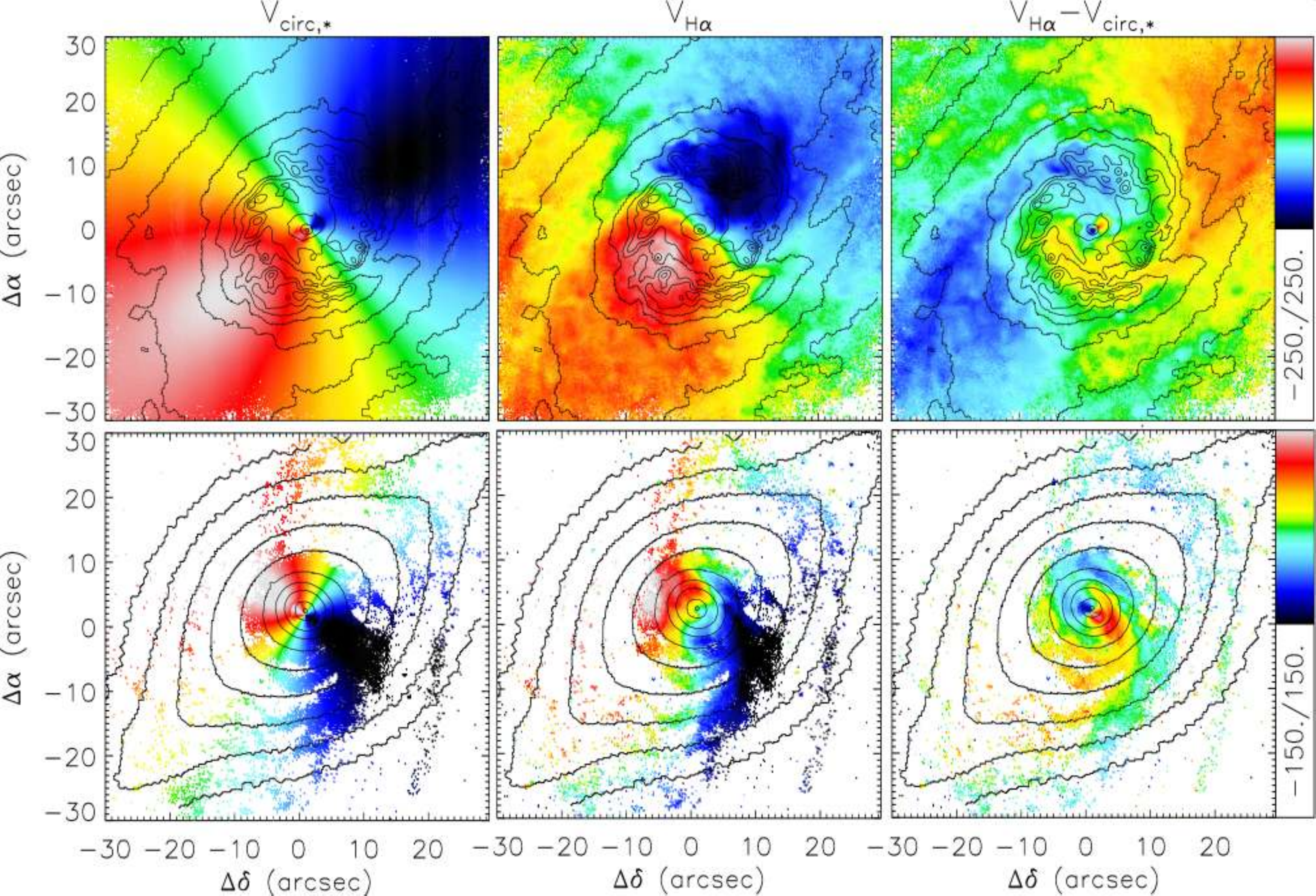}
\end{center}
    \caption{Jeans stellar dynamical model circular velocity field (left), un-binned H$\alpha$ velocity field (centre) and the difference between both (right) for NGC\,1097 (top) and NGC\,4643 (bottom). The colour bars on the side of each panel indicate the plotted range of the velocity in km s$^{-1}$. The isophotes shown are derived from the MUSE data cube reconstructed intensities and are equally spaced in steps of about 0.5 magnitudes. At the distances shown in Table \ref{tab:sample}, $1\arcsec$ corresponds to $\approx100\,\rm{pc}$ for NGC\,1097 and $\approx125\,\rm{pc}$ for NGC\,4643. North is up, east to the left.}
    \label{fig:gaskin}
\end{figure*}

Important inferences can be made already from the kinematic and stellar population maps of NGC\,1097 and NGC\,4643 (see Figs. \ref{fig:kin} and \ref{fig:agemet}), as well as from the star formation histories shown in Fig. \ref{fig:sfhs}. Firstly, all kinematic properties analysed are consistent with the picture in which the inner discs in these galaxies are built from gas brought to the central region by the bar, where it fuels the formation of a new stellar structure. This is inferred from the $V_*-\rm{h}_3$ anti-correlation, indicating the presence of near-circular orbits, and the elevated observed velocities, above those of the main underlying disc, indicating a stellar structure with high angular momentum and separate from the main disc. NGC\,4371 shows as well the same properties \citep[see][]{GadSeiSan15}. These properties are not expected in a scenario in which the stars (or gas) in the inner disc are accreted through galaxy mergers alone (i.e., without the effects of non-axisymmetric components in the disc such as bars). However, the lack of star formation in NCG\,4643 suggests a cessation in the flow of gas through the bar, which could be because the bar has formed sufficiently early that it had time already to push most of the available gas inwards. In addition, given that this galaxy sits at the outskirts of the Virgo cluster, environmental effects may have prevented later gas infall onto the disc. If the dense environment had removed the gas before the formation of the bar there would be no inner disc. On the other hand, the ongoing star formation in the inner disc of NGC\,1097 clearly shows that there is gas available for the bar to push it into the inner disc.

Secondly, the maps in Fig.~\ref{fig:agemet} show that in NGC\,1097 the oldest stars tend to be the most metal-rich. This can be seen in both luminosity- and mass-weighted age and metallicity maps. In fact, the young nuclear ring in NGC\,1097 shows the lowest metal content in the region covered by our MUSE field, which suggests that the gas forming stars in the ring, brought to the ring by the bar, was not pre-processed within the galaxy. In fact, NGC\,1097 is currently interacting with a low-mass satellite galaxy, which is likely the origin of the metal-poor gaseous material or helps pushing gas from the surroundings of the system towards NGC\,1097. This is consistent with the results from \citet{KewRupZah10}, who found that gas metallicity gradients are shallower in close galactic pairs, as a result of the dynamical effects of the interaction on the gas. If this material falls onto the disc of NGC\,1097, at a galactocentric distance within the bar radius, then it is pushed by the bar to the nuclear ring, fuelling the observed starburst. We note that \citet{SeiCacRui15} found that the nuclear ring in NGC\,7552, another interacting galaxy, within a galaxy group, also has very low metal content compared to the rest of its inner region.

Interestingly, NGC\,1097 is a massive galaxy and thus one in which the downsizing picture predicts that a bar forms early on in cosmic history. Therefore, the nuclear ring formed by the bar should not contain a dominant young stellar population, unless the disc within the bar radius is replenished with gas. This replenishment is certainly happening now, as indicated by the intense dust lanes seen in Fig. \ref{fig:rgb_cm} and the star-bursting ring. Alternatively, the disc of NGC\,1097 could have formed dynamically stable against the formation of a bar, and the bar was actually only recently formed, triggered by the interaction with the companion galaxy. A further possibility is that a bar formed earlier and has dissolved and the currently observed bar was formed more recently. However, note that bars seem to be robust structures \citep[see discussion in][]{GadSeiSan15}. Therefore, if the bar in NGC\,1097 formed at early cosmic epochs as expected from the downsizing picture, we expect to detect, within the nuclear ring of NGC\,1097, an underlying population of old stars formed when the bar pushed gas towards the centre for the first time, shortly after its formation. Indeed, the star formation history derived with the spectra sampled from the nuclear ring, as shown in Fig. \ref{fig:sfhs}, shows a peak at early times, which could indicate an old age for the first generation of stars formed within the nuclear ring, and therefore also for the bar. However, at this point it is not yet clear if these old stars belong to the ring or the underlying structures (see discussion in the Introduction). Thorough analyses are necessary to address this question properly.

Thirdly, it is very interesting to see that, despite the violent burst of star formation in the nuclear ring of NGC\,1097, the very centre of the galaxy is the oldest stellar component. This indicates that the nuclear ring indeed acts as a powerful barrier to the gas down-streaming along the bar and that most, if not all, gas gets trapped in the orbits that make up the ring. This is consistent with the results from \citet{DavMacHic09}, who found that the inflow of gas inside the inner disc in NGC\,1097 is two orders of magnitude lower than the inflow in the bar towards the inner disc. Supernovae in the ring could provide a reservoir of gas interior to the ring by pushing ejecta into this region -- which incidentally could in principle feed AGN activity -- but our results show that such reservoir, if present, does not induce star formation in that region. Either the conditions for the formation of stars are not currently met interior to the ring or supernovae simply blow the gas away from the disc. The fact that the region interior to the ring is also the most metal-rich is consistent with this picture. A central peak in metallicity is seen not only in NGC\,1097 but also in NGC\,4643 (see Fig.~\ref{fig:agemet}) and NGC\,4371 \citep[see][]{GadSeiSan15}.

The young, metal-poor nuclear ring in NGC\,1097, and the old, metal-rich inner disc in NGC\,4643 testify to the complexity and diversity present in bar-built nuclear structures today. Further, they preclude straightforward interpretations of the physical nature of stellar structures in galaxies exclusively from measurements of the mean stellar age and metallicity. Instead, these examples show that caution and thoroughness are required.

One remarkable result shown in Fig. \ref{fig:sfhs} is that a significant fraction of the stellar mass comes from stars older than 10\,Gyr in the three depicted regions in both NGC\,1097 and NGC\,4643. Examining the star formation histories of the whole sample across their entire MUSE fields, we find that in about half of the sample there is a significant peak above 10\,Gyr. The other half of the sample shows more structure in their star formation histories, with peaks at different epochs, or a rather smooth variation. For six galaxies, 50\% of the stellar mass or more in their central region resides in stars older than 10\,Gyr. Interestingly, \citet{McDAlaBli15} and \citet{GonPerCid17} also find large fractions of stars older than 10\,Gyr in the ATLAS$^{\rm 3D}$ sample of early-type galaxies, and in the CALIFA sample, respectively (the latter also includes late-type galaxies, as TIMER). These results will be explored in a forthcoming paper.

To conclude this section, we show an example of the powerful combination provided by analysing concomitantly the stellar and gaseous kinematics in the TIMER galaxies. The stellar kinematic fields derived above can be used to build stellar dynamical models. The left panels in Fig.~\ref{fig:gaskin} show circular velocity fields ($V_{\rm{circ},*}$) derived from Jeans modelling of the observed $V_*$ and $\sigma_*$ fields in Fig.~\ref{fig:kin} for NGC\,1097 and NGC\,4643. Combined with the observed velocity fields from H$\alpha$ emission in the MUSE data cubes -- from gas ionised by hot stars, similar to the ones shown in Figs. \ref{fig:maps_NGC0613} and \ref{fig:maps_NGC7140} for NGC\,613 and NGC\,7140 -- these models can be used to provide information on the non-gravitational motion of the ISM. The central panels in Fig.~\ref{fig:gaskin} show the un-binned H$\alpha$ velocity fields of NGC\,1097 and NGC\,4643, whereas the right panels show the difference between the Jeans $V_{\rm{circ},*}$ model and the H$\alpha$ velocities. The latter reveal deviations from circular motion in the dynamics of the gas, which can be caused by torques from non-axisymmetric structures such as bars, producing streaming motions, or non-gravitational motion, such as that caused by stellar winds or shocks.
The figure shows that, indeed, the ISM in both galaxies display non-circular motion due to the presence of the bar. Furthermore, the central region of NGC\,1097 shows residuals suggesting feedback mechanisms associated with the starburst in the nuclear ring or, alternatively, simply a result from the feeding of gas through the bar towards the ring. A proper understanding of these residuals requires further investigation and modelling, which will be presented in a forthcoming paper.

\section{Outlook}
\label{sec:conc}

The MUSE data cubes obtained and produced as a result of the effort from the TIMER team constitute a rich dataset with which a number of important astrophysical questions can be addressed, expanding further from the main scientific goals of the project, described above. Given the high physical spatial resolution and high SNR of the data, we find in several data cubes signatures of energy and/or momentum feedback mechanisms into the ISM, such as those shown in Fig. \ref{fig:maps_NGC0613} for NGC\,613. These mechanisms appear to be associated with either intense star formation (stellar feedback) or AGN activity (AGN feedback). The spatial and spectral coverage and sampling of the data allow us to obtain a detailed view on the physical processes governing the interactions between radiation and the different phases of the ISM. This creates a unique opportunity to investigate such feedback process in the context of their implementation in numerical simulations.

The expansion of the TIMER science case from its original proposal led to the initial team growing to include colleagues with the required expertise. It also led to the necessity of including ancillary data from other facilities in our analyses. The policies governing the project are such that collaborations with experts outside the team are encouraged to enhance the scientific exploration of the rich MUSE data cubes.

Further investigations being currently conducted by the TIMER team include:

\begin{enumerate}
\item The dynamics and stellar population content of nuclear bars;
\item Star formation and stellar populations in primary bars;
\item The star formation desert in barred galaxies;
\item The connection between box/peanuts and barlenses;
\item Stellar migration in disc galaxies;
\item The location of nuclear rings with respect to the Inner Lindblad Resonance;
\item Shear and shocks along bars;
\item The excitation states of the ionised gas in the ISM; and
\item The IMF across disc galaxies.
\end{enumerate}

Furthermore, the information contained in the cubes regarding the stellar and ionised gas dynamics, and the stellar population content of the different stellar structures present in the TIMER sample -- such as primary discs and bars, spiral arms, rings, inner discs and nuclear bars -- creates the opportunity to connect the formation and evolution histories of these different components to create a coherent and thorough history of the individual galaxies as a whole. Finally, this dataset also poses a challenge from the point of view of the analysis techniques and tools. The rich level of detail allows for envisaging new techniques to extract physical information.

The complete set of TIMER high-level data products and maps will be made available in the future at a central on-line repository for public access. Maps such as those in Figs. \ref{fig:kin} and \ref{fig:agemet}, describing the stellar kinematics and stellar population properties, will be published for the whole sample we present here in dedicated papers already in preparation. At a suitable point in time, the data corresponding to such maps will be made available along with other data products, such as those concerning the emission line properties. The data repository will eventually be updated once we finalise the data products corresponding to the three galaxies that remain to be observed.

The policies governing the TIMER project are such that collaborations with external experts are encouraged. Interested colleagues should contact the team to discuss the possibility of working together on additional topics.

\section*{Acknowledgements}

Based on observations collected at the European Organisation for Astronomical Research in the Southern Hemisphere under ESO programmes 097.B-0640(A), 095.B-0532(A), 094.B-0321(A) and 060.A-9313(A). The TIMER team is indebted to the MUSE Instrument Operation Team (IOT) and the Paranal Science Operations (PSO) team, in particular the MUSE Instrument Scientists and the UT4 Telescope Operators and Support Astronomers for their excellent support. We thank the anonymous referee for the positive and very constructive feedback. AdLC acknowledges support from grant AYA2016-77237-C3-1-P from the Spanish Ministry of Economy and Competitiveness (MINECO). GvdV acknowledges funding from the European Research Council (ERC) under the European Union's Horizon 2020 research and innovation programme under grant agreement No 724857 (Consolidator Grant ArcheoDyn). This research has made extensive use of the NASA Extragalactic Database (NED), the Lyon Extragalactic Data Archive (LEDA) and the NASA Astrophysics Data System (ADS).

\bibliographystyle{mnras}
\bibliography{../../gadotti_refs}

\begin{thebibliography}{}
\makeatletter
\relax
\def\mn@urlcharsother{\let\do\@makeother \do\$\do\&\do\#\do\^\do\_\do\%\do\~}
\def\mn@doi{\begingroup\mn@urlcharsother \@ifnextchar [ {\mn@doi@}
  {\mn@doi@[]}}
\def\mn@doi@[#1]#2{\def\@tempa{#1}\ifx\@tempa\@empty \href
  {http://dx.doi.org/#2} {doi:#2}\else \href {http://dx.doi.org/#2} {#1}\fi
  \endgroup}
\def\mn@eprint#1#2{\mn@eprint@#1:#2::\@nil}
\def\mn@eprint@arXiv#1{\href {http://arxiv.org/abs/#1} {{\tt arXiv:#1}}}
\def\mn@eprint@dblp#1{\href {http://dblp.uni-trier.de/rec/bibtex/#1.xml}
  {dblp:#1}}
\def\mn@eprint@#1:#2:#3:#4\@nil{\def\@tempa {#1}\def\@tempb {#2}\def\@tempc
  {#3}\ifx \@tempc \@empty \let \@tempc \@tempb \let \@tempb \@tempa \fi \ifx
  \@tempb \@empty \def\@tempb {arXiv}\fi \@ifundefined
  {mn@eprint@\@tempb}{\@tempb:\@tempc}{\expandafter \expandafter \csname
  mn@eprint@\@tempb\endcsname \expandafter{\@tempc}}}

\bibitem[\protect\citeauthoryear{{Athanassoula}}{{Athanassoula}}{1992}]{Ath92b}
{Athanassoula} E.,  1992, \mnras, \href
  {http://adsabs.harvard.edu/abs/1992MNRAS.259..345A} {259, 345}

\bibitem[\protect\citeauthoryear{{Athanassoula}}{{Athanassoula}}{2005}]{Ath05b}
{Athanassoula} E.,  2005, \mn@doi [\mnras] {10.1111/j.1365-2966.2005.08872.x},
  \href {http://adsabs.harvard.edu/abs/2005MNRAS.358.1477A} {358, 1477}

\bibitem[\protect\citeauthoryear{{Athanassoula} \& {Misiriotis}}{{Athanassoula}
  \& {Misiriotis}}{2002}]{AthMis02}
{Athanassoula} E.,  {Misiriotis} A.,  2002, \mn@doi [\mnras]
  {10.1046/j.1365-8711.2002.05028.x}, \href
  {http://adsabs.harvard.edu/abs/2002MNRAS.330...35A} {330, 35}

\bibitem[\protect\citeauthoryear{{Athanassoula}, {Machado}  \&
  {Rodionov}}{{Athanassoula} et~al.}{2013}]{AthMacRod13}
{Athanassoula} E.,  {Machado} R.~E.~G.,   {Rodionov} S.~A.,  2013, \mn@doi
  [\mnras] {10.1093/mnras/sts452}, \href
  {http://adsabs.harvard.edu/abs/2013MNRAS.429.1949A} {429, 1949}

\bibitem[\protect\citeauthoryear{{Baldwin}, {Phillips}  \&
  {Terlevich}}{{Baldwin} et~al.}{1981}]{BalPhiTer81}
{Baldwin} J.~A.,  {Phillips} M.~M.,   {Terlevich} R.,  1981, \mn@doi [\pasp]
  {10.1086/130766}, \href {http://adsabs.harvard.edu/abs/1981PASP...93....5B}
  {93, 5}

\bibitem[\protect\citeauthoryear{{Bender}, {Saglia}  \& {Gerhard}}{{Bender}
  et~al.}{1994}]{BenSagGer94}
{Bender} R.,  {Saglia} R.~P.,   {Gerhard} O.~E.,  1994, \mnras, \href
  {http://adsabs.harvard.edu/abs/1994MNRAS.269..785B} {269, 785}

\bibitem[\protect\citeauthoryear{{Buta} \& {Combes}}{{Buta} \&
  {Combes}}{1996}]{ButCom96}
{Buta} R.,  {Combes} F.,  1996, Fundamentals of Cosmic Physics, \href
  {http://adsabs.harvard.edu/abs/1996FCPh...17...95B} {17, 95}

\bibitem[\protect\citeauthoryear{{Buta} et~al.,}{{Buta}
  et~al.}{2015}]{ButSheAth15}
{Buta} R.~J.,  et~al., 2015, \mn@doi [\apjs] {10.1088/0067-0049/217/2/32},
  \href {http://adsabs.harvard.edu/abs/2015ApJS..217...32B} {217, 32}

\bibitem[\protect\citeauthoryear{{Cappellari} \& {Copin}}{{Cappellari} \&
  {Copin}}{2003}]{CapCop03}
{Cappellari} M.,  {Copin} Y.,  2003, \mn@doi [\mnras]
  {10.1046/j.1365-8711.2003.06541.x}, \href
  {http://adsabs.harvard.edu/abs/2003MNRAS.342..345C} {342, 345}

\bibitem[\protect\citeauthoryear{{Cappellari} \& {Emsellem}}{{Cappellari} \&
  {Emsellem}}{2004}]{CapEms04}
{Cappellari} M.,  {Emsellem} E.,  2004, \mn@doi [\pasp] {10.1086/381875}, \href
  {http://adsabs.harvard.edu/abs/2004PASP..116..138C} {116, 138}

\bibitem[\protect\citeauthoryear{{Cenarro}, {Cardiel}, {Gorgas}, {Peletier},
  {Vazdekis}  \& {Prada}}{{Cenarro} et~al.}{2001}]{CenCarGor01}
{Cenarro} A.~J.,  {Cardiel} N.,  {Gorgas} J.,  {Peletier} R.~F.,  {Vazdekis}
  A.,   {Prada} F.,  2001, \mn@doi [\mnras] {10.1046/j.1365-8711.2001.04688.x},
  \href {http://adsabs.harvard.edu/abs/2001MNRAS.326..959C} {326, 959}

\bibitem[\protect\citeauthoryear{{Cid Fernandes}, {Stasi{\'n}ska}, {Mateus}  \&
  {Vale Asari}}{{Cid Fernandes} et~al.}{2011}]{CidStaMat11}
{Cid Fernandes} R.,  {Stasi{\'n}ska} G.,  {Mateus} A.,   {Vale Asari} N.,
  2011, \mn@doi [\mnras] {10.1111/j.1365-2966.2011.18244.x}, \href
  {http://adsabs.harvard.edu/abs/2011MNRAS.413.1687C} {413, 1687}

\bibitem[\protect\citeauthoryear{{Coccato}, {Fabricius}, {Saglia}, {Bender},
  {Erwin}, {Drory}  \& {Morelli}}{{Coccato} et~al.}{2018}]{CocFabSag18}
{Coccato} L.,  {Fabricius} M.~H.,  {Saglia} R.~P.,  {Bender} R.,  {Erwin} P.,
  {Drory} N.,   {Morelli} L.,  2018, \mn@doi [\mnras] {10.1093/mnras/sty705},
  \href {http://esoads.eso.org/abs/2018MNRAS.477.1958C} {477, 1958}

\bibitem[\protect\citeauthoryear{{Cole}, {Debattista}, {Erwin}, {Earp}  \&
  {Ro{\v s}kar}}{{Cole} et~al.}{2014}]{ColDebErw14}
{Cole} D.~R.,  {Debattista} V.~P.,  {Erwin} P.,  {Earp} S.~W.~F.,   {Ro{\v
  s}kar} R.,  2014, \mn@doi [\mnras] {10.1093/mnras/stu1985}, \href
  {http://adsabs.harvard.edu/abs/2014MNRAS.445.3352C} {445, 3352}

\bibitem[\protect\citeauthoryear{{Combes} \& {Gerin}}{{Combes} \&
  {Gerin}}{1985}]{ComGer85}
{Combes} F.,  {Gerin} M.,  1985, \aap, \href
  {http://adsabs.harvard.edu/abs/1985A%26A...150..327C} {150, 327}

\bibitem[\protect\citeauthoryear{{Cowie}, {Songaila}, {Hu}  \& {Cohen}}{{Cowie}
  et~al.}{1996}]{CowSonHu96}
{Cowie} L.~L.,  {Songaila} A.,  {Hu} E.~M.,   {Cohen} J.~G.,  1996, \mn@doi
  [\aj] {10.1086/118058}, \href
  {http://adsabs.harvard.edu/abs/1996AJ....112..839C} {112, 839}

\bibitem[\protect\citeauthoryear{{Davies}, {Maciejewski}, {Hicks}, {Tacconi},
  {Genzel}  \& {Engel}}{{Davies} et~al.}{2009}]{DavMacHic09}
{Davies} R.~I.,  {Maciejewski} W.,  {Hicks} E.~K.~S.,  {Tacconi} L.~J.,
  {Genzel} R.,   {Engel} H.,  2009, \mn@doi [\apj]
  {10.1088/0004-637X/702/1/114}, \href
  {http://esoads.eso.org/abs/2009ApJ...702..114D} {702, 114}

\bibitem[\protect\citeauthoryear{{Elmegreen}, {Bournaud}  \&
  {Elmegreen}}{{Elmegreen} et~al.}{2008}]{ElmBouElm08}
{Elmegreen} B.~G.,  {Bournaud} F.,   {Elmegreen} D.~M.,  2008, \mn@doi [\apj]
  {10.1086/592190}, \href {http://adsabs.harvard.edu/abs/2008ApJ...688...67E}
  {688, 67}

\bibitem[\protect\citeauthoryear{{Emsellem}, {Renaud}, {Bournaud}, {Elmegreen},
  {Combes}  \& {Gabor}}{{Emsellem} et~al.}{2015}]{EmsRenBou15}
{Emsellem} E.,  {Renaud} F.,  {Bournaud} F.,  {Elmegreen} B.,  {Combes} F.,
  {Gabor} J.~M.,  2015, \mn@doi [\mnras] {10.1093/mnras/stu2209}, \href
  {http://adsabs.harvard.edu/abs/2015MNRAS.446.2468E} {446, 2468}

\bibitem[\protect\citeauthoryear{{Epinat} et~al.,}{{Epinat}
  et~al.}{2012}]{EpiTasAmr12}
{Epinat} B.,  et~al., 2012, \mn@doi [\aap] {10.1051/0004-6361/201117711}, \href
  {http://adsabs.harvard.edu/abs/2012A%26A...539A..92E} {539, A92}

\bibitem[\protect\citeauthoryear{{Eskridge}, {Frogel}, {Pogge}  \& {et
  al.}}{{Eskridge} et~al.}{2000}]{EskFroPog00}
{Eskridge} P.~B.,  {Frogel} J.~A.,  {Pogge} R.~W.,   {et al.} 2000, \mn@doi
  [\aj] {10.1086/301203}, \href
  {http://adsabs.harvard.edu/abs/2000AJ....119..536E} {119, 536}

\bibitem[\protect\citeauthoryear{{Falc{\'o}n-Barroso}
  et~al.,}{{Falc{\'o}n-Barroso} et~al.}{2006}]{FalBacBur06}
{Falc{\'o}n-Barroso} J.,  et~al., 2006, \mn@doi [\mnras]
  {10.1111/j.1365-2966.2006.10261.x}, \href
  {http://adsabs.harvard.edu/abs/2006MNRAS.369..529F} {369, 529}

\bibitem[\protect\citeauthoryear{{Falc{\'o}n-Barroso},
  {S{\'a}nchez-Bl{\'a}zquez}, {Vazdekis}, {Ricciardelli}, {Cardiel}, {Cenarro},
  {Gorgas}  \& {Peletier}}{{Falc{\'o}n-Barroso} et~al.}{2011}]{FalSanVaz11}
{Falc{\'o}n-Barroso} J.,  {S{\'a}nchez-Bl{\'a}zquez} P.,  {Vazdekis} A.,
  {Ricciardelli} E.,  {Cardiel} N.,  {Cenarro} A.~J.,  {Gorgas} J.,
  {Peletier} R.~F.,  2011, \mn@doi [\aap] {10.1051/0004-6361/201116842}, \href
  {http://adsabs.harvard.edu/abs/2011A%26A...532A..95F} {532, A95}

\bibitem[\protect\citeauthoryear{{F{\"o}rster Schreiber} et~al.,}{{F{\"o}rster
  Schreiber} et~al.}{2006}]{ForGenLeh06}
{F{\"o}rster Schreiber} N.~M.,  et~al., 2006, \mn@doi [\apj] {10.1086/504403},
  \href {http://esoads.eso.org/abs/2006ApJ...645.1062F} {645, 1062}

\bibitem[\protect\citeauthoryear{{F{\"o}rster Schreiber} et~al.,}{{F{\"o}rster
  Schreiber} et~al.}{2009}]{ForGenBou09}
{F{\"o}rster Schreiber} N.~M.,  et~al., 2009, \mn@doi [\apj]
  {10.1088/0004-637X/706/2/1364}, \href
  {http://adsabs.harvard.edu/abs/2009ApJ...706.1364F} {706, 1364}

\bibitem[\protect\citeauthoryear{{Fragkoudi}, {Athanassoula}  \&
  {Bosma}}{{Fragkoudi} et~al.}{2016}]{FraAthBos16}
{Fragkoudi} F.,  {Athanassoula} E.,   {Bosma} A.,  2016, \mn@doi [\mnras]
  {10.1093/mnrasl/slw120}, \href
  {http://esoads.eso.org/abs/2016MNRAS.462L..41F} {462, L41}

\bibitem[\protect\citeauthoryear{{Gadotti}}{{Gadotti}}{2009}]{Gad09a}
{Gadotti} D.~A.,  2009, in {Contopoulos} G.,  {Patsis} P.,  eds, {Chaos in
  Astronomy}. Springer Berlin Heidelberg, pp 159--172 (arXiv:0802.0495)
  (\mn@eprint {} {0802.0495})

\bibitem[\protect\citeauthoryear{{Gadotti}}{{Gadotti}}{2011}]{Gad11}
{Gadotti} D.~A.,  2011, \mn@doi [\mnras] {10.1111/j.1365-2966.2011.18945.x},
  \href {http://adsabs.harvard.edu/abs/2011MNRAS.415.3308G} {415, 3308}

\bibitem[\protect\citeauthoryear{{Gadotti}, {Seidel},
  {S{\'a}nchez-Bl{\'a}zquez}, {Falc{\'o}n-Barroso}, {Husemann}, {Coelho}  \&
  {P{\'e}rez}}{{Gadotti} et~al.}{2015}]{GadSeiSan15}
{Gadotti} D.~A.,  {Seidel} M.~K.,  {S{\'a}nchez-Bl{\'a}zquez} P.,
  {Falc{\'o}n-Barroso} J.,  {Husemann} B.,  {Coelho} P.,   {P{\'e}rez} I.,
  2015, \mn@doi [\aap] {10.1051/0004-6361/201526677}, \href
  {http://adsabs.harvard.edu/abs/2015A%26A...584A..90G} {584, A90}

\bibitem[\protect\citeauthoryear{{Gallo}, {Treu}, {Marshall}, {Woo}, {Leipski}
  \& {Antonucci}}{{Gallo} et~al.}{2010}]{GalTreMar10}
{Gallo} E.,  {Treu} T.,  {Marshall} P.~J.,  {Woo} J.-H.,  {Leipski} C.,
  {Antonucci} R.,  2010, \mn@doi [\apj] {10.1088/0004-637X/714/1/25}, \href
  {http://adsabs.harvard.edu/abs/2010ApJ...714...25G} {714, 25}

\bibitem[\protect\citeauthoryear{{Genzel} et~al.,}{{Genzel}
  et~al.}{2006}]{GenTacEis06}
{Genzel} R.,  et~al., 2006, \mn@doi [\nat] {10.1038/nature05052}, \href
  {http://adsabs.harvard.edu/abs/2006Natur.442..786G} {442, 786}

\bibitem[\protect\citeauthoryear{{Genzel} et~al.,}{{Genzel}
  et~al.}{2008}]{GenBurBou08}
{Genzel} R.,  et~al., 2008, \mn@doi [\apj] {10.1086/591840}, \href
  {http://esoads.eso.org/abs/2008ApJ...687...59G} {687, 59}

\bibitem[\protect\citeauthoryear{{Gonz{\'a}lez Delgado} et~al.,}{{Gonz{\'a}lez
  Delgado} et~al.}{2017}]{GonPerCid17}
{Gonz{\'a}lez Delgado} R.~M.,  et~al., 2017, \mn@doi [\aap]
  {10.1051/0004-6361/201730883}, \href
  {http://esoads.eso.org/abs/2017A%26A...607A.128G} {607, A128}

\bibitem[\protect\citeauthoryear{{Kauffmann}, {Heckman}, {Tremonti}  \& {et
  al.}}{{Kauffmann} et~al.}{2003}]{KauHecTre03}
{Kauffmann} G.,  {Heckman} T.~M.,  {Tremonti} C.,   {et al.} 2003, \mn@doi
  [\mnras] {10.1111/j.1365-2966.2003.07154.x}, \href
  {http://adsabs.harvard.edu/abs/2003MNRAS.346.1055K} {346, 1055}

\bibitem[\protect\citeauthoryear{{Kewley}, {Dopita}, {Sutherland}, {Heisler}
  \& {Trevena}}{{Kewley} et~al.}{2001}]{KewDopSut01}
{Kewley} L.~J.,  {Dopita} M.~A.,  {Sutherland} R.~S.,  {Heisler} C.~A.,
  {Trevena} J.,  2001, \mn@doi [\apj] {10.1086/321545}, \href
  {http://adsabs.harvard.edu/abs/2001ApJ...556..121K} {556, 121}

\bibitem[\protect\citeauthoryear{{Kewley}, {Groves}, {Kauffmann}  \&
  {Heckman}}{{Kewley} et~al.}{2006}]{KewGroKau06}
{Kewley} L.~J.,  {Groves} B.,  {Kauffmann} G.,   {Heckman} T.,  2006, \mn@doi
  [\mnras] {10.1111/j.1365-2966.2006.10859.x}, \href
  {http://adsabs.harvard.edu/abs/2006MNRAS.372..961K} {372, 961}

\bibitem[\protect\citeauthoryear{{Kewley}, {Rupke}, {Zahid}, {Geller}  \&
  {Barton}}{{Kewley} et~al.}{2010}]{KewRupZah10}
{Kewley} L.~J.,  {Rupke} D.,  {Zahid} H.~J.,  {Geller} M.~J.,   {Barton} E.~J.,
   2010, \mn@doi [\apjl] {10.1088/2041-8205/721/1/L48}, \href
  {http://esoads.eso.org/abs/2010ApJ...721L..48K} {721, L48}

\bibitem[\protect\citeauthoryear{{Kim}, {Seo}, {Stone}, {Yoon}  \&
  {Teuben}}{{Kim} et~al.}{2012}]{KimSeoSto12}
{Kim} W.-T.,  {Seo} W.-Y.,  {Stone} J.~M.,  {Yoon} D.,   {Teuben} P.~J.,  2012,
  \mn@doi [\apj] {10.1088/0004-637X/747/1/60}, \href
  {http://adsabs.harvard.edu/abs/2012ApJ...747...60K} {747, 60}

\bibitem[\protect\citeauthoryear{{Knapen}}{{Knapen}}{2007}]{Kna07}
{Knapen} J.~H.,  2007, {Barred Galaxies and Galaxy Evolution}.
p.~175, \mn@doi{10.1007/978-1-4020-5573-7_29}

\bibitem[\protect\citeauthoryear{{Krajnovi{\'c}} et~al.,}{{Krajnovi{\'c}}
  et~al.}{2015}]{KraWeiUrr15}
{Krajnovi{\'c}} D.,  et~al., 2015, \mn@doi [\mnras] {10.1093/mnras/stv958},
  \href {http://esoads.eso.org/abs/2015MNRAS.452....2K} {452, 2}

\bibitem[\protect\citeauthoryear{{Kroupa}}{{Kroupa}}{2001}]{Kro01}
{Kroupa} P.,  2001, \mn@doi [\mnras] {10.1046/j.1365-8711.2001.04022.x}, \href
  {http://adsabs.harvard.edu/abs/2001MNRAS.322..231K} {322, 231}

\bibitem[\protect\citeauthoryear{{Kuntschner} et~al.,}{{Kuntschner}
  et~al.}{2010}]{KunEmsBac10}
{Kuntschner} H.,  et~al., 2010, \mn@doi [\mnras]
  {10.1111/j.1365-2966.2010.17161.x}, \href
  {http://adsabs.harvard.edu/abs/2010MNRAS.408...97K} {408, 97}

\bibitem[\protect\citeauthoryear{{Law}, {Steidel}, {Erb}, {Larkin}, {Pettini},
  {Shapley}  \& {Wright}}{{Law} et~al.}{2007}]{LawSteErb07}
{Law} D.~R.,  {Steidel} C.~C.,  {Erb} D.~K.,  {Larkin} J.~E.,  {Pettini} M.,
  {Shapley} A.~E.,   {Wright} S.~A.,  2007, \mn@doi [\apj] {10.1086/521786},
  \href {http://esoads.eso.org/abs/2007ApJ...669..929L} {669, 929}

\bibitem[\protect\citeauthoryear{{Law}, {Steidel}, {Erb}, {Larkin}, {Pettini},
  {Shapley}  \& {Wright}}{{Law} et~al.}{2009}]{LawSteErb09}
{Law} D.~R.,  {Steidel} C.~C.,  {Erb} D.~K.,  {Larkin} J.~E.,  {Pettini} M.,
  {Shapley} A.~E.,   {Wright} S.~A.,  2009, \mn@doi [\apj]
  {10.1088/0004-637X/697/2/2057}, \href
  {http://esoads.eso.org/abs/2009ApJ...697.2057L} {697, 2057}

\bibitem[\protect\citeauthoryear{{Maoz}, {Barth}, {Ho}, {Sternberg}  \&
  {Filippenko}}{{Maoz} et~al.}{2001}]{MaoBarHo01}
{Maoz} D.,  {Barth} A.~J.,  {Ho} L.~C.,  {Sternberg} A.,   {Filippenko} A.~V.,
  2001, \mn@doi [\aj] {10.1086/321080}, \href
  {http://adsabs.harvard.edu/abs/2001AJ....121.3048M} {121, 3048}

\bibitem[\protect\citeauthoryear{{Marino} et~al.,}{{Marino}
  et~al.}{2013}]{MarRosSan13}
{Marino} R.~A.,  et~al., 2013, \mn@doi [\aap] {10.1051/0004-6361/201321956},
  \href {http://adsabs.harvard.edu/abs/2013A%26A...559A.114M} {559, A114}

\bibitem[\protect\citeauthoryear{{Masters} et~al.,}{{Masters}
  et~al.}{2011}]{MasNicHoy11}
{Masters} K.~L.,  et~al., 2011, \mn@doi [\mnras]
  {10.1111/j.1365-2966.2010.17834.x}, \href
  {http://adsabs.harvard.edu/abs/2011MNRAS.411.2026M} {411, 2026}

\bibitem[\protect\citeauthoryear{{McDermid} et~al.,}{{McDermid}
  et~al.}{2015}]{McDAlaBli15}
{McDermid} R.~M.,  et~al., 2015, \mn@doi [\mnras] {10.1093/mnras/stv105}, \href
  {http://esoads.eso.org/abs/2015MNRAS.448.3484M} {448, 3484}

\bibitem[\protect\citeauthoryear{{Men{\'e}ndez-Delmestre}, {Sheth},
  {Schinnerer}, {Jarrett}  \& {Scoville}}{{Men{\'e}ndez-Delmestre}
  et~al.}{2007}]{MenSheSch07}
{Men{\'e}ndez-Delmestre} K.,  {Sheth} K.,  {Schinnerer} E.,  {Jarrett} T.~H.,
  {Scoville} N.~Z.,  2007, \mn@doi [\apj] {10.1086/511025}, \href
  {http://adsabs.harvard.edu/abs/2007ApJ...657..790M} {657, 790}

\bibitem[\protect\citeauthoryear{{Mu{\~n}oz-Mateos} et~al.,}{{Mu{\~n}oz-Mateos}
  et~al.}{2013}]{MunSheGil13}
{Mu{\~n}oz-Mateos} J.~C.,  et~al., 2013, \mn@doi [\apj]
  {10.1088/0004-637X/771/1/59}, \href
  {http://adsabs.harvard.edu/abs/2013ApJ...771...59M} {771, 59}

\bibitem[\protect\citeauthoryear{{Mu{\~n}oz-Mateos} et~al.,}{{Mu{\~n}oz-Mateos}
  et~al.}{2015}]{MunSheReg15}
{Mu{\~n}oz-Mateos} J.~C.,  et~al., 2015, \mn@doi [\apjs]
  {10.1088/0067-0049/219/1/3}, \href
  {http://adsabs.harvard.edu/abs/2015ApJS..219....3M} {219, 3}

\bibitem[\protect\citeauthoryear{{Ocvirk}, {Pichon}, {Lan{\c c}on}  \&
  {Thi{\'e}baut}}{{Ocvirk} et~al.}{2006a}]{OcvPicLan06a}
{Ocvirk} P.,  {Pichon} C.,  {Lan{\c c}on} A.,   {Thi{\'e}baut} E.,  2006a,
  \mn@doi [\mnras] {10.1111/j.1365-2966.2005.09182.x}, \href
  {http://adsabs.harvard.edu/abs/2006MNRAS.365...46O} {365, 46}

\bibitem[\protect\citeauthoryear{{Ocvirk}, {Pichon}, {Lan{\c c}on}  \&
  {Thi{\'e}baut}}{{Ocvirk} et~al.}{2006b}]{OcvPicLan06b}
{Ocvirk} P.,  {Pichon} C.,  {Lan{\c c}on} A.,   {Thi{\'e}baut} E.,  2006b,
  \mn@doi [\mnras] {10.1111/j.1365-2966.2005.09323.x}, \href
  {http://adsabs.harvard.edu/abs/2006MNRAS.365...74O} {365, 74}

\bibitem[\protect\citeauthoryear{{Osterbrock} \& {Pogge}}{{Osterbrock} \&
  {Pogge}}{1985}]{OstPog85}
{Osterbrock} D.~E.,  {Pogge} R.~W.,  1985, \mn@doi [\apj] {10.1086/163513},
  \href {http://adsabs.harvard.edu/abs/1985ApJ...297..166O} {297, 166}

\bibitem[\protect\citeauthoryear{{Peletier} et~al.,}{{Peletier}
  et~al.}{2012}]{PelKutvan12}
{Peletier} R.~F.,  et~al., 2012, \mn@doi [\mnras]
  {10.1111/j.1365-2966.2011.19855.x}, \href
  {http://adsabs.harvard.edu/abs/2012MNRAS.419.2031P} {419, 2031}

\bibitem[\protect\citeauthoryear{{P{\'e}rez} et~al.,}{{P{\'e}rez}
  et~al.}{2017}]{PerMarRui17}
{P{\'e}rez} I.,  et~al., 2017, \mn@doi [\mnras] {10.1093/mnrasl/slx087}, \href
  {http://esoads.eso.org/abs/2017MNRAS.470L.122P} {470, L122}

\bibitem[\protect\citeauthoryear{{Pietrinferni}, {Cassisi}, {Salaris}  \&
  {Castelli}}{{Pietrinferni} et~al.}{2004}]{PieCasSal04}
{Pietrinferni} A.,  {Cassisi} S.,  {Salaris} M.,   {Castelli} F.,  2004,
  \mn@doi [\apj] {10.1086/422498}, \href
  {http://esoads.eso.org/abs/2004ApJ...612..168P} {612, 168}

\bibitem[\protect\citeauthoryear{{Pietrinferni}, {Cassisi}, {Salaris}  \&
  {Castelli}}{{Pietrinferni} et~al.}{2006}]{PieCasSal06}
{Pietrinferni} A.,  {Cassisi} S.,  {Salaris} M.,   {Castelli} F.,  2006,
  \mn@doi [\apj] {10.1086/501344}, \href
  {http://esoads.eso.org/abs/2006ApJ...642..797P} {642, 797}

\bibitem[\protect\citeauthoryear{{Pietrinferni}, {Cassisi}, {Salaris},
  {Percival}  \& {Ferguson}}{{Pietrinferni} et~al.}{2009}]{PieCasSal09}
{Pietrinferni} A.,  {Cassisi} S.,  {Salaris} M.,  {Percival} S.,   {Ferguson}
  J.~W.,  2009, \mn@doi [\apj] {10.1088/0004-637X/697/1/275}, \href
  {http://esoads.eso.org/abs/2009ApJ...697..275P} {697, 275}

\bibitem[\protect\citeauthoryear{{Pietrinferni}, {Cassisi}, {Salaris}  \&
  {Hidalgo}}{{Pietrinferni} et~al.}{2013}]{PieCasSal13}
{Pietrinferni} A.,  {Cassisi} S.,  {Salaris} M.,   {Hidalgo} S.,  2013, \mn@doi
  [\aap] {10.1051/0004-6361/201321950}, \href
  {http://esoads.eso.org/abs/2013A%26A...558A..46P} {558, A46}

\bibitem[\protect\citeauthoryear{{Piner}, {Stone}  \& {Teuben}}{{Piner}
  et~al.}{1995}]{PinStoTeu95}
{Piner} B.~G.,  {Stone} J.~M.,   {Teuben} P.~J.,  1995, \mn@doi [\apj]
  {10.1086/176075}, \href {http://adsabs.harvard.edu/abs/1995ApJ...449..508P}
  {449, 508}

\bibitem[\protect\citeauthoryear{{Planck Collaboration} et~al.,}{{Planck
  Collaboration} et~al.}{2015}]{AdeAghArn15}
{Planck Collaboration} et~al., 2015, ArXiv e-prints:1502.01589, \href
  {http://adsabs.harvard.edu/abs/2015arXiv150201589P} {}

\bibitem[\protect\citeauthoryear{{Querejeta} et~al.,}{{Querejeta}
  et~al.}{2015}]{QueMeiSch15}
{Querejeta} M.,  et~al., 2015, \mn@doi [\apjs] {10.1088/0067-0049/219/1/5},
  \href {http://esoads.eso.org/abs/2015ApJS..219....5Q} {219, 5}

\bibitem[\protect\citeauthoryear{{Querejeta} et~al.,}{{Querejeta}
  et~al.}{2016}]{QueMeiSch16}
{Querejeta} M.,  et~al., 2016, \mn@doi [\aap] {10.1051/0004-6361/201527536},
  \href {http://esoads.eso.org/abs/2016A%26A...588A..33Q} {588, A33}

\bibitem[\protect\citeauthoryear{{Rautiainen} \& {Salo}}{{Rautiainen} \&
  {Salo}}{2000}]{RauSal00}
{Rautiainen} P.,  {Salo} H.,  2000, \aap, \href
  {http://adsabs.harvard.edu/abs/2000A%26A...362..465R} {362, 465}

\bibitem[\protect\citeauthoryear{{Regan} \& {Teuben}}{{Regan} \&
  {Teuben}}{2003}]{RegTeu03}
{Regan} M.~W.,  {Teuben} P.,  2003, \mn@doi [\apj] {10.1086/344721}, \href
  {http://adsabs.harvard.edu/abs/2003ApJ...582..723R} {582, 723}

\bibitem[\protect\citeauthoryear{{Regan} \& {Teuben}}{{Regan} \&
  {Teuben}}{2004}]{RegTeu04}
{Regan} M.~W.,  {Teuben} P.~J.,  2004, \mn@doi [\apj] {10.1086/380116}, \href
  {http://adsabs.harvard.edu/abs/2004ApJ...600..595R} {600, 595}

\bibitem[\protect\citeauthoryear{{Rodrigues}, {Hammer}, {Flores}, {Puech}  \&
  {Athanassoula}}{{Rodrigues} et~al.}{2017}]{RodHamFlo17}
{Rodrigues} M.,  {Hammer} F.,  {Flores} H.,  {Puech} M.,   {Athanassoula} E.,
  2017, \mn@doi [\mnras] {10.1093/mnras/stw2711}, \href
  {http://adsabs.harvard.edu/abs/2017MNRAS.465.1157R} {465, 1157}

\bibitem[\protect\citeauthoryear{{Ruiz-Lara} et~al.,}{{Ruiz-Lara}
  et~al.}{2015}]{RuiPerGal15}
{Ruiz-Lara} T.,  et~al., 2015, \mn@doi [\aap] {10.1051/0004-6361/201526752},
  \href {http://esoads.eso.org/abs/2015A%26A...583A..60R} {583, A60}

\bibitem[\protect\citeauthoryear{{Sakamoto}, {Okumura}, {Ishizuki}  \&
  {Scoville}}{{Sakamoto} et~al.}{1999}]{SakOkuIsh99}
{Sakamoto} K.,  {Okumura} S.~K.,  {Ishizuki} S.,   {Scoville} N.~Z.,  1999,
  \mn@doi [\apj] {10.1086/307910}, \href
  {http://adsabs.harvard.edu/abs/1999ApJ...525..691S} {525, 691}

\bibitem[\protect\citeauthoryear{{S{\'a}nchez-Bl{\'a}zquez}, {Ocvirk},
  {Gibson}, {P{\'e}rez}  \& {Peletier}}{{S{\'a}nchez-Bl{\'a}zquez}
  et~al.}{2011}]{SanOcvGib11}
{S{\'a}nchez-Bl{\'a}zquez} P.,  {Ocvirk} P.,  {Gibson} B.~K.,  {P{\'e}rez} I.,
   {Peletier} R.~F.,  2011, \mn@doi [\mnras]
  {10.1111/j.1365-2966.2011.18749.x}, \href
  {http://adsabs.harvard.edu/abs/2011MNRAS.415..709S} {415, 709}

\bibitem[\protect\citeauthoryear{{Sarzi} et~al.,}{{Sarzi}
  et~al.}{2006}]{SarFalDav06}
{Sarzi} M.,  et~al., 2006, \mn@doi [\mnras] {10.1111/j.1365-2966.2005.09839.x},
  \href {http://adsabs.harvard.edu/abs/2006MNRAS.366.1151S} {366, 1151}

\bibitem[\protect\citeauthoryear{{Seidel} et~al.,}{{Seidel}
  et~al.}{2015}]{SeiCacRui15}
{Seidel} M.~K.,  et~al., 2015, \mn@doi [\mnras] {10.1093/mnras/stu2295}, \href
  {http://adsabs.harvard.edu/abs/2015MNRAS.446.2837S} {446, 2837}

\bibitem[\protect\citeauthoryear{{Sellwood} \& {Wilkinson}}{{Sellwood} \&
  {Wilkinson}}{1993}]{SelWil93}
{Sellwood} J.~A.,  {Wilkinson} A.,  1993, Reports of Progress in Physics, \href
  {http://adsabs.harvard.edu/abs/1993RPPh...56..173S} {56, 173}

\bibitem[\protect\citeauthoryear{{S{\'e}rsic} \& {Pastoriza}}{{S{\'e}rsic} \&
  {Pastoriza}}{1965}]{SerPas65}
{S{\'e}rsic} J.~L.,  {Pastoriza} M.,  1965, \pasp, \href
  {http://adsabs.harvard.edu/abs/1965PASP...77..287S} {77, 287}

\bibitem[\protect\citeauthoryear{{S{\'e}rsic} \& {Pastoriza}}{{S{\'e}rsic} \&
  {Pastoriza}}{1967}]{SerPas67}
{S{\'e}rsic} J.~L.,  {Pastoriza} M.,  1967, \pasp, \href
  {http://adsabs.harvard.edu/abs/1967PASP...79..152S} {79, 152}

\bibitem[\protect\citeauthoryear{{Shapiro} et~al.,}{{Shapiro}
  et~al.}{2008}]{ShaGenFor08}
{Shapiro} K.~L.,  et~al., 2008, \mn@doi [\apj] {10.1086/587133}, \href
  {http://adsabs.harvard.edu/abs/2008ApJ...682..231S} {682, 231}

\bibitem[\protect\citeauthoryear{{Sheth}, {Vogel}, {Regan}, {Thornley}  \&
  {Teuben}}{{Sheth} et~al.}{2005}]{SheVogReg05}
{Sheth} K.,  {Vogel} S.~N.,  {Regan} M.~W.,  {Thornley} M.~D.,   {Teuben}
  P.~J.,  2005, \mn@doi [\apj] {10.1086/432409}, \href
  {http://adsabs.harvard.edu/abs/2005ApJ...632..217S} {632, 217}

\bibitem[\protect\citeauthoryear{{Sheth}, {Regan}, {Hinz}  \& {et al.}}{{Sheth}
  et~al.}{2010}]{shereghin10}
{Sheth} K.,  {Regan} M.,  {Hinz} J.~L.,   {et al.} 2010, \mn@doi [\pasp]
  {10.1086/657638}, \href {http://adsabs.harvard.edu/abs/2010PASP..122.1397S}
  {122, 1397}

\bibitem[\protect\citeauthoryear{{Sheth}, {Melbourne}, {Elmegreen},
  {Elmegreen}, {Athanassoula}, {Abraham}  \& {Weiner}}{{Sheth}
  et~al.}{2012}]{SheMelElm12}
{Sheth} K.,  {Melbourne} J.,  {Elmegreen} D.~M.,  {Elmegreen} B.~G.,
  {Athanassoula} E.,  {Abraham} R.~G.,   {Weiner} B.~J.,  2012, \mn@doi [\apj]
  {10.1088/0004-637X/758/2/136}, \href
  {http://adsabs.harvard.edu/abs/2012ApJ...758..136S} {758, 136}

\bibitem[\protect\citeauthoryear{{Simmons} et~al.,}{{Simmons}
  et~al.}{2014}]{SimMelLin14}
{Simmons} B.~D.,  et~al., 2014, \mn@doi [\mnras] {10.1093/mnras/stu1817}, \href
  {http://adsabs.harvard.edu/abs/2014MNRAS.445.3466S} {445, 3466}

\bibitem[\protect\citeauthoryear{{Sormani}, {Binney}  \& {Magorrian}}{{Sormani}
  et~al.}{2015}]{SorBinMag15}
{Sormani} M.~C.,  {Binney} J.,   {Magorrian} J.,  2015, \mn@doi [\mnras]
  {10.1093/mnras/stv441}, \href
  {http://adsabs.harvard.edu/abs/2015MNRAS.449.2421S} {449, 2421}

\bibitem[\protect\citeauthoryear{{Soto}, {Lilly}, {Bacon}, {Richard}  \&
  {Conseil}}{{Soto} et~al.}{2016}]{SotLilBac16}
{Soto} K.~T.,  {Lilly} S.~J.,  {Bacon} R.,  {Richard} J.,   {Conseil} S.,
  2016, \mn@doi [\mnras] {10.1093/mnras/stw474}, \href
  {http://adsabs.harvard.edu/abs/2016MNRAS.458.3210S} {458, 3210}

\bibitem[\protect\citeauthoryear{{Stasi{\'n}ska}, {Tenorio-Tagle},
  {Rodr{\'{\i}}guez}  \& {Henney}}{{Stasi{\'n}ska} et~al.}{2007}]{StaTenRod07}
{Stasi{\'n}ska} G.,  {Tenorio-Tagle} G.,  {Rodr{\'{\i}}guez} M.,   {Henney}
  W.~J.,  2007, \mn@doi [\aap] {10.1051/0004-6361:20065675}, \href
  {http://adsabs.harvard.edu/abs/2007A%26A...471..193S} {471, 193}

\bibitem[\protect\citeauthoryear{{Storey} \& {Zeippen}}{{Storey} \&
  {Zeippen}}{2000}]{StoZei00}
{Storey} P.~J.,  {Zeippen} C.~J.,  2000, \mn@doi [\mnras]
  {10.1046/j.1365-8711.2000.03184.x}, \href
  {http://adsabs.harvard.edu/abs/2000MNRAS.312..813S} {312, 813}

\bibitem[\protect\citeauthoryear{{Tacconi} et~al.,}{{Tacconi}
  et~al.}{2013}]{TacNerGen13}
{Tacconi} L.~J.,  et~al., 2013, \mn@doi [\apj] {10.1088/0004-637X/768/1/74},
  \href {http://adsabs.harvard.edu/abs/2013ApJ...768...74T} {768, 74}

\bibitem[\protect\citeauthoryear{{Thomas}, {Maraston}, {Schawinski}, {Sarzi}
  \& {Silk}}{{Thomas} et~al.}{2010}]{ThoMarSch10}
{Thomas} D.,  {Maraston} C.,  {Schawinski} K.,  {Sarzi} M.,   {Silk} J.,  2010,
  \mn@doi [\mnras] {10.1111/j.1365-2966.2010.16427.x}, \href
  {http://adsabs.harvard.edu/abs/2010MNRAS.404.1775T} {404, 1775}

\bibitem[\protect\citeauthoryear{{Valdes}, {Gupta}, {Rose}, {Singh}  \&
  {Bell}}{{Valdes} et~al.}{2004}]{ValGupRos04}
{Valdes} F.,  {Gupta} R.,  {Rose} J.~A.,  {Singh} H.~P.,   {Bell} D.~J.,  2004,
  \mn@doi [\apjs] {10.1086/386343}, \href
  {http://adsabs.harvard.edu/abs/2004ApJS..152..251V} {152, 251}

\bibitem[\protect\citeauthoryear{{Vazdekis} et~al.,}{{Vazdekis}
  et~al.}{2015}]{VazCoeCas15}
{Vazdekis} A.,  et~al., 2015, \mn@doi [\mnras] {10.1093/mnras/stv151}, \href
  {http://adsabs.harvard.edu/abs/2015MNRAS.449.1177V} {449, 1177}

\bibitem[\protect\citeauthoryear{{Walcher}, {Coelho}, {Gallazzi}, {Bruzual},
  {Charlot}  \& {Chiappini}}{{Walcher} et~al.}{2015}]{WalCoeGal15}
{Walcher} C.~J.,  {Coelho} P.~R.~T.,  {Gallazzi} A.,  {Bruzual} G.,  {Charlot}
  S.,   {Chiappini} C.,  2015, \mn@doi [\aap] {10.1051/0004-6361/201525924},
  \href {http://adsabs.harvard.edu/abs/2015A%26A...582A..46W} {582, A46}

\bibitem[\protect\citeauthoryear{{Weilbacher}, {Streicher}, {Urrutia}, {Jarno},
  {P{\'e}contal-Rousset}, {Bacon}  \& {B{\"o}hm}}{{Weilbacher}
  et~al.}{2012}]{WeiStrUrr12}
{Weilbacher} P.~M.,  {Streicher} O.,  {Urrutia} T.,  {Jarno} A.,
  {P{\'e}contal-Rousset} A.,  {Bacon} R.,   {B{\"o}hm} P.,  2012, in Society of
  Photo-Optical Instrumentation Engineers (SPIE) Conference Series. p.~0,
  \mn@doi{10.1117/12.925114}

\bibitem[\protect\citeauthoryear{{Wisnioski} et~al.,}{{Wisnioski}
  et~al.}{2015}]{WisForWuy15}
{Wisnioski} E.,  et~al., 2015, \mn@doi [\apj] {10.1088/0004-637X/799/2/209},
  \href {http://adsabs.harvard.edu/abs/2015ApJ...799..209W} {799, 209}

\bibitem[\protect\citeauthoryear{{Wozniak}}{{Wozniak}}{2007}]{Woz07}
{Wozniak} H.,  2007, \mn@doi [\aap] {10.1051/0004-6361:20067020}, \href
  {http://adsabs.harvard.edu/abs/2007A%26A...465L...1W} {465, L1}

\bibitem[\protect\citeauthoryear{{van den Bosch} \& {Emsellem}}{{van den Bosch}
  \& {Emsellem}}{1998}]{vanEms98}
{van den Bosch} F.~C.,  {Emsellem} E.,  1998, \mn@doi [\mnras]
  {10.1046/j.1365-8711.1998.01616.x}, \href
  {http://adsabs.harvard.edu/abs/1998MNRAS.298..267V} {298, 267}

\bibitem[\protect\citeauthoryear{{van der Marel} \& {Franx}}{{van der Marel} \&
  {Franx}}{1993}]{vanFra93}
{van der Marel} R.~P.,  {Franx} M.,  1993, \mn@doi [\apj] {10.1086/172534},
  \href {http://adsabs.harvard.edu/abs/1993ApJ...407..525V} {407, 525}

\makeatother
\end{thebibliography}

\appendix
\section{Uncertainties in the derivation of mean stellar ages and metallicities}
\label{sec:app_err}

\begin{figure}
\begin{center}
	\includegraphics[width=\columnwidth]{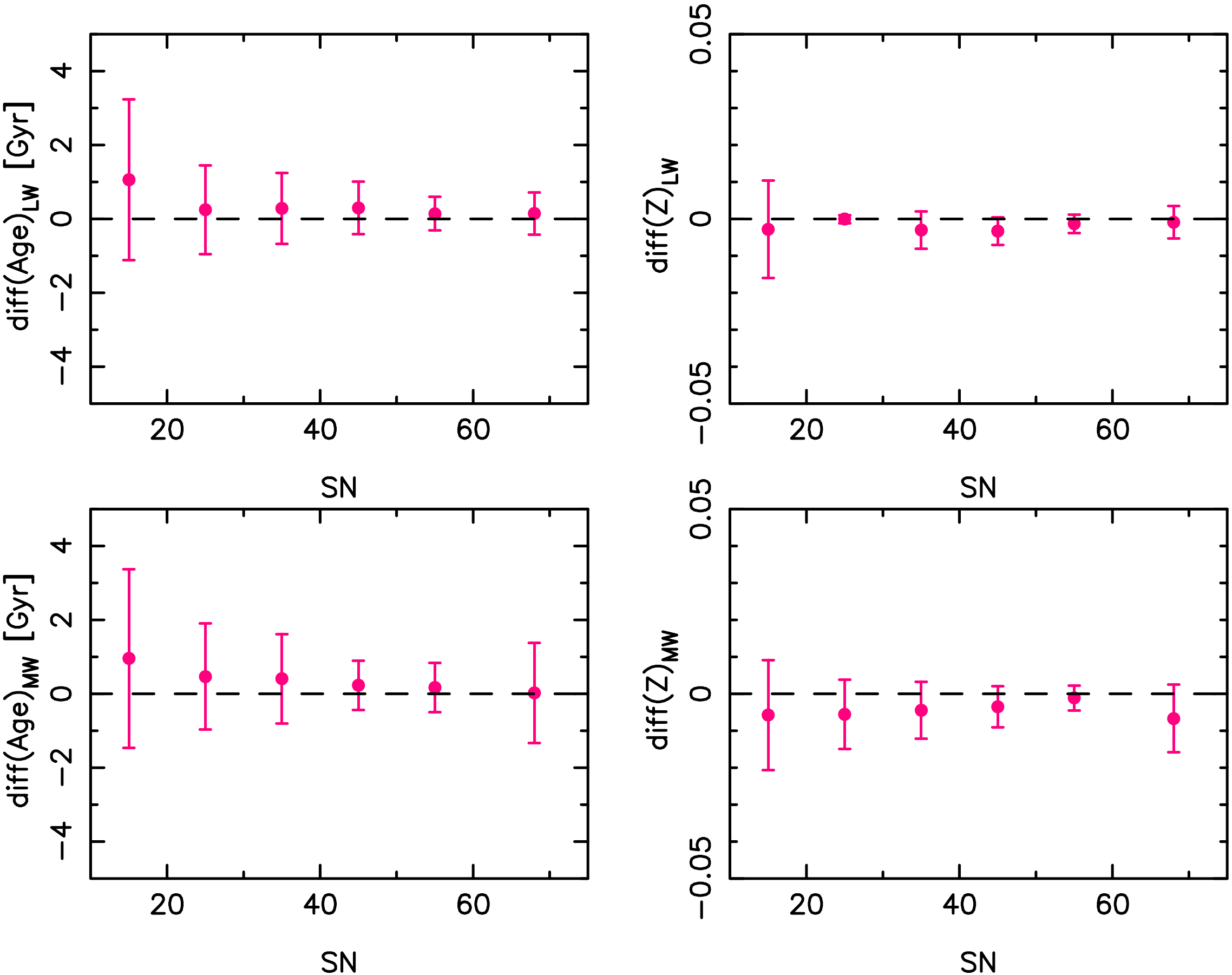}
\end{center}
    \caption{Differences between estimates of mean stellar age and metallicity (luminosity-weighted and mass-weighted, as indicated) when using a range of input parameters in {\tt STECKMAP}, and their variation with the signal-to-noise ratio of the spectrum. The filled circles indicate the mean values obtained with the different fits, whereas the error bars indicate the standard deviation of the different outputs. The dashed line indicates the position of the results obtained when using our canonical initial conditions employed in the fitting of all TIMER spectra.}
    \label{fig:err_guess}
\end{figure}

\begin{figure*}
\begin{center}
	\includegraphics[width=\columnwidth]{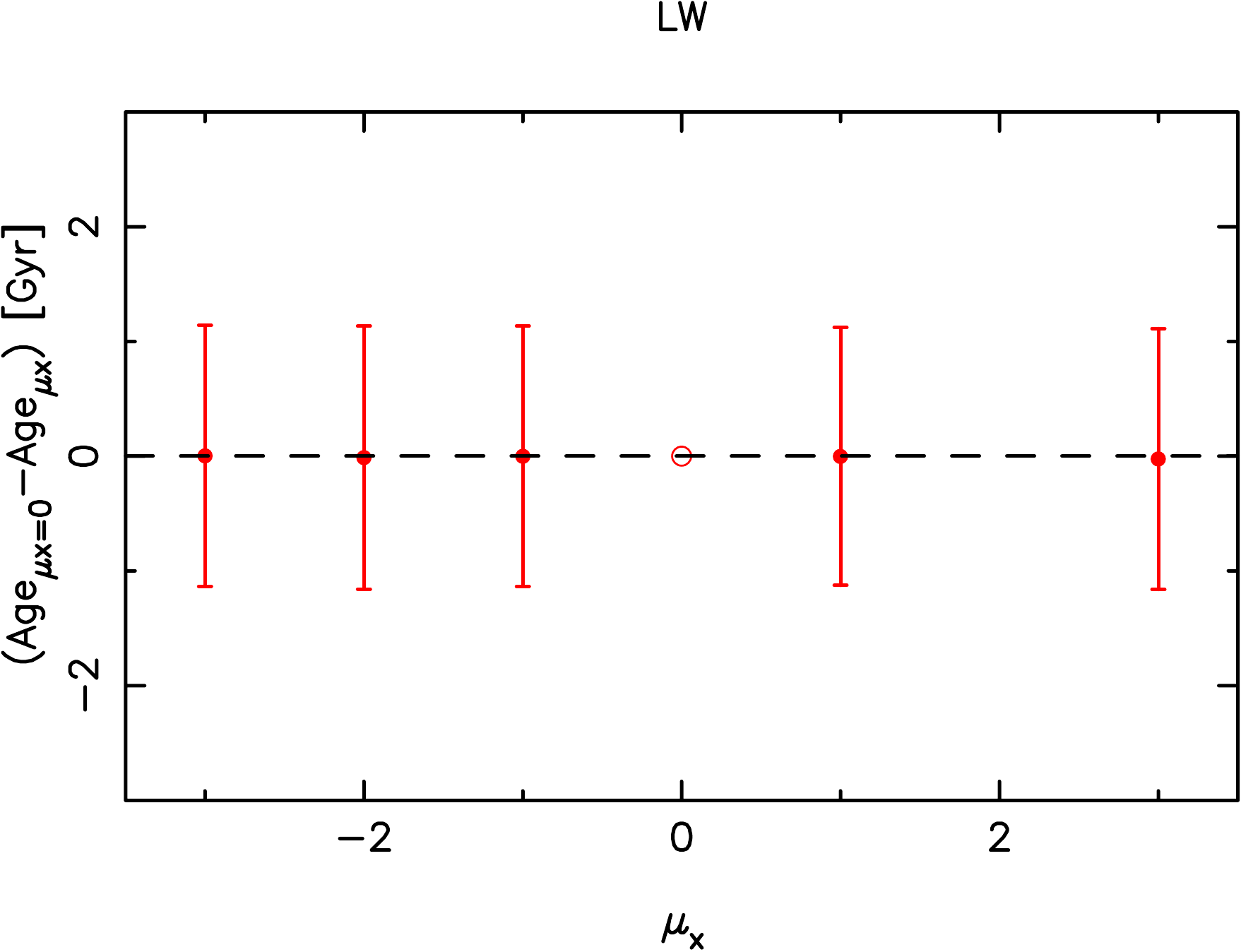}\hskip 0.5cm\includegraphics[width=\columnwidth]{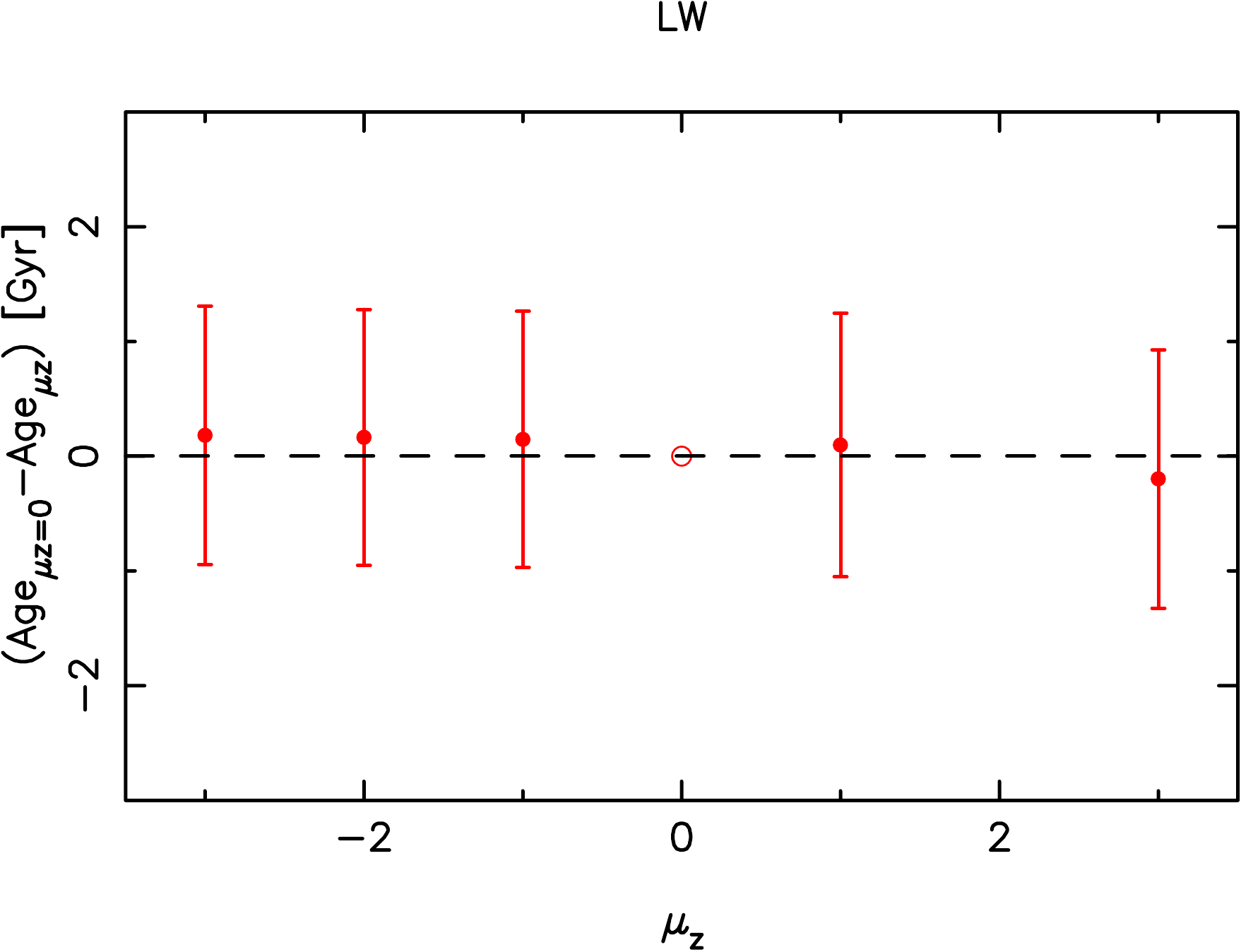}\\
	\vskip0.25cm
	\includegraphics[width=\columnwidth]{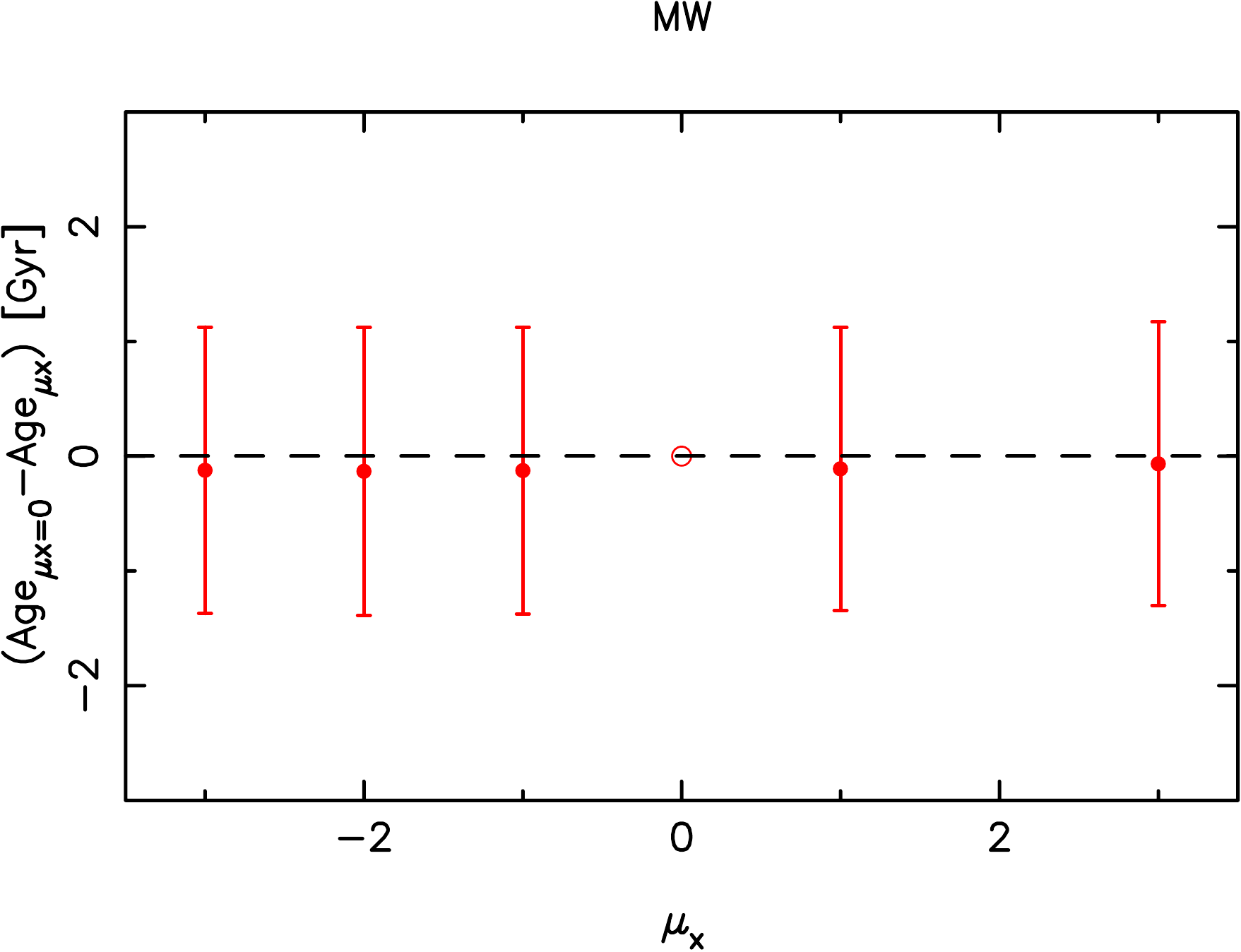}\hskip 0.5cm\includegraphics[width=\columnwidth]{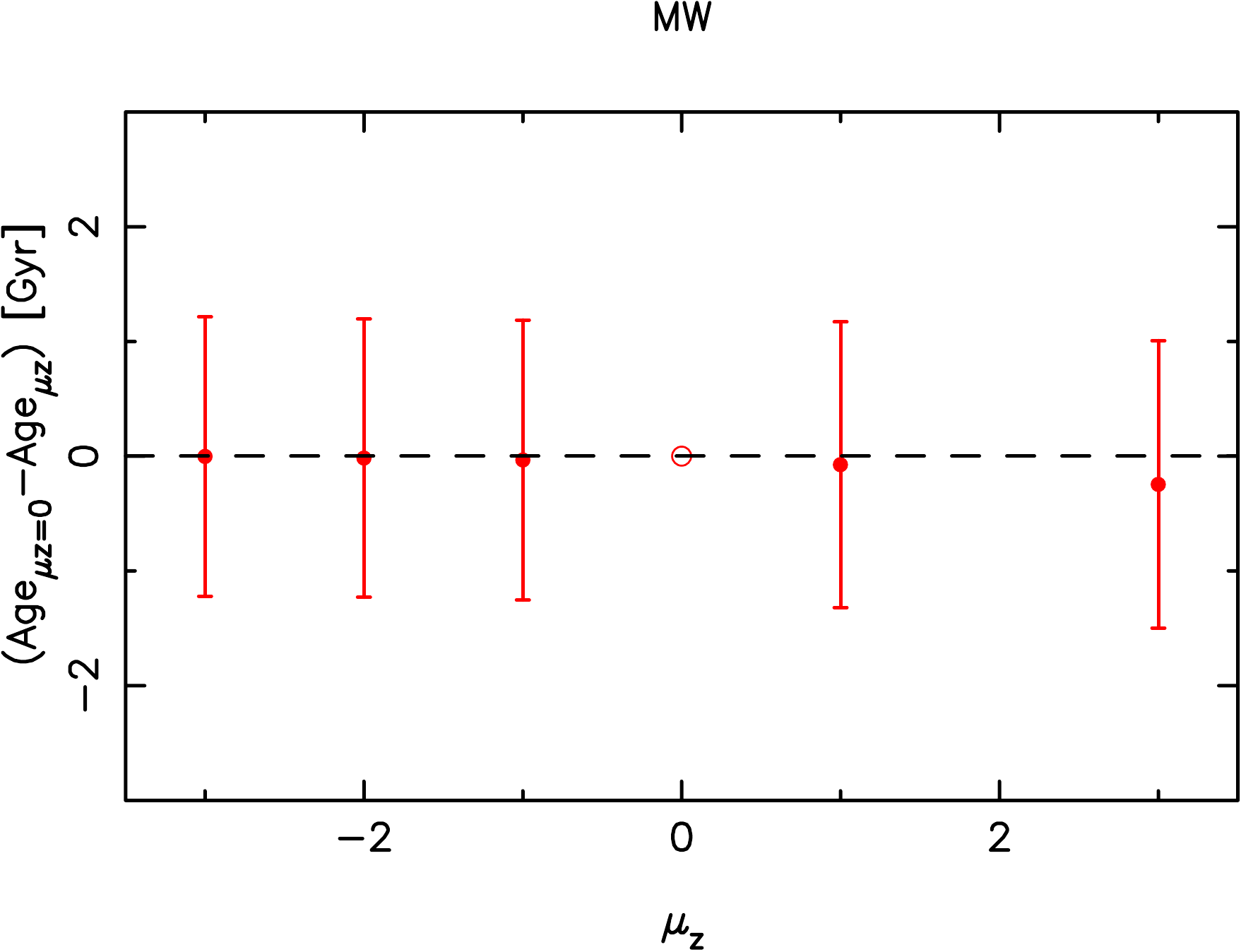}\\
\end{center}
    \caption{Difference in the measurement of mean stellar age when using a smoothness parameter compared to not using a smoothness parameter. Results for luminosity-weighted measurements are shown in the top panels, whereas mass-weighted measurements are shown in the bottom panels. The dependences on the value of $\mu_X$ (the age smoothness parameter) are shown in the left panels, whereas those on $\mu_Z$ (metallicity) are shown in the right panels. The panels show for each value of the smoothness parameters tested the mean (red filled circles) and the 1-$\sigma$ values (error bars) of the distributions of the results from the different fits.}
    \label{fig:err_smooth_a}
\end{figure*}

\begin{figure*}
\begin{center}
	\includegraphics[width=\columnwidth]{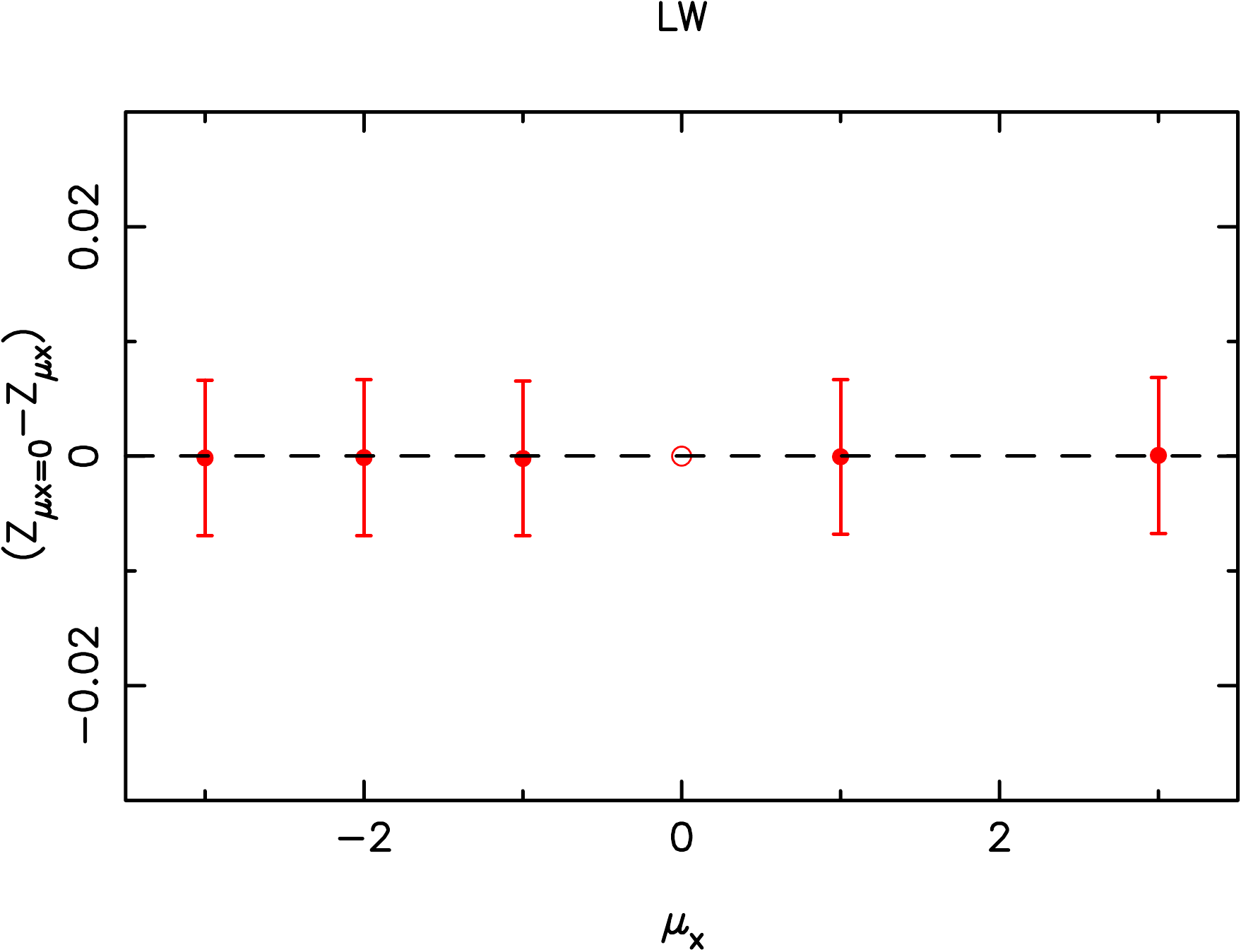}\hskip 0.5cm\includegraphics[width=\columnwidth]{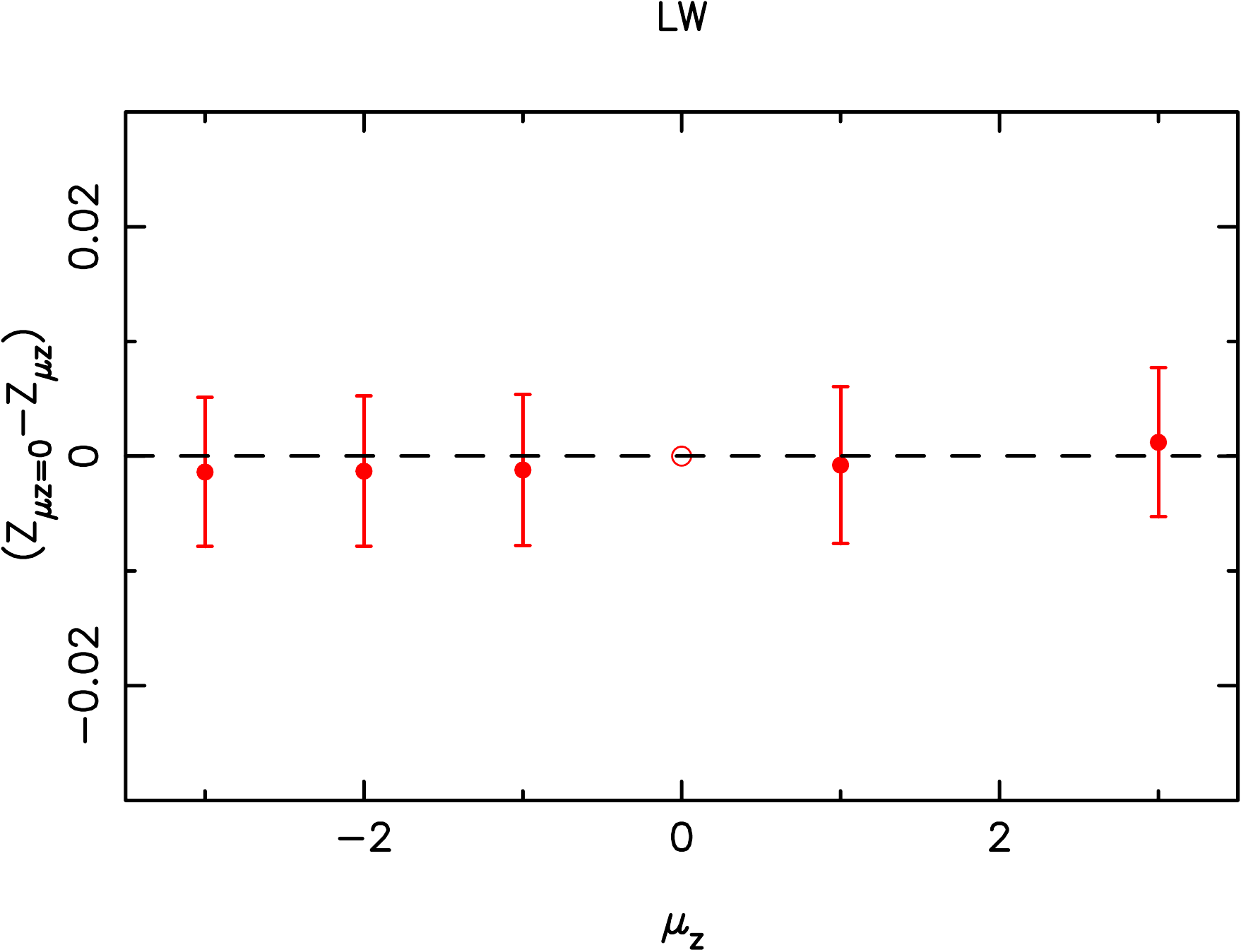}\\
	\vskip0.25cm
	\includegraphics[width=\columnwidth]{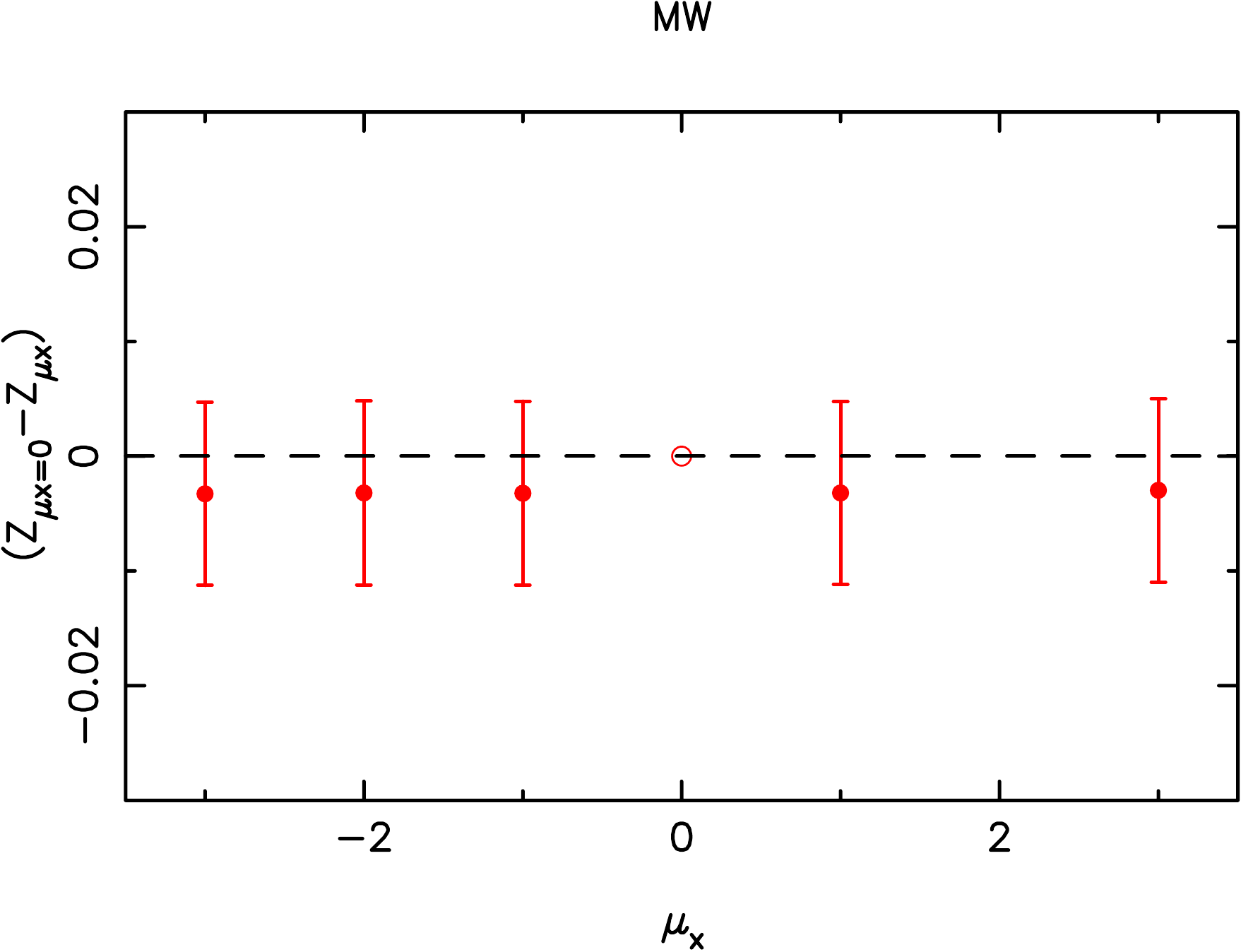}\hskip 0.5cm\includegraphics[width=\columnwidth]{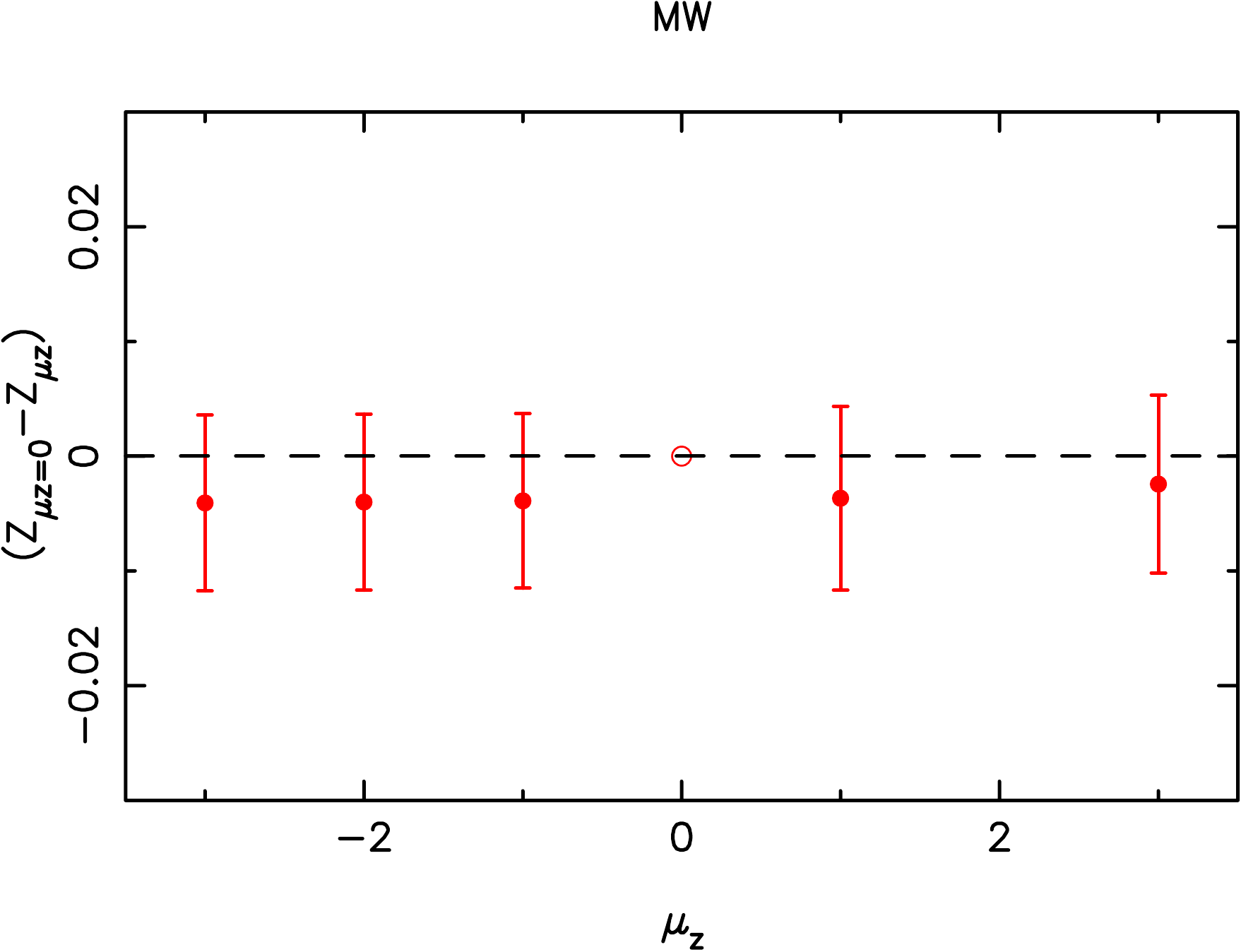}
\end{center}
    \caption{Same as Fig. \ref{fig:err_smooth_a} but for the measurements of mean stellar metallicity.}
    \label{fig:err_smooth_z}
\end{figure*}

\begin{figure*}
\begin{center}
	\includegraphics[width=2\columnwidth]{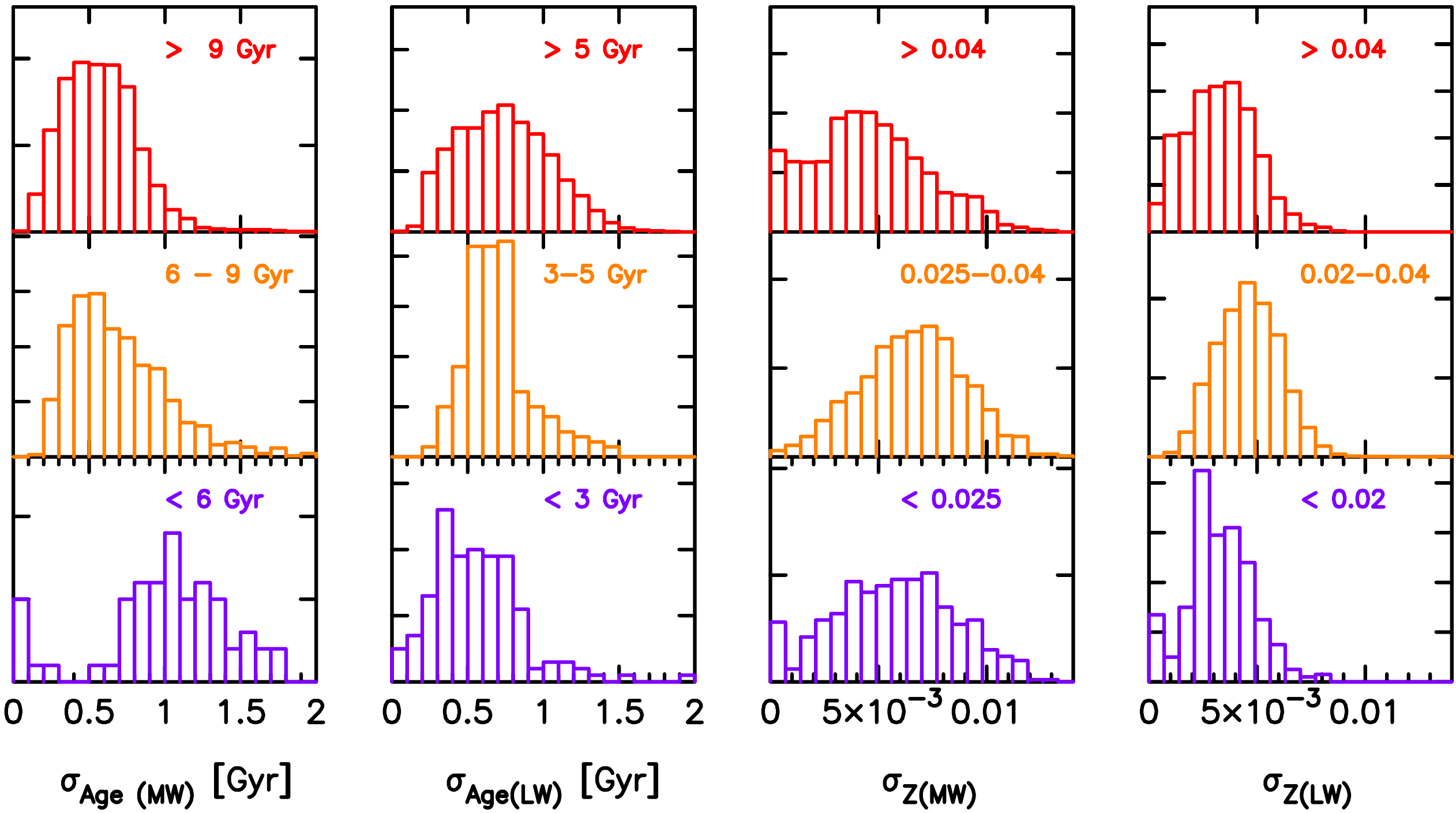}
\end{center}
    \caption{For the 5\,000 TIMER spectra used in these analyses we show histograms of the 1-$\sigma$ values of the distributions in mean stellar age and metallicity, mass-weighted and luminosity weighted, as indicated. The distributions correspond to the fits performed using 100 Monte Carlo realisations of each spectra and show what can be considered as an estimate of the error in each measurement. Since these uncertainties on the ages and metallicities depend on the mean age and metallicity, each panel shows the distributions of the 1-$\sigma$ values for different ranges of stellar age and metallicity.}
    \label{fig:err_hists}
\end{figure*}

To understand how robust are our measurements of mean stellar age and metallicity derived from the star formation histories produced from full spectral fitting using {\tt STECKMAP} (see Sect. \ref{sec:stelpop}), we performed a long series of tests. These tests employed 5\,000 spectra from the TIMER dataset with signal-to-noise ratios varying from about 20 per pixel to about 60 per pixel, and complement previous tests done by ourselves (see above) and \citet{OcvPicLan06a} and \citet{OcvPicLan06b} using synthetic spectra.

We first tested the impact of using different initial conditions for the minimisation routine. Each of the 5\,000 spectra were fitted 100 times using initial conditions that were randomly generated. The distribution of the resulting measurements of mean stellar age and metallicity are shown in Fig. \ref{fig:err_guess}. Noisier spectra result in a wider distribution of output values, but this quickly narrows down with increasing signal-to-noise ratio, and when this value reaches 40 per pixel the statistical improvement on the robustness of the fit with increasing signal-to-noise ratio is minor. This justifies our choice for the canonical value of 40 per pixel for the targeted signal-to-noise ratio during the creation of the Voronoi bins. Furthermore, all mean output values show very good agreement across the different bins of signal-to-noise ratio, with the exception of the age estimates from the noisiest spectra, which yield slightly larger values. In addition, the mean output values agree very well with the output using the canonical initial conditions we employed across this work (which is represented by the dashed line). Our canonical initial conditions for the minimisation routine follow the guidelines given by \citet{OcvPicLan06a}. These tests thus show that by using adequate initial conditions, one ensures that the solutions found are at (or close to) the global minimum of the $\chi^2$ distribution.

Next, we tested the impact of using different smoothness parameters in the regularisation of the {\tt STECKMAP} solution. The results are shown in Figs. \ref{fig:err_smooth_a} and \ref{fig:err_smooth_z}, for mean stellar age and metallicity, respectively, and for a range of values for the age and metallicity smoothness parameters (respectively, $\mu_X$ and $\mu_Z$). In the fits described above we used a range of values for the smoothness parameters, which are binned in the figures for visualisation purposes. The results show that, on average, using smoothness parameters have no effect as compared to {\em not} using smoothness parameters (i.e., with $\mu_X=\mu_Z=0$) in what concerns the measurements of mean stellar age and metallicity. There is a slight overestimation of the mass-weighted metallicity when regularisation is applied but this is well within the errors (see also the following paragraph). The results also show that the choice for the values of the smoothness parameters has on average no impact on the measurements. In this work, we used $\mu_X=\mu_Z=1$, following our tests with synthetic spectra mentioned in Sect. \ref{sec:stelpop}. We note, however, that the smoothness parameters have an effect on the {\em shape} of the star formation histories. Investigating the effects of using different smoothness parameters in this context is beyond the scope of the present analysis.

Finally, we ran 100 Monte Carlo realisations in each of the 5\,000 TIMER spectra and produced a new fit with {\tt STECKMAP} for each realisation. This allows us to derive the uncertainties in the measurements of mass-weighted and luminosity-weighted mean stellar ages and metallicities, which are shown in Fig. \ref{fig:err_hists}. One can see that typical values for these uncertainties are $0.5-1$ Gyr for age, and $0.005-0.010$ for metallicity ($Z$). In addition, these uncertainties depend in a complex fashion on the values of age and metallicity themselves, as well as on whether the measurements are mass-weighted or luminosity-weighted. For example, in absolute terms, mass-weighted measurements of mean stellar age are more robust for older ages, but luminosity-weighted measurements of age are more robust for younger ages. We point out again that these errors concern the mean values produced by the resulting star formation history, but the uncertainty in the star formation history itself -- which includes information on how many star-forming bursts occurred, and how intense and how long each of these bursts were, as well as when did they happen -- is substantially more difficult to assess, and beyond the scope of this paper.

\bsp	
\label{lastpage}
\end{document}